\documentclass[10pt,pra,aps,twocolumn,floatfix,superscriptaddress,nofootinbib]{revtex4-1}
\usepackage{graphicx}
\usepackage{amssymb}
\usepackage{amsmath}

\begin{document}

\title{Progress on stochastic analytic continuation of quantum Monte Carlo data}

\author{Hui Shao}
\email{huishao@bnu.edu.cn}
\affiliation{Center for Advanced Quantum Studies, Department of Physics, Beijing Normal University, Beijing 100875, China}

\author{Anders W. Sandvik}
\email{sandvik@bu.edu}
\affiliation{Department of Physics, Boston University, 590 Commonwealth Avenue, Boston, Massachusetts 02215}
\affiliation{Beijing National Laboratory for Condensed Matter Physics, Institute of Physics, Chinese Academy of Sciences, Beijing 100190, China}

\begin{abstract}
   We report multipronged progress on the stochastic averaging approach to numerical analytic continuation of
  imaginary-time correlation functions computed by quantum Monte Carlo simulations. After reviewing the conventional
  maximum-entropy approach and established stochastic analytic continuation methods, we present several new developments
  in which the configurational entropy of the sampled spectrum plays a key role. Parametrizing the spectrum as a large
  number of $\delta$-functions in continuous frequency space, an exact calculation of the entropy lends support to
  a simple goodness-of-fit criterion for the optimal sampling temperature. We also compare spectra sampled in
  continuous frequency with those from amplitudes sampled on a fixed frequency grid. Insights into the functional
  form of the entropy in different cases allow us to demonstrate equivalence in a generalized thermodynamic limit (large
  number of degrees of freedom) of the average spectrum and the maximum-entropy solution, with different parametrizations
  corresponding to different
  forms of the entropy in the prior probability. These results revise prevailing notions of the maximum-entropy method
  and its relationship to stochastic analytic continuation. In further developments of the sampling
  approach, we explore various adjustable (optimized) constraints that allow sharp low-temperature spectral features to
  be resolved, in particular at the lower frequency edge. The constraints, e.g., the location of the edge or the spectral weight
  of a quasi-particle peak, are optimized using a statistical criterion based on entropy {\it minimization} under the condition
  of optimal fit. We show with several examples that this method can correctly reproduce both narrow and broad
  quasi-particle peaks. We next introduce a parametrization for more intricate spectral functions with sharp edges, e.g.,
  power-law singularities. We present tests with synthetic data as well as with real simulation data for the spin-1/2 Heisenberg
  chain, where a divergent edge of the dynamic structure factor is due to deconfined spinon excitations. 
  Our results demonstrate that distortions of sharp edges or quasi-particle peaks, which arise with other analytic continuation
  methods, propagate and cause artificial spectral features also at higher energies. The constrained sampling methods overcome this
  problem and allow analytic continuation of spectra with sharp edge features at unprecedented fidelity. We present
  results for $S=1/2$ Heisenberg 2-leg and 3-leg ladders to illustrate the ability of the methods to resolve spectral
  features arising from both elementary and composite excitations. Finally, we also propose how the methods developed here could be
  used as ``pre processors'' for analytic continuation by machine learning. Edge singularities and narrow quasi-particle peaks being
  ubiquitous in quantum many-body systems, we expect the new methods to be broadly useful and take numerical
  analytic continuation to a new quantitative level in many applications.
  \null\vskip1mm\null\noindent  Published open-access version: Physics Reports {\bf 1003}, 1-88 (2023)
\end{abstract}

\date{Januari 9, 2023}

\maketitle

\tableofcontents

\vskip5mm

\section{Introduction}

\begin{figure}[t]
\centering
\includegraphics[width=75mm, clip]{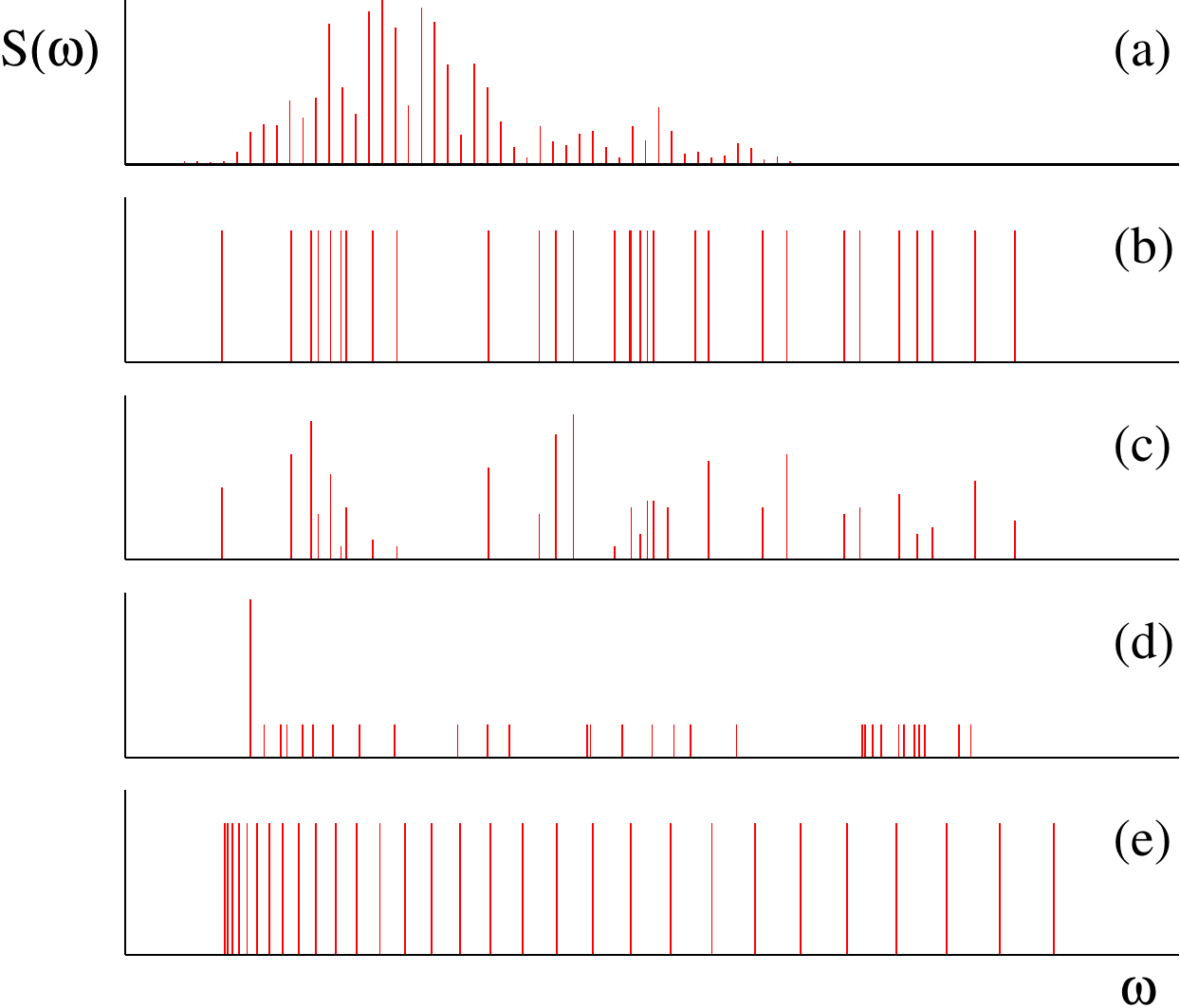}
\vskip-1mm
\caption{Parametrizations of spectra in terms of a large number $N_\omega$ of $\delta$-functions: (a) Variable (sampled) amplitudes
  on a fixed frequency grid. (b) Identical amplitudes and sampled frequencies in the continuum. (c) Variable frequencies and amplitudes.
  (d) A ``macroscopic'' $\delta$-function with amplitude $A_0$ at $\omega_0$, followed by $N_\omega$ ``microscopic'' $\delta$-functions
  at $\omega_i > \omega_0$ with uniform amplitudes $A_i=(1-A_0)/N_\omega$.
  The amplitude $A_0$ is optimized but held fixed in a given sampling run, while $\omega_0$ is sampled.
  (e) Equal amplitudes with monotonically increasing spacing $d_i \equiv \omega_{i+1}-\omega_i$. The lowest frequency $\omega_1$ is sampled along
  with all other frequencies with the constraint $d_{i+1} > d_{i}$. The final spectrum in all cases is
  the mean amplitude density accumulated in a histogram.}
\label{fig:spec}
\end{figure}

In order to extract dynamic response functions (spectral functions) from quantum Monte Carlo (QMC) simulations, 
the challenging task of numerical analytic continuation of imaginary-time dependent correlation functions to real 
frequency has to be performed. While significant progress has been made on this inverse problem during 
the past three decades, there are still questions remaining on what the best approach is for extracting the maximal 
amount of spectral information (frequency resolution) from a given set of QMC data. The maximum entropy (ME) method, of
which there are several variants \cite{gull84,silver90,gubernatis91,jarrell96,boninsegni96,bergeron16}, has dominated the
field for some time, but alongside it an alternative line of stochastic analytic continuation (SAC) methods (also called 
average spectrum methods) \cite{white91,sandvik98,beach04,vafay07,reichman09,syljuasen08,fuchs10,sandvik16,qin17,shao17,ghanem20a,ghanem20b}
have been gaining ground and show promise to outperform the ME method. Here the goodness-of-fit $\chi^2(S)$ is used as
an analogy of the energy in a statistical mechanics problem, and the spectrum $S(\omega)$, suitably parametrized, is Monte Carlo
sampled at a fictitious temperature $\Theta$ using the Boltzmann-like weight function ${\rm exp}{(-\chi^2(S)/2\Theta})$.
Some parametrizations in terms of $\delta$-functions are illustrated in Fig.~\ref{fig:spec} and will be explained
in detail later.

Building on the observation that suppression of configurational entropy with constraints on the sampled spectrum can
significantly improve the fidelity of the average spectrum \cite{sandvik16}, we here make further progress on the SAC scheme.
We primarily focus on low-temperature spectral functions with sharp features, e.g., narrow quasi-particle peaks and power-law
edge singularities (the latter of which often arise in systems with fractionalized excitations). Through a series of systematic studies of
different optimized constraints, with tests based on QMC and synthetic imaginary-time data, we demonstrate how spectra with
generic, sharp features can be reconstructed almost perfectly using parametrizations such as those illustrated
in Figs.~\ref{fig:spec}(d) and \ref{fig:spec}(e). These developments open opportunities for
studies of quantitative features of spectral functions that have so far been out of reach of QMC studies, not only in the most
obvious setting of condensed matter physics but also in lattice field theory \cite{ding18,aarts21,horak21}.

To prepare for these developments of constrained SAC, we first study the role of the parametrization of unconstrained spectral functions,
comparing $\delta$-functions on a fixed grid [Fig.~\ref{fig:spec}(a)], where the amplitudes constitute the sampled configuration space, and
in the frequency continuum (with the spectral weight distribution collected in a discrete histogram). In the continuum, either only the locations
of fixed-amplitude $\delta$-functions are sampled [Fig.~\ref{fig:spec}(b)] or the amplitudes are also updated along with the frequencies
[Fig.~\ref{fig:spec}(c)]. The average spectrum exhibits significant differences between the three parametrizations, which we
explain quantitatively by non-universal entropies originating from stochastic processes with different degrees of freedom. The fidelity
of the continuous frequency approach in reproducing known spectral functions is invariably better, and, moreover, the sampling is
also much more efficient in that case, typically requiring only of the order of a few minutes to achieve a smooth spectrum.

We discuss a simple method of selecting the sampling temperature $\Theta$ in such a way that the resolution is maximized while still
not being affected by over-fitting. The optimal sampling temperature can be related to the entropy of the spectral function in a given
parametrization. We derive the entropy in the case of $\delta$-functions in the continuum and confirm its role in the dependence of the
optimal sampling temperature on the number $N_\omega$ of $\delta$-functions used; $\Theta \propto 1/N_\omega$. The limit $\Theta \to 0$,
corresponding to fitting by pure $\chi^2$ minimization, is also of relevance to our arguments. For a given set of imaginary-time data,
we demonstrate that the positive-definite spectrum with lowest $\chi^2$ value defines a noise-limited effective number of fitting parameters.
In this context we also study how a small fraction (as low as $10^{-4}$) of negative spectral weight can favorably affect the average spectrum
at low $\Theta$.

The main purpose of the present work is to introduce various constraints imposed within the continuous frequency-space parametrization.
Such constraints, which represent some kind of auxiliary information, e.g., the existence of a sharp edge, can significantly improve the
ability of the SAC method to resolve spectral details also at frequencies far away from the feature(s) directly associated with
the constraint. In previous work on sharp edges \cite{sandvik16}, a fixed frequency grid was used and the constraints amounted to
enforcing and optimizing lower and upper bounds, outside which there is no spectral weight. This methods worked surprisingly
well, e.g., when a single-peak condition was also imposed (without any further information on the location or shape of the peak) it was possible
to closely reproduce the edge divergence at temperature $T=0$ of the structure factor of the Heisenberg chain---a feat that had been
impossible with previous approaches. In practice, the optimization of a constraint, which is based on a simple statistical criterion of
minimum $\langle \chi^2\rangle$ at fixed $\Theta > 0$, can be very time consuming. Here we introduce a variety of useful constraints within the
continuous-frequency representation, where either no further optimization is required or the optimization process is much faster than in the
previous approach.

In a previous work with collaborators \cite{shao17}, we already implemented an SAC method incorporating a spectral edge consisting of a single
$\delta$-function, whose relative weight $A_0$ was optimized and with the remaining weight $1-A_0$ divided over hundreds or thousands of
``microscopic'' $\delta$-functions to model a continuum; see Fig.~\ref{fig:spec}(d).
We here further explore the ability of the statistical optimization scheme to find
the correct value of $A_0$. In particular, we investigate how the optimal value converges when the statistical errors of the underlying
imaginary-time data are reduced. We also generalize the approach to a quasi-particle peak of finite width by splitting the weight $A_0$
over several sampled edge $\delta$-functions. This way, both broad and narrow quasi-particle peaks can be resolved to a degree
far exceeding what is possible with conventional methods.

Moving then to power-law and similar edge singularities, we introduce a constraint on the distances between the sampled $\delta$-functions,
such that the mean density of $\delta$-functions must increase monotonically when the edge is approached. This parametrization, illustrated in
Fig.~\ref{fig:spec}(e), most naturally describes a divergent edge. However, with different amplitude profiles and further constraints, both divergent
and convergent spectral edges can
be reproduced. We discuss the entropic pressures of the distance-monotonic parametrization and test its ability to reliably reproduce different
types of edges. Again, we find a remarkable improvement in the fidelity of the method in resolving not just the edge, but the entire spectral
function up to its high-frequency bound. We also generalize this approach to an arbitrary (not necessarily monotonically decaying) continuum
above the edge.

A key message of our study is that removal by constraints of distortions at the lower edge of a spectrum can also reveal features at higher
frequencies at unexpected level of detail. Thus, the imaginary-time data contain ``hidden information'' that is masked when a sharp edge is
not treated correctly but is revealed once the primary distortions are removed by appropriate constraints.

Our comprehensive series of tests of different parametrizations of increasingly complex spectral functions (also beyond those in Fig.~\ref{fig:spec})
build up to a scheme capable of resolving a broad range of common edge features, with only minimal input beyond the imaginary-time data. The method
in effect is a generic curve-fitting machinery, where the type of curve is only specified minimally (e.g., the edge takes an asymptotic power-law form,
with the exponent not necessarily specified but optimized in the process) and beyond this information the statistically best average
spectrum consistent with the input data is produced.

As an application to systems for which previous numerical approaches have provided only very limited information, we use constrained SAC methods to
study the dynamic structure factor of 2- and 3-leg spin-$1/2$ Heisenberg ladders \cite{dagotto96}. For the gapped 2-leg ladder, we extract not only
the gap $\Delta \approx 0.5025$ and spectral weight (about $96.7\%$) of the dominant $\delta$-function at the gap, but also resolve the
three-triplon contributions at $\omega > 3\Delta$. For the gapless 3-leg ladder we study the divergent spectral edge arising from deconfined spinons (as in the
Heisenberg chain considered in many of our test examples) and also obtain the profile at higher energies (which is different from that of the single
chain). These examples demonstrate how constraints corresponding to known spectral features at the lower edge (a $\delta$-function or a spinon
edge) can deliver previously unknown details of the spectrum away from the edge, including features arising from composite excitations.

The work reported here on SAC also led us to new insights into the ME method. In particular, we finally establish the
complete formal mapping between the ME and SAC methods, which in some important respects is at variance with previous attempts
\cite{beach04,bergeron16,ghanem20a}. We also introduce an SAC-inspired criterion for fixing the entropic weighting factor $\alpha$
in the ME method. With the proper relationship between the optimal $\alpha$ and $\Theta$ values, we demonstrate explicitly that SAC with different
parametrizations of the spectrum corresponds exactly, when $N_\omega \to \infty$ (a generalized thermodynamic limit of SAC),
to the ME approach with different functional forms of the entropy (i.e., the conventional
Shannon information entropy is not universal). Thus, unrestricted SAC can often be used interchangeably with the ME method, and some of our
constrained parametrizations can also in principle be translated to the ME framework. The sampling approach can still be necessary or
preferred, e.g., with some types of constraints and in cases where an analytical form of the entropy is not available.

The treatise is organized as follows:

In Sec.~\ref{sec:problem} we define the analytic continuation problem in detail and discuss QMC computed imaginary-time correlation functions
and the importance of taking covariance into account (which we do in practice by a basis transformation). We also explain how to generate realistic
synthetic data, including covariance, for testing and benchmarking purposes.

In Sec.~\ref{sec:methods} we briefly review some of the common analytic continuation methods, in particular data fitting based purely on
$\chi^2$ minimization, the ME method, as well as conventional SAC methods. We also describe the different parametrizations of the spectrum suitable
for SAC sampling and outline the further developments of optimized constraints that we present in detail in the later sections.

In Sec.~\ref{sec:plainsampling} we discuss Monte Carlo sampling techniques for spectral functions in the different parametrizations without
constraints [those illustrated in Figs.~\ref{fig:spec}(a)-(c)], including moment conserving updates used to increase the sampling efficiency.

In Sec.~\ref{sec:theta} we motivate the statistical criterion for the optimal sampling temperature $\Theta$ and present several illustrative examples,
using QMC data for the dynamic structure factor of the Heisenberg chain as well as synthetic data. We use a lower bound on the spectrum as an example of
a constraint that reduces the sampling entropy and can be optimized by a simple criterion. We argue that an effective number of fitting parameters
($\ll N_\omega$) can be defined by considering the low-$\Theta$ limit of the spectrum obtained with a given set of the noisy imaginary-time data, and
also explore an alternative way to regulate the entropy by optimizing the number of sampled $\delta$-functions at $\Theta=1$.

In Sec.~\ref{sec:entropy} we first derive the entropy of the spectrum explicitly in the case of equal-amplitude $\delta$-functions in the frequency
continuum and discuss related results from the literature. We then discuss and make use of a recent result for the entropy in fixed-grid SAC \cite{ghanem20a}.
Based on these two entropy forms, we conjecture a new ``mixed entropy'' when both frequencies and amplitudes are sampled. We also demonstrate how the
optimal sampling temperature delivered by our criterion relates to the extensive form of the entropy by $\Theta \propto 1/N_\omega$.

In Sec.~\ref{sec:deltapeak} we discuss constrained SAC applied to a spectrum with a sharp peak at the lower edge, either a $\delta$-function as
in Fig.~\ref{fig:spec}(d) or a generalization to a multi-$\delta$ parametrization for sampling a quasi-particle peak of finite width. We test the ability of the 
optimization method to converge to the correct values of the amplitude and width of the peak using synthetic spectral functions. We also discuss results 
for the dynamic structure factor of the two-dimensional (2D) $S=1/2$ Heisenberg antiferromagnet, where a sharp peak at the lower edge of the spectrum 
represents the dominant single-magnon excitation.

In Sec.~\ref{sec:contedge1} we apply the monotonicity constraint on the distance between the $\delta$-functions, Fig.~\ref{fig:spec}(e),
which is appropriate for reproducing a continuum with an edge singularity. We present test results for the dynamic structure factor of the Heisenberg chain,
where deconfined spinons lead to a divergent power-law singularity, which can be resolved to a remarkable degree with our method. We also consider a synthetic
spectrum with non-divergent sharp edge and show how a parameter regulating the $\delta$-function spacing can be optimized and leads to high-fidelity results
without any other input.

In Sec.~\ref{sec:contedge2} we discuss the entropic pressures existing within the parametrization used in the previous section and introduce an
optimized parameter to further improve the resolution of the edge singularity, specifically enabling determination of the exponent governing an
asymptotic power-law divergent or convergent form. Here we also discuss a combination of the parametrization used for the sharp edge with that
for a generic continuum, with which a completely arbitrary spectral function with power-law edge can be modeled with no other input.

In Sec.~\ref{sec:ladders} we compute the dynamic structure factor of $S=1/2$ Heisenberg 2- and 3-leg ladders. In the former case, we optimize the
leading isolated $\delta$-function at the gap, which arises from the triplon quasi-particle at momentum $q=\pi$, and further improve the results by
imposing also the three-triplon gap. In the latter case, we employ the parametrization with the expected divergent spinon edge followed by an arbitrary
continuum.

In Sec.~\ref{sec:maxent} we demonstrate the exact relationships between SAC with different parametrizations of the spectrum and the ME method with
the corresponding functional forms of the entropy. By comparing ME and SAC results in the proper way following from our exact mapping, we demonstrate
the correctness of the three forms of the entropy discussed in Sec.~\ref{sec:entropy}. We also discuss sampling versus probability maximization within
the ME framework and point out a previously overlooked problem arising from extensive sampling entropy. Readers who are interested in these topics and who
are familiar with the basic aspects of the SAC and ME methods can read Sec.~\ref{sec:maxent} essentially independently of the other parts of the paper (with just a
few jumps back to referenced results of earlier sections).

In Sec.~\ref{sec:discussion} we conclude with a brief summary as well as further comments and conclusions. For future prospects, we discuss more general
constrained parametrizations and present a proposal for machine learning to identify the best spectrum in a large set of SAC or ME spectra. We also suggest
potential advantages of including a small fraction of negative spectral weight in SAC.

In \ref{app:lowtheta} we report new insights into the $\chi^2$ minimization procedure corresponding to $\Theta \to 0$,
explaining why the ultimate best-fit spectrum should consist of a small number of $\delta$-functions. We also discuss how this limit
defines an effective number of fitting parameters for noisy data, when positive definiteness is enforced.

In \ref{app:low2} we further discuss how the SAC spectrum at very low sampling temperatures changes in the presence of a small fraction of negative
spectral weight, as a result of additional entropy contributions. Our preliminary results indicate that the SAC method may some times be further improved by
exploiting negative spectral weight.

In \ref{app:fluct} we discuss the fluctuations of the sampled spectral weight within a fixed frequency window; specifically arguing that
these fluctuations cannot be translated into statistical errors on the average spectrum. We also explicitly demonstrate the additivity of amplitude
and frequency fluctuations of a spectrum sampled with $\delta$-functions.

In \ref{app:statmech} we compare and contrast conventional statistical mechanics and the unrestricted SAC sampling problem, providing further
arguments for an unusual thermodynamic limit ($N_\omega \to \infty$) of SAC, where the fluctuations of the spectrum about the maximum-probability ME solution
vanish.

\section{The numerical analytic continuation problem}
\label{sec:problem}

In Sec.~\ref{sec:defs} we outline the mathematical formalism of the analytic continuation problem, establishing definitions and notation
used in the later sections. We discuss QMC generated imaginary-time data in Sec.~\ref{sec:qmcdata}, e.g., the choice of time grid and the
characterization of the statistical errors and covariance. In Sec.~\ref{sec:syntdata} we discuss synthetic data, i.e.,
imaginary-time correlations generated for testing purposes from an artificial model spectrum, with correlated noise added to mimic the
statistical fluctuations in typical QMC data.

\subsection{Definition of the problem}
\label{sec:defs}

The correlation function computed in a QMC simulation is defined with some operator $O$ 
of interest (typically corresponding to some experimental probe) as
\begin{equation}
G(\tau) = \langle O^\dagger(\tau) O(0)\rangle,
\label{gtaudef1}
\end{equation}
where the imaginary-time dependence is defined in the Heisenberg representation as (working
in dimensionless units where $\hbar=1$)
\begin{equation}
O(\tau) = {\rm e}^{\tau H} O {\rm e}^{-\tau H},
\label{otau}
\end{equation}
with the Hamiltonian $H$ of the system under study. In the basis of eigenstates $|n\rangle$ and
eigenvalues $E_n$ of $H$, the spectral function of $O$ at temperature $T=\beta^{-1}$
(setting $k_B=1$) is given by
\begin{equation}
S(\omega)=\frac{\pi}{Z} \sum_{m,n}{\rm e}^{-\beta E_n}|\langle m|O|n\rangle|^2 \delta(\omega - [E_m-E_n]),
\label{somegasum}
\end{equation}
where $Z$ is the partition function. The relationship between this spectral
function and the imaginary-time correlation function in Eq.~(\ref{gtaudef1}) is
\begin{equation}
G(\tau) = \frac{1}{\pi}\int_{-\infty}^\infty d\omega S(\omega){\rm e}^{-\tau \omega}.
\label{contrel1}
\end{equation}
For a bosonic operator $O$, $G(\beta-\tau)=G(\tau)$ and we need only $\tau \in [0,\beta/2]$. Further, in the case of a bosonic
function the spectral weight distributions at negative and positive frequencies are related according to 
\begin{equation}
S(-\omega)={\rm e}^{-\beta \omega}S(\omega).
\label{sminus}
\end{equation}
The relationship between $S(\omega)$ and $G(\tau)$ can therefore be modified so that integration is required only over positive
frequencies, with Eq.~(\ref{contrel1}) written as
\begin{equation}
G(\tau) = \int_{0}^\infty d\omega S(\omega) K(\tau,\omega),
\label{contrel2}
\end{equation}
where the kernel is given by
\begin{equation}
K(\tau,\omega) = \frac{1}{\pi}(e^{-\tau\omega} + e^{-(\beta-\tau)\omega}).
\label{kernel1}
\end{equation}
We will here consider only bosonic spectral functions, but the fermionic case can be studied with very minor modifications of the methods. For examples
and applications, we will study quantum spin models and synthetic spectral functions.

A well known example of a bosonic spectral function is the dynamic spin structure factor $S^\alpha(q,\omega)$, measured, e.g., by magnetic inelastic
neutron scattering as the cross section for momentum ($q$) and energy ($\omega$) transfer. In this case the operator in Eq.~(\ref{somegasum}) is
$O=S_q^\alpha$, the Fourier transform of the $\alpha$-component $S_r^\alpha$ $(\alpha=x,y,z)$ of the real-space spin operators.
We will test our improved (as well as previous) SAC schemes on the dynamic structure factor of  spin-isotropic
Heisenberg models in one and two dimensions, where the function $S(q,\omega)$ is independent of the direction $\alpha$.
In addition, we will also consider synthetic imaginary-time data to test the ability to resolve a variety of spectral features.

A QMC simulation delivers a statistical estimate $\bar G_i \equiv\bar G(\tau_i)$ of the true correlation function $G(\tau)$ for a set of
imaginary times $\tau_i$, $i=1,\ldots,N_\tau$ (or one can work in Matsubara space with a set of Fourier components at frequencies
$\omega_n = n2\pi/\beta$ \cite{bergeron16,smakov05}, but here we will consider only formulations in the original time space). Often a uniform grid of
$\tau$ points is used, but other grids are some times preferable, as we will discuss further below in Sec.~\ref{sec:qmcdata}. Ideally, for a given
system there are no other approximations than the unavoidable statistical errors of $\bar G_i$, the magnitudes of which depend inversely on
the square-root of the length of the QMC run as usual. 

We denote by $\sigma_i$ one standard deviation of the mean $\bar G_i$. Importantly, the statistical errors of different data points $i$
are correlated, and their full characterization requires the covariance matrix \cite{jarrell96}. With the QMC data divided up into bins
$b=1,2,\ldots,N_B$ (assumed to be statistically independent, which in practice is essentially satisfied if the bins represent sufficiently
long simulation times), the covariance matrix is given by
\begin{equation}
C_{ij} = \frac{1}{N_B(N_B-1)}\sum_{b=1}^{N_B} (G^b_i-\bar G_i)(G^b_j-\bar G_j),
\label{cijdef}
\end{equation}
where $G^b_i$ is the mean of the correlation function computed with the data of bin $b$ [or $b$ could represent bootstrap samples, as will be
discussed further below, in which case $N_B$ is the number of samples and the factor $1/(N_B-1)$ should be removed above]. The diagonal elements
of $C$, the variances, are the squares of the conventional statistical errors; $\sigma_i^2 = C_{ii}$.

When applying a numerical analytic continuation procedure, some suitable (sufficiently flexible) parametrization of the spectral function
$S(\omega)$ is optimized for compatibility with the QMC data. Alternatively, in the SAC approach, statistically acceptable instances of $S(\omega)$
are sampled. In either case, given $S(\omega)$ the corresponding $G_i$ values can be computed according to Eq.~(\ref{contrel2}) and the
overall closeness of these to the QMC-computed values $\bar G_i$ is quantified in the standard way by the ``goodness of the fit'',
\begin{equation}
\chi^2 = \sum_{i=1}^{N_\tau}\sum_{j=1}^{N_\tau} (G_i-\bar G_i)C^{-1}_{ij}(G_j-\bar G_j).
\label{chi2}
\end{equation}
Some times only the diagonal elements of $C$ are included, and the goodness of the fit then reduces to
\begin{equation}
\chi_d^2 = \sum_{i=1}^{N_\tau} \left ( \frac{G_i-\bar G_i}{\sigma_i} \right )^2.
\label{chi2d}
\end{equation}
Here we will always use the full covariance matrix, which is necessary for the SAC method to be statistically sound.

\subsection{QMC correlation functions}
\label{sec:qmcdata}
      
It is convenient to work with normalized spectral functions. According to Eq.~(\ref{contrel1}) the normalization
$\int S(\omega)d\omega$ is just the value $\pi G(0)$ and we therefore divide the QMC data $\bar G(\tau_i)$ by $\bar G(0)$ for a spectrum normalized
to $\pi$. We point out that dividing by $\bar G(0)$ also cancels out some covariance and makes the standard statistical errors $\sigma(\tau)$
smaller for $\tau$ close to $0$ (vanishing as $\tau \to 0$). We use the bootstrap method for computing the covariance matrix, i.e.,
with a large number $M$ of samples of $N_B$ randomly chosen bins among all the $N_B$ bins. As mentioned, Eq.~(\ref{cijdef}) then holds if
the sum is taken over the $M$ bootstrap samples and the denominator $N_B(N_B-1)$ is replaced by $M$. Normalizing each bootstrap sample, the normalization
$\int S(\omega)d\omega=\pi$ is also enforced exactly in the sampled spectra, and we do not use the $\tau=0$ data point explicitly. The original
normalization is put back in after the analytic continuation by just multiplying the spectrum by the original pre-normalization value of $\bar G(0)$.

For temperatures $T>0$, the integral $\int S(\omega)d\omega$ includes spectral weight at both negative and positive frequencies, and with
Eq.~(\ref{contrel2}) $G(0)$ does not correspond directly to a fixed normalization of $S(\omega)$ in the corresponding frequency
range $\omega \in [0,\infty)$. The Monte Carlo sampling can be simplified by working with a different spectral function that is normalized to
unity on the positive frequency axis. We therefore define a modified spectral function to use internally in the computer program,
\begin{equation}
A(\omega) = S(\omega)(1 + {\rm e}^{-\beta\omega})/\pi,
\label{barelation}
\end{equation}
for which Eq.~(\ref{contrel1}) implies [with the convention $G(0)=1$] the desired normalization 
\begin{equation}
\int_{0}^\infty d\omega A(\omega) = 1.
\end{equation}
Then, in place of Eq.~(\ref{contrel2}) and Eq.~(\ref{kernel1}) we use 
\begin{equation}
G(\tau) = \int_{0}^\infty d\omega A(\omega) \bar K(\tau,\omega).
\label{eir}
\end{equation}
where the kernel is given by
\begin{equation}
\bar K(\tau,\omega) = \frac{e^{-\tau\omega} + e^{-(\beta-\tau)\omega}}{1 + e^{-\beta\omega}}.
\label{kernel2}
\end{equation}
We always convert back from $A(\omega)$ to $S(\omega)$ after the analytic continuation.

Several considerations are involved in choosing the set of points $\{\tau_i\}$ at which to evaluate $G(\tau)$ in a QMC simulation. The behavior of
$S(\omega)$ at high frequencies is predominantly reflected at short times, because of the kernel $\rm{e}^{-\tau\omega}$ in Eq.~(\ref{contrel1})
and the typical increase in the relative statistical errors with $\tau$. The low-frequency behavior is conversely most cleanly
``filtered out'' at the longest accessible time, before the relative errors increase so much that the data become useless in practice. Thus, we would like to have
many $\tau$ points at very short times as well as at long times. However, we do not want too many points in total, because each one requires QMC computation
time, and, moreover, the covariance matrix can become unstable (its inverse may become noise dominated) if the matrix is too large for given statistical
quality of the data. It should be noted that the computational effort of the analytic continuation procedure also increases with the number
of $\tau$ points (linearly, as we will see). 

To sufficiently cover both short and long times $\tau \in [0,\beta/2]$ for large $\beta$, while limiting the total number of points, it is often convenient
to use a non-linear $\tau$ grid, e.g., a quadratic grid where $\tau_i \propto i^2$. It is also sometimes useful to start with a linear grid for short times and
switch to a quadratic grid at longer times. Overall, our experience is, thankfully, that the end result for $S(\omega)$ is not very sensitive to exactly what
$\tau$ grid is chosen, as long as a reasonably large number of points is used (typically tens of points) and they are spread over the range where the statistical
errors are relatively small (which we typically take as below $10\%$ or $20\%$ relative error).

In practice, when using the covariance matrix $C$, instead of inverting it and using Eq.~(\ref{chi2}) to compute $\chi^2$, it is better to
diagonalize $C$ and transform $\bar G(\tau)$ and the kernel to the resulting eigenbasis. Then $\chi^2$ is given by a formula like Eq.~(\ref{chi2d}),
with $\sigma^2_i$ replaced by the $i$th eigenvalue $\epsilon_i$ of the covariance matrix and $\{ \bar G_i\}$ being the elements of the vector obtained
by transforming the original $\tau$ data, arranged in a vector, to the eigenbasis of $C$.

To summarize the change of basis, with $U$ denoting the orthogonal matrix that transforms the covariance matrix $C$ to its diagonal form,
$\epsilon = U^{\rm T} CU$, we make these substitutions when applying the above formulas:
\begin{equation}
\bar G \to U^{\rm T} \bar G,~~~~\bar K(\omega) \to U^{\rm T}\bar K(\omega), 
\label{basistransf}
\end{equation}  
where on the right-hand side $\bar G$ denotes the vector of $N_\tau$ QMC-computed $G(\tau)$ points and $\bar K(\omega)$ similarly is
the vector containing the $N_\tau$ kernel points evaluated at frequency $\omega$. After the transformation, the invariant goodness-of-fit, Eq.~(\ref{chi2}),
is a single sum,
\begin{equation}
\chi^2 = \sum_{i=1}^{N_\tau} \left ( \frac{G_i-\bar G_i}{\sqrt{\epsilon_i}} \right )^2,
\label{chi2eps}
\end{equation}
where now of course the vector $G$ is automatically produced in the transformed basis because it is computed with the transformed kernel according
to Eq.~(\ref{eir}).

To compute the correlation functions with QMC simulations, we here use the stochastic series expansion (SSE) method \cite{sandvik10}. It should be noted
that there are no other approximations in this scheme beyond the statistical errors. We will not discuss how the correlation functions are calculated
with the SSE method; see Refs.~\cite{sandvik92,sandvik96,dorneich01}. We refer to the recent review Ref.~\cite{sandvik19} for some further
discussion and references.

\begin{figure*}[t]
\centering
\includegraphics[width=105mm]{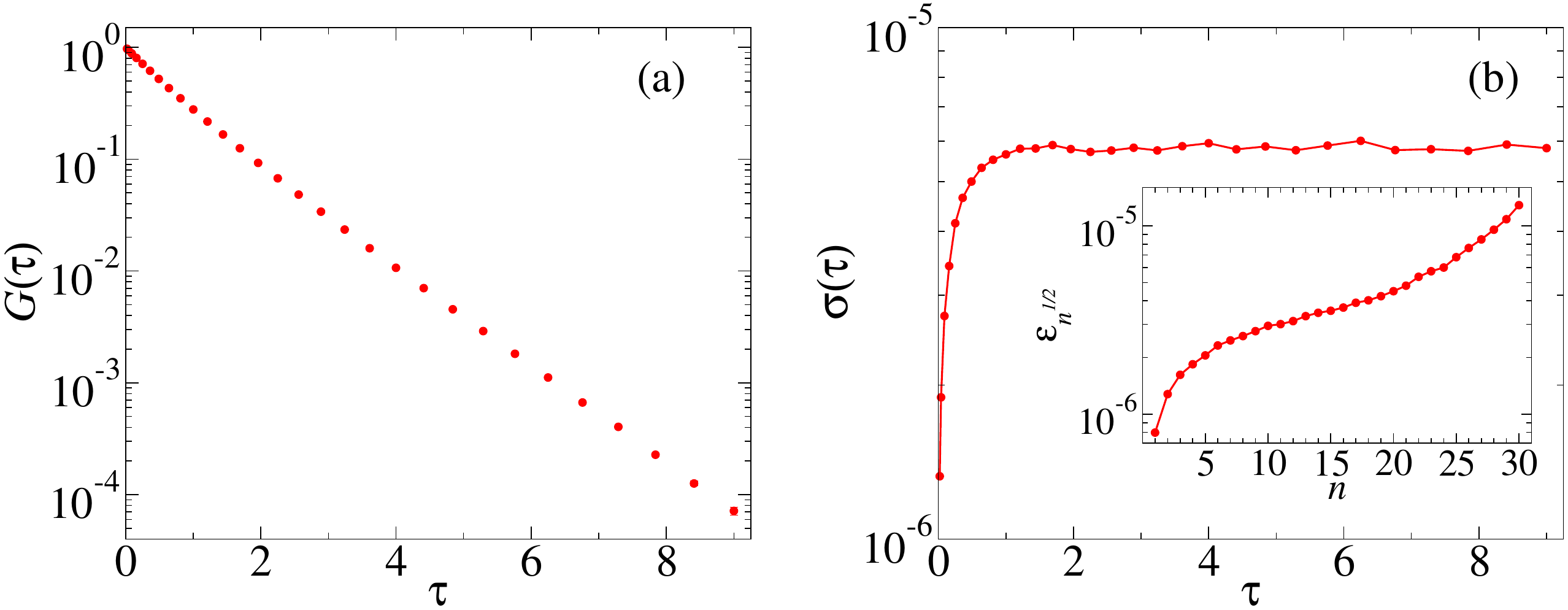}
\vskip-1mm
\caption{(a) Normalized imaginary-time spin correlations at momentum $q=4\pi/5$  generated in SSE QMC simulations of the Heisenberg chain of size $L=500$
at a temperature low enough ($\beta=1000$) to produce ground state results. The grid is quadratic, with $\tau_i=i^2/100$ for $i =1,\ldots,30$, with the
cut-off representing a relative error exceeding $20\%$.  (b) The statistical errors, same as those shown with error bars in (a). The square-root of the
eigenvalues of the covariance matrix are graphed in increasing order in the inset.}
\label{fig:gtau}
\end{figure*}

Fig.~\ref{fig:gtau}(a) shows an example of a QMC-computed correlation function on a quadratic grid of $\tau$ values, with data included up
to the point before the relative error (the conventional standard deviation) exceeds $20\%$. Fig.~\ref{fig:gtau}(b) shows the corresponding statistical
error versus $\tau$, as well as the eigenvalues of the covariance matrix graphed in ascending order (where we have taken the square-root so
that the eigenvalues can be directly compared with the conventional error bars $\sigma_i$).

The statistical errors $\sigma_i$ are typically only weakly dependent on $\tau_i$ except when $\tau_i$ are close to zero, as can be seen in
Fig.~\ref{fig:gtau}(b). We will refer to the approximate overall
magnitude of $\sigma_i$ [with the normalization $\bar G(0)=1$] close to the cut-off value of $\tau$ as the ``error level''. Thus, in Fig.~\ref{fig:gtau}
the error level is about $6\times 10^{-6}$. An alternative simple definition of the error level could be the largest eigenvalue of the covariance matrix
(which is a bit above $10^{-5}$ in the present case). With the SSE method, it is often possible to
reach error levels of $10^{-5}$, or even $10^{-6}$, without too much effort.

\begin{figure*}
\centering
\includegraphics[width=107mm]{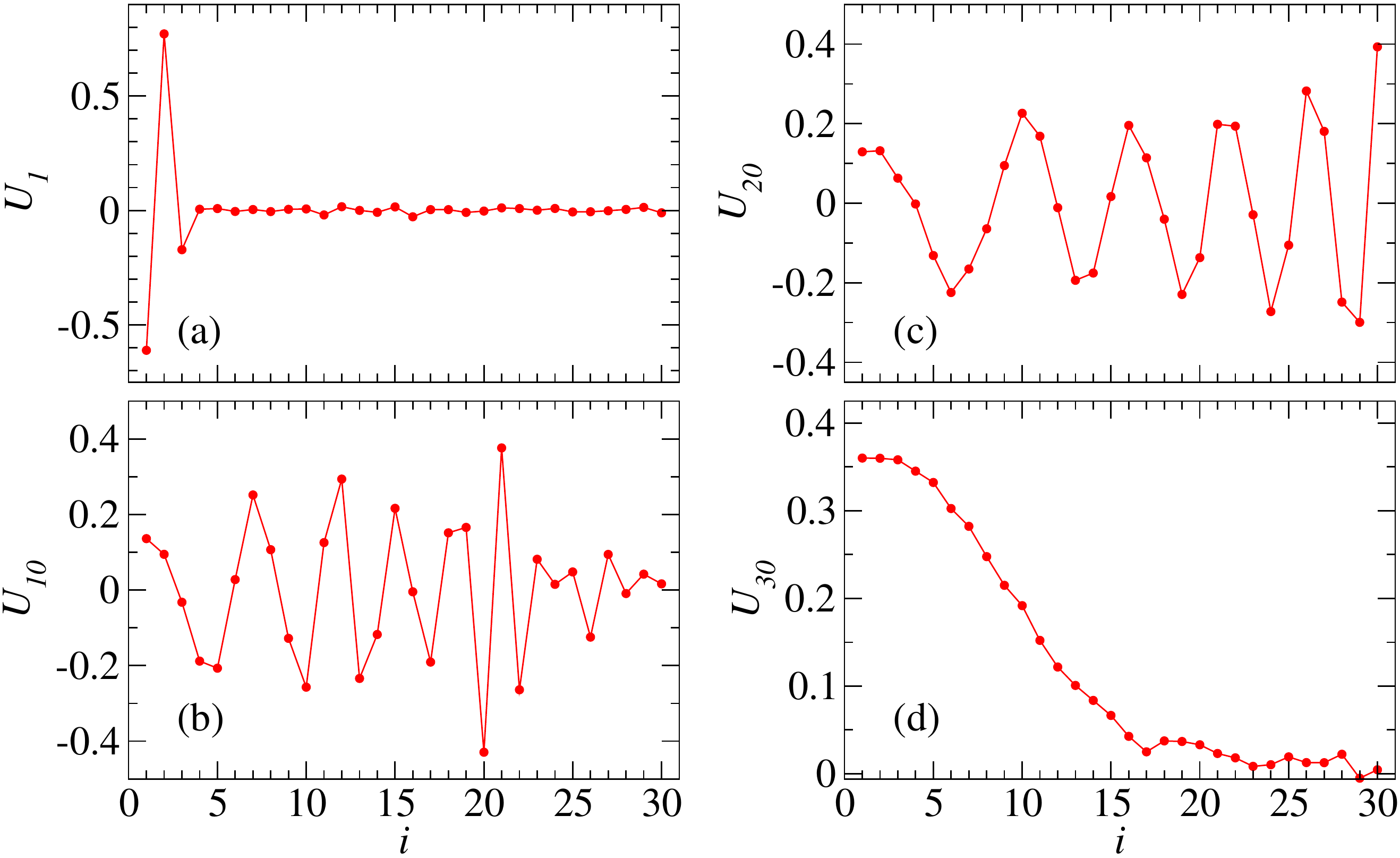}
\vskip-1mm
\caption{Selected eigenfunctions corresponding to the data in Fig.~\ref{fig:gtau}. Panels (a)--(d) show the eigenfunctions $U_1$, $U_{10}$, $U_{20}$,
and $U_{30}$, respectively (with smallest to largest eigenvalues).}
\label{fig:geigen}
\end{figure*}

Some examples of eigenfunctions of $C$ are shown in Fig.~\ref{fig:geigen}. The eigenvector $U_1$ in Fig.~\ref{fig:geigen}(a) corresponds to the
smallest eigenvalue $\epsilon_1$ in the inset of  Fig.~\ref{fig:gtau}(b) and involves essentially only $\bar G(\tau_i)$ for the three shortest times, $i=1,2,3$,
added out-of-phase. The subsequent eigenvectors
involve most of the original $\tau$ points, with the weights exhibiting oscillations with increasing distance between maxima and minima. The
eigenvector $U_{30}$ for the largest eigenvalue, Fig.~\ref{fig:geigen}(d), has only positive contributions, with significant weight for about the first
half of the $\tau$ points. Thus, even after normalizing the data by $\bar G(0)$, which automatically eliminates some of the uniform covariance, the largest
fluctuations are still in-phase, involving the points with $\tau  \lesssim 3$ in this case. While these details of the covariance depend on the model
and the spectral function considered, the behaviors observed in Figs.~\ref{fig:gtau} and \ref{fig:geigen} are qualitatively fairly typical for a wide range
of quantum spin models and observables that we have studied (e.g., the 2D and 3D system in Refs.~\cite{shao17} and \cite{qin17}, respectively).

It has been argued that only data corresponding to ``large''eigenvalues (or singular values) should be used \cite{jarrell96}, but we do not see any reason for
truncation at this stage. Some of the smaller eigenvalues correspond to the very small error bars at small $\tau$, e.g., in the case of the
eigenvector $U_1$ in Fig.~\ref{fig:geigen}(a), and throwing away these data may be detrimental. It is also difficult to identify a clear boundary
between ``large'' and ``small'' eigenvalues, as seen in the inset of Fig.~\ref{fig:gtau}(b). We here only make sure that the eigenvalues are stable
with respect to the amount of QMC data used. Often problems with the covariance matrix can be detected at the analytic continuation stage, if a
statistically acceptable $\chi^2$ value cannot be obtained (and we will later quantify the meaning of ``acceptable''). In such cases, we prune the data
set by, e.g., only using every second $\tau$ point or by increasing the amount of QMC data until the eigenvalues and eigenvectors become reliable and
$\chi^2$ is acceptable.

\subsection{Synthetic data}
\label{sec:syntdata}

In order to rigorously test analytic continuation methods, models with exactly known spectral functions are invaluable. However, models that are
exactly solvable,  amenable to QMC simulations, and also have non-trivial spectral functions with interesting features are rare. Even in the case of
the $S=1/2$ Heisenberg chain, which we will study frequently in the following sections, the dynamic spin structure factor has not been calculated entirely
without approximations, though the results can still serve as very useful benchmarks \cite{caux05a,caux05b,pereira06}.

Given the need to test methods for many different types of spectral functions, it is necessary to consider synthetic data, by which we mean imaginary-time
correlation functions computed directly using Eq.~(\ref{contrel1}) with an artificial spectral function $S(\omega)$ constructed with the desired
features to be tested. Several examples will be considered in the tests presented in the later sections. 

To mimic the statistical errors present in QMC data, noise should be added to the values $G(\tau_i)$ obtained from Eq.~(\ref{contrel1}). Since results
for different $\tau$ points are always strongly correlated when they are computed on the basis of the same QMC configurations, we construct synthetic
correlated noise in the way already described in Refs.~\cite{qin17} and \cite{shao17} and repeated here for completeness.

We begin by generating normal-distributed values $\sigma^0_i$ with zero mean and the same standard deviation $\sigma^0$ for all $i$ (for simplicity). To build
in correlations between the noise for different $i$, we compute weighted averages over several points using an exponentially decaying weight function;
\begin{equation}
\sigma_i = \sum_{j} \sigma^0_j {\rm e}^{-|\tau_i-\tau_j|/\xi_\tau},
\label{corrnoise}
\end{equation}
which we then add to $G(\tau_i)$. We generate a large number of such noisy data sets (corresponding to QMC data bins) and run these through the
same bootstrapping code that we use for real QMC data to compute the mean values and the covariance matrix.

The correlation time $\xi_\tau$ and the common (in the simplest case) standard deviation $\sigma^0$ of the generated noise instances $\sigma^0_j$
in Eq.~(\ref{corrnoise}) are adjusted so that the eigenvalues of the covariance matrix are similar to those of real QMC eigenvalues, e.g., those in
Fig.~\ref{fig:gtau}(b). For all results presented in this work, we used $\xi_\tau=1$. The overall error level $\sigma$ (i.e., the approximate
statistical errors for $\tau\gg \xi_\tau$, where the $\tau$ dependence is weak, as in Fig.~\ref{fig:gtau}) depends on the original (uncorrelated)
standard deviation $\sigma^0$ and the number of bins, and these parameters can be adjusted for a desired error level $\sigma$.

We note here that covariance should not necessarely be regarded as a detriment to analytic continuation. In fact, covariance is what causes the small eigenvalues,
e.g., in the inset of Fig.~\ref{fig:geigen}(b), and some of these small eigenvalues can translate into better frequency resolution compared to data with
uniform statistical errors of magnitude similar to the larger eigenvalues. Using tests with synthetic data with and without covariance, but otherwise the
same number of $\tau$ points and noise level, we have indeed found that the presence of covariance actually improves the outcome of analytic continuation.
We will not discuss this aspect of covariance further, but it is important to keep it in mind when benchmarking methods with synthetic spectral functions.

In our tests with synthetic data in this work, we set the temperature in Eq.~(\ref{contrel1}) to a low value in all cases (often $T=0$), since the improved
SAC methods considered here are mainly intended for applications to systems at low temperatures, where sharp spectral features often appear.

\section{Review of methods and outline of stochastic analytic continuation with constraints}
\label{sec:methods}

In perturbative diagrammatic many-body calculations, where imaginary-time correlations can be obtained to machine precision, Pad\'e approximants
\cite{vidberg77,beach00} are commonly used for analytic continuation (though not always successfully---an apparently superior method was developed
recently \cite{fei21}). The Pad\'e approach was applied also with QMC data, with some success for simple cases, e.g., in early work to extract
single-mode information \cite{thirumvalai83}. In general this method is not stable when attempting to reproduce more complicated spectral
functions, though some progress has been made in recent years \cite{schott16,han17,motoyama22}. The new Nevanlinna analytic continuation method
\cite{fei21} has not yet been broadly explored with QMC data.

To set the stage for the new developments presented in this treatise, we here first briefly review the most commonly used numerical methods for
analytic continuation of QMC data. In Sec.~\ref{sec:methods_a} we summarize some of the maximum-likelihood techniques explored early on and
also mention more recent related developments. We then focus on the commonly used ME method in Sec.~\ref{sec:methods_b}. In Sec.~\ref{sec:methods_c}
we discuss the main ideas behind the conventional (unrestricted) SAC method (with more details and new insights to follow in Sec.~\ref{sec:plainsampling}).
In Sec.~\ref{sec:methods_d} we discuss the motivations for restricted sampling (with optimized constraints) and give a preview of the further progress
that will be presented in more detail in Secs.~\ref{sec:deltapeak}, \ref{sec:contedge1}, and \ref{sec:contedge2}. We defer discussion of
the more recently explored machine learning methods for analytic continuation \cite{arsenault17,fournier20} to Sec.~\ref{sec:discussion}.

\subsection{Data fitting based on $\chi^2$}
\label{sec:methods_a}

An early approach to analytic continuation of QMC data
was to represent the spectrum as a positive-definite histogram with a small number of bins (four to eight) and optimize the
distribution of weight by minimizing $\chi^2$ \cite{schuttler85a,schuttler85b} (where sum rules and computed moments of the spectrum were also
included in addition to the correlation function). It was noted that the amount of information contained in the noisy QMC data did not warrant a
larger number of histogram bins. If one attempts to use a large number of bins (or closely spaced $\delta$-functions with optimized amplitudes)
the result is in general not smooth, but sharp spikes appear. Perhaps surprisingly the spectrum truly minimizing $\chi^2$ typically consists of a
few isolated $\delta$-functions \cite{sandvik98}, as we will also discuss further below when studying the dependence of a sampled spectrum on the
fictitious temperature $\Theta$ in the SAC method. In \ref{app:lowtheta} we confirm that the ultimately best-fit spectrum is a small set
of $\delta$-functions. The reason for this behavior can be found in the imposed positive-definiteness of the spectrum.

In a further development of the $\chi^2$ fitting approach, a discrete gradient-squared term was maximized along with the minimization
of $\chi^2$, thus leading to smooth spectral functions \cite{white89}. While this represented an improvement and some useful results
were obtained, the lack of a general criterion for the relative degree of smoothness and goodness of the fit made it hard
to judge the validity of the results. Around the same time, the mathematically more appealing ME method \cite{gull84} was adapted to the
QMC context \cite{silver90,gubernatis91,jarrell96} and quickly became the dominant analytic continuation technique.

In principle, a very good solution to the analytic continuation problem would be to just use $\chi^2$ minimization with a suitable functional form,
not based on a dense histogram or a large set of $\delta$-functions but still with enough flexibility to describe the expected spectrum. This approach
has been applied with some success with functions depending on a small number of parameters motivated by physical insights, e.g.,
in Refs.~\cite{sandvik98b,sandvik01,katz14}. However, it is very difficult to construct a generic form with sufficient flexibility to
reproduce {\it a priori} unknown spectral functions. Increasing the flexibility beyond some rather small number of parameters leads to problems similar
to those encountered with histograms, unless the solution can be sufficiently constrained (regularized).
Some progress has been made recently along these lines, by combining rather complex, flexible functional forms
with a Bayesian method to discriminate between results of different parametrizations \cite{linden19}. The spectral functions here are still
not completely arbitrary but should be constructed based on prior knowledge.

Analytic continuation based on singular-value decomposition is also possible \cite{gazit13}. As an alternative to minimizing $\chi^2$ (the $L_2$ norm)
other norms defining the ``best'' solution can also be used in sparse modeling. Here the goal is to find a minimal number of parameters, in some
representation of the spectrum, to model noisy imaginary-time (or Matsubara frequency) data. So far we have not seen any advantages of these approaches
relative to the other methods reviewed here and further below, though the goal is also exactly to solve problems
analogous to the spike histograms discussed above. We refer to the recent review Ref.~\cite{otsuki20} for further details on sparse modeling.

\subsection{Maximum entropy methods}
\label{sec:methods_b}

In the ME method the spectrum is normally parametrized as amplitudes on a dense frequency grid, regularized by the information entropy $E(S)$
defined with respect to a default model $D(\omega)$. In the limit of a continuum
\begin{equation}
E(S)=-\int_{-\infty}^\infty d\omega S(\omega) \ln{\left (\frac{S(\omega)}{D(\omega)} \right )},
\label{esdef}
\end{equation}
which is maximal ($E=0$) when $S(\omega)=D(\omega)$.

The mathematical foundation of the ME method is Bayes' theorem, with
\begin{equation}
P(S|\bar G)P(\bar G)=P(\bar G|S)P(S),
\label{bayes}
\end{equation}
postulated. Here the prior probability (i.e., the result in the absence of data) is given by
\begin{equation}
P(S) \propto {\rm exp}[\alpha E(S)],
\label{meprior}  
\end{equation}
where $\alpha$ is a constant to be determined [or in some cases integrated over with a suitable prior $P(\alpha)$]. The conditional probability
$P(\bar G|S)$ of the data set $\{\bar G(\tau_i)\}$ given a spectrum $S$ (implicitely defined by the model simulated) is well known from the
statistics of Gaussian fluctuations;
\begin{equation}
P(\bar G|S) \propto {\rm exp} \left (-\frac{\chi^2(\bar G,S)}{2} \right ).
\label{psgme}  
\end{equation}
Finally, $P(\bar G)$ in Eq.~(\ref{bayes}) acts as an irrelevant normalization, since $\bar G$ has already been generated when solving
for $S$. Maximizing the probability $P(S|\bar G) \propto P(\bar G|S)P(S)$ amounts to minimizing the functional
\begin{equation}
F(S)=\chi^2(S)/2-\alpha E(S),
\label{fsdef}
\end{equation}
which delivers the default model if $\alpha$ is large and corresponds to pure $\chi^2$ fitting when $\alpha \to 0$. These limits
represent poor solutions to the problem for completely different reasons, and the ``best'' spectrum should be obtained at some intermediate value
of $\alpha$.

The entropy serves a similar purpose as the gradient-squared smoothing discussed above in Sec.~\ref{sec:methods_a}, but in the ME method the
ambiguity in the choice of the parameter $\alpha$ has been partially resolved by using Bayesian inference arguments \cite{silver90,jarrell96}.
The basic premises of those arguments are themselves not rigorous, however, and there are differing opinions on exactly how to best
determine $\alpha$ or integrate results over $\alpha$. Recently a completely different criterion, originating in the SAC approach
\cite{beach04}, was put forward \cite{bergeron16} where the ``best'' $\alpha$ corresponds to a maximum of the derivative of $\ln(\chi^2)$
with respect to $\ln(\alpha)$ in the region where $\chi^2(\alpha)$ is close to its minimum value. However, as we will show in Sec.~\ref{sec:thetacrit}
in the context of the SAC method, and also in Sec.~\ref{sec:maxent} when further discussing the ME method, this criterion cannot guarantee an
acceptable fit to the data. 

To improve the performance of the ME method in general, all prior available information should be incorporated to constrain the solution \cite{linden95}.
A default model $D(\omega)$ in the entropy definition, Eq.~(\ref{esdef}), does not impose any hard constraints, but a good model (i.e., close to the
true spectrum) often helps significantly
to achieve a solution with minimal distortions. In the QMC context, frequency moments from sum rules can be imposed when
constructing $D(\omega)$ by maximizing the entropy \cite{jarrell96}. Approximate results, e.g., from perturbation theory \cite{diamantis14,fanto21}, can be also used,
but this approach can produce biased or misleading results, given that the perturbative results in most cases would be far from correct \cite{jarrell12}.

Often, in the absence of strong constraints or a default model known to be very close to the solution sought, the best option is to use a flat default
extending from $\omega=0$ [in the case where the $\omega \ge 0$ formulation with the kernel in Eq.~(\ref{kernel2}) is used] up to some frequency beyond which
the remaining spectral weight is negligible. If the lower bound $\omega_0>0$ is known, incorporating it can significantly improve also the resolution at higher
frequencies, as was noted in the context of the SAC method \cite{sandvik16}.

We will further discuss the ME method in Sec.~\ref{sec:maxent}, where we present transferable insights from the SAC approach. We propose a new
SAC-inspired way of fixing the entropy factor $\alpha$. Most importantly, we formulate equivalences (not invoking any mean-field arguments \cite{beach04}
or other approximations) between SAC sampling with different parametrizations and ME methods with associated functional forms of the entropy---not
always the Shannon information entropy.

\subsection{Stochastic averaging}
\label{sec:methods_c}

It was early on realized \cite{white91} that a different way to achieve a smooth spectrum is to average over many solutions with reasonable $\chi^2$
values, though this method initially fell in the shadows of the ME method. The SAC approach (also called stochastic analytic inference or the average
spectrum method) was introduced independently in a slightly different form \cite{sandvik98} several years later. Subsequent works further developed and
explored the method \cite{beach04,syljuasen08,fuchs10,sandvik16,qin17,shao17,ghanem20a,ghanem20b}. Applications of SAC methods to various quantum lattice models
abound; we list a representative sample of mostly recent works as
Refs.~\cite{syljuasen08,qin17,shao17,feldner11,voll15,lohofer15,lohofer17,becker17,becker18,ying19,shu18,sun18,ma18,xu19,li20,raczkowski20,zhou21,sato21,cheng22,liu22}.

In most SAC methods, the averaging is carried out by importance sampling of the parameters of a fully flexible positive (semi) definite spectrum $S(\omega)$,
e.g., the amplitudes of $\delta$-functions on a dense grid of frequencies or with some large number of $\delta$-functions residing in continuous frequency space,
as illustrated in Fig.~\ref{fig:spec}. Such a parametrization with large $N_\omega$ can be regarded as the configuration space of a statistical-mechanics problem,
with $\chi^2(S)/2$ playing the role of the energy in a Boltzmann-like weight function. One can then carry out a Metropolis Monte Carlo simulation in this space
at a fictitious temperature $\Theta$, whence the probability density of the spectrum $S(\omega)$, given the data $\bar G$ (and the covariance matrix), is
\begin{equation}
P(S|\bar G) \propto {\rm exp} \left (-\frac{\chi^2(S)}{2\Theta} \right ),
\label{psg}  
\end{equation}
where $\chi^2$ also depends on $\bar G$
and the covariance matrix according to Eq.~(\ref{chi2}). With $\Theta=1$, which was used by White~\cite{white91} and also advocated
by Sylju{\aa}sen \cite{syljuasen08}, this conditional probability is exactly the likelihood function arising from Bayes' theorem, Eq.~(\ref{bayes}),
if one assumes a prior probability $P(S)$ independent of $S$ (i.e., all combinations of the parameters defining the spectrum are a priori equally
likely). Leaving out the unimportant (and unknown) normalization $P(\bar G)$ we then have $P(S|\bar G) \propto P(\bar G|S)$, i.e., Eq.~(\ref{psg}).

As an alternative to fixing $\Theta=1$, it can be argued that $\Theta$ should be adjusted to make sure that the sampled average goodness of the
fit $\langle \chi^2\rangle$ is close to its minimum value (obtained when $\Theta \to 0$). The fluctuations of the spectrum still must be large enough
to produce a smooth average \cite{sandvik98,beach04}, and then some criterion balancing $\langle \chi^2\rangle$ minimization and sampling entropy must
be devised. It has been shown explicitly \cite{sandvik16} that configurational entropy leads to a deteriorating spectrum (increasing $\langle \chi^2\rangle$)
as the number of degrees of freedom of the spectrum is increased if the temperature is held fixed at $\Theta=1$. This effect can be
counteracted by imposing constraints \cite{sandvik16}, as we will also discuss here, but in most cases it is still necessary to also suppress
the entropy by reducing $\Theta$.

The need to lower the sampling temperature below $\Theta=1$ leads to a problem similar to the selection of the entropy weighting factor $\alpha$ in the ME
method. A commonly used method has been to identify the point of sharpest drop (peak in the derivative) of $\ln\langle \chi^2(\Theta)\rangle$ versus $\ln(\Theta)$,
which, in a loose analogy with the specific heat  \cite{beach04} can be regarded as signaling a transition into a ``glassy'' state in which the
spectrum becomes difficult to sample and is affected by the statistical errors. Bayesian inference can also be applied to fix $\Theta$ \cite{fuchs10}, with
the same caveats as in the case of $\alpha$ in the ME method. As demonstrated by Beach, the ME and SAC methods are in fact closely related \cite{beach04}---treating
the SAC setup within a mean-field-like approximation gives exactly the ME method, with $\alpha=\Theta$. This fact initially appeared to lend further credence
to SAC as superior to the ME method, as the fluctuations neglected in the ME solution can contain additional spectral structure \cite{fuchs10}. Here, in
Sec.~\ref{sec:entropy} and \ref{sec:maxent}, we will further quantify the relationship between the two methods, which in some important aspects is at odds
with the previous notions.

As mentioned in the previous section, the likelihood function Eq.~(\ref{psg}) with $\Theta=1$ arises from Bayes' theorem with a constant prior $P(S)$.
Alternatively, this likelihood function, with $\Theta=1$ or $\Theta \not= 1$, can be heuristically interpreted as a reasonable way to carry out averaging
over a class of spectral functions compatible with the QMC data set $\{ \bar G_i\}$. Ultimately, however, a given parametrization of the spectrum also can be
regarded as an entropic prior, i.e., there is some bias toward certain shapes of the spectrum due to specific entropic pressures associated with the
sampled degrees of freedom (i.e., the details of the stochastic process). Investigating these entropic effects, formally quantifying them and counteracting
them with constraints will be the major themes of this paper.

There is another line of average-spectrum methods where the sampling is not carried out with a Boltzmann-like weight function, but some other mechanism
is used to generate an ensemble of spectra with reasonable $\chi^2$ values \cite{mishchenko00,mishchenko02,mishchenko12,goulko17,bao16,krivenko2019}.
We will here only discuss sampling with Eq.~(\ref{psg}), which has the advantage of many analogies with statistical mechanics. Specifically, regularizing
the average spectrum with the fictitious temperature $\Theta$, there is an adjustable competition between ``internal energy'', i.e., minimization of
$\langle \chi^2\rangle$, and entropy, the latter of which can distort the spectrum but also counteracts overfitting (preventing freezing into an undesirable
ground state corresponding to a noise dominated spectrum). Further analogies appear when constraints are imposed, whence the sampling maximizes the
entropy within a restricted space. Optimizing the parameter regulating a constraint corresponds to $\langle \chi^2\rangle$ (internal energy) minimization
by suppression of entropy at fixed $\Theta$. The non-thermal regularization parameters used in
Refs.~\cite{mishchenko00,mishchenko02,mishchenko12,goulko17,bao16,krivenko2019} imply other, less understood, relationships between
$\langle \chi^2\rangle$ and entropy.

Sampling of the spectrum is also sometimes done within the ME approach \cite{boninsegni96,kora18}, using $e^{-F(S)}$ with $F(S)$ given by Eq.~(\ref{fsdef})
as the weight function. Thus, an entropic prior is used to weight the configurations, and at the same time there is also an often overlooked native entropy
of the sampling space (normally the spectral weight distribution stored as a discrete histogram). The latter in principle causes diverging fluctuations
in the continuum limit of the frequency space, as we will demonstrate explicitly in Sec.~\ref{sec:maxent}.

\subsection{Stochastic analytic continuation; parametrizations and constraints}
\label{sec:methods_d}

It seems very difficult to completely avoid some bias in any numerical analytic continuation method. A main goal of the work presented here is to show
how bias arising from entropic pressures can be removed by constraints. A simple example would be to fix a lower bound of a gapped spectrum at some frequency
$\omega_0>0$, thus avoiding ``leaking'' of spectral weight below the bound, which inevitably takes place with unrestricted sampling. While the gap
would typically not be known in advance, it can be determined by a statistical goodness-of-fit measure produced by SAC runs at different values of
$\omega_0$. Constraints can also be helpful for reproducing more intricate sharp spectral features, especially at the lower edge, which will be our main
focus in the later sections of this paper.

In some cases, a certain prominent feature  may be known, e.g., an edge $\omega_0>0$ at which the spectrum diverges asymptotically as a
power law, $(\omega - \omega_0)^{-p}$ when $\omega \to \omega_0$ from above, with  a known or unknown exponent $p$. It is then clearly legitimate to
incorporate a constraint that implies such an asymptotic form but does not further hard-impose any details, i.e., allowing deviations from the  native
constrained form away from the asymptotic $\omega \to \omega_0$ limit. We will demonstrate that SAC procedures with such constraints, exemplified
in the simplest case in Fig.~\ref{fig:spec}(e), can deliver previously unknown information, e.g., the value of $\omega_0$ and, ideally (if the imaginary-time
data are sufficiently good) the exponent $p$ controlling the asymptotic divergence if it is not a priori known. Another example is a quasi-particle peak with
some weight and width, for which Fig.~\ref{fig:spec}(d) is the simplest parametrization when the peak is extremely sharp. Provided that the spectrum beyond
the edge feature imposed is not too complex, it can also be resolved in its entirety to a much better degree once the edge has been treated correctly.

In cases where an edge feature is not known with certainty, imposing a specific constraint can be thought of as testing a hypothesis, and the optimization
process should then also ideally be able to signal the inapplicability of an incorrect constraint. Though significant work may be required for selecting,
testing, and optimizing constraints, the efforts can pay off with results that far exceed in quality what can be expected with other approaches. A main
goal of this work is to provide some examples of what can be achieved with suitable constraints and parametrizations of the spectral function within
the SAC approach.

\subsubsection{Role of configurational entropy}

As already mentioned, if one takes the Bayesian point of view, the natural assumption would be to keep $\Theta=1$ in the sampling weight, Eq.~(\ref{psg})
\cite{syljuasen08}. However, if the number of parameters of the spectrum is large, e.g., when using $\delta$-functions on a dense frequency grid, this leads
to poor data fitting, i.e., the sampled average goodness of the fit $\langle \chi^2\rangle$ is much larger than would be statistically expected for a
good fit \cite{sandvik16}. The root cause of this effect can be found in the analogy with statistical mechanics, where $\chi^2$ would be an unusual form of the
energy. Most notably, $\chi^2$ is not extensive in the number of degrees of freedom sampled with, e.g., the $\delta$-functions as in Fig.~\ref{fig:spec}.
This fact is most clearly reflected by the minimum value $\chi^2_{\rm min}$, which is positive and
cannot change significantly when more $\delta$-functions are added after
some threshold number. Note that $\chi^2_{\rm min}=0$ cannot be reached when positive-definiteness of the spectrum is imposed, as discussed in detail in
\ref{app:lowtheta}. Therefore, unlike conventional statistical mechanics, when more degrees of freedom are added, to make the spectrum more flexible, the
entropy will eventually, for any fixed value of $\Theta$, drive the spectrum toward some limit that maximizes the entropy while taking $\langle \chi^2\rangle$
further away from $\chi^2_{\rm min}$. We also note here that the form of the entropy depends on the parametrization used (as we discuss in detail in
Sec.~\ref{sec:entropy1} and also when relating SAC to the ME method in Sec.~\ref{sec:maxent}) but it is always extensive in the sampled degrees of freedom.

While the ``entropic catastrophe'' can be counteracted by choosing $\Theta < 1$ (adapted to $N_\omega$, as will be discussed in
Sec.~\ref{sec:plainsampling}), in SAC methods with restricted sampling \cite{sandvik16,shao17} the entropy associated with distortions of sharp spectral
features is also impeded by suitable optimized constraints. Then sampling at $\Theta=1$ can be appropriate, but, if the number of parameters used
to sample the spectrum is large, it will still be necessary to also reduce $\Theta$.

Some entropic effects on the average spectrum were investigated in Ref.~\cite{sandvik16}. In particular, for a spectrum with known lower
and upper bounds, it was shown that spectral weight gradually leaks out further beyond these bounds when the number of $\delta$-functions is increased.
This deteriorating effect of configurational entropy can then also be used to identify the correct frequency bounds, by monitoring
$\langle \chi^2\rangle$ at fixed $\Theta > 0$ when the constraints are varied.

Prohibiting spectral weight from certain frequency regions implies an entropy reduction. For a spectrum with a sharp lower edge, a clear
minimum in $\langle \chi^2\rangle$ was found when the lower bound was close to the true location of the edge \cite{sandvik16}.
The upper bound is normally less sharp, and the $\langle \chi^2\rangle$ signal is weaker but still useful.
Further, in cases where the true spectrum has a single peak, it was shown that imposing a single peak in the spectrum at each stage of the sampling 
can dramatically improve the ability to resolve the shape of the spectrum (and one can expect this to generalize to two or more peaks). This
quality boost was also explained by a reduction of configurational entropy when the fluctuations of the sampled spectrum are constrained.

The above example of frequency bounds can be regarded as a certain prior probability $P(S)$. With no other restrictions imposed, sampling with one of the
three parametrizations in Fig.~\ref{fig:spec}(a)-(c) in the absence of QMC data clearly results in a uniform spectral density. While this flat average spectrum
can in some sense be regarded as a default model, as in the ME method, the different entropic pressures of the three parametrizations will still produce different
outcomes when the spectra are weighted by $\chi^2$, as discussed in detail in Secs.~\ref{sec:theta} and \ref{sec:entropy}. These differences can also be
reproduced with the ME method with a flat default model and different forms of the entropic prior, as we will demontrate in Sec.~\ref{sec:maxent}. 

In other cases, it may be less useful to think of constraints as analogues to ME default models. For instance, imposing a single peak at the level of
each sampled configuration can have a large impact in SAC, as mentioned above, even if the average spectrum has a single peak also in the absence of
such a constraint (and that peak would then be broader than when the sampling constraint is imposed). The imposition of the sampling-level constraint
has no direct translation (at least not an obvious one) into a default model for the most probable ME spectrum.

The difference between the SAC and ME approaches in this regard is that the SAC spectrum is an average, and constraints on how the average is taken (what
configurations are sampled over) will affect the outcome, while in the ME method a single function is normally optimized and constraints can only be
applied to this one function. An exception is the version \cite{boninsegni96,kora18} of the ME method in which the histogram representing the spectrum
is not determined by minimizing $F(S)$ but is averaged over samples with probability $\propto {\rm e}^{-F(S)}$. Then a constraint on the individual sampled
spectra could also in principle be built in exactly as in the SAC method implemented with $\delta$-functions on a frequency grid \cite{sandvik16}.
However, as will be discussed in Sec.~\ref{sec:maxent2}, such a method may suffer from problems related to the extensive sampling entropy and
``double counting'' of entropy; in the continuum limit the average becomes ill-defined and cannot be sampled properly with the above ME probability
distribution.

In Ref.~\cite{shao17}, together with collaborators we showed how a the amplitude $A_0$ of an isolated $\delta$-function at the lower edge of a continuum
(i.e., a sharp quasi-particle peak) can be optimized within the SAC parametrization in Fig.~\ref{fig:spec}(d). In this case a default model analogy within
the standard ME scheme would also be rather contrived, though a optimized $\delta$-function could certainly also be incorporated in a rather similar way
within the ME approach, especially in light of our new results for the relationships between the SAC and ME method, detailed in Sec.~\ref{sec:maxent}.

\subsubsection{Different parametrizations}
\label{sec:parametr}

One issue that has not been addressed in sufficient detail previously is how the averaged spectral function depends on the parametrization. A 
commonly used SAC parametrizations is a set of $\delta$-functions on a fixed frequency grid $\{\omega_i\}$ \cite{sandvik98,syljuasen08,sandvik16},
Fig.~\ref{fig:spec}(a), where the amplitudes $A_i$ are sampled. A non-uniform grid (not considered here) can also be used, in which case the
local density of the grid points can be regarded as a default model \cite{ghanem20a}.

With the spectrum defined in the frequency continuum, we
consider either sampling of only the locations $\omega_i$ of equal-$A_i$ $\delta$-functions as in Fig.~\ref{fig:spec}(b) \cite{qin17}, or also including
updates of the amplitudes as in Fig.~\ref{fig:spec}(c) \cite{beach04}. A related ``grid point sampling'' approach was developed in Ref.~\cite{ghanem20b}.
The parametrization with fluctuating frequencies and amplitudes has been the most commonly used in SAC applications, e.g.,
Refs.~\cite{fuchs10,feldner11,voll15,lohofer15,lohofer17,becker17,becker18,ying19,raczkowski20,zhou21,sato21}, and the fixed-amplitude
parametrization has also been used in several works, e.g., Refs.~\cite{shao17,ma18,cheng22,liu22}. As an alternative to all-equal amplitudes
in Fig.~\ref{fig:spec}(b), some other predetermined distribution of amplitudes can also be used; the values $A_i$ are then still held fixed but
they reside at unrestricted frequencies $\omega_i$ \cite{qin17}.

Here we will describe all the five different parametrizations illustrated in Fig.~\ref{fig:spec}, and further variations of these cases will be
introduced in later sections. In this section we focus on qualitative aspects of configurational entropy and constraints, and defer practical details
on the Monte Carlo sampling algorithms to Secs.~\ref{sec:plainsampling} and \ref{sec:contedge1}. While we will perform some tests with the fixed-grid
parametrization illustrated in Fig.~\ref{fig:spec}(a), in most of our tests and applications we consider $\delta$-functions in the frequency
continuum. The average spectrum is then defined as the mean amplitude density collected in a histogram with bin width suitably chosen to
accomodate the details of the spectrum.

With a large number $N_\omega$ of $\delta$-functions, the three parametrizations illustrated in Figs.~\ref{fig:spec}(a)--(c) are generic and
suitable in principle to describe any kind of spectrum. They are, however, associated with different types of entropic pressures, i.e., for
given $\bar G(\tau)$ data and a fixed value of $\Theta$, they will each favor different spectral profiles.
Importantly, a result obtained with one parametrization can typically not be reproduced at any value of $\Theta$ with another parametrization.
In Sec.~\ref{sec:plainsampling} we will compare results for several examples of spectral functions, following the average spectrum as a function of
the sampling temperature $\Theta$. Using a moderately large number ($N_\omega=1000$) of $\delta$-functions at, say, $\Theta=1$, we find that $\langle \chi^2\rangle$
is significantly larger with the fixed-grid parametrization, Fig.~\ref{fig:spec}(a), i.e., the entropic pressures pushing the solution away from
the correct one is larger than it is with $\delta$-functions in the continuum, Fig.~\ref{fig:spec}(a) and \ref{fig:spec}(b). In all cases, when
$N_\omega$ is large [and exactly how large depends on the error level of the $\bar G(\tau)$ data], $\Theta$ has to
be lowered from $1$ in order for the value of $\langle \chi^2\rangle$ to be statistically sound (and we will later define our criterion for
statistical soundness). Unless the error level $\sigma$ is extremely low, the average spectra obtained with different parametrizations are all
different.

Based on our systematic investigations, we conclude that SAC in the frequency continuum has higher fidelity, and sampling such parametrizations
is also much faster than on the fixed grid. The fixed-grid results in general have excessively sharp peaks.
There are also differences between continuous-frequency spectra sampled with or without amplitude updates, and we will show that sampling with amplitude
updates often leads to better frequency resolution. However, only sampling frequencies may be safer from the standpoint of the standard information
entropic arguments, according to which a spectrum with a larger Shannon entropy would be preferred. Indeed, we will also find (Sec.~\ref{sec:maxent})
that SAC with only frequency updates [Fig.~\ref{fig:spec}(b)] is exactly equivalent to the ME method with the conventional Shannon entropy in
the limit $N_\omega \to \infty$, if $\Theta$ and $\alpha$ are chosen such that  $\langle \chi^2(\Theta)\rangle$ in SAC equals $\chi^2(\alpha)$ of
the ME spectrum. For other parametrizations, the equivalence between the methods requires different functional forms of the entropy in the
prior probability used in ME method.

There are many possibilities to build in specific known or expected features of the spectrum, and doing so can lead to substantial
reduction or elimination of entropic pressures that distort the averaged spectrum in the absence of such constraints. Imposing a constraint of
course means that some piece of information beyond the imaginary-time data is supplied to the solution, but this information can be of a very
generic form, e.g., the spectrum should have a sharp lower edge (which the constraints discussed in this paper will mostly be
focused on). With a corresponding constraint, it should be possible to extract unknown information on the properties of the edge, e.g., its
location and details on the shape (e.g., the width of a quasi-particle peak or the exponent governing a power-law singularity).
As we will show, imposing a constraint on the edge
can also greatly improve the fidelity of the other parts of the average spectrum. In other words, without the constraint the broadened edge 
will be reflected in compensating distortions also at higher frequencies, in order for the spectrum to fit the imaginary-time data. Once the
edge is treated correctly, the other parts of the spectrum are also better reproduced, often in surprising detail.

\begin{figure}
\centering
\includegraphics[width=7.5cm, clip]{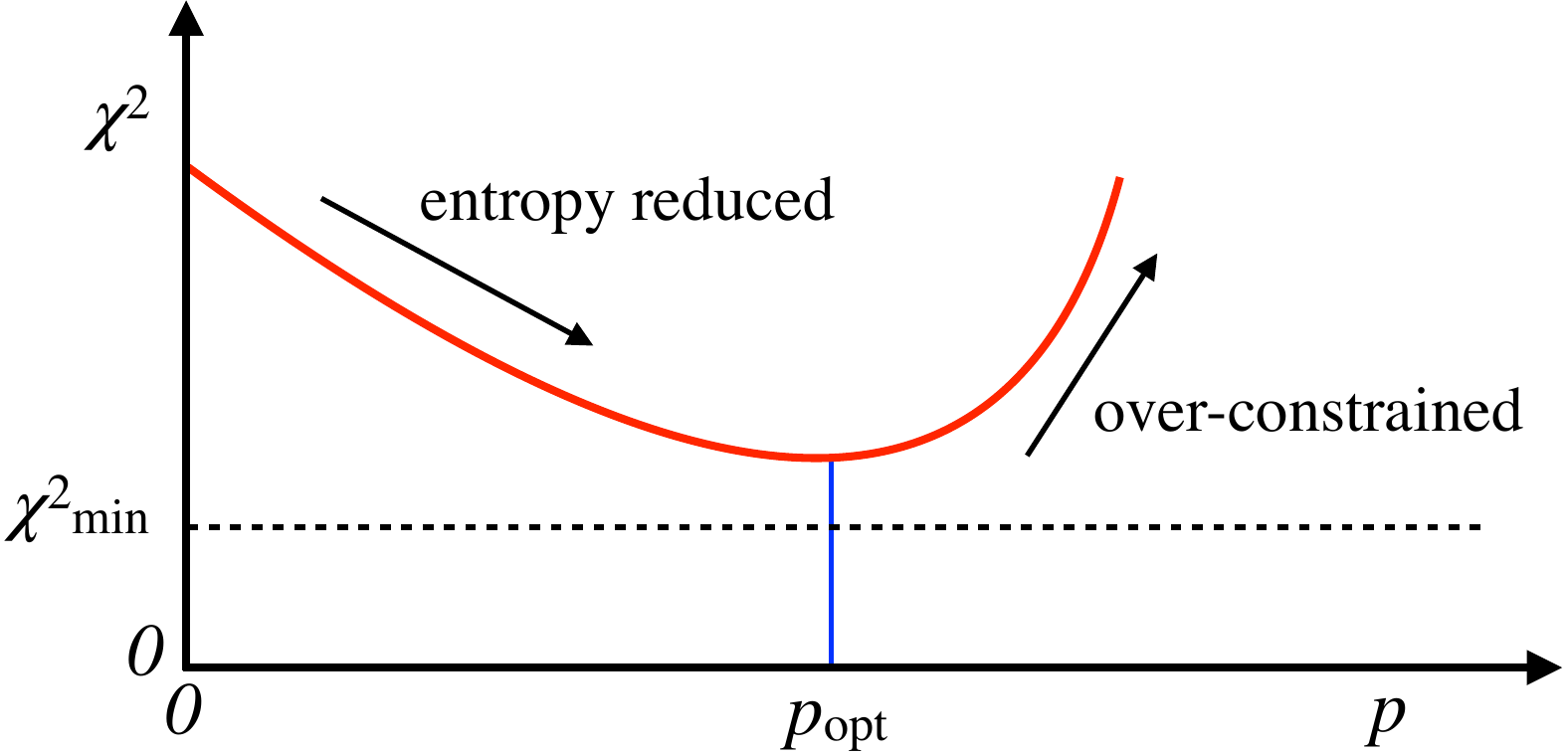}
\vskip-1mm
\caption{Schematic illustration of the optimization of a parameter $p$ regulating an appropriate constraint, with $p=0$ corresponding
  to unconstrained sampling using, e.g., the parametrization in Fig.~\ref{fig:spec}(b). When $p$ is increased, the configurational entropy
  of typical spectra is reduced, thereby allowing $\langle \chi^2\rangle$ to decrease even though the temperature
$\Theta$ is kept fixed [at a value for which $\langle \chi^2(p=0)\rangle$ is clearly above the minimum value $\chi^2_{\rm min}$ attainable].
When $p$ exceeds its correct value, the spectrum becomes too constrained and the fit deteriorates rapidly. The two competing effects lead to an
optimal value $p_{\rm opt}$ (minimizing $\langle \chi^2\rangle$), which is close to the correct $p$ of the spectrum provided that
the constraint is applicable to the spectrum sought.}
\label{fig:optim}
\vskip-1mm
\end{figure}

The perhaps simplest constraint is just an imposed lower frequency bound $\omega_0$ within the parametrizations in Figs.~\ref{fig:spec}(a)--(c),
optimized in the generic way illustrated in Fig.~\ref{fig:optim} by running the SAC procedure for a range of values of the parameter $p=\omega_0$.
Unless the true spectrum has a sharp finite step at the edge, this constraint is in general not sufficient to reproduce the actual shape of the
edge, however, though the results can still be much better than without any constraints (as we will show with examples).

The first more sophisticated constraint studied here, following Ref.~\cite{shao17}, imposes a macroscopic $\delta$-function at the edge frequency $\omega_0$,
with an adjustable (optimized) amplitude $A_0$. The edge is the lower bound for a set of $N_\omega$ microscopic $\delta$-functions with total weight
$1-A_0$, i.e., their individual amplitudes are $A_i=(1-A_0)/N_\omega$. This constrained parametrization is illustrated in Fig.~\ref{fig:spec}(d), where
all the frequencies $\omega_i$ are sampled, including $\omega_0$ (unlike the simpler edge constraints discussed above, where $\omega_0$ is optimized
but held fixed during the sampling process). If the $\delta$-edge ansatz is appropriate, $\omega_0$ will fluctuate relatively weakly
close to the correct position once $A_0$ has been optimized in the way illustrated in Fig.~\ref{fig:optim} (with the generic parameter $p=A_0$).

A different type of edge-tailored constraint, depicted in Fig.~\ref{fig:spec}(e), is intended to model a sharp edge at $\omega=\omega_1$ followed by a continuum
with monotonically decreasing spectral weight. The constraint on the equal-weight $\delta$-functions is a monotonically increasing spacing
$d_i \equiv \omega_{i+1}-\omega_i$. If the number $N_\omega$ of $\delta$-functions is sufficiently large, the average density becomes essentially continuous and, as
we will show, can well reproduce arbitrary monotonically decreasing spectral weight. The entropic pressures in this parametrization naturally
favor a divergent peak, and in certain cases no further optimization has to be carried out. The sampling automatically adapts the spectral weight to
the correct frequency window. A non-divergent edge can be obtained by modifications of this parametrization, with a single parameter optimized for
the edge shape (including arbitrary divergent or non-divergent power law).

We will also combine the parametrizations in Figs.~\ref{fig:spec}(b) and \ref{fig:spec}(e) for a generic spectrum consisting of a sharp edge followed
by an arbitrary (i.e., not necessarily monotonically decaying) continuum. We will also consider some modifications of these parametrizations and comment
on other potentially useful constraints. As test examples, in addition to synthetic spectral functions we will present results for the dynamic structure
factor of the $S=1/2$ Heisenberg models in one and two dimensions. In Sec.~\ref{sec:ladders} we present initial results of a study of Heisenberg ladders,
which have less well known dynamic structure factors that are expected to host sharp edge features suited for constrained SAC.

Other parametrizations in the continuous frequency space have also been used in SAC and related methods, e.g., a number of boxes whose locations
and shapes are sampled \cite{mishchenko00,mishchenko02,mishchenko12,bao16,goulko17}. SAC with an orthogonal-polynomial expansion has also been
tested \cite{wu13}. In this work we only consider parametrizations based on $\delta$-functions, but it would be interesting to further explore
also other parametrizations.

\subsubsection{Optimization of constraints}

An important question is of course how to optimize a constraint. The general principle was laid out in Ref.~\cite{sandvik16} and is illustrated
schematically in Fig.~\ref{fig:optim}. It was shown that the removal of entropy by the imposition of an appropriate constraint leads to a reduction of
the mean goodness-of-fit $\langle \chi^2\rangle $ if the sampling temperature $\Theta$ has been suitably chosen (in the way further discussed below).
Eventually, when the parameter regulating the constraint, e.g., the weight $A_0$ of the leading $\delta$-function in Fig.~\ref{fig:spec}(d), becomes
too large, a good fit is no longer possible (i.e., the constraint is too constraining) and then $\langle \chi^2\rangle$ begins to increase. Thus, there
will be a specific value of $A_0$ for which the fit is optimal. This method was used in Ref.~\cite{shao17} to study the magnon pole of the dynamic
structure factor of the 2D Heisenberg model, as well as a generalized model in which the magnon pole eventually vanishes due to fractionalization 
into spinons (see also Ref.~\cite{ma18}). Here we will further characterize the convergence of the $\langle \chi^2\rangle$ minimum to the correct
value of the constraining parameter $A_0$ as the data quality increases (the error level $\sigma$ decreases) and the sampling temperature is adjusted.
We will also investigate the behavior when the edge peak has finite width, which we model by dividing the peak weight $A_0$ among $N_p > 1$ edge
$\delta$-functions. In this case two parameters $A_0$ and $N_p$ must be optimized, which is more time consuming but still feasible with the
techniques we have developed.

Using the parametrization with monotonically increasing distance between the $\delta$-functions in Fig.~\ref{fig:spec}(e), we will study the dynamic structure
factor of the Heisenberg chain. We achieve excellent agreement with Bethe ansatz (BA) results---better than Ref.~\cite{sandvik16} and without the
time-consuming work to locate the edges (optimizing the lower and upper frequency bounds) that was necessary there. Both the lower edge and the high-frequency
bound are automatically found by the sampling, thanks to the very minor tendency to artificial broadening with the new  constrained parametrization.
We will show how the entropic
pressures within the space of increasing distances between the $\delta$-functions instead naturally favor a divergent peak, which has a specific shape in
the absence of data but is morphed into a shape close to the true spectral profile when sampled according to $\chi^2$ with good data. If the actual peak is
not divergent, an optimized constraint on the distance $\omega_2-\omega_1$ between the first two $\delta$-functions can be used. As we will show with a
synthetic-data example, when the edge feature has been properly treated, also the other parts of the spectrum are reproduced with remarkable fidelity.

We will also discuss a modification of the parametrization in Fig.~\ref{fig:spec}(e), in which the amplitudes $A_i$ depend on the index $i$ of the
increasing frequencies $\omega_i$ in a way that can be directly related to the exponent controlling an asymptotic power-law behavior $(\omega-\omega_0)^p$
with positive or negative $p$. The exponent can be extracted surprisingly precisely when $p<0$ (using the generic approach in Fig.~\ref{fig:optim}) in tests
with synthetic data of reasonable statistical quality, while the case $p>0$ is more challenging (for reasons that we also explain).
Finally, we will use synthetic data to study spectral functions with edge divergencies followed by arbitrary continua, using combinations
of the parametrizations depicted in Figs.~\ref{fig:spec}(b) and \ref{fig:spec}(e). The relative weight of the two groups of $\delta$-functions
then has to be optimized, along with the exponent $p$ if need be.

We stress here that our method for optimizing constraints [Fig.~\ref{fig:optim}] is purely based on $\chi^2$ minimization at fixed $\Theta$, motivated by
analogies with statistical mechanics and considering the
obvious effects of over-constraining. In principle, Bayesian inference could also be used, in a way similar to
the treatment of the sampling temperature in Ref.~\cite{fuchs10}. However, then a prior probability for the optimized parameter ($p$ in Fig.~\ref{fig:optim})
also has to be postulated, and the procedures become much more complicated computationally. We will show numerous examples of success with the much the simpler
maximum-likelihood approach.

Naturally, an applied constraint must be suitable for the spectrum sought, and a clearly unsutable constraint will be reflected in the inability to
reach an acceptable goodness-of-fit. For a satisfactory result, the constrained spectrum must not have additional strong entropic pressures
that push the sampled configurations away from the correct solution. To the extent additional features
of the spectrum are known (or suspected), further constraints can be applied in principle to improve the results. In the examples to be
reported here, we will optimize up to two different parameters regulating constraints.

\section{SAC with unrestricted sampling}
\label{sec:plainsampling}
 
Here we discuss simple Monte Carlo algorithms for spectral function with the fixed-grid and continuous frequency parametrizations illustrated
in Figs.~\ref{fig:spec}(a), \ref{fig:spec}(b), and \ref{fig:spec}(c). The fixed-grid case is considered in Sec.~\ref{sec:gridsamp}, and in  
Sec.~\ref{sec:contsamp} we discuss continuous frequencies without and including amplitude updates. We focus on the principles of
unrestricted sampling and defer results and incorporation of constraints to the later sections.

\subsection{Fixed-grid spectrum}
\label{sec:gridsamp}

At first sight, a fixed frequency grid, with a number $N_\omega$ $\delta$-functions at frequencies $\omega_i=i\Delta_\omega$, $i=1,\ldots,N_\omega$,
and variable amplitudes $A_i$ corresponding to Eq.~(\ref{barelation}) is perhaps the most natural parametrization of the spectrum. It was indeed used
in many of the previous works on the SAC method \cite{white91,sandvik98,vafay07,fuchs10,syljuasen08,sandvik16}. Non-uniform frequency grids have also
been considered \cite{ghanem20a,ghanem20b}.

While importance sampling of the amplitudes may seem like a trivial task formally,
in practice simple moves of weight among the amplitudes $A_i$ can evolve the spectrum only slowly, and collecting smooth averages is time
consuming. When the quality of the imaginary-time data is good, say, with the error level $\sigma$ (defined in Sec.~\ref{sec:qmcdata})
of order $10^{-6} \sim 10^{-5}$ or less, only very small changes will be accepted at a good rate. In Ref.~\cite{sandvik98} multi-amplitude moment
conserving updates were developed that help considerably in this regard, but still a smooth spectrum may require sampling from several minutes to many
hours, depending on the number of frequencies used, the sampling temperature $\Theta$, and the error level of the imaginary-time data (with better data
implying less efficient sampling, thus requiring longer runs). The nature of the spectrum sought will implicitly affect the sampling efficiency as well.

Since a fixed normalization is used here with the kernel in Eq.~(\ref{kernel2}), the sum of the amplitudes is conserved, $\sum_i A_i=1$, and the simplest Monte
Carlo move corresponds to a random re-distribution of the sum $A_i+A_j$ of the amplitudes at two chosen grid points $i$ and $j$. The move is accepted or
rejected using the standard Metropolis algorithm with the weight function in Eq.~(\ref{psg}). The change in the weight, which is needed for
the acceptance probability, involves the changes in $G(\tau_k)$ at all time points $\tau_k$, and the new value of $\chi^2$ is computed according to 
Eq.~(\ref{chi2eps}) in the basis where the covariance matrix is diagonal. The change in $G(\tau_k)$ only depends on the affected amplitudes $A_i$ and $A_j$
and can easily be obtained from Eq.~(\ref{eir}) with the kernel pre-computed.

In updates involving a set of $n$ amplitudes $\{A_i\}$, $n-1$ moments of the spectrum can be conserved. Unless some moments are exactly known (e.g., because
of some sum rule) or approximately known to much higher precision than what is implicit in the imaginary-time data, the purpose here is not for the
sampling to actually conserve any moments (beyond the fixed normalization), but to speed up the sampling. The idea \cite{sandvik98} is that an update
conserving several moments may reduce the changes in $G(\tau)$ while still involving significant spectral weight re-distribution, thus improving the acceptance
rate.

It is in principle easy to compute the line in an $n$-dimensional hypercube corresponding to conserving $n-1$ moments, but ensuring positive
definiteness complicates the expressions. In the tests here, we only carried out simultaneous updates of two or three amplitudes, conserving only the normalization 
in the first case and also the first frequency moment in the second case.

In moves of weight between two amplitudes, we use two ways of selecting the grid points $i$ and $j$: (a) with $i$ chosen at random among
$i \in \{1,\ldots N_\omega-1\}$ and $j=i+1$ or (b) with both $i$ and $j \not=i$ chosen at random among all possibilities. We redistribute the total weight $A_i+A_j$
within a window of the allowed values, i.e, the updated amplitudes are
\begin{equation}
A'_i=A_i+\delta_A,~~~A'_j=A_j-\delta_A,
\label{aprimedelta}
\end{equation}
with $\delta_A$ uniformly generated with
zero mean within a window adjusted so that the acceptance rate is close to $0.5$. Moves violating the positivity constraint are rejected, while those
satisfying the constraint are accepted using the standard Metropolis probability with the weight function in Eq.~(\ref{psg}).
In principle, the frequency window for the attempted shifts $\delta_A$ can
depend on the frequencies but we here use the same window for all $i,j$. It is useful to also attempt some updates with the full allowed range
$\delta_A \in [-A_i,A_j]$ in Eq.~(\ref{aprimedelta}), even when the acceptance rate is low.

When simultaneously updating three amplitudes $A_i$, $A_j$, $A_k$, the indices $i,j,k$ are here assumed to be ordered so that $\omega_i < \omega_j < \omega_k$
(which is necessary for some of the formulas below to be correct). We again choose either consecutive grid points or randomly select them from the set of
$N_\omega$ points. The conserved quantities are
\begin{subequations}
\begin{eqnarray}
m_0 & = & A_i+A_j+A_k \\
m_1 & = & A_i\omega_i+A_j\omega_j+A_k\omega_k,
\end{eqnarray}
\end{subequations}
and the procedure for uniformly sampling among all possible new amplitudes $A'_i$, $A'_j$, $A_k'$ is as follows: First, for positive definiteness of all three updated
amplitudes, $A_i'$ is allowed ony within the window $[a_i,b_i]$, where
\begin{subequations}
\begin{eqnarray}
a_i & = & {\rm max}\left [0,\frac{m_0\omega_j-m_1}{\omega_j-\omega_i}\right ],\\
b_i & = & {\rm min}\left [m_0,\frac{m_0\omega_k-m_1}{\omega_k-\omega_i}\right ].
\end{eqnarray}
\end{subequations}
The new amplitudes are then constructed in order $i,j,k$ according to
\begin{subequations}
\begin{eqnarray}
A'_i & \in & [a_i,b_i],\\
A'_j & = & [m_0\omega _k-m_1-A'_i(\omega_k-\omega_i)]/(\omega_k-\omega_j),~~\\
A'_k & = & m_0 - A'_i - A'_j.
\end{eqnarray}
\label{amp3updates}
\end{subequations}
The choice of $A_i'$ in the allowed window $[a_i,b_i]$ is made at random and the expressions for $A_j'$ and $A'_k$ ensure conservation of $m_0$
and $m_1$. If the acceptance rate is low, it can be improved by choosing $A'_i$ in a smaller window centered at the present value $A_i$, instead of selecting
the new value from the full allowed range. As in the $n=2$ case, moves within a fixed symmetric window around $A_i$ may
in some cases generate negative values, which are rejected immediately. The initial convergence of the spectrum, in particular, benefits from the
full-range moves, which can be accepted at a good rate before the sampling has fully equilibrated.

The above $n=2$ and $n=3$ amplitude updates are sufficient for the simple tests presented here. The formal scaling of the computational effort of
one updating sweep, each with $\propto N_\omega$ attempts with $n=2$ and $n=3$, is linear in both $N_\tau$ and $N_\omega$. The overall sampling efficiency
also depends in a less quantifiable way on the  frequency spacing $\Delta_\omega$ (thus on $N_\omega)$, as larger relative weight can be transferred
between the $\delta$-functions at a good acceptance rate if $\Delta_\omega$ is smaller, but the individual amplitude fluctuations are then also larger.
The smoothness of the average spectrum can be improved post-sampling by averaging the mean amplitudes over several of the native
grid points of the sampled spectrum.

In Ref.~\cite{ghanem20a} an improved sampling algorithm for grid-based spectra was constructed, with a singular-value decomposition used to transform
the kernel to the principal axes of $\chi^2$. The different frequency modes still cannot be sampled independently, as they are coupled through the
condition of positive definiteness. A scheme of sampling in different ``blocked'' bases was developed and appears to be much more efficient than
the above simple updates.

In our further developments and practical applications of SAC, we do not use the fixed grid but prefer to sample $\delta$-functions
in continuous frequency space, for the entropic reasons that were briefly mentioned in Sec.~\ref{sec:parametr} and which will become apparent in our examples.
The Monte Carlo sampling in the continuous representations, Figs.~\ref{fig:spec}(b) and \ref{fig:spec}(c), are also simple and in practice much more
efficient than sampling amplitudes on the fixed grid (without the basis transformation used in Ref.~\cite{ghanem20a}, which we have not implemented).
In Sec.~\ref{sec:maxent} we will use the fixed-grid updates described here for stochastic optimization of spectra with the ME method,
where the inefficiencies are less severe.

\subsection{Continuous frequency space}
\label{sec:contsamp}

With the continuous frequency space, the spectrum is the amplitude weighted mean density of a number $N_\omega$ of $\delta$-functions whose frequencies 
are sampled. The amplitudes $A_i$ can be kept fixed, as in Fig.~\ref{fig:spec}(b), or importance sampled along with the frequencies as in as in
Fig.~\ref{fig:spec}(c). The $\omega$ space can be approximated with a very fine ``micro grid'' of many (e.g., millions of) equally spaced frequencies
$\omega_j$ up to a cut-off. The much smaller number $N_\omega$ of actual $\delta$-functions (typically thousands, in some cases more than
$10^5$) occupy only a small fraction of the grid. The kernel can be precomputed on the micro grid points, and the integer-valued
grid indices $j$ can be used in the sampling procedure and easily converted to the actual frequencies $\omega_j$ when needed (which is only at the stage
of final output of the results).

Alternatively, floating-point numbers can be used for the frequencies, but then the kernel, which involves time consuming
exponential functions, cannot be precomputed exactly. More importantly, the transformation in Eq.~(\ref{basistransf}) to the eigenbasis of the covariance
matrix would have to be performed on the fly with the current frequencies $\{\omega_i\}$, or else $\chi^2$ has to be computed in the original
basis using the double-sum over $N_\tau$ in Eq.~(\ref{chi2}), instead of performing the single sum, Eq.~(\ref{chi2eps}), in the transformed basis.
In either case, the the computational effort will then scale as $N_\tau^2$ instead of $N_\tau$ with the discrete method.

To achieve the best of both worlds, realizing the frequency continuum in practice, we still use a grid of points on which the kernel and its
frequency derivative are precomputed. The kernel can then be evaluated rapidly for the continuous (double-precision floating-point) frequencies $\omega_i$
by interpolation using the two closest grid points, corrected with the aid of the stored derivatives. With a fine grid, a practically exact kernel can be
evaluated faster in this manner than with its mathematically exact expression. Both the kernel and its derivatives are transformed to the eigenbasis
of the covariance matrix, and the $O(N_\tau)$ scaling is maintained.

\subsubsection{Sampling frequencies}

In the simplest case, all the $\delta$-functions have the same amplitude, $A_i=1/N_\omega$. At a given instant, the frequencies of the $\delta$-functions
are $\omega_i$, $i \in \{1,\ldots,N_\omega\}$, and these frequencies are integer multiples of a micro grid spacing $\delta_\omega$ (typically we use
$10^{-5}$ or $10^{-6}$). In most cases, we have not observed any differences between the discrete and full floating-point representations, which is
natural if the spectrum is featureless on the micro grid scale $\delta_\omega$. When the spectrum has fine structure in the form of a very sharp edge and
we use the monotonicity constraint in Fig.~\ref{fig:spec}(e) (sampling of which is discussed in Sec.~\ref{sec:hbergedge}), the discrete representation
can introduce anomalies, however, and it is then better to use floating-point frequencies. The results in this section and Secs.~\ref{sec:deltapeak} and
\ref{sec:contedge1} were obtained with $\delta_\omega=10^{-5}$, which represent the continuum sufficiently well according to our tests.
The double-precision floating point representation was used in Sec.~\ref{sec:contedge2}.

In a single-frequency update, an index $i \in \{1,\ldots,N_\omega\}$ is selected at random and the corresponding frequency is changed, $\omega_i \to \omega_i + R$, 
by a random amount $R \in [-\epsilon,\epsilon]$ with $\epsilon$  adjusted to give an acceptance rate of approximately $0.5$ (in the discrete representation with
the index itself, $i \to i + I$ with an integer $I \not=0$ similarly drawn within a suitably adjusted window---in the following we use only the frequency
representation for clarity). The calculation of the weight ratio for the Metropolis acceptance probability again requires re-computation of the contributions to
all $G(\tau_i)$ from the moved frequency $\omega_i$, demanding an $O(N_\tau)$ computational effort for a single update and $O(N_\tau N_\omega)$ for an entire sweep of
$\propto N_\omega$ updates.

With two frequencies updated simultaneously, indices $i \not= j$ are randomly selected and the corresponding
frequencies are changed as
\begin{equation}
\omega'_i = \omega_i + R,~~~ \omega'_j = \omega_j - R,
\label{omegaupdate}
\end{equation}
with a random displacement $R$ with zero mean in a window adjusted for good acceptance rate. This update of course conserves the first frequency moment
since we here consider equal amplitudes. In most of the cases we have considered here, updating one and two frequencies suffices for effective sampling
when $N_\omega$ is large. In some cases, there are signs of poor ergodicity; specifically when sampling narrow quasi-particle peaks with the method
presented in Sec.~\ref{sec:deltanp}. Simultaneous updates of three frequencies resolves this issue and can speed up the sampling also in other cases.

To conserve the first moment when three frequencies are changed, we generalize the two-frequency update Eq.~(\ref{omegaupdate}) to
\begin{equation}
\omega'_i = \omega_i + 2R,~~~ \omega'_j = \omega_j - R,~~~ \omega'_k = \omega_k - R,
\label{omegaupdate2}
\end{equation}
and then take $R$ such that also the second moment is conserved;
\begin{equation}
R = \frac{1}{3}(\omega_k + \omega_j - 2\omega_i).
\label{romegaupdate2}
\end{equation}  
Here it should be noted that $R$ may not correspond exactly to an integer number of the units $\delta_\omega$ of the frequency grid, but, the spacing being
very small, the second moment is still conserved to a high degree. As long as the rounding to a the nearest higher or lower grid point is applied
consistently, so that an update with a rounded $R$ has a reverse update with exactly $-R$, detailed balance is not affected. Of course Eq.~(\ref{omegaupdate2})
is not the only way to conserve the first moment, but it is a simple and efficient way in combination with Eq.~(\ref{romegaupdate2}). 

A Monte Carlo updating sweep consists of a sequence of $N_\omega$ single-frequency moves followed by $N_\omega/2$ two-frequency updates and, optionally
$N_\omega/3$ three-frequency moves. After each sweep, the
amplitudes are added to the histogram in which the spectral density is accumulated. During the sampling stage, in the discrete representation the working
histogram is one with exactly the same frequency spacing $\delta_\omega$ as in the sampling space. At the output stage, the small bins are combined into 
much larger bins, of size $\Delta_\omega$ small enough so that all features of the spectrum can be well represented while large enough for sampling noise
ro be averaged out sufficiently. In the tests reported here we used $\Delta_\omega=0.01$ or $\Delta_\omega=0.005$ in most cases.

In a modification of the fixed-amplitude scheme, the amplitudes $A_i$ are initialized with some dispersion, e.g, $A_i = c + di$ with $c$ and $d$ 
constants such that the normalization $\sum_i A_i = 1$ is obeyed. These amplitudes are kept constant during the sampling process and, because there 
are no constraints on the locations of the $\delta$-functions, smaller amplitudes can migrate to areas with lower spectral weight. The two-frequency
move of the form Eq.~(\ref{omegaupdate}) does not conserve the first moment if the two amplitudes are different, but we can still use it (optionally
in combination with amplitude swaps).

As long as $N_\omega$ is sufficiently large and the dispersion of $A_i$ is not too large, we find no significant differences in the final averaged spectrum between
the methods of uniform or varying (but not updated) amplitudes. A more significant dispersion leads to increased fluctuations and changes in the entropic
pressures of the method and slightly different averaged spectra. The sampling efficiency can be better in a parametrization with varying amplitudes if swaps
of two amplitudes are included, $A'_i = A_j$, $A'_j = A_i$, with $i$ and $j$ chosen randomly among all the $\delta$-functions. When sampling spectra with two
or more peaks with little weight between them, this update leads to considerable improvements in the efficiency of transferring weight between the peaks
(especially when the spectrum is not yet equilibrated), as was found in Ref.~\cite{qin17}. Here we will use equal amplitudes when only the
fequencies are sampled.

\subsubsection{Sampling amplitudes and frequencies}

If also the amplitudes are sampled, i.e., with the parametrization in Fig.~\ref{fig:spec}(c) (discussed in the previous literature, e.g., in
Refs.~\cite{beach04,fuchs10,ghanem20b}), we supplement the frequency moves with the updates described in Sec.~\ref{sec:gridsamp} for the case of amplitudes
on the fixed grid. For updating two amplitudes, with the corresponding frequencies unchanged, it is clear that Eq.~(\ref{aprimedelta}) works. The algorithm
for changing three amplitudes is also the same as before, as long as the frequencies $\omega_i,\omega_j,\omega_k$ are labeled in increasing order in
Eq.~(\ref{amp3updates}). 

With the individual amplitudes not conserved, we can also modify the two-frequency update to conserve the first frequency moment. After the proposed move
of the frequencies in Eq.~(\ref{omegaupdate}), we change the amplitudes with the normalization conserved as in Eq.~(\ref{aprimedelta}), but now with the
unique $\delta_A$ that conserves also the first moment;
\begin{equation}
\delta_A = \frac{R(A_i-A_j)}{2R + \omega_i - \omega_j}.
\label{daconserving}
\end{equation}
If this $\delta_A$ leads to negative $A'_i$ or $A'_j$ the move is rejected, otherwise it is accepted with the Metropolis probability. We mix these
moves with updates of only the frequencies as in Eq.~(\ref{omegaupdate}), with the windows for selecting $R$ adapted individually in the two cases
for good acceptance rates.

In multi-frequency updates with $n>2$ $\delta$-functions, which we have constructed explicitly only for $n=3$ in Eqs.~(\ref{omegaupdate2}) and
(\ref{romegaupdate2}), the corresponding amplitudes can also be updated along with the frequencies to ensure conservation of moments. Then $n$ new frequencies
are first according to Eq.~(\ref{omegaupdate2}) with $R$ in Eq.~(\ref{romegaupdate2}) (or generalized formulas for larger $n$), after which conservation of
$n-1$ frequency moments is ensured by changing the corresponding amplitudes according to a more complicated generalization of Eq.~(\ref{daconserving}).
In practice, the $n\le 3$ updates we have presented above allow rather effective sampling already and we do not go to higher $n$.

In general, when the amplitudes are also sampled the number $N_\omega$ of $\delta$-functions can be smaller than when only frequency moves
are performed, as some amplitudes can always become very small and reach far into thin spectral tails.
However, the sampling efficiency appears to be generally worse, likely because some amplitudes occasionally become large and
more difficult to move. Not performing amplitude updates also of course saves time. The amplitude updates are helpful when sampling at very low $\Theta$,
to obtain a good approximation for the best goodness of fit $\chi^2_{\rm min}$ (which we need for setting the optimal sampling temperature, as discussed
below in Sec.~\ref{sec:theta}). Sufficiently good values can also be easily obtained with only frequency updates if $N_\omega$ is large enough.

Whether or not to include amplitude moves is not in the end only a matter of sampling efficiency, since the entropy is also affected. We will demonstrate this
fundamental difference between the two parametrizations both in practice, in the test cases presented below in Sec.~\ref{sec:theta}, and formally
when comparing the functional forms of the entropies in Sec.~\ref{sec:entropy}. We will also argue that better results (in particular, better resolved
peak widths) are in general obtained when amplitude moves are included. 

If the averaged spectrum sampled in the absence of QMC data is regarded as a default model, then the continuum representations considered here correspond
to an infinitely stretched out spectrum when the $\delta$-functions can migrate without bounds (at least in principle). In contrast, the fixed-grid spectrum
always has an upper bound, and then the default average spectrum is flat with the amplitude equal to the inverse of the frequency range. In practice, the formal
difference in the frequency bound on the grid and in the continuum should not be very important; once the sampling is restricted by the imaginary-time data,
$\chi^2$ imposes a similar de facto upper bound in both cases---under the assumption that $\Theta$ is correctly chosen (as explained in the next section)
in each case.

It is worth noting that a frequency cut-off has to be imposed in the computer program only for the purpose of precomputing the kernel and for storing
the histogram. If the cut-off is high enough, it does not imply any restriction on the migration of the $\delta$-functions. The high-frequency
tail is in practice dictated by the QMC data, as a $\delta$-function moving up very high in frequency will ruin $\chi^2$ and not be accepted. One might then
worry that increasing $N_\omega$ (i.e., reducing the individual amplitudes) would affect the tail, as the $\chi^2$ cost of migrating high up diminishes,
especially in the fixed-amplitude case where the unit of spectral weigh is $1/N_\omega$. However, in practice we have not noticed any issues with the tail
of the spectrum even for very large $N_\omega$ (more than $10^5$ in some tests) when the sampling temperature $\Theta$ is fixed according to the scheme discussed
below in Sec.~\ref{sec:theta}. There is also no fundamental reason to expect any problems when $N_\omega \to \infty$, because of the exact mapping
of the SAC to the ME method in this limit (Sec.~\ref{sec:maxent}). We will show examples of the effect of varying $N_\omega$ in Sec.~\ref{sec:nomega}.

\section{Optimal sampling temperature}
\label{sec:theta}

An important aspect of SAC is how to select the sampling temperature $\Theta$. The general situation \cite{sandvik98,beach04} prevailing with unrestricted
sampling is that the spectrum at low $\Theta$ freezes into a stable or metastable $\chi^2$ minimum. In the $\Theta \to 0$ limit a few sharp peakst,
for reasons that will be mentioned below and further elucidated in \ref{app:lowtheta}. High $\Theta$
values lead to smooth featureless spectra with large $\langle \chi^2\rangle$. There is a range of $\Theta$ over which $\langle \chi^2\rangle$ is small,
close to its minimum value $\chi^2_{\rm min}$, but the fluctuations are still significant and smoothen the average spectrum. It is not possible to reach
$\chi^2=0$, even in principle, when positive definiteness is imposed; again see \ref{app:lowtheta}, and also \ref{app:low2}, where we include a small fraction of negative spectral weight.

There is still no consensus on exactly how $\Theta$ should best be chosen, but overall the different schemes proposed in the literature produce rather
similar results in most cases. The $\Theta$-fixing issue is similar to the various ways in which the entropic weighting parameter $\alpha$
of the spectrum can be chosen in the ME method \cite{jarrell96,bergeron16}. Some criteria proposed for optimal $\Theta$ were
inspired  by the physics analogy of a phase transition between ``data fitting'' and ``noise fitting'' \cite{sandvik98,beach04}. However, there is 
no rigorous motivation for such physics inspired criteria, and statistically acceptable $\langle \chi^2\rangle$ values cannot be guaranteed.
In this section we discuss our alternative optimal-$\Theta$ criterion, show examples of its application, and also discuss related issues that further motivate
and support the criterion. We begin by a brief outline of the subsections to follow.

The criterion we advocate for the optimal value of $\Theta$ involves a balance between entropy and goodness-of-fit, formulated in a simple way motivated
by the properties of the $\chi^2$ distribution, namely, $\Theta$ is adjusted so that
\begin{equation}
\langle \chi^2(\Theta) \rangle \approx \chi^2_{\rm min} + a\sqrt{2\chi^2_{\rm min}},
\label{eq:chi2}
\end{equation}
where $\chi^2_{\rm min}$ is the minimum value of $\chi^2$, which can be reached in a simulated annealing \cite{kirkpatrick83}
process to very low $\Theta$, and $a$ is of order $1$. This criterion was first applied in Ref.~\cite{qin17} (though it was stated in a
slightly different way) and will be further motivated here in Sec.~\ref{sec:thetacrit}. In the later subsections we present various results
illustrating the applicability of the criterion when used with the different parametrizations of the spectrum.

We stress already here that finding $\chi_{\rm min}$ to sufficient precision is not difficult with the sampling methods
we have developed in continuous frequency space [Sec.~\ref{sec:contsamp}]; in typical cases annealing runs of a few minutes suffice, and we have
not encountered any severe problems with getting stuck in local minima. In principle, conventional grid-based non-negative least squares fitting methods
could also be used for this purpose \cite{bergeron16,Koch18,Ghanemthesis}. Given that simulated annealing works well and uses the same programming elements
as the subsequent main sampling runs, we do not incorporate any other optimization methods here.

Before providing a detailed motivation for Eq.~(\ref{eq:chi2}), we stress that it is a criterion based purely on the goodness of the fit and the
properties of the $\chi^2$ distribution. It is also further justified by the existence of an effective number $N_{\rm para}$ of parameters that the
positive definite spectrum can provide to fit the noisy $\bar G(\tau)$ data. We define $N_{\rm para}$ below and demonstrate its origin in more detail
in \ref{app:lowtheta}. Bayesian inference arguments can also be used \cite{fuchs10}, but then a postulated prior $\Theta$-distribution 
is also required. We here stay away from such ambiguous input, instead favoring more concrete statistical arguments. 

Many of the SAC results in this section will be based on SSE-QMC data for the antiferromagnetic Heisenberg spin-$1/2$ chain, defined by the Hamiltonian
\begin{equation}
H = J \sum_{i=1}^L {\bf S}_i \cdot {\bf S}_{i+1},
\label{heisenberg}
\end{equation}
with periodic boundary conditions and $J=1$. In the first test, in Sec.~\ref{sec:example1} we study the same $16$-spin system that was used in 
Ref.~\cite{sandvik98}, where SAC results for the dynamic structure factor were compared with exact diagonalization results at
rather high temperatures. For the second example, in Sec.~\ref{sec:example2} we consider a Heisenberg chain with $L=500$ spins at inverse temperature
$\beta=500$, which effectively corresponds to $T=0$ for the structure factor computed. The results will be compared with a  numerical BA calculation
for the same $L$ \cite{caux05a,cauxdata}.

In Sec.~\ref{sec:example3} we consider a case of a synthetic spectrum with non-trivial shape to further illustrate the dependence of the average
spectrum on the parametrization used. We also discuss the convergence toward the correct profile with decreasing error level $\sigma$ of the
imaginary-time data. We study $\sigma$ as low $\sigma=10^{-8}$, which would in most cases be unrealistic in QMC simulations but may be of interest
in potential applications of SAC to problems where there are no statistical errors in the imaginary-time data. We also discuss the convergence of the
spectrum with increasing $N_\omega$ at fixed error level.

In Sec.~\ref{sec:nomega} we will show how spectra sampled in continuous frequency space evolve with $N_\omega$ when using the Bayesian sampling
temperature, $\Theta=1$. The main question here is whether an optimal $N_\omega$ can be identified, as a potential alternative to optimizing $\Theta$ for
arbitrary (normally large) $N_\omega$. We conclude that there is no obvious practical way to choose an optimal $N_\omega$, and it is better to aim for the
$N_\omega \to \infty$ limit by using sufficiently large $N_\omega$ with $\Theta$ appropriately chosen. We already now point out that
Eq.~(\ref{eq:chi2}) implies $\Theta \propto 1/N_\omega$, which will be demonstrated formally in Sec.~\ref{sec:entropy}.

We will not show any statistical errors on the spectral functions presented in the tests in this section or elsewhere in the paper. In fact, in
Sec.~\ref{sec:maxent} (and further in \ref{app:fluct} and \ref{app:statmech}) we argue that there are no fluctuations in the average spectrum
in the limit $N_\omega \to \infty$, and that the fluctuations for finite $N_\omega$ also do not correspond to meaningful statistical errors reflecting the errors
in $\bar G(\tau)$. We will also further discuss the issue of statistical errors in Sec.~\ref{sec:errors}. For now, we note that the statistical errors related to
typical noise levels of $\bar G(\tau)$, preferably obtained using bootstrapping, are very small compared to the differences between spectra sampled with 
different parametrizations.

\subsection{Optimal-$\Theta$ criterion}
\label{sec:thetacrit}

The imposed ``thermal noise'' level implied by Eq.~(\ref{eq:chi2}) is in spirit similar to the cut-off in Tikhonov regularization \cite{tikhonov63}, which has
recently been discussed in the context of analytic continuation of QMC data by singular-value decomposition \cite{Ghanemthesis}. In practice, the criterion is
implemented in SAC as follows: A simulated annealing procedure is first carried out to find $\chi^2_{\rm min}$. The value only needs to be reasonably close to the true
minimum [considering the fact that the factor $a \lesssim 1$ in Eq.~(\ref{eq:chi2}) should not have to be specified exactly], which is difficult to reach exactly
but which can be rather easily approached to the degree needed here. After this initial step, a second annealing procedure is carried out where $\Theta$ is
lowered until the mean value $\langle \chi^2\rangle$ satisfies Eq.~(\ref{eq:chi2}), with $a \approx 0.5$ typically. Alternatively, the $\Theta$ criterion can be
applied with saved data from the first stage, but we often do the second run at a slower annealing rate in order to obtain a more precise
$\langle \chi^2(\Theta)\rangle$ form (i.e., with smaller statistical errors). With a single slow annealing run, the average spectra accumulated at all
$\Theta$ values can also be saved for later examination, e.g., to study in detail the sensitivity of the spectrum to the value of $\Theta$.

In some cases, e.g., when the number of $\delta$-functions is small or when constraints are imposed, the optimum according to Eq.~(\ref{eq:chi2}) may be for
$\Theta > 1$ (though never much above $1$ in our experience). In such cases it could be argued that $\Theta=1$ should be used, since there would be no reason
to introduce more fluctuations than with the original probability from Bayes' theorem. However, the need to use $\Theta < 1$ when $N_\omega$ is large
already demonstrates the failure of the application of Bayes' theorem in this context (as further elaborate in Sec.~\ref{sec:pathology}), and then it is also
legitimate to use $\Theta > 1$ when $\Theta=1$ leads to an overly constrained spectrum.

To further motivate the way in which $\langle \chi^2\rangle$ is fixed above the minimum value $\chi^2_{\rm min}$ in Eq.~(\ref{eq:chi2}), we note that the mean $E(\chi^2)$
and variance $V(\chi^2)$ of the $\chi^2$ distribution are
\begin{equation}
E(\chi^2) = N_{\rm dof},~~~~V(\chi^2) = 2N_{\rm dof},
\label{evchi2}
\end{equation}
which applies to a best-fit procedure (with an applicable fitting function) where the number of independent degrees of freedom is
\begin{equation}
N_{\rm dof} = N_{\rm data} - N_{\rm para},
\label{ndofdef}
\end{equation}
where $N_{\rm data}$ is the number of fitted data points and $N_{\rm para}$ the number of optimized parameters in the fitting function. In the case of
SAC, $N_{\rm data}=N_\tau$, but the number of sampled frequencies or amplitudes (or both) cannot be regarded as $N_{\rm para}$. First, since
typically $N_\omega \gg N_\tau$, formally $N_{\rm dof}=N_\tau-N_\omega$ would be negative and nonsensical. Second, with positive definite amplitudes,
when more $\delta$-functions (or other parameters defining the positive definite spectrum) are added, at some point $\chi^2 \to \chi^2_{\rm min}>0$
and the value can no longer be reduced. Thus, while a larger number of $\delta$-functions implies a more flexible spectrum, these cannot all
independently contribute to minimizing $\chi^2$---instead, one can regard them as collectively forming an effective (rather small) number of
parameters. Indeed, when minimizing $\chi^2$ by, e.g., simulated annealing, the optimal spectrum is essentially the same as that of data-fitting with
some small number (2-6 typically) of $\delta$-functions (and we argue in \ref{app:lowtheta} that no better positive definite fitting function
exists), only with some small broadening of the peaks or small spurious peaks, with both types of flaws diminishing with improved optimization.

A number $N_\delta$ of $\delta$-functions functions (or sharp peaks) corresponds to $N_{\rm para}=2N_\delta$ effective fitting parameters (frequencies
and amplitudes).
Further, according to Eq.~(\ref{evchi2}) we can regard $\chi^2_{\rm min}$ as a proxy for the effective number of degrees of freedom, and similarly
$2\chi^2_{\rm min}$ as a proxy for the variance of the $\chi^2$ distribution with this number of degrees of freedom. In practice, a well defined
number $N_\delta$ of sharp peaks appear only when $\chi^2$ is extremely close to its perfect minimum, and our method of fixing $\Theta$ only needs
a reasonably good approximation to $\chi^2_{\rm min}$, say to within $1\%$ of the optimal value (where the average spectrum may still look very ill-converged
unless the annealing process is extremely slow). We can then still consider $\chi^2_{\rm min} \approx N_{\rm dof}$. According to Eq.~(\ref{ndofdef}), the
effective number of parameters should then also be given by $N_{\rm para} \approx  N_\tau - \chi^2_{\rm min}$, which can be used as a statistical
test when comparing with $2N_\delta$ (after a sufficiently slow annealing); we expect
\begin{equation}
\chi^2_{\rm min} \approx N_{\tau} - 2N_{\delta},
\label{chi2expected}
\end{equation}
to be satisfied within a well defined uncertainty stemming from the variance of the $\chi^2$ distribution, which is sufficiently well approximated
by $2\chi^2_{\rm min}$. It is not necessary to always perform such tests, but we will present examples in \ref{app:lowtheta} when discussing
the limit $\Theta \to 0$ in detail.

As mentioned above, for typical QMC data quality $N_\delta$ is normally small ($2-6$), and if $N_\tau \gg 2N_\delta$ we should of course simply have
$\chi^2_{\rm min}/N_\tau \approx 1$. However, since $N_\delta$ is also itself related to the data quality and the data points are highly correlated, in practice
we have found that there is no gain in using very large $N_\tau$ (especially considering the increase in computational effort with $N_\tau$). In our tests,
we typically take $N_\tau$ in the range $10$ to $50$, and $\chi^2_{\rm min}/N_\tau$ then only very rarely exceeds $1$.

With $a=1$ in Eq.~(\ref{eq:chi2}) the sampled spectrum, on average, has $\chi^2$ deviating from $\chi^2_{\min}$ by one standard deviation.
This level of fluctuations should remove distortions arising from ``fitting to the noise'' (a concept that we also make more precise in the case
of SAC in \ref{app:lowtheta}). We will show below that the criterion Eq.~(\ref{eq:chi2}) with $a$ of order $1$ indeed produces excellent
spectra in tests both on synthetic imaginary-time data and on actual QMC results for systems with known spectral functions. We typically use $a=0.5$,
which further reduces the rare fluctuations of $\chi^2$ up to values corresponding to several standard deviations away from the mean. With $a=0.25$, we
some times have obtained spectra with clear signs of overfitting.

We will illustrate the optimal-$\Theta$ criterion here in SAC with both the fixed-grid and continuum parametrizations of the spectrum.
We often normalize the goodness of the fit by the number $N_\tau$ of $\tau$ points. Because $N_\tau$ is always larger than the effective number of
degrees of freedom, Eq.~(\ref{ndofdef}), the statistical expectation is that $\chi_{\rm min}^2/N_\tau < 1$. Here it is worth pointing out that the rarely
used ``historical'' variant of the ME method imposes $\chi^2/N_\tau = 1$ (in that case there is no averaging) as the criterion for determining the entropic
coefficient $\alpha$. As has been pointed out \cite{jarrell96}, it is not always possible, because of ``unlucky'' statistical fluctuations, to reach this
$\chi^2$ value, and, moreover, by arguments similar to those above, in general $\chi^2=N_\tau$ will be statistically too high
(though only marginally so). In principle, a failsafe criterion like Eq.~(\ref{eq:chi2}) can also be used with the ME method, but we
are not aware of any works using it so far. We will demonstrate its applicability with the ME method in Sec.~\ref{sec:maxent}.

\subsection{Example 1: 16-spin Heisenberg chain}
\label{sec:example1}

The exact spectrum $S(q,\omega)$ for a finite system is a sum of $\delta$-functions, Eq.~(\ref{somegasum}), and for only $16$ spins the number
of $\delta$-functions with significant weight is not very large (decreasing to just a few when $T \to 0$). No realistic numerical analytic
continuation method can resolve many individual $\delta$-functions, and considerable smoothing has to be imposed in order to compare
results. As in Ref.~\cite{sandvik98}, we here use a histogram with relatively large bins to represent the exact spectral weight
distribution of $S(q,\omega)$, in this case for $q=\pi/2$ at $T=J/2$.

\begin{figure}[t]
\centerline{\includegraphics[width=8.25cm, clip]{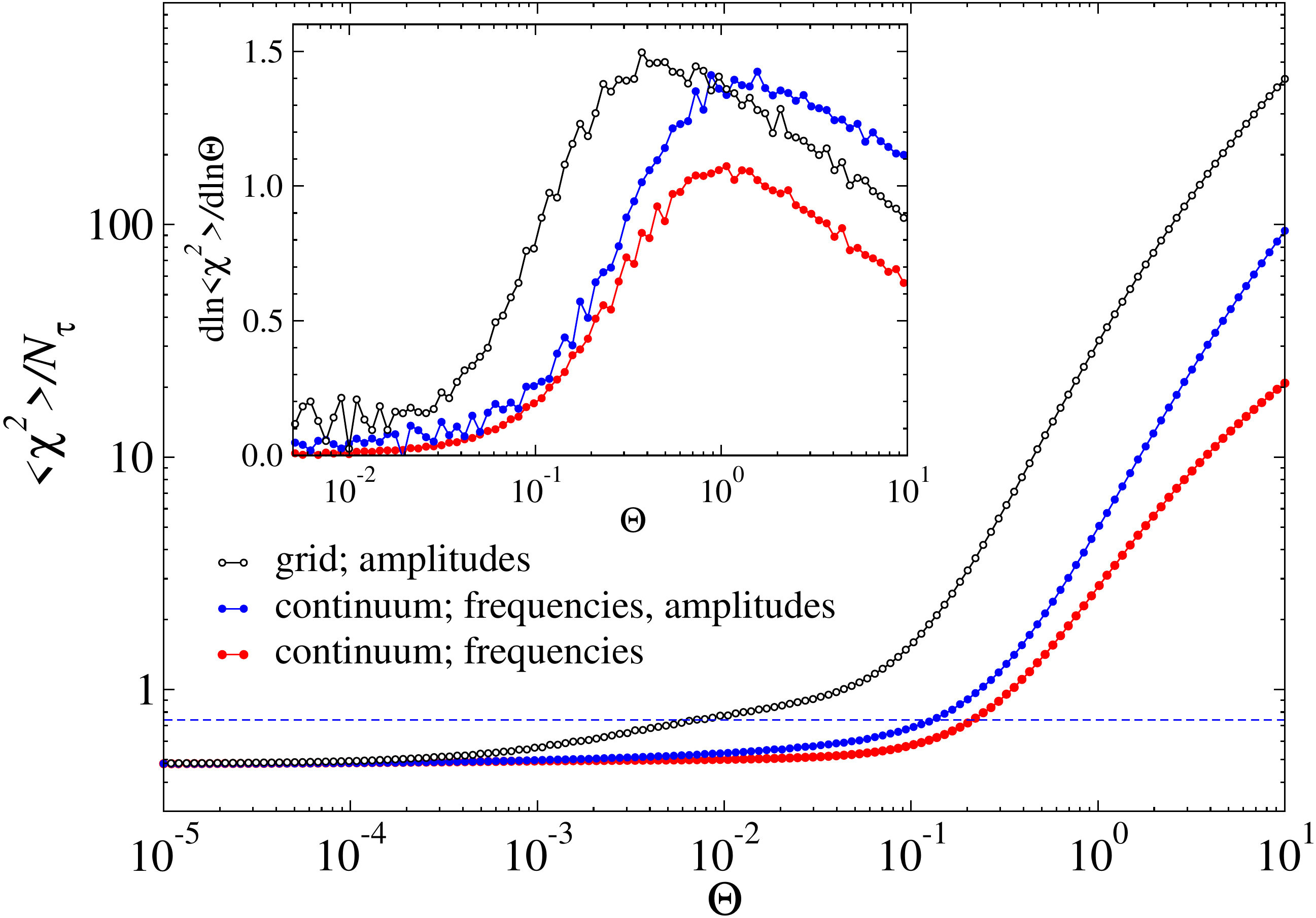}}
\vskip-1mm
\caption{Examples of the evolution of the goodness of the fit with the sampling temperature, shown here on a log-log plot for three
different parametrizations of the spectrum [Figs.~\ref{fig:spec}(a)--(c)]; on a fixed frequency grid (open black circles), in the continuum with both
frequencies and amplitudes sampled (blue solid circles), and in the continuum with fixed amplitudes $A_i=1/N_\omega$ and sampled frequencies (red
solid circles). The system is the $16$-site Heisenberg chain at inverse temperature $\beta=2$, with time spacing $\Delta_\tau=0.125$ in the
QMC data (i.e., $N_\tau=8$) and statistical error level $\sigma \approx 10^{-5}$. The dashed blue line corresponds to the criterion
for optimal sampling temperature, Eq.~(\ref{eq:chi2}) with the constant $a=0.5$. The inset shows the logarithmic derivatives extracted from
differences between data at $\Theta_i$ and $\Theta_{i+1}$.}
\label{fig:x2}
\end{figure}

Figure \ref{fig:x2} shows how $\langle \chi^2\rangle/N_\tau$ evolves with $\Theta$ in SAC simulated annealing runs. In the fixed-grid case, a frequency
spacing $\Delta_\omega=0.005$ was used for $\omega \in (0,5)$ and with the continuous-$\omega$ parameterizations $1000$ $\delta$-functions were sampled.
In all cases, the annealing processes started at $\Theta=10$ and the temperature was gradually lowered by dividing $\Theta$ by $1.1$ between each of the
points in Fig.~\ref{fig:x2}. For each $\Theta$, of the order of $10^6$ full Monte Carlo updating sweeps were carried out and all the accumulated spectra
were saved. With all three  parameterizations, $\langle \chi^2\rangle$ decreases monotonically with decreasing sampling temperature,
eventually converging to, for all practical purposes, the same minimum value, $\langle \chi^2\rangle/N_\tau \approx 0.479$, indeed well
below $1$ as expected. It can be noted that the three $\langle \chi^2(\Theta)\rangle$ curves differ considerably at higher $\Theta$, pointing to
different entropy contents of the spectra defined with these parameterizations.

\begin{figure*}
\centerline{\includegraphics[width=16cm, clip]{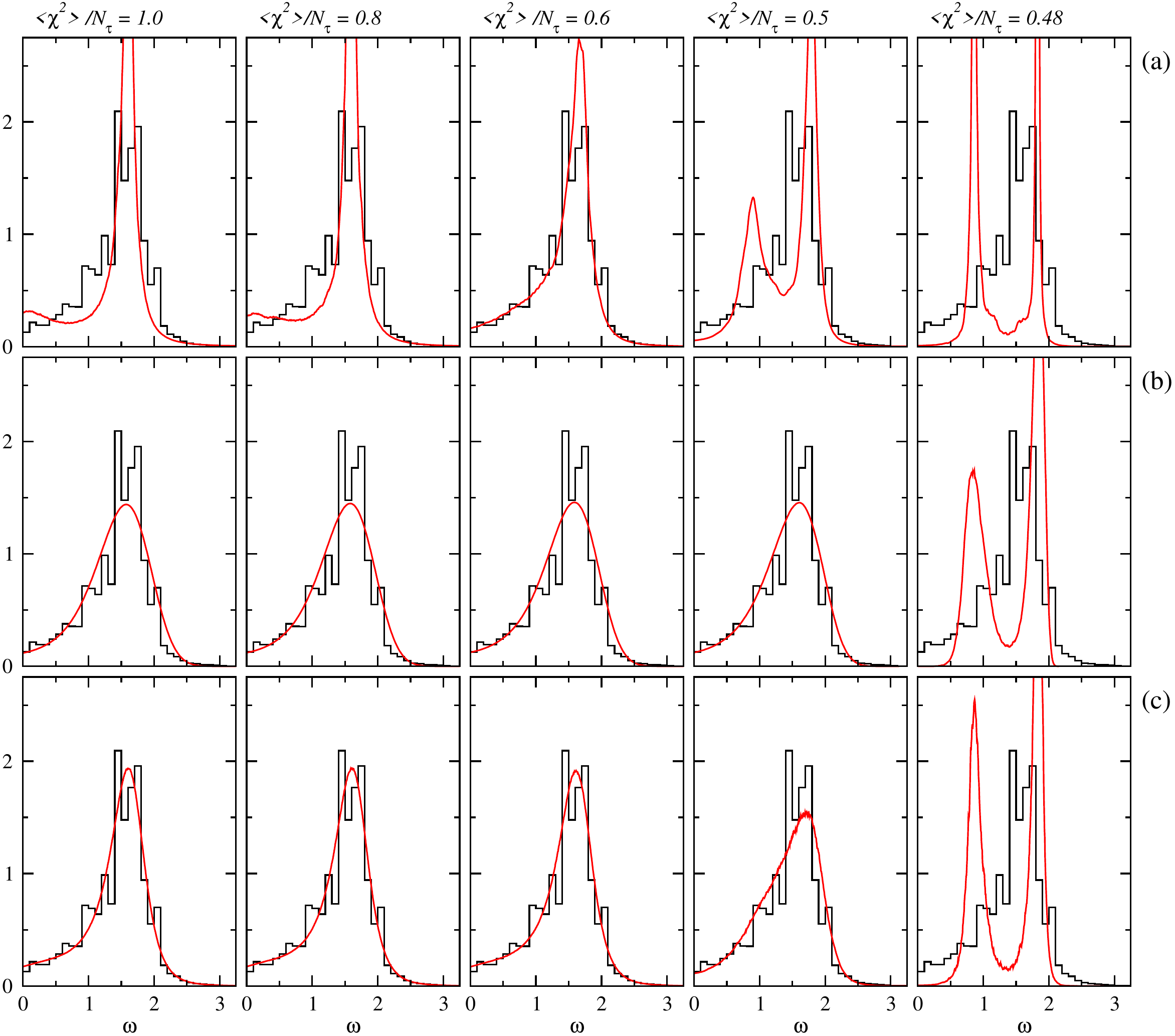}}
\vskip-1mm
\caption{Spectral functions (dynamic structure factor at $q=\pi/2$) corresponding to Fig.~\ref{fig:x2} for selected values of $\langle \chi^2\rangle/N_\tau$.
The three rows correspond to the three different $\delta$-function parameterizations illustrated in Fig.~\ref{fig:spec}(a)--(c); a fixed frequency grid with 
sampled amplitudes in (a), a continuum with only frequencies sampled (equal fixed  amplitudes) in (b), and with both frequencies and amplitudes sampled in (c). 
The columns correspond to different goodness-of-fit values, indicated on top of each column, which are realized at different $\Theta$ values according
to Fig.~\ref{fig:x2}. The histograms 
(shown with black lines) represent the spectral weight distribution calculated according to the exact form, Eq.~(\ref{somegasum}), for the same $16$-spin 
Heisenberg chain at the same temperature $T=1/2$.}
\label{sw1}
\vskip-2mm
\end{figure*}

The inset of Fig.~\ref{fig:x2} shows the $\ln(\Theta)$ derivative of $\ln \langle \chi^2(\Theta)/N\tau \rangle$, where we observe broad maxima in all
three cases. Such maxima have been likened to heat capacity peaks \cite{beach04}, and it was argued that the peak location is a good way to choose $\Theta$
(arguably representing the ``phase transition'' between data fitting and noise fitting). However, in Fig.~\ref{fig:x2} the peak locations correspond to rather
large $\langle \chi^2\rangle$ values, well above $\chi^2_{\rm min}$ and outside the range that would be considered acceptable based on the width of the
$\chi^2$ distribution. Therefore, the peak location should not, in general, be an optimal criterion, though it has often produced good results
\cite{feldner11,voll15,lohofer15,lohofer17,becker17,becker18,ying19,raczkowski20,sato21}. In contrast, adjusting $\Theta$ according to our criterion in
Eq.~(\ref{eq:chi2}), indicated by a dashed line corresponding to the factor $a=0.5$ in Fig.~\ref{fig:x2}, guarantees that the spectrum represents a
statistically good fit to $\bar G(\tau)$.

Figure \ref{sw1} shows examples of average spectra obtained at different sampling temperatures in the same runs as above, with
the columns labeled by the corresponding normalized $\langle \chi^2\rangle$ values, with fixed-grid and continuum-frequency results shown at matching
values, i.e., at different values of $\Theta$ according to Fig.~\ref{fig:x2}. With the $\delta$-functions in
continuous frequency space, the spectral weight density was accumulated in histograms of the same bin size, $\Delta_\omega=0.005$, as in the fixed-grid
spectrum. For comparison, each panel in Fig.~\ref{sw1} also shows exact diagonalization results represented by a histogram with bin width $\Delta_\omega=0.1$.
Even at the rather high (physical) temperature $T=J/2$ used here, the histogram has a jagged structure that one should not expect
to reproduce (and of course the true spectral function is even less smooth, consisting of some hundreds of $\delta$-functions). A realistic hope
would be to reproduce the overall shape of the frequency distribution without the fine details.

For given $\langle \chi^2\rangle/N_\tau$ value in Fig.~\ref{sw1}, the SAC spectra obtained with the different parameterizations look distinct in their details,
though the overall distribution of spectral weight is rather similar and close to the exact histogram when $\langle \chi^2\rangle/N_\tau  = 0.6$. The
fixed grid produces the sharpest peaks, while the continuum with only frequencies sampled leads to the overall broadest spectra. The continuous-$\omega$
results clearly have better overall shapes when $\langle \chi^2\rangle/N_\tau > 0.6$, where the peak of the fixed-grid spectra is too narrow and does not
represent satisfactory envelopes of the histogram. There is also a broad spurious maximum at low freqiency. The continuous-$\omega$ results are overall
broader and represent more satisfactory weight distributions. The fact that the spectra sampled with all three parametrizations differ significantly from
each other, even for matching values of $\langle \chi^2\rangle$, implies that their entropies do not just differ by constant factors but are affected by
their respective stochastic processes in some more fundamental way, which we will investigate in detail in Secs.~\ref{sec:entropy} and \ref{sec:maxent}.

Without a quantitative criterion for the match to the exact result, which would not be very useful here given that the exact spectrum has fine
structure that can never be reproduced, it is not possible to say definitely which parametrization is the best one here. Overall, it appears that the
fixed-amplitude spectra for $1 \ge \langle \chi^2\rangle/N_\tau \ge 0.6$ [Fig.~\ref{sw1}(b)] represent somewhat broadened envelopes of the histograms, and that
the added amplitude sampling [Fig.~\ref{sw1}(c)] causes the peak to narrow significantly and visually brings it closer to the histogram. However,
the bin size of the histogram is largely arbitrary. Preceding a peak-split at lower temperatures, a broadening is seen in Fig.~\ref{sw1}(c)
at $\langle \chi^2\rangle/N_\tau = 0.5$, and a precursor to the splitting is also present in Fig.~\ref{sw1}(a) at $\langle \chi^2\rangle/N_\tau = 0.6$.

With all three parametrizations, the spectrum eventually splits into two peaks below a non-universal value of $\Theta$, which can be regarded as a
consequence of overfitting with a spectrum constrained to be positive definite, as further discussed in \ref{app:lowtheta} (and also in \ref{app:low2}, where we relax the constraint and include negative spectral weight). As already mentioned, we also establish that each sharp peak
corresponds to two effective parameters that the positive definite spectrum provides for fitting the noisy imaginary-time data. Thus, in the present
case there are four parameters, and only with significantly improved data quality would this number at some point increase to six.

As $\Theta$ is lowered, the splitting into two peaks occurs in Fig.~\ref{sw1} at the highest value of $\langle \chi^2\rangle$ in the fixed-grid case
(though at a much lower value of $\Theta$, as seen in Fig.~\ref{fig:x2}), followed by the continuum with both frequency and amplitude updates.
In Ref.~\cite{sandvik98}, it was shown that the fixed-grid spectrum actually is the best right before the point where the split occurs, when
the  single maximum broadens out before the peak at lower frequency emerges; this broadening is manifested in the fixed-grid panel for 
$\langle \chi^2\rangle/N_\tau =  0.6$ in Fig.~\ref{sw1}(a). A broadening before the split also takes place with the combined amplitude and frequency
sampling, as seen in the $\langle \chi^2\rangle/N_\tau =  0.5$ panel of Fig.~\ref{sw1}(c). The shape of the spectrum here is very different from the
fixed-grid case in Fig.~\ref{sw1}(a), however. In Ref.~\cite{sandvik98}, the peak splitting was related to a local entropy maximum, which was suggested as
a criterion for selecting the best $\Theta$ value. It has proven difficult, however, to clearly identify such an entropy maximum in general. The $\Theta$
fixing based on $\langle \chi^2\rangle$ in Eq.~(\ref{eq:chi2}) is both easier to satisfy in practice and better motivated by statistical arguments.

With $a=1$, the optimal sampling temperature according to Eq.~(\ref{eq:chi2}) and Fig.~\ref{fig:x2} corresponds to $\langle \chi^2\rangle/N_\tau \approx 0.82$.
In practice, our experience is that somewhat smaller $a$ values produce better results (though the differences are normally only marginal). We typically use
$a=0.5$ (corresponding to the horizontal line in Fig.~\ref{fig:x2}), which in this present case corresponds to $\langle \chi^2\rangle/N_\tau \approx 0.65$.
Even $a=0.25$ (here $\langle \chi^2\rangle/N_\tau \approx 0.55$) typically produces good spectra without obvious effects of overfitting in the frequency
continuum (while the grid-bases spectra seem to suffer earlier from overfitting). If the underlying
imaginary-time data are good enough, this detail of the $\Theta$ criterion is not critical, as the spectrum changes very little over a wide range of
temperatures before the effects of overfitting (peak splitting) become visible. This insensitivity to $\Theta$ variations is apparent in the set of
continuous-frequency results in Figs.~\ref{sw1}(b) and \ref{sw1}(c) for $\langle \chi^2\rangle/N_\tau > 0.5$, while the corresponding fixed-grid results
in Figs.~\ref{sw1}(c) evolve more significantly with $\Theta$.

Overall, a reasonable conclusion that can be drawn from the results in Fig.~\ref{sw1} is that frequency-only sampling is the safest way to avoid too much
structure in the spectrum when the $\Theta$ criterion Eq.~(\ref{eq:chi2}) is applied with a reasonable value of the factor $a\lesssim 1$. However, the profiles
obtained when also the amplitudes are sampled visually appear closer to the true weight distribution, not only in the peak shape but also in the
way the tail of the spectrum is well reproduced. The generality of these behaviors is of course not clear based on just this test, but we have
found very similar differences between spectra in other cases where the correct spectrum consists of a single broad maximum.

It is interesting that the spectra for $\langle \chi^2\rangle/N_\tau > 0.5$ in Fig.~\ref{sw1}(c) are very close to an analytical result based
on a high-temperature expansion in Ref.~\cite{starykh97} (Fig.~4, $T/J=0.5$ panel). While that result is for the thermodynamic limit, at this
high temperature the spectrum obtained with SAC evolves very little from the $L=16$ form when $L$ increases. This excellent agreement with a presumably very
good analytical form provides additional support for the favorable effects of amplitude fluctuations on the average spectrum. Moreover, the results in
Fig.~\ref{sw1}(b) are closer to the ME result also shown in Fig.~4 of Ref.~\cite{starykh97}. The closeness of the ME result to that of frequency-only
sampling will be explained in Sec.~\ref{sec:maxent} based on a mapping between the two methods.

The sampling time was a few minutes for each $\Theta$ point in Fig.~\ref{fig:x2} (several hours for a total of 200 $\Theta$ values, down to $\Theta$ much
lower than shown in the figure). This is longer than typically required to just establish $\chi^2_{\rm min}$ to sufficient accuracy with the continuous
frequency parametrizations. It was done in this way here in order to obtain the smooth goodness-of-fit curves in Fig.~\ref{fig:x2} and sufficiently
smooth spectra when  $\langle\chi^2\rangle\approx \chi^2_{\rm min}$ in Fig.~\ref{sw1}. All 200 spectra were saved in this process and some of them
were selected for Fig.~\ref{sw1}.

If such detailed information is not needed, the entire first annealing process can often be carried out in a few
minutes and still produces sufficiently converged $\chi^2_{\rm min}$ estimates. It is clear that a slight over-estimation of $\chi^2_{\rm min}$ in practice just
corresponds to the effective value of $a$ in Eq.~(\ref{eq:chi2}) being slightly higher than the target value, and, as we have seen, the end
result is not sensitive to minor variations in $a$. For the second annealing procedure, which stops when the criterion in Eq.~(\ref{eq:chi2}) is satisfied,
it is better to sample a bit longer for each $\Theta$ (and also the rate of lowering $\Theta$ can be slower), so that the error bars on $\langle \chi^2\rangle$
are sufficiently small for reliably applying the optimal-$\Theta$ criterion. For the final sampling when $\Theta$ has been fixed (and the sampling continues
from the last configuration of the annealing process), the sampling time can be adapted to the desired smoothness of the spectrum.

\begin{figure*}[t]
\centerline{\includegraphics[width=16cm,clip]{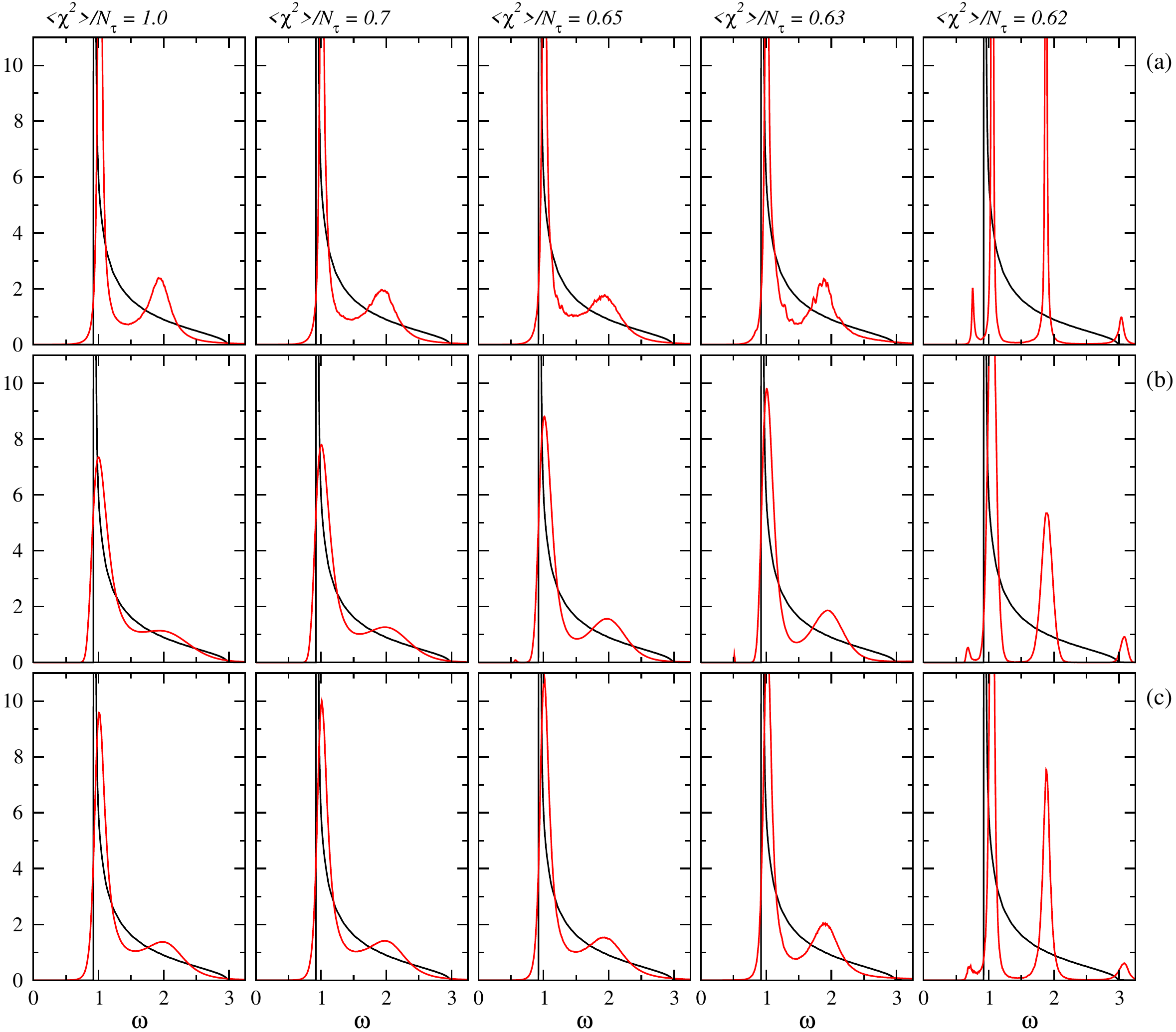}}
\vskip-1mm
\caption{SAC results for the dynamic spin structure factor (red curves) at $q=4\pi/5$ organized as in Fig.~\ref{sw1} but for $L=500$ spins at $T=J/500$,
  sufficiently low for ground state results. The spectra are compared with a $T=0$ numerical BA calculation \cite{caux05a,cauxdata} for the same system size
  (black curves). Results are shown at several values of the goodness of the fit based on simulated annealing with gradually lowered 
$\Theta$ values, similar to Fig.~\ref{fig:x2}. In this case $\chi^2_{\rm min}/N_\tau=0.619$.}
\label{sw2}
\vskip-1mm
\end{figure*}

\subsection{Example 2: 500-spin Heisenberg chain}
\label{sec:example2}

Next, we present a similar study of a much larger Heisenberg chain: $L=500$ spins simulated with the SSE method at $T=J/500$. We consider
the dynamic structure factor at momentum $q=4\pi/5$ and compare with $T=0$ results calculated numerically using the BA wave function
\cite{caux05a,caux05b,pereira06,cauxdata} for the same system size. It should be noted again that these BA based results are not exact, but include the
contributions from only two- \cite{muller81,bougourzi96,karbach97} and four-spinon \cite{caux05a} processes. The total spectral weight being known
exactly, it can be inferred that the BA spectrum contains about $98\%$ of the total weight in the case $q=4\pi/5$ that we consider here.
In Sec.~\ref{sec:contedge1} we will also study other $q$ values (with a different SAC parametrization), where the captured weight is
slightly different. The lower edge of the BA spectrum is the exact spinon dispersion relation \cite{cloizeaux62}.

The correlation function $\bar G(\tau)$ was computed on a uniform imaginary-time grid with spacing $\Delta_\tau=0.25$ between points. For the particular
momentum considered here, good data (using  20\% relative error as the cut-off) was obtained up to $\tau=9.5$; a total of $33$ data points and the
corresponding covariance matrix was thus used in the SAC runs. Data obtained at $T=J/1000$ with a similar level of error on a nonlinear $\tau$ grid are shown in
Fig~\ref{fig:gtau}. For this calculation, the two different $\tau$ grids produced essentially identical results, thus demonstrating that the choice of
$\tau$ grid is not critical when a reasonably large number of points is used. In general, once a certain number of $\tau$ points are used, additional
points do not contribute substantially more information, due to strong covariance, unless the data quality is further improved. There are also no signs
of differences in the $\tau$-dependence at the level of the error bars with these two data sets calculated at two different very low temperatures, and we
expect that there are no finite-temperature effects in the spectral functions presented below.

\begin{figure*}[t]
\centerline{\includegraphics[width=115mm,clip]{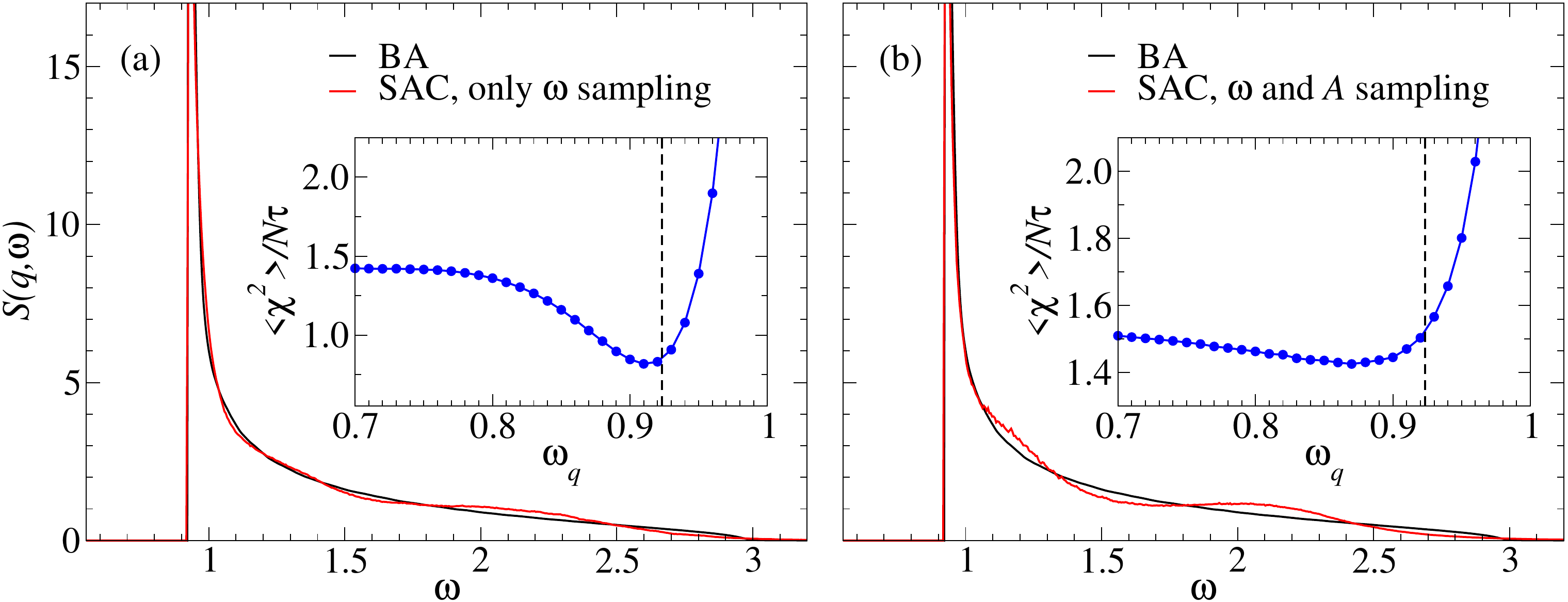}}
\vskip-1mm
\caption{Continuous frequency SAC results (red curves) for the dynamic structure factor of the $L=500$ Heisenberg chain (same as in Fig.~\ref{sw2}) with the
lower edge of the spectrum fixed at the known frequency $\omega_q \approx 0.923$ for $q=4\pi/5$. The BA result is shown in black. In (a) only frequency updates were
carried out while in (b) also the amplitudes were updated. In both (a) and (b), $\Theta$ was adjusted to give $\langle \chi^2\rangle/N_\tau \approx 0.75$ corresponding
to $a = 0.5$ in the criterion in Eq.~(\ref{eq:chi2}) when applied with the lower bound imposed (in which case $\chi^2_{\rm min}/N_\tau\approx 0.65$). The insets
show results for the goodness-of-fit vs the edge location, where in both cases $\Theta$ was fixed at a higher value than in the main graphs, so that
$\langle \chi^2\rangle/N_\tau \approx 1.5$ in both cases at $\omega_q=0$ (and there are no appreciable changes until $\omega_q \approx 0.7$). The correct
edge location is indicated by the vertical dashed lines.}
\label{w0fix}
\vskip-2mm
\end{figure*}

Simulated annealing runs were carried out and produced consistent goodness-of-fit curves similar to those in Fig.~\ref{fig:x2}, with convergence to
$\chi^2_{\rm min}/N_\tau \approx 0.619$ for all three parametrizations. Fig.~\ref{sw2} shows spectral functions obtained at five different values of the mean
goodness-of-fit, graphed together with the BA result. The exact dynamic structure factor of the Heisenberg chain in the thermodynamic limit has a power-law 
singularity at the lower edge $\omega_q$, of the form $S(\omega) \propto (\omega-\omega_q)^{-1/2}$ (with a logarithmic correction that we will discuss later 
in Sec.~\ref{sec:contedge2}). The BA results also being calculated on a finite chain, a broadening of the $\delta$-functions has been imposed
\cite{caux05a,cauxdata} and the divergence is thus quenched.

Unrestricted SAC sampling naturally cannot reproduce the sharp edge at $\omega_q$. Then, as is apparent in Fig.~\ref{sw2}, because of the spectral weigh appearing below
the true edge there is a compensatory effect by the sampling procedure (in order to produce a good fit to the imaginary-time data) that shifts the tip of
the rounded peak to higher frequency. The optimal $\Theta$ criterion in Eq.~(\ref{eq:chi2}) with $a =0.5 \sim 1$ corresponds to 
$\langle \chi^2\rangle/N_\tau = 0.7 \sim 0.8$, where it can be seen in Fig.~\ref{sw2} that there is always a second spurious broad maximum (often 
referred to as ``ringing'' in the literature) at $\omega\approx 2$. Clearly, overall the results are unsatisfactory with all parametrizations.

When $\Theta$ is taken very low, a spectrum with a few sharp peaks emerges with all parametrizations. With the fixed grid, it is very difficult to 
reach the ultimate spectrum with true minimum $\chi^2$ value, though the peaks seen in Fig.~\ref{sw2} still sharpen considerably when $\Theta$
is further lowered. While it is difficult to reach the true $\chi^2_{\rm min}$ value also with the other parametrizations, the narrow peaks at
low $\Theta$ are again suggestive of the actual $\chi^2$-minimizing positive definite spectrum consisting of four $\delta$-functions. In
\ref{app:lowtheta} we show further evidence of this behavior in results at much lower $\Theta$ than in Fig.~\ref{sw2}.

To demonstrate that the spurious second maximum, which is present at all reasonable values of $\Theta$ in Fig.~\ref{sw2}, indeed is caused by the
inability of the method to resolve the sharp edge, in Fig.~\ref{w0fix} we show results obtained with the continuous frequency parametrizations when the
lower edge of the spectrum has been fixed at its known value, which is \cite{cloizeaux62} $\omega_q = \pi\sin(q)/2$ in the thermodynamic limit (and not
significantly different for $L=500$ \cite{caux05a}). Here we observe a very sharp peak at the edge and a far less pronounced second maximum. Overall,
the results are much closer to the BA spectrum, though some distortions are still visible, and more so when both frequencies and amplitudes are sampled.
The distortions in the broad tail portion of the spectrum can still be thought of as induced by the primary distortions close to the edge. With no weight
below the true edge, the secondary distortions are also much milder.

As a first example of optimization of a constraint according to the principles illustrated in Fig.~\ref{fig:optim}, the insets of Fig.~\ref{w0fix}
show how the goodness of the fit changes with the location
$\omega_q$ of the imposed lower bound. Here the sampling temperature was held fixed at a value higher than prescribed by our $\Theta$ criterion, but
still low enough to give reasonable fits of the spectral functions to the data. With the slightly elevated $\Theta$ value, there is room for the fit to improve
when entropy is removed by the constraint, until the fit again deteriorates when the spectrum becomes over-constrained by an excessively high lower bound.

In the inset of Fig.~\ref{w0fix}(a)
we observe that the $\langle \chi^2 \rangle$ minimum is well developed at $\omega_q$ close to the correct value, while in Fig.~\ref{w0fix}(b) the minimum is 
much shallower and further away from the correct value. Moreover, in Fig.~\ref{w0fix}(a) the best value of $\langle \chi^2 \rangle/N_\tau$ is well below $1$, in the
realm of satisfying our $\Theta$ criterion, while in Fig.~\ref{w0fix}(b) the value always stays high above $1$. In both cases (in particular in the latter case),
we can of course improve the goodness of the fit by lowering $\Theta$, but then the minimum becomes shallower.

\begin{figure*}[t]
\centerline{\includegraphics[width=135mm, clip]{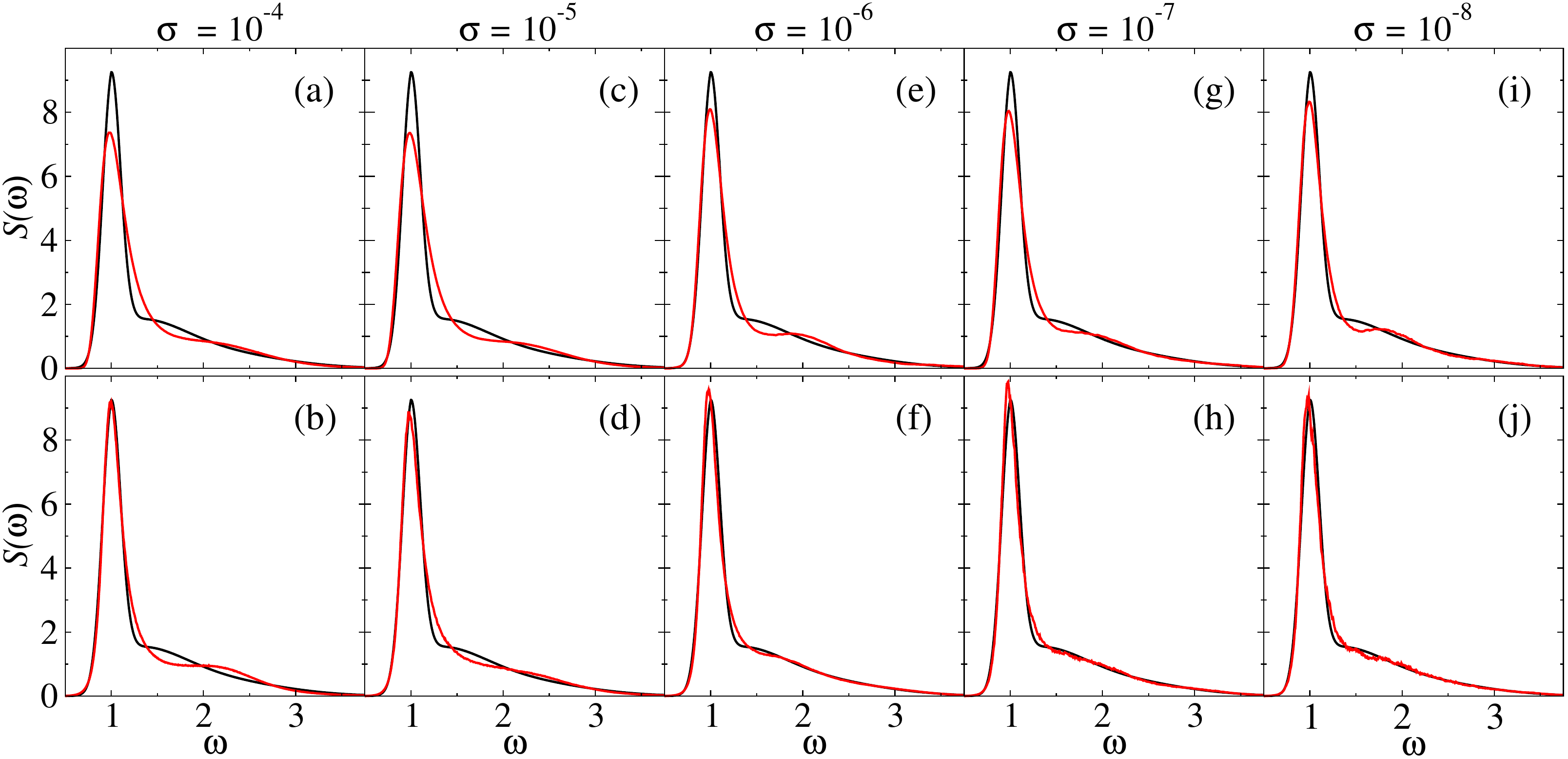}}
\vskip-1mm
\caption{A synthetic $T=1/16$ spectrum (black curves) reproduced by continuous-frequency SAC with only frequency moves in the upper row and
also including amplitude moves in the lower row. In both cases, results are shown for five different error levels; from $\sigma=10^{-4}$ in (a) and (b)
to $10^{-8}$ in (i) and (j), as indicated on top of each column. The $\sigma=10^{-4}$ and $\sigma=10^{-5}$ $\bar G(\tau)$ data sets included
40 $\tau$ points at spacing $\Delta_\tau=0.2$, while 80 points with $\Delta_\tau=0.1$ were used in the other cases. The number of sampled
$\delta$-functions ranged from $10^3$ with both parametrizations at $\sigma=10^{-4}$ to $2\times 10^5$ with frequency updates only at $\sigma=10^{-8}$
($5\times 10^{4}$ with the amplitude updates included).}
\label{syntcomp}
\vskip-2mm
\end{figure*}

Here it should be noted that just imposing the lower bound still does not enable the sampling to reproduce the correct edge shape. This inability
to resolve a very sharp feature, because of further entropic pressures not impeded by the simplest edge constraint, should be responsible for the imperfect
determination of the bound $\omega_q$ in this case. The
$\langle \chi^2 \rangle$ minimum should also only be expected to reflect the correct value of the constraining parameter in the limit of very low error
level $\sigma$ of the $\bar G(\tau)$ data, as we will discuss further in Sec.~\ref{sec:delta2} (and we will also further explain how to choose $\Theta$ when
scanning over a parameter). Nevertheless, even if it is not yet a perfect constraint, imposing the lower bound (at the known $\omega_q$ or at the
imperfect optimized value) clearly improves the fidelity of the method, especially with the equal-amplitude parametrization. The less favorable effects
of the constraint when amplitude updates are included (quantitatively seen in $\langle \chi^2 \rangle/N_\tau$ in the insets of Fig.~\ref{w0fix})
indicate that there are other entropic effects, beyond leakage of weight below $\omega_q$, that distort the spectrum in this case, more so than when only
the frequencies are sampled. This conclusion is supported by the differences in the spectral functions in Fig.~\ref{w0fix}, where the amplitude
updates cause more ringing in the tail of the spectrum.

In Ref.~\cite{sandvik16}, the fidelity of the SAC method for edge-divergent spectral functions was further improved significantly by imposing the
constraint of a single maximum in the amplitudes versus frequency in the fixed-grid parametrization (i.e., only amplitude updates maintaining the
single-peak structure were carried out). Such a constraint clearly impedes the entropic effects tending to flatten the peak and also suppresses the
ringing behavior, e.g., non-monotonic undulations are no longer possible. Tests were carried out using the same $L=500$ chain studied here. In
Secs.~\ref{sec:contedge1} and \ref{sec:contedge2}, we will present even more powerful methods for treating edge singularities with continuous-frequency
SAC representations.

\subsection{Example 3: Synthetic spectrum}
\label{sec:example3}

We next discuss an example based on synthetic data, a spectrum constructed from three Gaussians, shown as the black curves in Fig.~\ref{syntcomp}.  
Two of the Gaussians are broad and close to each other, so that their individual maxima do not appear and instead a single flatter maximum forms.
The third peak is narrower and located clearly below the other two. The low-frequency parts of the broader Gaussians are further damped by a fast
exponential decay below the maximum of the taller peak. The peak width is
not extremely narrow, so that there is some hope of reconstructing the entire spectrum from imaginary-time data at error levels achievable in QMC 
calculations. When converting the spectrum to $G(\tau)$ according to Eq.~(\ref{contrel1}), we here set  the inverse temperature to $\beta=16$, in light
of the fact that the tall peak of this spectrum could mimic a temperature broadened quasi-particle peak or a $T>0$ version of a singular dynamic
structure factor such as the one in Example 2 above.

\begin{figure*}[t]
\centerline{\includegraphics[width=150mm, clip]{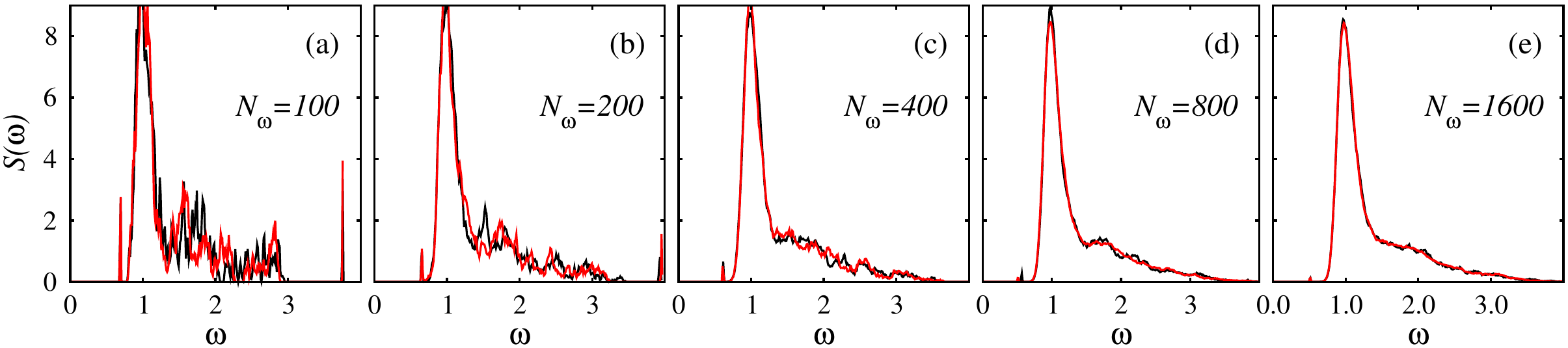}}
\vskip-1mm
\caption{Results for the same synthetic spectrum as in Fig.~\ref{syntcomp}, obtained with data of error level $\sigma=10^{-7}$ with sampling only
of the frequencies.  Results are shown for several different choices of the number $N_\omega$ of $\delta$-functions, and in each case
two independent runs were carried out (red and black curves). The spectra were sampled at roughly equal computational effort, using
$4 \times 10^{9}/N_\omega$ Monte Carlo sweeps (as defined in Sec.~\ref{sec:contsamp}). In all cases, $\Theta$ was adjusted to give
$\langle \chi^2\rangle/N_\tau \approx 0.79$, corresponding to the criterion Eq.~(\ref{eq:chi2}) with $a=0.5$.} 
\label{nconv}
\end{figure*}

\subsubsection{Dependence on the data quality}

Having already concluded that the continuous-frequency parametrizations are better than the fixed grid, we here only consider the former, comparing the case 
of frequency-only updates (equal weight $\delta$-functions) with that of both frequency and amplitude updates. We consider noise levels from
$\sigma=10^{-4}$ down to $10^{-8}$. When sampling at very low noise levels, the advantage of using large $N_\omega$ are apparent, and we went as
high as $N_\omega=2\times 10^5$ in the case of frequency-only updates at $\sigma=10^{-8}$, for which the complete SAC process needed many hours
of annealing and final sampling at $\Theta=0.025$. At the smallest error levels the process took only of the order ten minutes.

The results are shown in Fig.~\ref{syntcomp} for both continuous frequency parametrizations and may be considered acceptable already at the highest noise
levels, especially when both frequency and amplitude moves are used. The rather flat region where the broad maximum merges with the peak is the most challenging
feature to reproduce, with the SAC results instead exhibiting excessive broadening of the right side of the peak followed by ringing behavior. Reducing the noise
level improves the fidelity of this part of the spectrum, though ringing persists in the case of frequency-only updates even at $\sigma = 10^{-8}$. The gradual
improvements with decreasing $\sigma$ appear more systematic when both frequencies and amplitudes are sampled, and even though some deviations persist the
spectra at $\sigma = 10^{-7}$ and $10^{-8}$ are very close to correct. Overall, it is clear that the amplitude updates have a favorable effect on the fidelity,
with the peak height very close to correct and smaller deviations overall also for the other parts of the spectrum for all $\sigma$ values. However, it should
be noted that the tip of the peak actually in some cases is slightly too tall, which may indicate that the sampling entropy in this parametrization overly
favors sharp peaks. Still, in Fig.~\ref{syntcomp} it is rather obvious that the spectra obtained with the amplitude updates are better.

In all cases, the remaining distortions of the broad maximum and the region between the two peaks can be traced back to imperfect resolution of the tall peak,
which leads to secondary distortions like ringing at higher frequencies. There then has to be a compensating effect in order for the spectrum to reproduce
$\bar G(\tau)$, which implies a distortion at higher frequency, in particular in the rather flat region where the broad maximum merges with the peak.
In Sec.~\ref{sec:deltanp} we will provide further, concrete evidence of this interpretation of distortions propagating from the sharpest spectral feature. 
The tail portion above $\omega=2$ is rather well reproduced even at the highest error levels and improves systematically as $\sigma$ is reduced with
both parametrizations. 

QMC simulations would not normally be able to produce data with errors $\sigma=10^{-8}$, or even $\sigma=10^{-7}$, but such data are is still relevant for
exploring the performance of the method in such extreme cases. There are potential applications of SAC also to analytic continuation of data without statistical
errors (but often with some systematical errors instead), e.g., results produced with density matrix renormalization group (DMRG) calculations in imaginary time
\cite{linden20}, which is a potentially promising alternative to calculations performed directly in real time \cite{barthel09,yang21}. To process such data
by SAC, artificial noise can be added and systematically reduced as far as possible to converge to the best possible positive definite spectrum. With some
more computational effort, sampling millions of $\delta$-functions, the methods used here can likely be pushed to error levels much below $\sigma=10^{-8}$;
we have not yet established the practical limitations.

\subsubsection{Dependence on $N_\omega$}

The above example also provides a good opportunity to test the dependence of the results and the sampling efficiency on $N_\omega$. Fig.~\ref{nconv} shows
results for the same synthetic spectrum as in Fig.~\ref{syntcomp}, now at fixed error level $\sigma=10^{-7}$ and sampled only with frequency updates.
The number of $\delta$-functions ranges from $N_\omega=100$ to $1600$, i.e., in all cases smaller than in Fig.~\ref{syntcomp}, where $N_\omega=10^4$ in the
case $\sigma=10^{-7}$. The number of sampling sweeps was chosen proportional to $1/N_\omega$, so that the computational effort is roughly the same in
all cases---about 30 single-core CPU minutes for each, after a simulated annealing from higher $\Theta$ of over an hour. We show results for
two independent runs in each case to check the consistency of the procedures. Three key observations can be made:

(i) When $N_\omega$ is too small for the given data quality, the relatively large units $1/N_\omega$ of spectral weight (in the normalized, sampled spectrum) 
cannot migrate sufficiently, either to the low or the high tail of the spectrum. As a consequence, a small number of them will form separate small
spikes to roughly account for the spectral weigh in the tails and, thus, produce a good $\langle \chi^2\rangle$ value [which is the same for all
the cases shown and the same as in Fig.~\ref{syntcomp}(e)]. Note that these spectral features for small $N_{\omega}$ do not simply result from poor
sampling efficiency (though indeed the sampling is also slow), but from the inability of a small number of relatively large-amplitude $\delta$-functions
to sufficiently approximate the spectral weight distribution. The artificial spikes gradually move and shrink away as $N_\omega$ is increased.

\begin{figure*}[t]
\centering
\includegraphics[width=100mm]{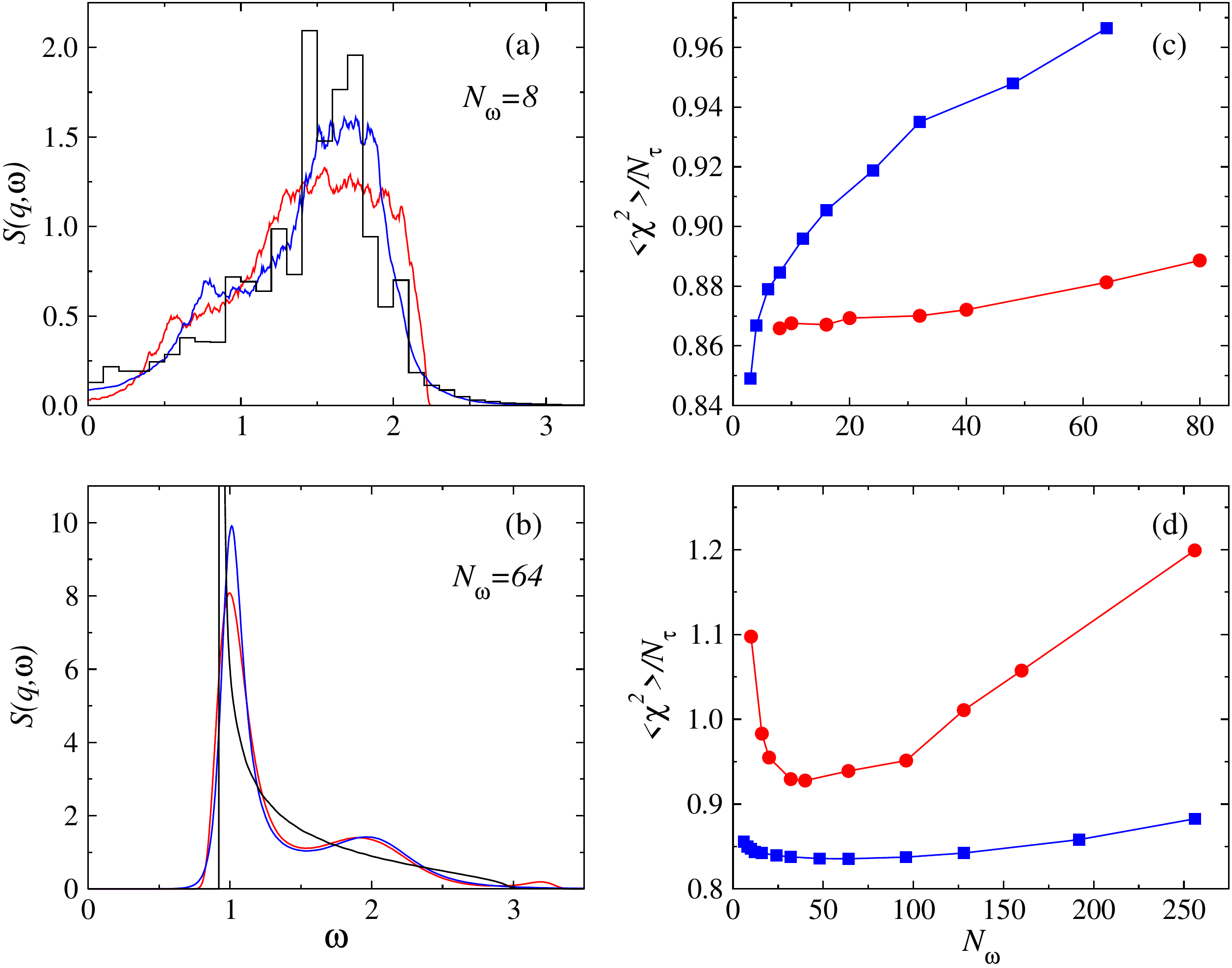}
\vskip-2mm
\caption{Examples of spectra obtained with a relatively small number $N_\omega$ of $\delta$-functions in the continuum, with the sampling
temperature fixed at $\Theta=1$ and using the same imaginary-time data underlying Figs.~\ref{sw1} and \ref{sw2}; (a) for $L=16,T=J/2$ with $N_\omega=8$ 
and (b) for $L=500,T=J/500$ with $N_\omega=64$. In both cases, the red curve was obtained solely with frequency updates, while for the blue curve 
also amplitude updates were performed. The dependence of $\langle \chi^2\rangle$ on $N_\omega$ is graphed in (c) for $L=16$ and in (d) for $L=500$.
Red and blue symbols correspond to sampling without and with amplitude updates, respectively.}
\vskip-2mm
\label{ndep}
\end{figure*}

(ii) With larger $N_\omega$ the sampling is much more efficient, leading to smoother curves.
For small $N_\omega$, it is very expensive in $\chi^2$ to move a unit
$1/N_\omega$ of spectral weight a substantial distance, while for large $N_\omega$ the small units can move significantly. There
are clear differences between the two independent runs for both $N_\omega=100$ and $200$, illustrating the very slow evolution of the spectrum and
trapping behavior (note, however, that the anomalous edge spikes are quite reproducible in all cases). It is clear from this example that the larger
moves in frequency for larger $N_\omega$ very well compensate for the fact that the unit of spectral weight $1/N_\omega$ is smaller and the number
of Monte Carlo sweeps is lower (though the total number of frequency update attempts was the same for all cases).

(iii) The profiles converge to a limiting shape for increasing $N_\omega$. Looking again at Fig.~\ref{syntcomp}(e), where a much larger $N_\omega=10^4$
was used, and comparing with the results for $N_\omega=800$ and $1600$ in Figs.~\ref{nconv}(d) and \ref{nconv}(e), the latter spectra still have a more
pronounced shoulder at $\omega\approx 2$. This feature smoothens out when $N_\omega$ increases, and for $N_\omega = 5000$ and larger this part is very close
to the exact spectrum and nothing changes when going to much larger $N_\omega$. It can also be noted that the height of the dominant peak systematically
decreases with $N_\omega$ and Fig.~\ref{syntcomp}(e) represents the converged profile also in this respect. Thus, as should be expected based on the analogy
with statistical mechanics, there is a ``thermodynamic limit'' of the spectrum, though in this case, because of the intensive property of $\chi^2$ and
extensive entropy (which we will demonstrate explicitly for this parametrization in Sec.~\ref{sec:entropy}), the parameter that regulates this limit
is not $\Theta$ but $\Theta/N_\omega$. The thermodynamic-limit analogy of $N_\omega \to \infty$, which has some unusual aspects related to $\chi^2$ as an
energy and the high-density limit of the $\delta$-functions (``particles''),
is discussed in detail in \ref{app:statmech}.

Based on this test, we can conclude that it should always be better to use relatively large $N_\omega$. If also the amplitudes are sampled, there are
no artificial spikes beyond the main part of the spectrum even for small $N_\omega$ and the spectrum converges faster with increasing $N_\omega$.
There is still no harm in using larger $N_\omega$, and the sampling efficiency is then improved also in this case. It should also be noted that there are
two sources of un-smoothness in the sampled spectral functions; i) when $N_\omega$ is fundamentally too small and ii) from insufficient averaging. Both effects
can be seen in Fig.~\ref{nconv} when comparing with Fig.~\ref{syntcomp}(e).

\subsection{Optimal number of $\delta$-functions at $\Theta=1$}
\label{sec:nomega}

Figure \ref{fig:x2} illustrates how SAC fails when using the Bayesian sampling temperature  $\Theta=1$ in Eq.~(\ref{psg}), because of entropic
effects when the number of sampled degrees of freedom is large. A useful remedy is to introduce the optimized sampling temperature $\Theta$ as explained
above, and from the extensive entropy (Sec.~\ref{sec:entropy}) we know that $\Theta \propto 1/N_\omega$ should be expected. In Ref.~\cite{sandvik16},
another remedy of the entropic catastrophe was proposed---the use of optimized constraints to prohibit the spreading of spectral weight that is the
root cause of the problem. 

Since the entropic catastrophe is related to large $N_\omega$, a different route to solving the problem may be to work with the smallest possible $N_\omega$.
The use of very large $N_\omega$ is largely motivated by improvements in sampling efficiency, as we discussed above, as well as to reach the
$N_\omega \to \infty$ limit if that is desired. Here we will present some results for the evolution of the spectrum with $N_\omega$ at $\Theta=1$ in
the regime where the changes with $N_\omega$ are still substantial. We will in particular investigate whether an optimal $N_\omega$ exists in the sense of
a minimum $\langle \chi^2(N_\omega)\rangle$.

In general, we find that this approach does not work well. Results for the same $L=16$ and $L=500$ Heisenberg chains as in Figs.~\ref{sw1} and
\ref{sw2} are shown in Fig.~\ref{ndep}, now with $\Theta=1$. The previous results were obtained with $N_\omega=1000$ in both cases, while here spectral
functions obtained with continuous frequency are graphed for $N_\omega=8$ ($L=16$) and $N_\omega=64$ ($L=500$) in Fig.~\ref{ndep}(a) and \ref{ndep}(b),
respectively. Figs.~\ref{ndep}(c) and \ref{ndep}(d) show plots of $\langle \chi^2\rangle/N_\tau$ versus $N_\omega$ for the two systems. In the case of
$L=16$ there is no discernible minimum in $\langle \chi^2\rangle$, while for $N=500$ there is in fact a rather well defined minimum.

Let us discuss the $L=16$ case first. For the smallest values of $N_\omega$, the spectrum (not shown) exhibits two sharp peaks, similar to the results 
in Fig.~\ref{sw1} for the smaller $\langle \chi^2\rangle$ values. The spectrum broadens out as $N_\omega$ increases, with $\langle \chi^2\rangle$ slowly 
increasing as well, as seen in Fig.~\ref{ndep}(c). Even with $N_\omega$ as small as $8$, the average spectral weight in Fig.~\ref{ndep}(a) reproduces 
quite well the exact diagonalization histogram, though the match at small $\omega$ is worse, with both parametrizations, than the optimal-$\Theta$ results 
in Fig.~\ref{sw1}. The result obtained with both frequency and amplitude sampling is clearly better, especially at the high-frequency tail.
With the fixed amplitudes, it is very costly in $\chi^2$ for even one unit of weight to migrate up to the thin-tail part of the spectrum, as in Fig.~\ref{nconv}
for the smaller values of $N_\omega$. However, in this case the consequence is just the sharp drop in mean spectral weight above $\omega=2$ in Fig.~\ref{ndep}(a),
with no artificial spike at higher frequency because of the very large amplitude $1/N_\omega$ when $N_\omega=8$. As $N_\omega$ increases, the sharp edge gradually
vanishes, and for $N_\omega \approx 50$ and higher the tail is well reproduced. Since there is no minimum in $\langle \chi^2\rangle$, the optimization
approach with $N_\omega$ as the parameter in Fig.~\ref{fig:optim} is not applicable in this case. 

The $L=500, N_\omega=64$ spectra in Fig.~\ref{ndep}(b) look very similar to the optimal ones in Figs.~\ref{sw2}, except for small third peak at
high frequency when only the frequencies are sampled. This peak is again a consequence of the rather large unit of spectral weight $1/N_\omega$, which
prohibits the density of $\delta$-functions to properly reproduce the tail. When the amplitudes are also sampled, the tail bump is not present.
In this case, $N_\omega=64$ is in the region of the $\langle \chi^2\rangle$ minimum observed in Fig.~\ref{ndep}(d), but the results are arguably still
worse than the (also far from satisfactory) results obtained with optimized $\Theta$ and large $N_\omega$ in Fig.~\ref{sw2}.

Even the best $\langle \chi^2\rangle$ values in Fig.~\ref{ndep}(c) exceed the optimum discussed in Sec.~\ref{sec:thetacrit}. Thus, with $\Theta=1$ the entropy
dominates too much over the goodness of the fit (though the $\langle \chi^2\rangle$ values still represent reasonable fits). This problem can be traced back
to the fact that the number of effective parameters, discussed above in Sec.~\ref{sec:thetacrit} and further in \ref{app:lowtheta}, is always much
smaller than the minimum number of $\delta$-functions required to obtain a smooth averaged spectrum. With unrestricted sampling it is in practice necessary to
optimize (lower) $\Theta$, while, as we will see further below (and as was also  found in Ref.~\cite{sandvik16}) in some cases constrained sampling
with $\Theta=1$ (or even $\Theta > 1$) is valid, with good $\langle \chi^2\rangle$ values even for large $N_\omega$.

In a related investigation of the optimal density of frequencies in "grid sampling'' \cite{ghanem20b}, a generic behavior of the goodness of the fit
similar to Fig.~\ref{ndep}(c) was noted (and a flatter goodness-of-fit can also be observed in Fig.~1 of Ref.~\cite{sandvik16} before a logarithmic
increase sets in at higher grid point density). It was argued \cite{ghanem20b} that the optimal density is in the flat region close to the minimum.
In that case the sampling temperature was also not adjusted ($\Theta=1$) and reasonable $\langle \chi^2\rangle$ values were still
obtained in the optimal regime. Though an optimal number of degrees of freedom at $\Theta=1$ can exist with some parametrizations, because of particular
$\chi^2$ versus entropy competition, in our experience with the parametrizations used here it is better to use large $N_\omega$ and adjust $\Theta$.

\section{Sampling entropy}
\label{sec:entropy}

In the case of sampled frequencies with constant equal amplitudes, the configurational entropy of a given spectral density profile (the $\delta$-function
amplitudes collected in a histogram) can be easily calculated exactly, which we will do in Sec.~\ref{sec:entropy1}. We also discuss the non-universality (dependence
on the parametrization) of the functional form of the entropy. In Sec.~\ref{sec:entropy2}, we demonstrate that the extensive property of the entropy
is consistent with our criterion Eq.~(\ref{eq:chi2}) for determining the sampling temperature.

\subsection{Different entropy forms}
\label{sec:entropy1}

Consider the situation in Fig.~\ref{fig:spec}(b), where $N$ ``particles'' labeled $i=1,\ldots,N$ occupy positions $\omega_i$ in the
continuum (where now, in this subsection, we suppress the $\omega$ subscript on $N$ for simplicity of the notation). It will not be necessary to provide any
bounds on these positions, and in principle they can be both positive and negative though in the present applications all $\omega_i \ge 0$. For a particle
configuration $(\omega_1,\ldots,\omega_N)$ and a histogram with bins $b=1,2,\ldots$ of width $\Delta$, let there be $n_b$ particles in bin $b$.
The number of identical histograms from combinatorics is
\begin{equation}
N_C = \frac{N!}{\prod_b n_b!}.
\end{equation}
Given that each particle can move within its window of size $\Delta$ without changing the bin assignment, the total configurational volume corresponding to
a set of occupation numbers is $V_C=\Delta^NN_C$, and the equal-amplitude (EA) entropy is 
\begin{equation}
E_{\rm EA}=\ln(V_C) = N \ln(\Delta)+\ln(N!) - \sum_b \ln(n_b!).
\end{equation}
Applying Stirling's formula in the form $\ln(n!)=n\ln(n)-n$ and using $\sum_b n_b = N$, we obtain
\begin{equation}
E_{\rm EA}= - \sum_b n_b \ln \left ( \frac{n_b}{N\Delta}\right ).
\label{eeaderiv1}  
\end{equation}
Stirling's formula is a good approximation when $n_b$ is large, but is also completely correct for $n_b=0$,
which is reassuring because bins beyond the tail of the spectrum will have $n_b=0$. For the regions where there is
finite spectral weight, $n_b \propto N$. Thus, the effects of the approximation will diminish for large $N$ and the above entropy becomes exact when
$N \to \infty$ for small $\Delta$. The mean occupation numbers will then also be proportional to $\Delta$, and we can define amplitudes $A_b$ such
that $n_b=N\Delta A_b$. Eq.~(\ref{eeaderiv1}) then becomes 
\begin{equation}
E_{\rm EA}= - N\Delta \sum_b A_b \ln (A_b).
\end{equation}
Converting to an integral, $\Delta A_b \to d\omega A(\omega)$, we have 
\begin{equation}
E_{\rm EA}= - N \int d\omega A(\omega) \ln [A(\omega)].
\label{eea}
\end{equation}
Apart from the factor $N$, the entropy then looks like the standard Shannon information entropy. Note that $\int d\omega A(\omega)=1$ follows from
the definitions above. The entropy clearly favors configurations with smooth distribution $A(\omega)$. Unlike the way the entropy is defined in the ME
method, Eq.~(\ref{esdef}), here there is no default model dividing $A(\omega)$ under the logarithm (though it can be incorporated as well if so desired
\cite{beach04}).

The above standard result for the entropy was also derived in the SAC context
by Bergeron and Tremblay \cite{bergeron16}, and previously also by Beach \cite{beach04}.
However, these works did not explicitly consider a specific SAC parametrization but analyzed a stochastic process that is in practice
very similar to our fixed-amplitude parametrization. The number $N$ was not regarded as specifically related to the SAC sampling space, but as a device
introduced for the purpose of the calculation that drops out when relating the entropy to the ME method. Apparently, the results were also regarded as
generic and suggestive \cite{bergeron16} of a relationship between the SAC and the ME method (in a mean-field sense in Ref.~\cite{beach04}).
The details of the stochastic process are in fact crucial, and different SAC parameterizations realize different stochastic processes with different
functional forms of the entropy.

Ghanem and Koch \cite{ghanem20a} recently calculated the entropy in the case of a fixed grid (including a default model), where a more sophisticated
functional-integral method was required. Interestingly, the GK entropy is formally equivalent to the Shannon entropy of the default model with respect
to the sampled spectrum. This difference from the conventional Shannon entropy with respect to the default model might seem insignificant, but actually
is quite dramatic. With a flat default, the GK entropy can be written as
\begin{equation}
E_{\rm GK}= N \int d\omega \ln [A(\omega)],
\label{egk}
\end{equation}
up to a factor which would only affect (implicitly) the value of $\Theta$ in our approach (fixing $\Theta$ based on $\langle\chi^2\rangle$). This entropic
form is quite different from Eq.~(\ref{eea}), and it was suggested that it imposes higher entropic penalties to deviations from the default model \cite{ghanem20a}.
This statement seems counter to our results in Fig.~\ref{sw1}, where the grid sampling produced the sharpest peaks (i.e., largest deviations from the flat
effective default). As pointed out already, overall scale factors are absorbed in our SAC formulation by the sampling temperature $\Theta$, and what matters
for determining the shape of the spectrum $A$ is the functional-integral form of the entropy $E(A)$ irrespective of overall factors. Without more detailed
analysis, it is difficult to predict what effects a given entropy will have in combination with the $\chi^2$ weighting when sampling the spectrum.

We next consider the most complicated of the three basic SAC parametrization, Fig.~\ref{fig:spec}(c), where both frequencies and amplitudes are
sampled. This is also the parametrization first considered by Beach within a mean-field approximation \cite{beach04} of the entire SAC problem
including the $\chi^2$ weighting. He assumed the applicability of the conventional Shannon information entropy, i.e., without the factor $N$
in Eq.~(\ref{eea}). We have not managed to compute the entropy exactly in this case but have constructed what should be a good approximation.

We take the approach used in \ref{app:fluct}, where, in a simplified example, we compute the variance of the spectral weight fluctuations
by summing the independent contributions from amplitude and frequency fluctuations. Similarly, we argue that the entropy should be the sum of the
entropies from frequency-only
updates, Eq.~(\ref{eea}), and amplitude-only updates, Eq.~(\ref{egk}), at least to a first approximation. In this case, we initially have to keep the
default model $D(\omega)$ in the sum of the two entropies, because it appears in different ways in the two forms. Thus, we propose the {\it mixed entropy}
$E_{\rm MX}$ defined as
\begin{equation}
E_{\rm MX} = - N \int d\omega [A(\omega)-D(\omega)] \ln \left ( \frac{A(\omega)}{D(\omega)} \right ).
\label{emxw}  
\end{equation}
Here it should be noted that both terms are negative semi-definite and, thus, the maximum mixed entropy is also zero. Taking a flat default model (within
the relevant range of $\omega$) and calling its constant value $\gamma$, we obtain a simper form of the mixed entropy that we will consider here;
\begin{equation}
E_{\rm MX} = - N \int d\omega [A(\omega)-\gamma] \ln[A(\omega)].
\label{emx}  
\end{equation}
It is interesting to note that the integrand here is closer to that for the conventional entropy when $A(\omega)>\gamma$, while being closer to the
GK form when $A(\omega) < \gamma$.

We conjecture that the mixed entropy for some value of $\gamma$ corresponds closely to the de facto entropic weight when sampling both the frequencies
$\omega_i$ and the associated amplitudes $A_i$ with the SAC method. In SAC, we do not use any explicit default model, and, as we have mentioned
before, in some sense the default model formally stretches out to $\omega=\infty$ when the continuous frequency space is used. However, the true spectrum
covers only a limited range of frequencies and $\gamma$ may then be close to the inverse of an effective width of the spectrum. The ambiguity in defining
such a width implies that $E_{\rm MX}$, for any value of $\gamma$, cannot be the exact expression for the entropy. There may be some effective non-flat
default model $D(\omega)$ that is generated by the ``confining potential'' imposed implicitly by the $\bar G(\tau)$ data in the SAC process.

In Sec.~\ref{sec:maxent} we will demonstrate exact mappings between the SAC method with different parametrizations and corresponding ME
methods with different forms of the entropy used in the prior probability $P(S)$ in Eq.~(\ref{meprior}). With these mappings, we can test the
three forms of the entropies discussed above. We will find that ME calculations with $E_{\rm EA}$ in Eq.~(\ref{eea}) and $E_{\rm GK}$ in Eq.~(\ref{egk}) 
match essentially perfectly results of SAC with equal-amplitude $\delta$-functions and fixed grid, respectively, when the goodness-of-fit values
of the two methods match. Moreover, SAC results obtained with
both amplitude and frequency updates included are well described by the mixed entropy, $E_{\rm MX}$ in Eq.~(\ref{emx}), when $\gamma$ is a
number closely corresponding to the inverse width of the spectrum.

The role of the parametrization was discussed qualitatively in our previous works with collaborators \cite{qin17,shao17}, where it was pointed out that
SAC with $\delta$-functions in the frequency continuum shows less ``entropic bias'' than the fixed-grid sampling. Ghanem and Koch \cite{ghanem20b}
likewise found advantages when ``relaxing the grid'' and sampling frequencies. Thanks to the now available explicit results for the entropy in different
parametrization, especially the non-Shannon $E_{\rm GK}$ and $E_{\rm MX}$ forms, we have a more quantitative understanding of the different
parametrizations. In particular, with the mixed entropy we can understand why spectra obtained with both amplitude and frequency updates, e.g., in
Figs.~\ref{sw1} and \ref{sw2}, fall between those when only amplitudes or only frequencies are sampled. More detailed insights will be presented
in Sec.~\ref{sec:maxent}.

The increase in entropy with $N$ was noted first in the fixed-grid parametrization in Ref.~\cite{sandvik16}, and Eqs.~(\ref{eea}) and
(\ref{egk}) demonstrate the extensive property explicitly. Any parametrization with $N$ sampled parameters (here with $N=N_\omega$ when only
frequencies are sampled and $N=2N_\omega$ when the amplitudes are also sampled) should have an extensive entropy
for large $N$. The extensive property is of course expected if we think of the $\delta$-functions as a set of particles. However, while we 
here treat the analytic continuation as a statistical mechanics problem, the energy in the form of $\chi^2/2$ does not have the normal extensive
property---for example, the minimum value $\chi^2_{\rm min}$ will not change much after some value of $N_\omega$ has been increased but approaches a
non-zero constant. Thus, when using the Boltzmann-like probability distribution Eq.~(\ref{psg}) with $\Theta=1$, the sampling will become
increasingly entropy dominated when $N_\omega$ grows, causing a gradually smoothing of the average spectrum whose ``internal energy'' $\langle \chi^2\rangle$
diverges with $N_\omega$. This deterioration of the goodness of the fit can be observed in the $\Theta=1$ tests in Fig.~\ref{ndep}.

To counteract the entropic catastrophe and keep the average spectrum $N_\omega$-independent for large $N_\omega$, the sampling temperature will have to be
reduced as $1/N_\omega$ (which we will explicitly confirm below), or constraints have to be introduced that sufficiently suppress the entropy as was done in
Ref.~\cite{sandvik16}. For any parametrization, the entropy eventually, for large $N_\omega$, has to be suppressed by lowering $\Theta$. With
$\Theta =\theta/N_\omega$, the SAC sampling problem looks very similar to a statistical mechanics problem at temperature $\theta$, though with some important
subtle differences that are spelled out in detail in \ref{app:statmech} and which impact the mapping between SAC and the ME method (Sec.~\ref{sec:maxent}).

In the grid case, an imposed density of points acts as a default model \cite{ghanem20a}.
In the continuum case, the a priori effective default model in the absence of imposed bound is a spectrum spread out to infinity, and if bounds
are imposed a flat default within those bounds is realized. In principle, other default models can be implemented in the continuum by adding a
potential on the ``particles'', which was done by Beach \cite{beach04} and later also by Ghanem and Koch \cite{ghanem20b} in their ``grid sampling''
method. We will not consider the conventional default model approach here and instead, starting from Sec.~\ref{sec:deltapeak}, we will introduce
various hard constraints that modify the entropic pressures affecting the average spectrum. This approach can also in some sense be regarded as
introduction of generalized, optimized default models, though we will not use that language. We will make connections to default models
when discussing future prospects in Sec.~\ref{sec:conc3}.

\subsection{Consistency with optimal $\Theta$}
\label{sec:entropy2}

In practice, use of the criterion Eq.~(\ref{eq:chi2}) for the sampling temperature guarantees a good fit though $N_\omega$ is not referenced directly. We next show
that this criterion indeed represents the optimal balance between $\chi^2$ and entropy, in the sense that it delivers $\Theta \propto 1/N_\omega$, so that the
factor $N_\omega$ in the entropy is in effect canceled.

\begin{figure}[t]
\centering
\includegraphics[width=70mm]{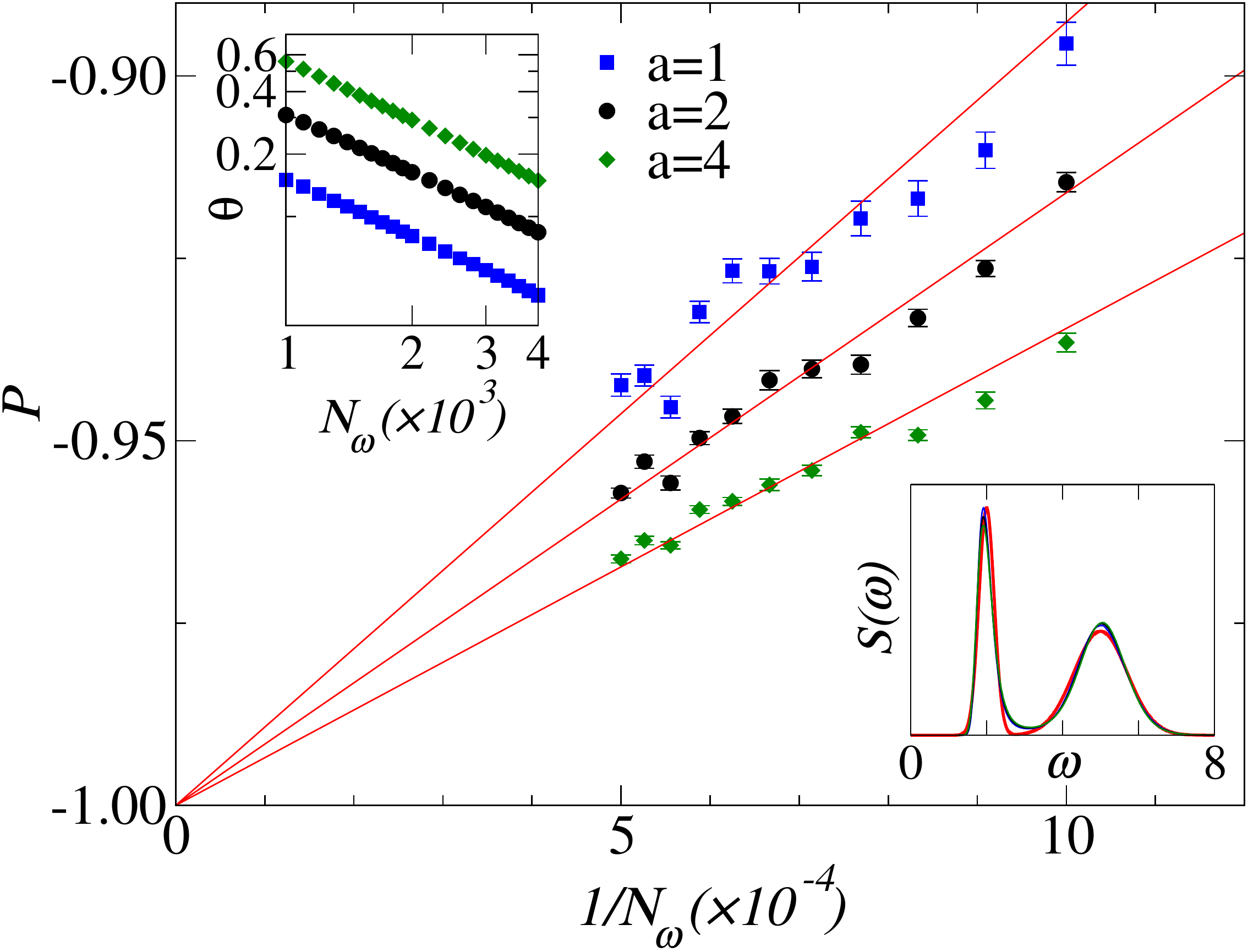}
\vskip-1mm
\caption{Slope of the $N_\omega$ dependence of the sampling temperature $\Theta$ determined using Eq.~(\ref{eq:chi2}) with $a=1,2$, and $4$.  A synthetic
spectrum consisting of two Gaussians, shown as the red curve in the low-right inset, was used to generate imaginary time correlations, to which
noise at level $\sigma=10^{-5}$ was added. The slope $P$, shown vs $1/N_\omega$ in the main graph, is defined on the basis of results for the number of frequencies
being $N_\omega=N$ and $N_\omega=2N$ with the results for $\Theta(N_\omega)$ shown in the upper-left inset. The red lines show consistency with $P=-1$ for
$N_\omega \to \infty$, thus confirming the predicted impact of the expected $\propto N_\omega$ scaling of the entropy. The spectrum obtained
with $N_\omega=4000$, $a=2$ is shown in the inset as well (black curve). The spectra obtained with $a=1$ and $a=4$ are almost indistinguishable
from the $a=2$ result.}
\vskip-1mm
\label{nw-th}
\end{figure}

We use a synthetic spectral function with two separated Gaussian peaks, as shown with the red curve in the inset in the low-right corner of Fig.~\ref{nw-th}.
We have tested prefactors $a=1$, $2$, and $4$ in Eq.~(\ref{eq:chi2}), with the number of $\delta$-functions $N_\omega$ ranging from from $1000$ to $4000$. With a
slow enough annealing process, the sampling temperature is determined for each given $N_\omega$, as shown in the upper left inset of Fig.~\ref{nw-th}. We then
calculated the slope $P$, the $N_\omega$ derivative of $\chi^2$, on the basis of pairs of points obtained with $N=N_\omega$ and $N=2N_\omega$ $\delta$-functions.

Results for the slope versus $1/N_\omega$ are shown in the main part of Fig.~\ref{nw-th}, where the error bars were obtained using bootstrapping of the
$\langle\chi^2\rangle$ data. The values deviate by at most $10\%$ from the expected slope $P=-1$, corresponding to the scaling $\Theta \propto 1/N_\omega$
for which the entropy and $\chi^2$ should be  balanced according to the discussion in Sec.~\ref{sec:entropy1}. The systematic corrections to the
exponent $P=-1$ are close to linear in $1/N_\omega$, as as shown with the red lines in Fig.~\ref{nw-th}, and the agreement can be further improved by
including small second-order corrections (not shown). Thus, the optimal $\Theta$ versus $N_\omega$ should have a power-series form
$\Theta = b_1/N_\omega + b_2/N_\omega^2 + \ldots$.

This demonstration of course only confirms that the optimal $\Theta$ should scale as $1/N_\omega$, and does not tell us what the factor $a$ should be in the fixing
criterion, Eq.~(\ref{eq:chi2}). However, the properties of the $\chi^2$ distribution dictate that $a$ should be of order $1$.  For reasonably good QMC data,
as we have demonstrated, e.g., in Fig.~\ref{sw1}, the final result does not depend significantly on $a$ as long as the value is reasonable; in practice we
often use $a=0.5$ as mentioned.

The insensitivity of the average spectrum to $a$ is also exemplified in Fig.~\ref{nw-th}, where in the low-right inset we graph the result for
$N_\omega=4000$ obtained with $a=2$. This spectrum, as well as those for $a=1$ and $a=4$ (not shown), falls almost on top of the synthetic spectrum. The
dependence of the average spectrum on $\Theta$ (and hence on $a$) can be expected to be more significant if the imaginary-time data are barely good enough
for resolving some specific feature of the spectrum.

\section{Quasi-particle peaks}
\label{sec:deltapeak}

In this section we begin a detailed discussion of how to adapt the SAC method to spectral functions with sharp features. The prototypical first example
is a $\delta$-function edge;
a quasi-particle peak which often appears at the lower edge of the spectrum (in most cases with some broadening). In Sec.~\ref{sec:delta1}
we review the procedures that we developed in previous work \cite{shao17} to optimize the amplitude of such a $\delta$-function. In Sec.~\ref{sec:delta2}
we present a systematic study of the statistical optimization criterion, focusing on the convergence properties as the data quality is improved. We also
explain how a finite width of a peak can be detected if the parametrization with a $\delta$-peak is imposed when this form is strictly not appropriate.
In Sec.~\ref{sec:deltanp} we propose a new high-resolution method for quasi-particle peaks of finite width.

In Sec.~\ref{sec:delta1} and \ref{sec:delta2} we use exclusively the continuum parametrization in Fig.~\ref{fig:spec}(d), with equal-amplitude
``microscopic'' $\delta$-functions to model the continuum and a larger ``macroscopic'' one at the lower edge; the quasi-particle peak with weight
$A_0$. In Sec.~\ref{sec:deltanp} we will generalize this parametrization by distributing the quasi-particle weight $A_0$ over a relatively small number
$N_p \ll N_w$ of $\delta$-functions (where $N_\omega$ is the total of number $\delta$-functions), with some feature of the peak (e.g., its lower or upper edge
or its mean frequency) serving as a lower bound for the continuum.

In principle, the microscopic $\delta$-functions could also have fluctuating amplitudes. However, in this case we have found that it is in general better
to use fixed equal amplitudes, which is the parametrization that produces the least spectral details in the case of unrestricted sampling (e.g., in Figs.~\ref{sw1}
and \ref{syntcomp}). We saw this already in Fig.~\ref{w0fix}, in the context of the lower spectral bound, where the equal amplitudes gave a better spectrum
when the correct bound was imposed. It is of course not clear how generic these observations are, and likely amplitude updates could also be useful if there is
significant structure (sharp peaks) within the continuum. Here we will consider rather smooth continua.

\begin{figure}[t]
\centering
\includegraphics[width=70mm]{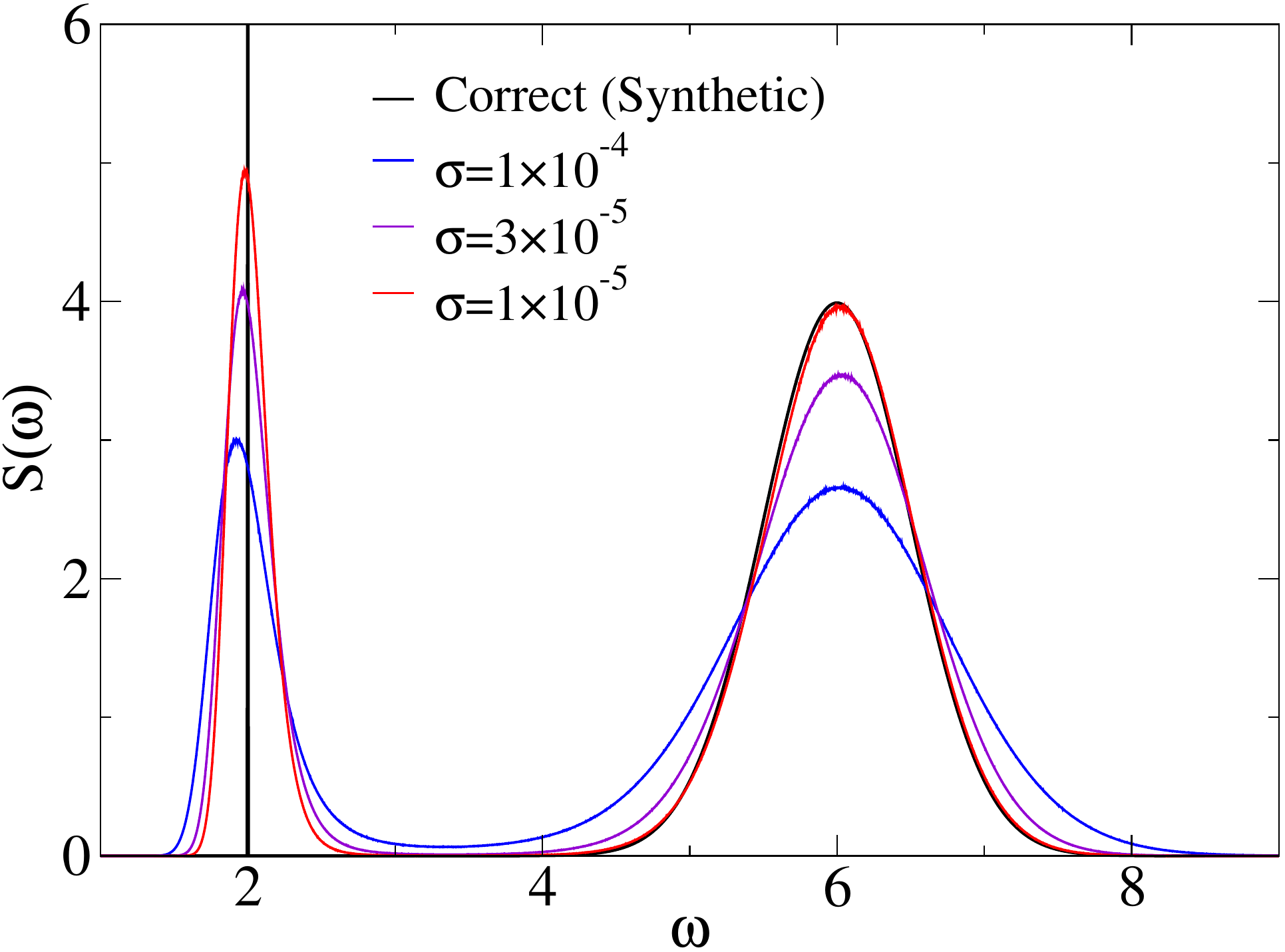}
\vskip-1mm
\caption{Results of unrestricted SAC with the frequency-continuum representation, Fig.~\ref{fig:spec}(b), applied to a synthetic
spectrum (shown in black) with a $\delta$-function containing $25\%$ of the total weight at $\omega=2$ and a broad Gaussian peak centered at
$\omega=6$. The other curves show SAC results obtained with noise added to the synthetic correlation function $G(\tau)$, with the levels
of noise $\sigma$ indicated in the legends. At $\sigma=10^{-5}$, the Gaussian maximum is reproduced almost perfectly and partially covers
this part of the synthetic spectrum.}
\vskip-1mm
\label{delta-free}
\end{figure}

\subsection{Delta-function and continuum}
\label{sec:delta1}

We first illustrate the limitations of the unrestricted SAC by tests with a synthetic spectral function, graphed in Fig.~\ref{delta-free},
where a dominant $\delta$-peak  containing $25\%$ of the total spectral weight is separated from a broad Gaussian continuum,
After calculating $G(\tau)$ on a linear grid with $\Delta_\tau=0.05$ from the synthetic spectrum, we added correlated noise as
described in Sec.~\ref{sec:syntdata}, using three different noise levels; $\sigma=1\times10^{-4}$, $3\times10^{-5}$ and $1\times10^{-5}$. With
the cut-off in $\tau$ set at relative error of $20\%$, the maximum $\tau$ values were $2.8$, $3.4$, and $4.0$, respectively.

We sampled $2000$ equal-weight $\delta$-functions, with $\Theta$ adjusted according to the criterion stated in Eq.~(\ref{eq:chi2}).
As seen in Fig.~\ref{delta-free}, when the data quality is good enough, in this case $\sigma=10^{-5}$, the continuum is resolved almost
perfectly even though the sharp $\delta$-peak can never be resolved with unrestricted sampling. The lower peak does narrow systematically
(likely logarithmically) with improving data quality.
When the data quality is insufficient, not only is the tip of the lower peak shifted down from the location of the
$\delta$-function (to compensate for the overall more substantial broadening on the high-$\omega$ side of the peak), but also the shape
of the continuum is excessively broadened. By including also amplitude updates, a narrower peak can be produced, and the resolution also improves
if the peak and continuum are further separated; an example of very narrow isolated peak produced by unrestricted sampling will be presented
in Sec.~\ref{sec:hladd2}; the ``triplon'' peak of the 2-leg Heisenberg ladder model.

The good resolution of the continuum in Fig.~\ref{delta-free} for $\sigma=10^{-5}$ is aided by the  gap between the two spectral features, which of course
is frequently not present in physical spectral functions. As an example, the dynamic structure factor of the square-lattice Heisenberg antiferromagnet is known
to have a dominant $\delta$-function representing the single-magnon excitation (likely with some extremely small broadening even at $T=0$ \cite{chernyshev09})
followed by a continuum. Unlike the spectrum in Fig.~\ref{delta-free}, the continuum apparently extends all the way down to the quasi-particle peak, with no
gap or significant reduction of spectral weight between the two features \cite{shao17}. In this case, unrestricted SAC cannot easily (unless the underlying
imaginary-time data are extremely good) resolve even the continuum part of the spectrum, because the entropic broadening at the edge is compensated by
distortions of the connected continuum (similar to the results in Fig.~\ref{sw2}). Examples of such distortions of synthetic spectral functions will
be given in Sec.~\ref{sec:delta2}. First we discuss our implementation of the $\delta$-edge spectrum and present an illustrative
result for the 2D Heisenberg model.

To resolve a $\delta$-edge followed by an arbitrary continuum, we parameterize the  spectrum as in Fig.~\ref{fig:spec}(d), with a single $\delta$-function
with amplitude $A_0$ and location $\omega_0$, followed by $N_\omega$ microscopic $\delta$-functions with equal amplitude $A_i=(1-A_0)/N_\omega$, with
the constraint $\omega_i > \omega_0$ for $i>0$. Typically we here use $N_\omega=500-1000$, and the results do not change significantly for larger $N_\omega$.
If the amplitude $A_0$ is sampled, its mean will not be close to the correct value, because its effect on $\chi^2$ is small. Even though, as we will demonstrate
explicitly below, increasing $A_0$ will reduce $\langle \chi^2\rangle$, the entropy will drive $A_0$ to small values and push $\omega_0$ below the true edge.
Therefore, $A_0$ has to be fixed in the sampling process, and its value must be optimized in some way.

The way to optimize a generic constraint, regulated by a parameter $p$, is illustrated in Fig.~\ref{fig:optim}, and we already gave a simple (though imperfect)
example of optimization of the lower spectral bound in Fig.~\ref{w0fix}. We applied the optimum-$\langle \chi^2\rangle$ approach to the $\delta$-edge, with
the optimized parameter $p=A_0$, in previous work \cite{shao17}. We review the details of the method in this case next, in preparation for the further
insights on the optimization protocol that we present in Sec.~\ref{sec:delta2}.

Before the special amplitude $A_0$ is introduced, we first perform a standard unrestricted SAC simulated annealing procedure (formally we set $A_0=0$),
as explained in Sec.~\ref{sec:plainsampling}, resulting in a minimum goodness-of-fit $\chi^2_{\rm min}$. For the next step, we fix the sampling
temperature $\Theta$ not according to the statistical criterion in Eq.~(\ref{eq:chi2}), but at a somewhat higher value. The elevated sampling
temperature is preferred in this context because the optimization procedure involves scanning over $A_0$ to find a minimum in $\langle \chi^2\rangle$,
which will be more pronounced if $\Theta$ is higher. The detailed effects of the sampling temperature will be discussed in Sec.~\ref{sec:delta2}, and
here we simply state that we use $\langle \chi^2(\Theta) \rangle \approx \chi^2_{\rm min} + aN_\tau$ instead of Eq.~(\ref{eq:chi2}). Then unrestricted sampling
still produces a reasonably good fit, $\langle \chi^2\rangle/N_\tau \lesssim 2$, if the prefactor $a=1$ is used, though strictly speaking the fit is suboptimal.
When $A_0$ is turned on gradually, $\langle \chi^2\rangle$ decreases and the minimum value (when $A_0$ is close to its optimum) often satisfies the
conventional $\Theta$ criterion.

With $\Theta$ being fixed, a scan of $A_0$ is carried out on a grid (running sequentially or in parallel), with typically tens of points, from $A_0=0$ to
up to $A_0=1$ (or some smaller window can be chosen if there is already some knowledge on the magnitude of $A_0$). For a given $A_0\neq0$, the sampling now
includes two steps of a simple modification of the method presented in Sec.~\ref{sec:contsamp}: (i) the standard updates of the $N_\omega$ equal-amplitude
$\delta$-functions with their lower bound being  $\omega_0$ (i.e., any move taking $\omega_i$ with $i>0$ below $\omega_0$ is rejected); and (ii) updating
the location $\omega_0$ of the edge $\delta$-function (where any move taking it above any of the other $\delta$-functions is rejected). The acceptance
rates for $\omega_0$ and $\omega_{i>0}$ are kept track of separately, and the size of the moves (frequency windows) are adapted to give rates close to $0.5$.

With the fixed $A_0$ value, the entropy-driven suppression of its value is avoided. Moreover, when $A_0>0$ entropy is removed from the spectrum.
Simply put, the larger $A_0$ the less can the spectrum fluctuate overall. In particular, the edge frequency $\omega_0$ will move up as $A_0$
increases, thereby restricting the fluctuations of the continuum.
As illustrated schematically in Fig.~\ref{fig:optim} for a general parameter $p$ with similar effect, 
this entropy suppression initially leads to a reduction of $\langle \chi^2\rangle$. However, when $p$ (here $A_0$)
becomes too large, a spectrum representing a good
fit to $\bar G(\tau)$ can no longer form and $\langle \chi^2\rangle$ then starts to grow sharply with $A_0$. Thus, there will be a minimum in
$\langle \chi^2\rangle$ that represents the point where a further entropy reduction leads to a deteriorated fit, and it is natural to posit $A_0$
at this point as the optimal value. Of course, this can be strictly true only in the limit of vanishing statistical errors in the imaginary-time data,
but we invariably obtain good results also with realistic error levels.

In a sense, this $\langle \chi^2\rangle$ minimization procedure could be referred to as a ``minimum entropy method'', as the entropy of the
statistical-mechanics system, consisting of the large number of $\delta$ functions, is minimized under the condition of also optimizing the
mean goodness-of-fit (which acts as the internal energy of the problem). Of course the entropy is still maximized by the sampling within a
fixed constraint (here the value of $A_0$).

Since $\omega_0$ is also sampled, the spectrum accumulated in a histogram may exhibit a slightly broadened peak instead of a
$\delta$-function. Unless $A_0$ is very small, the peak is typically still sharp, however, and the peak often occupies only a single bin
of the accumulated histogram. As shown in Ref.~\cite{shao17}, at the $\langle \chi^2\rangle$ minimum both the amplitude $A_0$ and the mean location
$\langle \omega_0\rangle$ are typically very close to their correct values (for good enough data, as will be further discussed in
Sec.~\ref{sec:delta2}). The lack of fluctuations of the peak location will be further explained in Sec.~\ref{sec:deltanp}.

After $A_0$ has been optimized, $\langle \chi^2\rangle$ is often statistically acceptable and the spectrum accumulated there
can simply be used. However, we often still carry out a second annealing step (starting from the $\Theta$ value used in the scan) to find a
new $\Theta$ according to the standard criterion in Eq.~(\ref{eq:chi2}). The differences in the results are typically minimal at
this stage.

\begin{figure}[t]
\centering
\includegraphics[width=7cm]{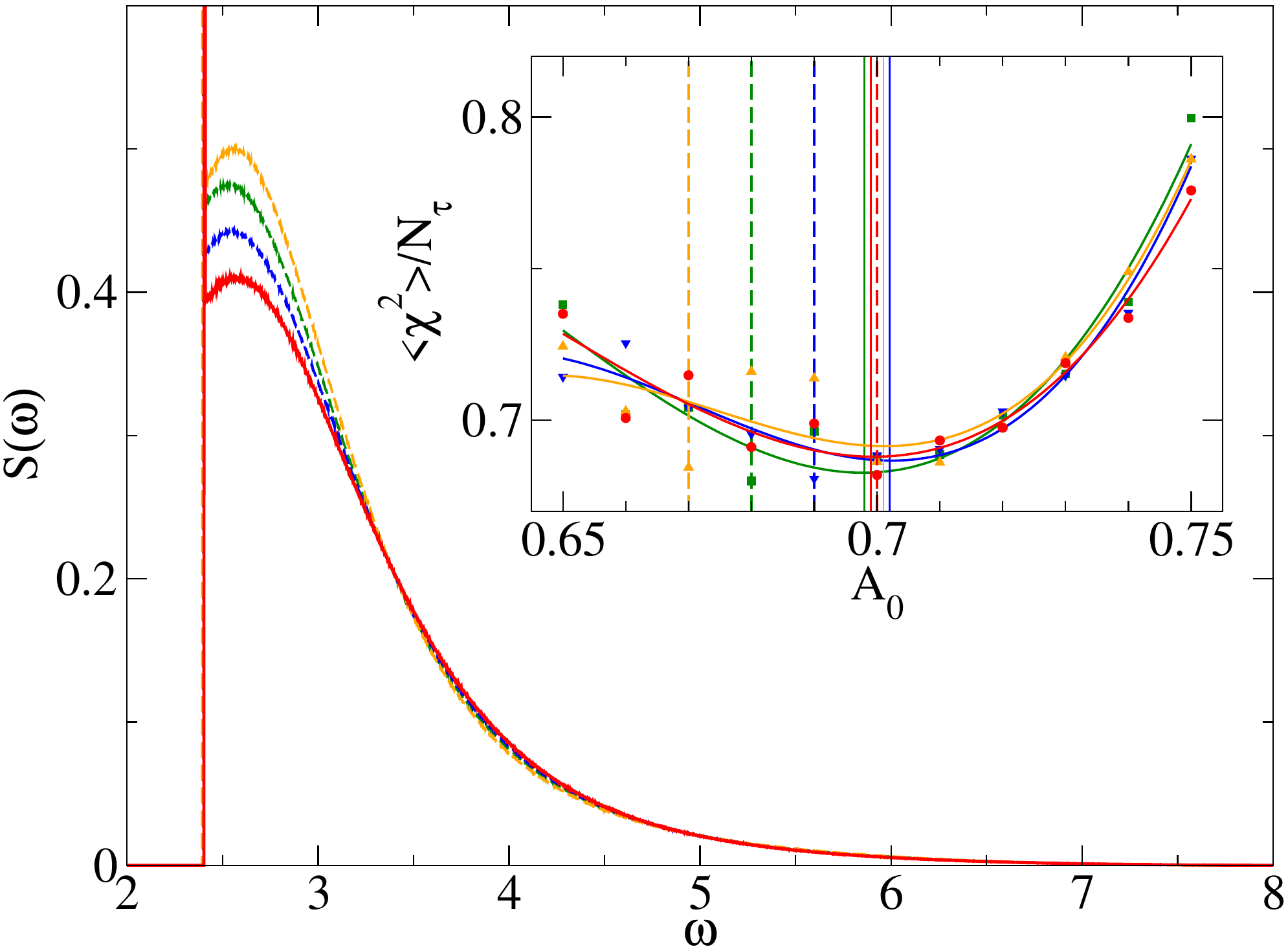}
\caption{Dynamic structure factor of the 2D Heisenberg model with $32 \times 32$ spins at $q=(\pi/2,\pi/2)$, determined using SAC with leading
  $\delta$-peak. The inset shows how $\langle\chi^2\rangle$ varies in four different scans over $A_0$, with cubic polynomial fits used to extract
  the optimal value at the $\langle \chi^2\rangle$ minimum; in all cases here $A_0 \approx 0.70$ as indicated by the solid colored lines matching
  the fitted curves. For clarity we have not plotted error bars on the data points; they are of order $0.01$ below $A_0=0.7$ and smaller for larger
  $A_0$. The dashed lines indicate other $A_0$ values at which the spectra in the main graph were sampled, shown with the same color coding.}
\label{2dsw}
\end{figure}

As an example, we here consider the dynamic structure factor of the 2D Heisenberg model at wave-vector $q=(\pi/2,\pi/2)$, with $\bar G(\tau)$ computed
with the SSE QMC method for a system with $32 \times 32$ spins at a temperature low enough to give ground state results in practice
(see Ref.~\cite{shao17} for further technical details). In this case we used a quadratic $\tau$ grid, similar to Fig.~\ref{fig:gtau}, from
$\tau=0.01$ to $\tau=3.61$. The error level was $\sigma \approx 10^{-5}$. To illustrate the stability of the optimization procedure, in the inset of
Fig.~\ref{2dsw} we show results of several independent scans of $A_0$ in the neighborhood of the shallow minimum produced in this case. Because of
fluctuations in the individual points, the minimum is best located by carrying out a polynomial fit (here of third order) to several data points and
extracting the minimum from the fitted function. The minima extracted in the four independent runs in Fig.~\ref{2dsw} are all very close to each
other at $A_0 \approx 0.70$, and of course the procedure can be further improved by sampling longer and using a finer grid of $A_0$ points. Here
we carried out short runs on a coarse grid in order to better illustrate the fluctuations and interpolation method.

A spectral function obtained with $A_0$ close to its optimal value is shown in red in Fig.~\ref{2dsw}; the results at the four different estimated
minima are all very close to this curve and are not shown separately. For reference, we also show results of sampling
with $A_0$ away from its estimated optimal value. As seen in the inset of the figure, for $A_0$ larger than the optimum the goodness-of-fit deteriorates
rapidly and the statistical fluctuations are small (which follows from the fact that $A_0$ over-constrains the spectrum here). Therefore, it is more likely
to underestimate $A_0$ with this method (as we will also discuss in more detail below in Sec.~\ref{sec:delta2}), and we show examples of suboptimal spectra
in this region. As expected, the spectral weight close to the $\delta$-edge increases as $A_0$ is reduced, but other parts of the continuum are little affected.

\begin{figure*}[t]
\includegraphics[width=150mm]{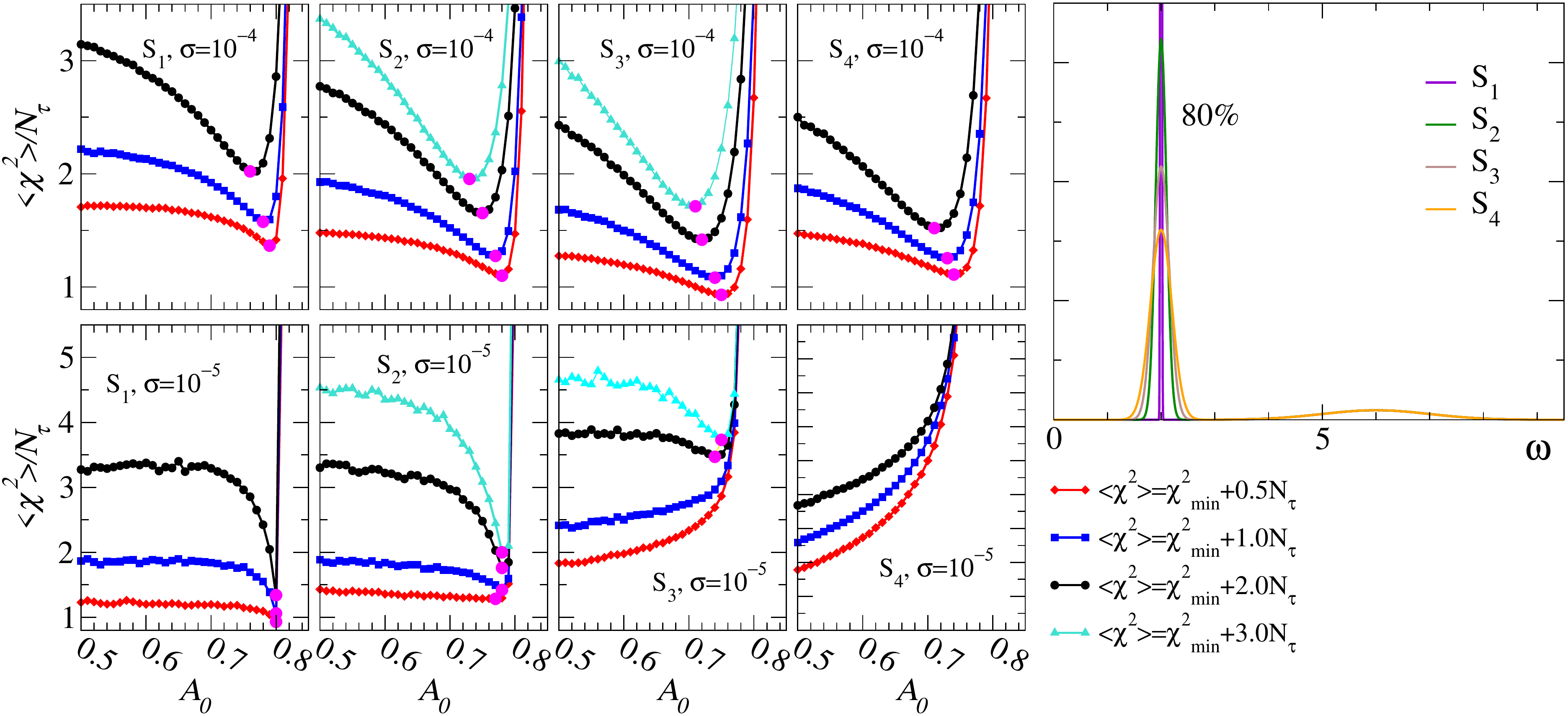}
\caption{Demonstration of the sensitivity of the optimal leading $\delta$-function amplitude to the error level $\sigma$ (with two cases, $\sigma=10^{-4}$
  and $\sigma=10^{-5}$ as indicated) and the sampling temperature $\Theta$. The results also demonstrate the characteristic signals of a broadened
  edge peak. The synthetic spectral functions with peaks of different widths, labeled $S_1,\ldots,S_4$ in order of increasing width (all with $80\%$
  of the weight in the edge peak) are shown in the rightmost panel. The other panels show $\langle \chi^2\rangle/N_\tau$ versus $A_0$ in SAC runs with a
  $\delta$-function edge, sampled at different $\Theta$ values corresponding to the initial unrestricted $\langle \chi^2\rangle$ values in the way
  stated in legends to the right. For $S_4$ with $\sigma=10^{-5}$ the $\langle \chi^2\rangle$ minimum is outside the figure, at $A_0=0$ for all values of
  $a$, and this is the case also for $S_3$ with $a=0.5$ and $1$, thus indicating that the data quality is good enough (and $a$ is low enough) to determine
  that the true spectra do not host a $\delta$-function edge.}
\label{broadened-1}
\end{figure*}

We can in principle also estimate error bars on $A_0$ by a bootstrapping approach, with optimized $A_0$ based on several random samples of the data set
(instead of repeating the optimization steps with the same data sets, as we did for Fig.~\ref{2dsw}). As will be discussed in Sec.~\ref{sec:delta2}, the
(rather small) dependence of the result on the value of $\Theta$ in the optimization step should also be considered for such a procedure to produce correct error
bars. It should also be noted, however, that statistical errors are not very meaningful, as there are always some systematic errors from various entropic pressures
as well. For example, as we will discuss further below in Sec.~\ref{sec:delta2}, $A_0 > 0$ extracted from the $\langle \chi^2\rangle$ minimum is typically somewhat
below the correct value.

\subsection{Detecting a finite peak width}
\label{sec:delta2}

Spectral functions with $\delta$-peaks, or extremely narrow peaks, at the lower edge are very common, e.g., spin wave excitation in magnetically
ordered states as in the example above. However, these peaks become broadened due to finite-temperature effects, and in most cases there would be
at least some broadening also at $T=0$,  due to the quasi-particle not being an exact eigenstate of the Hamiltonian. Nevertheless, the spectral
information carried by the quasi-particle  $\delta$-peak is often the most prominent and important feature of the spectrum.

In principle, the method described above for optimizing a $\delta$-edge can be extended to also optimizing its width---there should be a minimum in
$\langle \chi^2\rangle$ in the 2D space of amplitude and width of the peak. We will explore such an  approach in Sec.~\ref{sec:deltanp}.
Here we first show how a finite peak width can be detected within the framework of the $\delta$-edge parametrization. We will show this while also
providing further insights into the existence of the $\langle \chi^2\rangle$ minimum; how it evolves with the sampling temperature $\Theta$ and the data quality.

As mentioned in the previous section, to see a pronounced minimum value of $\langle \chi^2\rangle$ in the scan over $A_0$, $\Theta$ should be
chosen slightly above the value at which the free sampling ($A_0=0$) satisfies the criterion in Eq.~(\ref{eq:chi2}). At the minimum, $\langle \chi^2\rangle$
may then still satisfy Eq.~(\ref{eq:chi2}) with $a\lesssim 1$, but this cannot be guaranteed. It is therefore useful to carry out the optimization process at gradually
lower values of $\Theta$ and monitor the evolution of $A_0$ at the $\langle \chi^2\rangle$ minimum. In general, one should expect lower $\Theta$ to
produce better results, to the extent that a minimum can be clearly identified (which of course is easier with better data quality).

Here we will demonstrate that the scanning procedure can also signal the inapplicability of the $\delta$-edge when, at a given error level of the
$\bar G(\tau)$ data, there is enough broadening of the peak for our statistical criteria to signal deviations from the $\delta$-function.
We carry out $A_0$ scans with four different imaginary-time data sets, obtained from synthetic spectra $S_1$, $S_2$, $S_3$, and $S_4$, all consisting
of a dominant Gaussian centered at $\omega=2$, containing $80\%$ of the spectral weight, and a second broader Gaussian centered at $\omega=6$ and truncated
at $\omega=2$ (see the rightmost panel of Fig.~\ref{broadened-1}). The dominant Gaussian has width (standard deviation) $0.01$, $0.1$, $0.15$, and $0.2$,
in $S_1,\ldots,S_4$, respectively. For all these cases we use the same SAC parametrization with a $\delta$-edge, which is strictly not correct since all
the spectra have a finite-width peak. We set the inverse temperature to $\beta=32$ when converting the spectral functions to imaginary-time
functions $G(\tau)$ according to Eq.~(\ref{contrel1}), with $\Delta_\tau=0.1$ and noise added at levels $\sigma=10^{-4}$ and $10^{-5}$. The
maximum $\tau$ values were $\tau_{\rm max}=3.0$ and $4.2$, respectively.

\begin{figure*}[t]
\centering
\includegraphics[width=100mm]{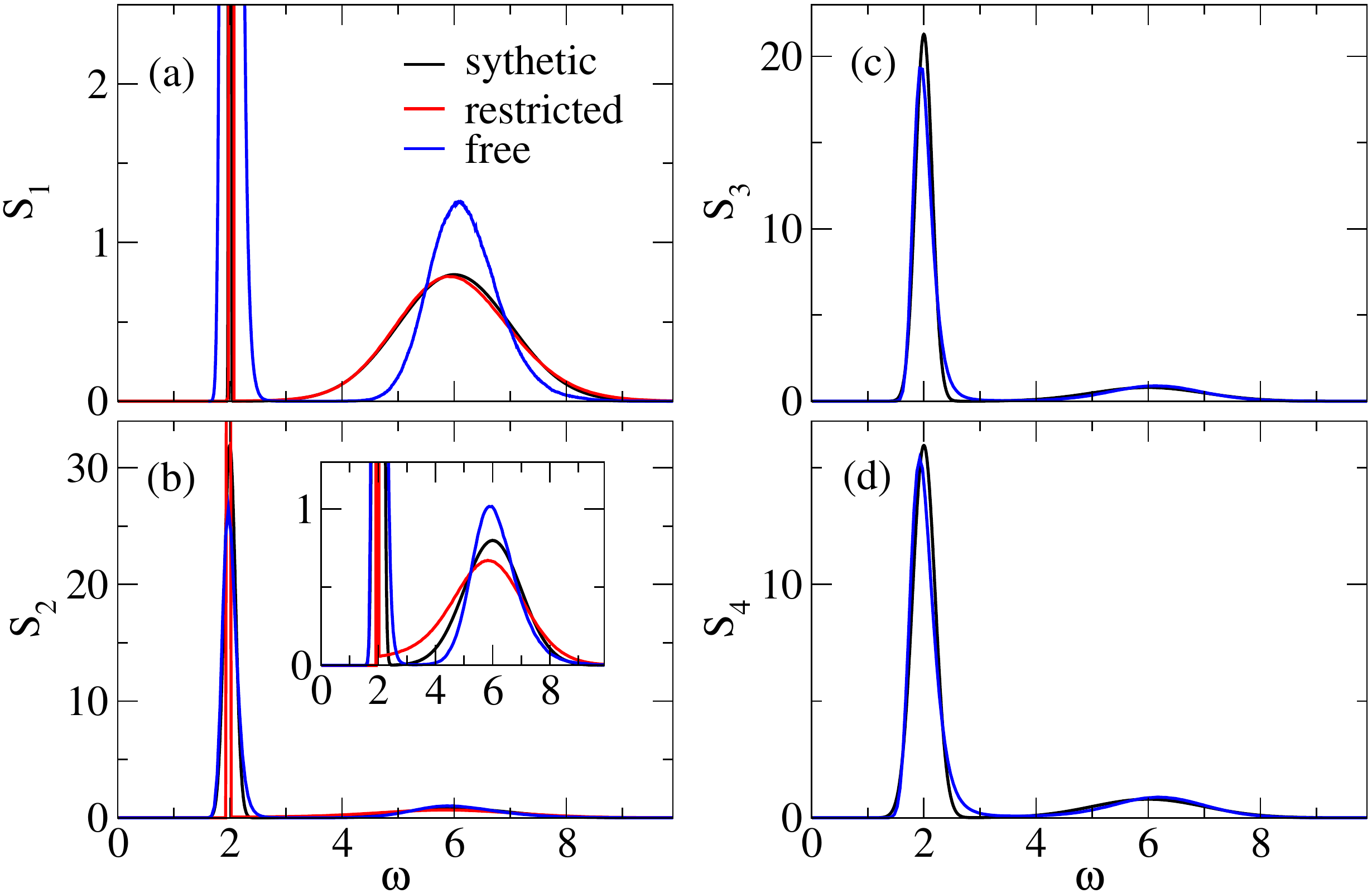}
\caption{The synthetic spectral functions $S_1,\ldots,S_4$, used for the tests in Fig.~\ref{broadened-1}, are shown as black curves in (a)--(d). These
  spectra are compared to SAC results at the corresponding optimal $A_0$ values at error level $\sigma=10^{-5}$ (red curves). In the cases $S_3$ and $S_4$,
  where the optimal weight is $A_0=0$ for $a=0.5$ in Fig.~\ref{broadened-1}, only the results of the equivalent unrestricted (free) sampling are shown
  (blue curves), while in the other cases both free-sampling and optimal-$A_0$ (determined at $a=0.5$) results are shown.}
\label{broadened-1-sw}
\end{figure*}

We carried out systematic scans over $A_0$ at three or four different sampling temperatures, fixing $\Theta$ such that
$\langle \chi^2\rangle = \chi^2_{\rm min} + aN_\tau$ at $A_0=0$ for different values of the factor $a$. Since the location of the edge-$\delta$ is not
fixed in the SAC procedure, there is some broadening of the peak in the mean sampled density accumulated in the histogram when $A_0 > 0$.
However, this peak width is very small
when the weight of the $\delta$-function is as large as it is in this example. The question we want to address here is how the finite width of the leading peak
of the synthetic spectrum is reflected in the functional form of $\langle \chi^2\rangle$ versus $A_0$, and how this form changes when the sampling temperature
and the data quality (noise level) are varied.

The test results are collected in Fig.~\ref{broadened-1}, where the top and bottom rows of panels correspond to the two error levels 
and the four panels in each row show $\langle \chi^2\rangle/N_\tau$ versus $A_0$ for the spectra
$S_1,\ldots,S_4$. The different data sets correspond to the factors $a=0.5,1,2,3$ used when fixing $\Theta$ so that
$\langle \chi^2\rangle = \chi^2_{\rm min} + aN_\tau$ at $A_0=0$. It is clear that sampling at a higher value of $\Theta$ (larger $a$) has a similar
effect as a higher noise level, since in either case the function $G(\tau)$ corresponding to the sampled spectra is pushed
further from its true value.
This intuitive understanding is confirmed by Fig.~\ref{broadened-1}, where for $\sigma=10^{-4}$ we see clearly that increasing $a$ leads to a sharper
$\langle \chi^2\rangle$ minimum, but if $a$ is large the minimum value of $\langle \chi^2\rangle$ is too high for a statistically sound fit. Moreover,
the value of $A_0$ at the minimum is consistently too small ($<0.8$) but shifts toward $0.8$ as $a$ is decreased. The $A_0$ value also moves
closer to $0.8$ when the error level is decreased from $\sigma=10^{-4}$ to $10^{-5}$ in the case of $S_1$ (the spectrum with the narrowest peak).

If the true spectrum indeed has a $\delta$-edge or a very narrow peak, a good fit cannot be obtained when $A_0$ is larger than its correct value
$A_{0,{\rm true}}$. Too small a value of $A_0$ can still in principle be compensated for by the distribution of the other sampled $\delta$-functions,
but entropic broadening leads to a suboptimal fit (though not to the same extent as does severe overconstraining). Thus, for small $\sigma$ one can expect
a minimum with a typically sharp increase for $A_0 > A_{0,{\rm true}}$ and a less dramatic increase for $A_0 < A_{0,{\rm true}}$. As $a$ is
reduced, the feature for $A_0 < A_{0,{\rm true}}$ should become more shallow, as the starting $A_0$ value of $\langle \chi^2\rangle$ is closer to
the minimum value $\chi^2_{\rm min}$, while for $A_0 > A_{0,{\rm true}}$ an even sharper increase should be
expected because of the higher sensitivity to excessive weight in the $\delta$-function. When $\sigma$ is decreased, both sides of the minimum should become
sharper, though the variations on the $A_0 < A_{0,{\rm true}}$ side are still limited by the fixing of $\Theta$ using $\langle \chi^2\rangle$ at $A_0=0$
(or some other value of $A_0$ well below the optimal value). While the above behavior is expected strictly for a $\delta$-edge, clearly a very narrow
peak will result in the same kind of features, as observed at $\sigma=10^{-5}$ in Fig.~\ref{broadened-1}, unless the error level $\sigma$ is extremely low.

\begin{figure*}[t]
\centering
\includegraphics[width=120mm]{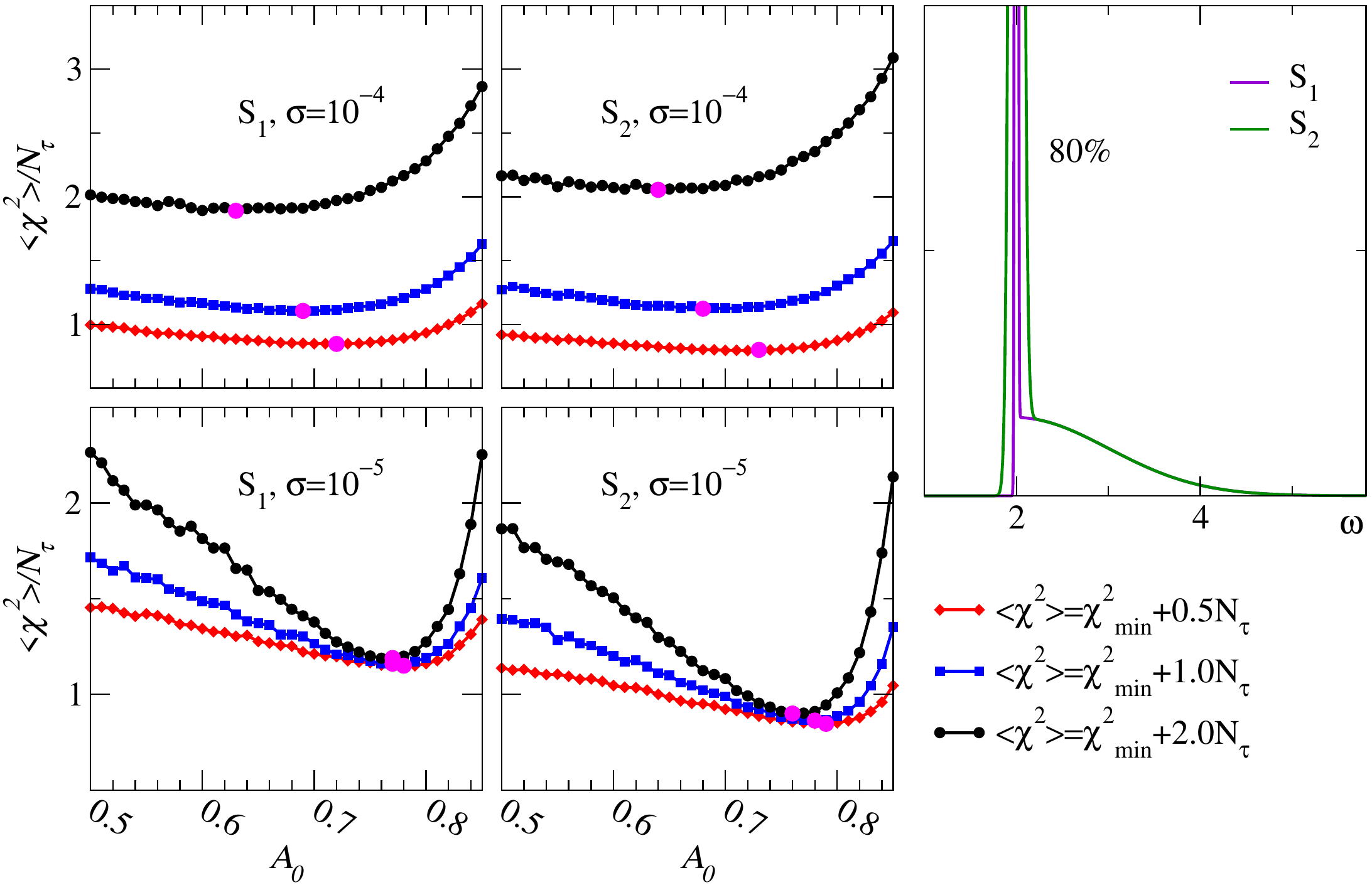}
\caption{Tests similar to those in Fig.~\ref{broadened-1} but for spectral functions (shown in the rightmost panel) where the continuum is
connected to the dominant peak. The other panels show $\langle \chi^2\rangle$ versus $A_0$ in scans with different values of the factor
$a$ used in fixing $\theta$ and two different error levels as in Fig.~\ref{broadened-1}.}
\label{broadened-2}
\end{figure*}

Turning now to cases where deviations from the $\delta$-edge shape can be detected, for $S_4$ (the broadest peak),
the optimal value of $A_0$ is $0$ (outside the graph
boundary) for all cases of $a$ in Fig.~\ref{broadened-1} when the error level is $\sigma^{-5}$, while at $\sigma=10^{-4}$ there is still a clear minimum
for $A_0$ between $0.7$ and $0.8$. Thus, with the better data quality the method can correctly detect the inapplicability of the parametrization with
$A_0>0$. In this case, the final sampling is equivalent to the unrestricted SAC, and, as shown in Fig.~\ref{broadened-1-sw}, the resulting spectral
function is very close to the correct (synthetic) spectrum.

In the case of $S_3$, the scans for error level $\sigma=10^{-5}$ with $a=2$ and $a=4$ in Fig.~\ref{broadened-1} produce minima, but the optimal $A_0$
value is smaller for $a=2$ than for $a=4$, and for $a=1$ and $a=0.5$ the minimum is again at $A_0=0$. This case shows clearly how choosing a larger $a$
has a similar effect as a larger error level $\sigma$, so that for $a=2$ and $a=4$ there is an apparent optimal $A_0>0$, as also for all values of $a$ in the case
$\sigma=10^{-4}$. It is only when the fit becomes tight enough, for sufficiently small $a$ and $\sigma$, that the procedure becomes sensitive to the
finite peak width and results in unrestricted sampling ($A_0=0$) being optimal.

In general, for this type of spectral function, once the inapplicability of the $\delta$-peak can be unambiguously detected, the unrestricted SAC produces
spectra that match reasonably well the correct profile, as observed in Fig.~\ref{broadened-1-sw} for both $S_3$ and $S_4$. In the case of $S_2$, the behavior
of the $\langle \chi^2\rangle$ minimum for $\sigma=10^{-5}$ in Fig.~\ref{broadened-1} also hints at the absence of a strict $\delta$-edge, since the
optimal $A_0$ value moves slightly to the left when $a$ is decreased and the minimum is very shallow for $a=0.5$. However, the unrestricted
sampling cannot resolve the correct peak width of $S_2$ at the levels of errors considered here, as shown in Fig.~\ref{broadened-1-sw}.

We next consider a harder case, where a sharp peak which still contains 80\% of the spectral weight is followed by a broad ``half Gaussian''
continuum, as shown in the rightmost panel of Fig.~\ref{broadened-2}. The peaks in the two synthetic spectral functions $S_1$ and $S_2$ are Gaussians of
width $0.01$ and $0.1$, respectively. We have again used a linear $\tau$ grid with $\Delta_\tau=0.1$ and set the inverse temperature to $\beta=32$.
The four other panels of Fig.~\ref{broadened-2} show $A_0$ scans in the same kind of arrangement as previously in Fig.~\ref{broadened-1}.

As we already discussed in Sec.~\ref{sec:delta1}, when there is no clear separation between the edge peak and the continuum the optimization of $A_0$
is more challenging. In Fig.~\ref{broadened-2}, the very shallow minima seen in the results for error level $\sigma=10^{-4}$ reflect this difficulty.
It is still clear that the location of the minimum moves toward the correct value $A_0=0.8$ as the factor $a$ is reduced. With the higher data
quality, $\sigma=10^{-5}$, the minima are quite sharp and better converged to $A_0\approx 0.8$. In these tests the two synthetic spectra both
have narrow Gaussian edge peaks and the data quality is not sufficient to detect the finite width. With increasing data quality we expect
behaviors similar to those seen in Fig.~\ref{broadened-1} when the optimal $A_0$ eventually starts to move toward $0$.

\begin{figure*}[t]
\centering
\includegraphics[width=110mm]{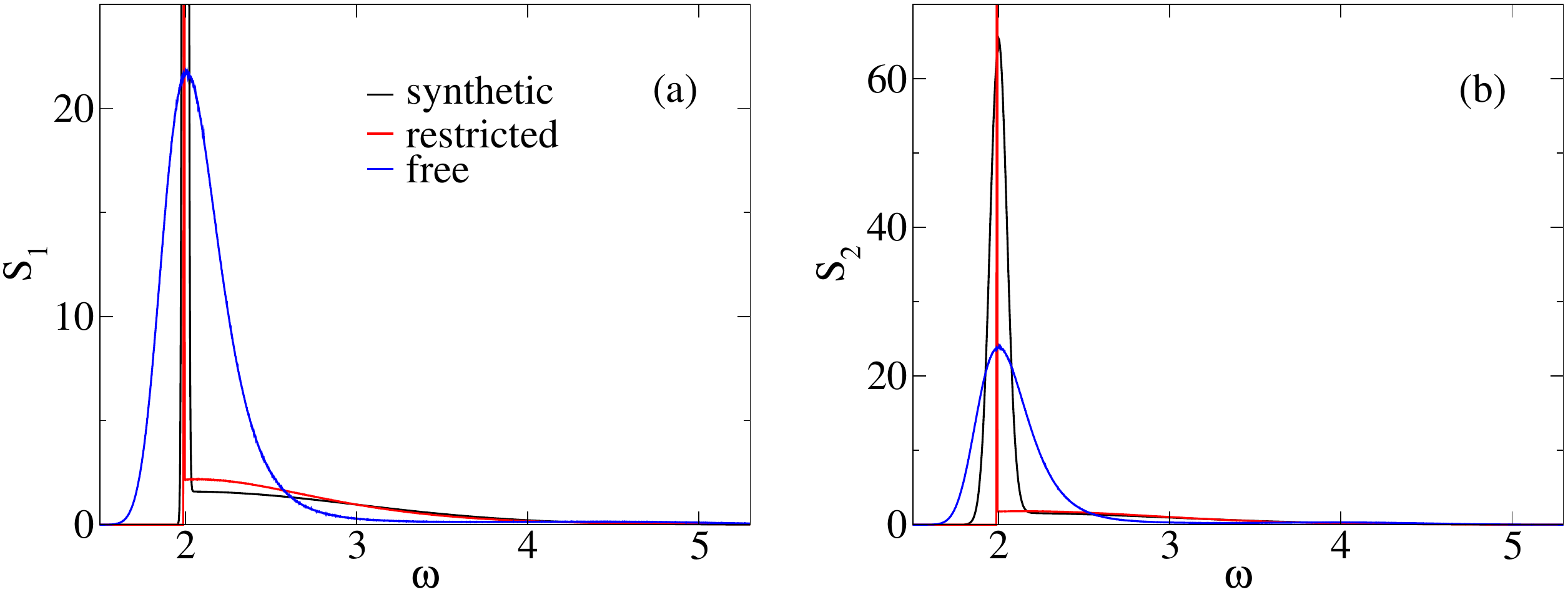}
\caption{The two synthetic spectral functions $S_1$ and $S_2$ (black curves) used in the tests in Fig.~\ref{broadened-2} compared with SAC results
at optimal $A_0$ (determined at $a=0.5$ in Fig.~\ref{broadened-2}) in those tests (red curves) as well as results of unrestricted sampling (blue).}
\label{broadened-2-sw}
\end{figure*}

We show results for the above two synthetic spectral functions obtained with the data of error level $10^{-5}$ in Fig.~\ref{broadened-2-sw}. 
With unrestricted sampling, the leading peak is much too broad in both cases and the continuum is poorly reproduced. With the $\delta$-edge,
the peak is too narrow but the continuum is much closer to the correct shape except very close to the peak. Thus, while the edge is not perfect,
the incorporation of a sharp peak still improves the resolution at higher frequency, because of the absence of distorting effects of spectral
weight below the peak when the sampling is unrestricted.

\subsection{Broadened quasi-particle peaks}
\label{sec:deltanp}

A macroscopic $\delta$-function at the lower edge of a spectrum is an extreme case of a quasi-particle peak. Typically such a peak has some finite width
due to decay processes at $T=0$ or $T>0$. It would clearly be of great value if also the width of a narrow peak could be reliably determined. As we
saw above, our method of optimizing the weight of the $\delta$-peak can also in principle, if the $\bar G(\tau)$ data are good enough, detect the unsuitability of
this imposed form if the peak actually has finite width. If the peak is broad enough, it can be reproduced by unrestricted sampling, and of course
the better the imaginary-time data the narrower the peak that can be properly resolved.

Since the location $\omega_0$ of the leading $\delta$-edge in the parametrization, Fig.~\ref{fig:spec}(d), is also sampled, one might imagine that
its fluctuations could result in a broadening of the peak collected in a histogram, and that the profile would then correspond to the actual width of a
narrow quasi-particle peak when it is not strictly a $\delta$-function. In practice, as observed in the tests in Sec.~\ref{sec:delta2}, the fluctuation
broadened peak always comes out too narrow, however (unless the true peak is extremely narrow). This lack of fluctuations fundamentally follows from the
entropic downward pressure on the single macroscopic $\delta$-peak from the microscopic continuum contributions. In the limit $N_\omega \to \infty$, these
pressures can be expected to completely localize the edge-$\delta$ at a frequency where the entropy is best balanced by $\langle \chi^2\rangle$. The
diminishing fluctuations of the SAC spectrum overall when $N_\omega \to \infty$ is discussed in detail in Sec.~\ref{sec:maxent} and \ref{app:statmech}.

The inability of the single macroscopic $\delta$-function to model a quasi-particle peak of finite width can be understood in view of the composition of
the individual sampled spectral configurations: If the true peak has significant width, then each individual configuration should be able to reproduce
this width, at least approximately. The single macroscopic $\delta$-function clearly does not allow such flexibility, given that the continuum contributions
are spread out by entropic pressures and cannot concentrate sufficiently within the peak region, unless the peak is very broad.

There are potentially many ways to construct suitable parametrizations generalizing the $\delta$-edge in Fig.~\ref{fig:spec}(d) to a finite peak width.
For example, we could replace the isolated $\delta$-function by a Gaussian, whose width $\sigma_0$ is also optimized. The SAC would then involve a scan of
$\langle \chi^2\rangle$ to optimize the peak in the two-dimensional parameter space $(A_0,\sigma_0)$. The way in which the Gaussian is merged with the
microscopic $\delta$-functions belonging to the continuum then also has to be specified. One option would be to let the center of the Gaussian serve as
the lower bound of the $\delta$-functions. Of course the real peak may not have a Gaussian shape, and imposing this form from the outset is then not ideal.

\subsubsection{Multi-$\delta$ peak parametrization}

\begin{figure}[t]
\centering
\includegraphics[width=70mm]{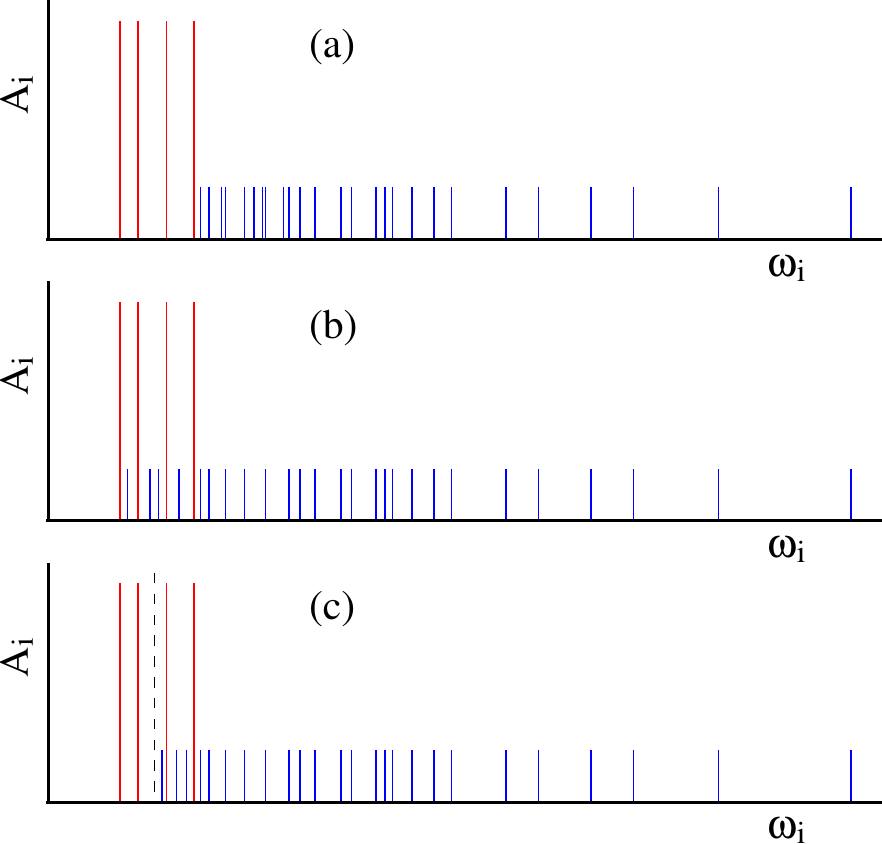}
\caption{Parametrizations of the spectrum in terms of two sets of $\delta$-functions, with $N_\omega = N_p + N_c$. The set of $N_p$ contributions at the
lower end of the spectrum, with amplitudes $A_0/N_p$, are intended to model the quasi-particle peak (here $N_p=4$, shown in red), while the remaining
$N_c$ $\delta$-functions (blue) with amplitudes $(1-A_0)/N_c$ will account primarily for an arbitrary continuum. Typically $N_p \ll N_c$. In (a) and (b),
the highest and lowest, respectively, of the peak frequencies acts as the lower bound for the continuum contributions, while in (c) the lower bound
is given by the first moment of the peak (vertical dashed line).}
\label{peakfig}
\end{figure}

We here proceed with a different generalization of the $\delta$-edge, illustrated in Fig.~\ref{peakfig}, using a relatively small number $N_p>1$ of $\delta$-functions
to model the peak, each of these contributions having amplitude $A_0/N_p$. For the continuum, we use $N_c$ contributions, each of weight $(1-A_0)/N_c$, for a total
of $N_\omega=N_p+N_c$ units of spectral weight. We again wish to model a relatively narrow peak located at the lower frequency edge and anticipate that
$N_p \ll N_c$, so that the entropic pressures broadening the leading peak will be relatively low. We will devise methods for optimizing the peak by
scanning over $A_0$ and $N_p$ for given $N_c$.

Constraining the continuum contributions in some way by a lower bound associated with the peak location will ensure that the continuum cannot migrate below
the peak. Such a constraint, depending on exactly how it is implemented, can also
impose entropic pressures that squeeze the peak from above and prohibit the individual pieces to migrate too much away from each other into the continuum.
However, it may also be beneficial to allow the continuum to at least partially penetrate down among the main peak contributions.

Figure \ref{peakfig} illustrates three slightly different constraining mechanisms regulating the merger of the peak and the continuum.
In Fig.~\ref{peakfig}(a), the highest of the peak $\delta$-functions serves as the lower bound for the continuum. Given that $N_p \ll N_c$,
the entire peak is pushed down in frequency by the entropy of the continuum. This entropic squeezing of the peak can become excessive
when $N_c$ is large, leading to a very abrupt boundary between the peak and the continuum; an asymmetric averaged peak that
is rounded at the lower edge but very sharp at the upper edge. We therefore do not use this constraint in practice, though we will show
some test results below.

The opposite extreme is represented by the constraint illustrated in Fig.~\ref{peakfig}(b), where the lowest peak frequency acts as
the constraining lower bound for the continuum contributions.
Except for the lowest $\delta$-function, the peak elements are not explicitly constrained in their migration up in frequency, which in some cases
causes them to split off from the lower edge and cause a double peak in the averaged spectrum.

A compromise between the two extreme constraints is shown in
Fig.~\ref{peakfig}(c), where the lower continuum bound is given by the mean frequency of the peak $\delta$-functions, thus, the overall peak location is
pushed down by entropic pressures but the peak width can fluctuate without explicit entropic coupling to the continuum.

Both the low-edge contraint in Fig.~\ref{peakfig}(b) and the peak-center constraint
in Fig.~\ref{peakfig}(c) allow the continuum to extend down into the peak region, which further contributes flexibility to the peak shape. While the
parametrizations with constraints discussed here do not assume that the true peak has a certain shape, as always there are implicit entropic pressures
that to some extent will be reflected in the average spectrum---less so for better data quality, so that an arbitrary peak shape can, in principle,
emerge in the limit of very small error bars on $\bar G(\tau)$. We will demonstrate that narrow quasi-particle peaks can be reproduced at a fidelity not
practically attainable with unconstrained SAC.

As indicated above, the $\delta$-functions within the peak and continuum groups will have fixed amplitudes in the tests reported here.
Varying amplitudes can of course also be used to generalize the parametrizations in Fig.~\ref{peakfig}. However, our tests on synthetic
spectral functions show that better resolution of the peak, in particular, is obtained when sampling only the frequencies and keeping fixed amplitudes.
The entropic pressures when the peak amplitudes are sampled along with the frequencies push the spectrum to develop an excessively sharp peak, which is
consistent with the tendencies found in Sec.~\ref{sec:theta} but is more pronounced here in typical cases when $N_p$ is small. It is still possible that
varying amplitudes could be advantageous in some cases.

\subsubsection{Peak optimization}

For given $N_c$, sufficiently large to model the continuum (depending on the QMC data quality), we have to  optimize $\langle \chi^2\rangle$ over the 2D
space $(A_0,N_p)$. We will later see that the task can be simplified by performing only a small number of line scans through limited portions of the space, but we
begin by demonstrating that a well-defined minimum can indeed be identified at which the average spectrum represents a close to optimal reproduction of the
correct shape. In these tests, we exclusively use synthetic spectral functions.

Sampling of the $\delta$-functions can be carried out in the same way as described in Sec.~\ref{sec:contsamp}, but with the two- and three-frequency moves
only involving $\delta$-functions within either the quasi-particle part or the continuum part. Moves violating the mutual peak-continuum constraint are of
course rejected. The three-frequency moves significantly improve the ergodicity of the peak part when $N_p \ge 3$, and without such moves there
are some times signs of meta-stability.\footnote{Meta-stability is reflected in the average spectral function changing when three-frequency updates are
included. We have not encountered meta-stability with in any of the parametrizations used in other sections, though the sampling efficiency can
often be improved significantly by including three-frequency updates.} It is easy to incorporate the constraints of Fig.~\ref{peakfig}.
Since moves involving more than one frequency conserve the first frequency moment, the lower bound of the continuum with the
constraint in Fig.~\ref{peakfig}(c) only changes in the single-frequency updates of the peak. The maximum attempted movement of the $\delta$-functions in
the single- and two-frequency updates should be adjusted separately for the peak and continuum groups, to keep the acceptance rate close to $0.5$ in either case.

\begin{figure*}[t]
\centering
\includegraphics[width=120mm]{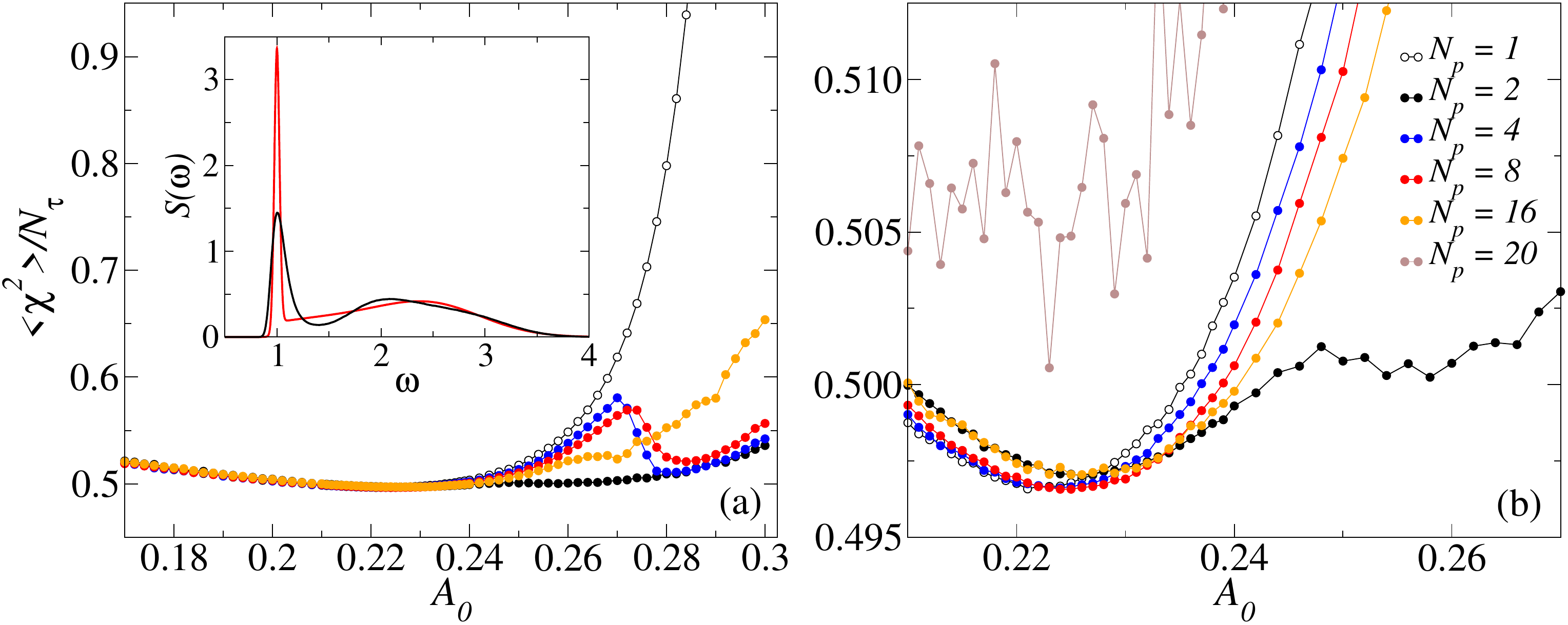}
\caption{Mean goodness-of-fit versus the peak weight $A_0$ obtained with the parametrization in Fig.~\ref{peakfig}(b) for the synthetic spectrum shown with
the red curve in the inset of (a), where the SAC result obtained with unrestricted frequency and amplitude sampling is also shown (black curve, obtained with
$N_\omega=5000$ $\delta$-functions at $\Theta=0.02$, where  $\langle \chi^2\rangle/N_\tau \approx 0.498$). The noise level of the imaginary time data
(53 points uniformly spaced at $\Delta_\tau=0.2$) was $10^{-6}$. Results for fixed $N_c=500$ and $N_p=1,2,4,8$ and $16$ are shown in (a), and in (b)
details of the same data are shown in the neighborhood of the minimum with matching symbol colors. In (b) results are also shown for $N_p=20$, where
$\langle \chi^2\rangle$ overall is elevated and the statistical noise is much higher. Corresponding spectral functions, displayed in Fig.~\ref{swnc500},
show that the qualitative change in behavior is caused by the main peak splitting above $N_p=18$. The sampling temperature in all cases was $\Theta=0.5$.
Unrestricted sampling gave $\chi_{\rm min}/N_\tau \approx 0.43$ and the same value can almost be reached also with all $N_p$
used here when $\Theta \to 0$.}
\label{x2np}
\end{figure*}

To test the method, we first consider a synthetic spectrum consisting of three Gaussians, with the lowest one much sharper than the other two and with the
higher ones exponentially damped for $\omega$ below the center of the sharp peak---similar to the spectrum considered before, e.g., in Fig.~\ref{syntcomp},
but now with a much narrower peak that cannot be reproduced by unrestricted SAC for reasonable imaginary-time data quality. The synthetic spectrum is shown in the
inset of Fig.~\ref{x2np}(a) along with the result of unrestricted (free) sampling. The peak is clearly too narrow to be properly resolved by free sampling even 
at the low error level $\sigma = 10^{-6}$ of the data used here.

We first consider the constraint of Fig.~\ref{peakfig}(b), where the continuum can extend down to the lowest frequency of the peak.
Fig.~\ref{x2np}(a) shows scans over $A_0$ with $N_c=500$ for several different choices of $N_p$ up to $N_p=16$. In all cases, a shallow minimum in
$\langle \chi^2\rangle$ is observed, which is shown in more detail in Fig.~\ref{x2np}(b). The location of the minimum is only weakly dependent 
on $N_p$, and the minimum is typically followed by a maximum at higher $A_0$. A significant change between $N_p=16$ and $N_p=20$ is seen in
Fig.~\ref{x2np}(b), with the overall goodness of the fit deteriorating significantly (relatively speaking) for $N_p=20$ and the statistical
fluctuations of $\langle \chi^2\rangle/N_\tau$ also increasing dramatically, as evidenced by the significant scattering of the data points.

\begin{figure*}[t]
\includegraphics[width=150mm]{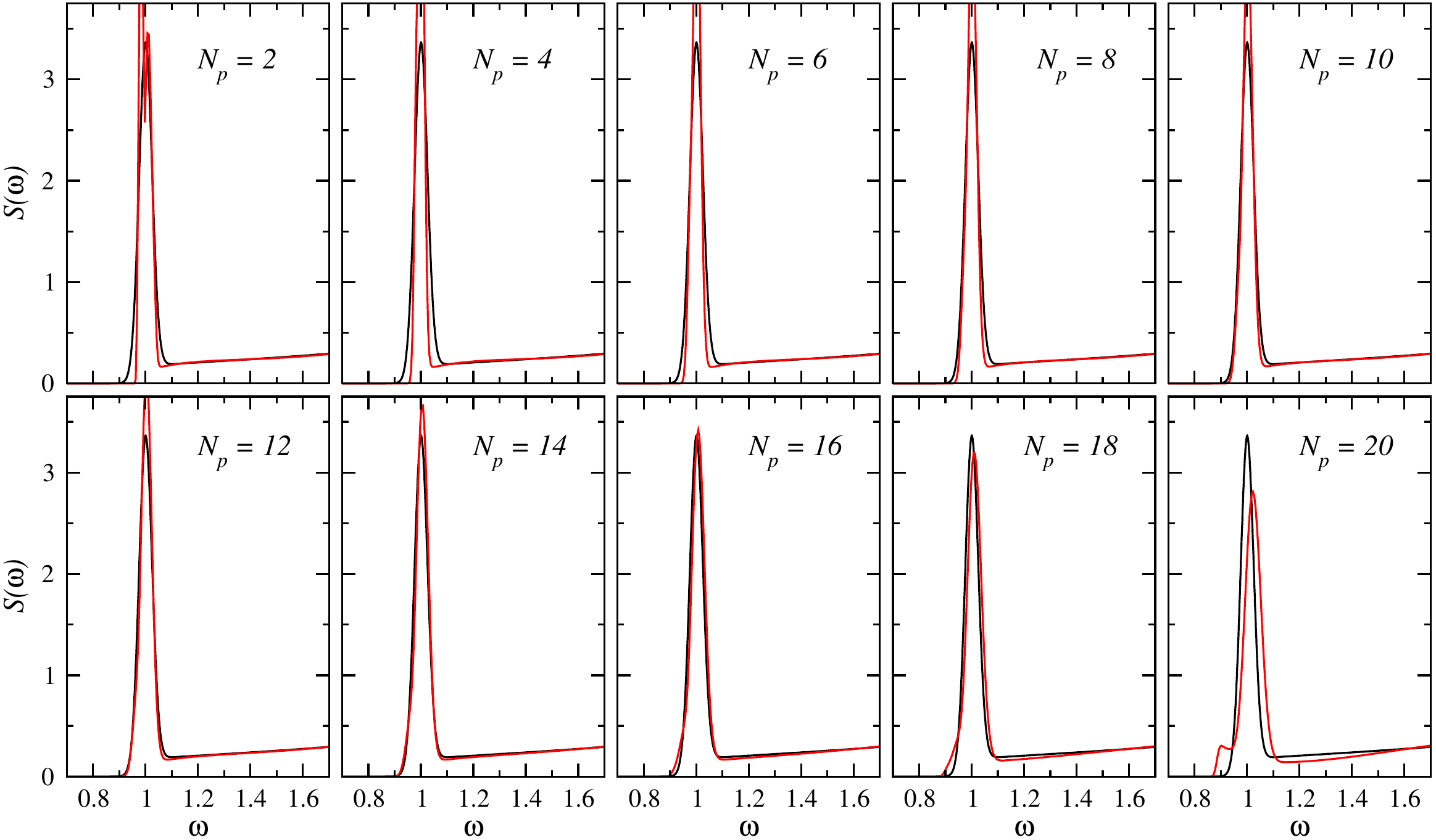}
\caption{SAC spectral functions (red curves) obtained with $N_p$ ranging from $2$ to $20$ in steps of $2$, compared with the exact synthetic spectrum (black
curves). The continuum is constrained by the lowest peak $\delta$-function, as illustrated in Fig.~\ref{peakfig}(b), with $A_0$ chosen according to the
$\langle \chi^2\rangle$ minima shown for selected cases of $N_p$ in Fig.~\ref{x2np}.}  
\label{swnc500}
\end{figure*}

The different behavior of the goodness of the fit when $N_p=20$ can be traced to a qualitative change in the average spectral function, which is shown
in Fig.~\ref{swnc500} for all even values of $N_p$ between $2$ and $20$, with $A_0$ fixed in each case at the respective $\langle \chi^2\rangle$ minimum.
The peak part of the spectrum has a single maximum up to $N_p=18$, while a small secondary peak at the lower edge has emerged for $N_p=20$. A precursor
of the splitting in the form of broadening at the lower edge can be seen already for $N_p=18$.

In general, peak splitting with the low-edge constraint is a consequence of competition between
different entropies: the continuum exerts a downward pressure on its constraining lower bound, i.e., the lowest $\delta$-function in the peak group.
The other elements of the peak are not subject to any such pressure but tend to spread out to higher frequencies by their own entropy. When $N_p$
is small and the total peak weight $A_0$ is well below the optimal value, the peak must be localized in a narrow region in order to keep $\chi^2$ low.
When $A_0$ increases, the weight at the lower edge, which is pushed down by the entropy of the continuum, becomes too high and it is then
advantageous for the rest of the peak to migrate up. This upward shift is further amplified by the entropy of the peak group when $N_p$ increases.
Eventually, these effects all conspire to cause the observed peak split when $N_p$ is large and $A_0$ exceeds its optimal value.
Indeed, as we will see below, for large $N_c$ the peak often splits even for rather small $N_p$ and $A_0$ below its optimal value.

The spectrum with the lowest $\langle \chi^2\rangle$ in the case under consideration here
is that for $N_p=15$, which is not included in Fig.~\ref{swnc500} but which naturally
falls between those for $N_p=14$ and $16$. The correct spectrum is reproduced remarkably well. There is some dependence of the result (including also
the optimal $A_0$ and $N_p$) on the exact value of the sampling temperature $\Theta$, but as long as the the standard $\langle \chi^2\rangle$ criterion
is satisfied the variations with $\Theta$ are again only mild. In the case at hand, at the sampling temperature used, Eq.~(\ref{eq:chi2}) is satisfied
with $a$ slightly below $0.5$.

\begin{figure*}[t]
\centering
  \includegraphics[width=150mm]{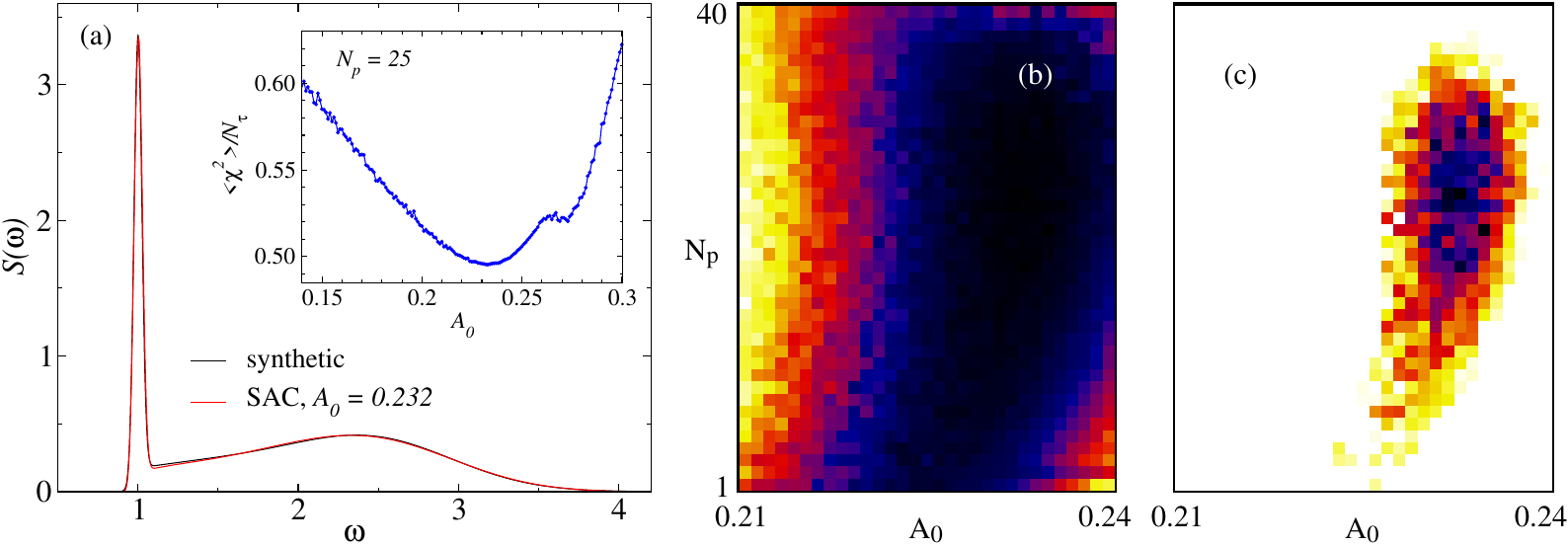}
  \caption{Test with the peak-center constraint of Fig.~\ref{peakfig}(c), using the same synthetic spectrum as in Fig.~\ref{x2np}; here shown
    in (a) as the black curve. The almost identical red curve shows the optimal SAC spectrum, obtained with $N_c=1000$, $N_p=25$, and $A_0=0.232$, the latter
    corresponding to the $\langle \chi^2\rangle$ minimum in the inset.
  Color maps of $\langle \chi^2\rangle$ in the relevant range of ($A_0,N_p$) close to the optimum are shown in (b) and (c), with black corresponding
  to the minimum and lighter colors showing gradually larger values up to a cut-off. The cut-off in (b) is the largest $\langle \chi^2\rangle$ value
  in the parameter window shown, while in (c) it is $10\%$ of the range from the minimum to maximum value.}
 \label{chicolor1}
\end{figure*}

Next, we test the peak-center constraint of Fig.~\ref{peakfig}(c). Fig.~\ref{chicolor1} shows results for the same synthetic spectrum as above
[the inset of Fig.~\ref{x2np}(a)], based on
runs with $N_c=1000$, for which the optimal number of peak elements is (as we will show below) $N_p \approx 25$. Again, a clear $\langle \chi^2\rangle$
minimum is observed [inset of Fig.~\ref{chicolor1}(a)] and at the corresponding $A_0$ value the resulting average spectrum is very close to the correct
profile [main Fig.~\ref{chicolor1}(a)].

To illustrate the well-defined optimum in the full space $(A_0,N_p)$, a 2D color-coded $\langle \chi^2\rangle$ map is shown in Fig.~\ref{chicolor1}(b)
in the region close to the minimum. Here black corresponds to the smallest $\langle \chi^2\rangle$ value within the window and lighter colors indicate
larger values on a linear (in a standard RGB coding) scale.
To see the minimum clearly, it is necessary to stretch the scale, as we have done in Fig.~\ref{chicolor1}(c) by assigning
the lightest color (white) to all cases where $\langle \chi^2\rangle$ exceeds the minimum by more than $1/10$ of the full $\langle \chi^2\rangle$ range spanned by
the points in the window. Here the individual runs were relatively short (about 2 hours of CPU time each) and statistical fluctuations make it impossible
to extract the optimum with absolute certainty (though function fitting, which we have not performed here, would clearly help in this regard). However, the
differences in the average spectrum are small within the region of the darkest colors. The spectrum in Fig.~\ref{chicolor1}(a) is the one corresponding
to the smallest $\langle \chi^2\rangle$ value obtained in these particular runs, where $N_p=25$ and the $\langle \chi^2\rangle$ minimum at $A_0=0.232$
is seen more clearly in the scan in the inset of Fig.~\ref{chicolor1}(a).

\begin{figure*}[t]
\centering  
\includegraphics[width=115mm]{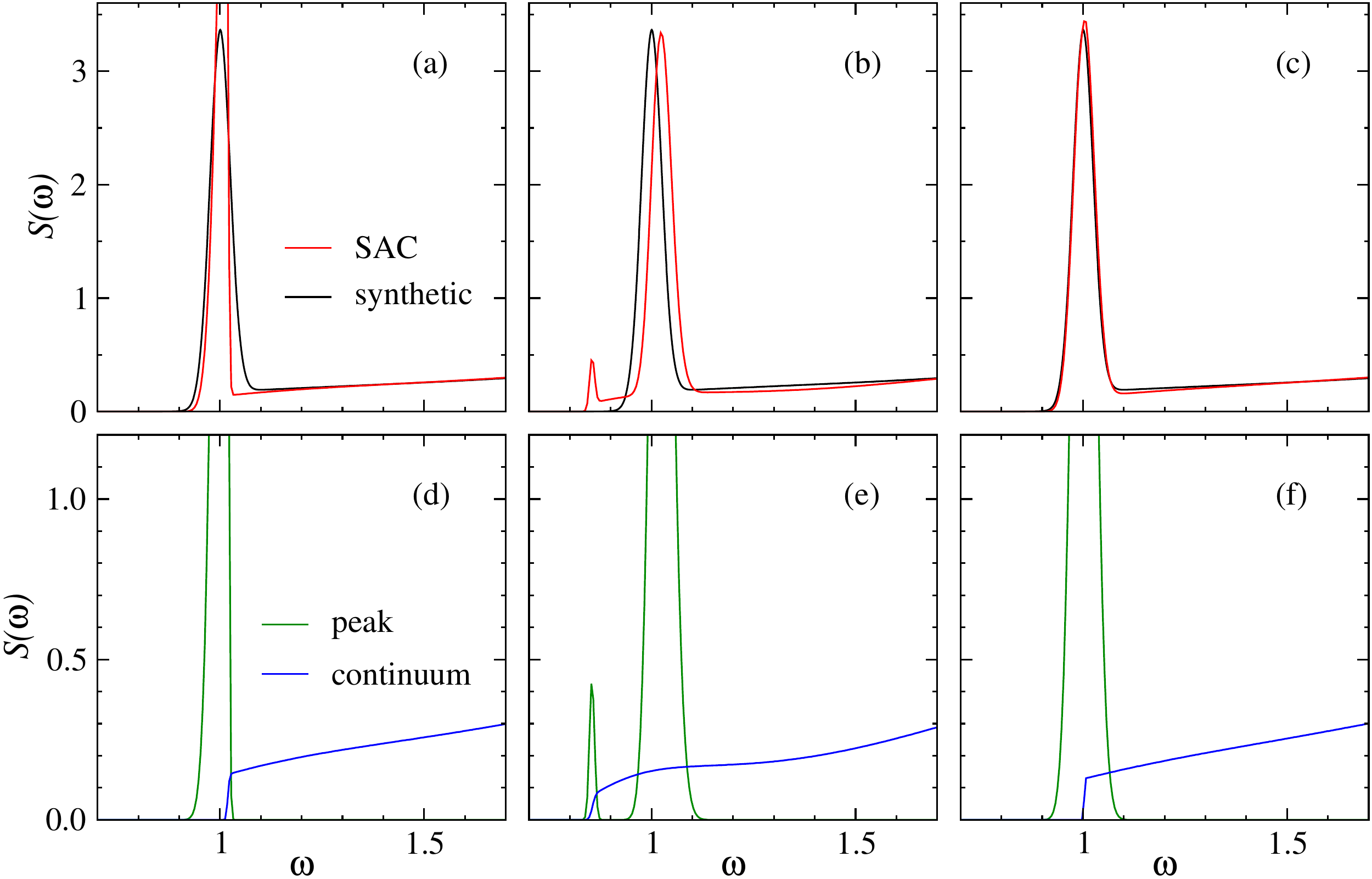}
\caption{SAC results for the synthetic spectrum shown as the black curves in (a)-(c) (the same as in Figs.~\ref{x2np} and \ref{chicolor1}), obtained
with the three different constraints in Figs.~\ref{peakfig}(a)-(c), correspondingly. In all cases $N_c=5000$ and $N_p=30$. The peak weight $A_0$
in each case corresponds to the $\langle \chi^2\rangle$ minimum for the given $N_p$ ($A_0 \approx 0.23$). Panels (d)-(f) show separately the peak
and continuum contributions to the results in (a)-(c).}
 \label{pcompare}
\end{figure*}

It should be pointed out here that the $\langle \chi^2\rangle$ minimum corresponding to the optimal spectrum is not always the global minimum in the
space $(A_0,N_p)$. As seen clearly in Fig.~\ref{x2np}(a), for some $N_p$ values there is a second minimum at $A_0$ above the more consistent (among a range
of $N_p$ values) minimum that we have associated with the optimal fit. A small second local minimum is also seen in the inset of Fig.~\ref{chicolor1}(a).
In some cases, with all three constraints, we have found that a second minimum at a larger $A_0$ value can be lower than the first one, but the corresponding
spectrum is nevertheless much worse, with the peak either excessively broadened or split. For $N_p=1$, there is never any second minimum above the correct one.
Based on many tests, we conclude that it is always the minimum for the lowest $A_0$ that should be used when $N_p>1$, even if it is not the global minimum. A good
strategy to approximately locate the correct minimum is to first perform a scan over $A_0$ with $N_p=1$, for which the global $\langle \chi^2\rangle$ minimum
is invariably close to that for small $N_p > 1$. Other approaches for rapid searches for the relevant region, in $A_0$ as well as $N_p$, will be discussed
further below.

The second $\langle \chi^2\rangle$ minimum for fixed $N_p$, whether local or global, reflects a locally optimal re-distribution of the peak weight when $A_0$
is larger than the true spectral weight of the peak. At this minimum the fluctuations of the spectrum are suppressed, and $\langle \chi^2\rangle$
can then be low for reasons similar to why the $\Theta \to 0$ spectrum in unrestricted sampling represents a good fit to the imaginary-time data even though
the spectral weight distribution is poor (as discussed in Sec.~\ref{sec:theta}).

Given the role of the entropic pressures exerted on the peak by the continuum, it should also be expected that the results to some extent depend on the number
$N_c$ of $\delta$-functions in the continuum part. While an appropriate value of the sampling temperature $\Theta$ regulates the overall entropy, the relative
entropies of the peak and the continuum are also clearly important. Though the optimal $N_p$ value increases with $N_c$, it is not simply proportional
to $N_p/N_c$ (at least for small $N_c$), and the optimal $A_0$ value also shows a weak dependence on $N_c$. It is difficult (not practical) to search
for the optimum in the space ($A_0,N_p,N_c$), especially since our strategy of determining $\Theta$ is based on fixed $N_c$, with $\Theta$ held fixed
in the subsequent scans over $A_0$ and $N_p$. While increasing $N_c$ initially can lead to lower $\langle \chi^2\rangle$ values at fixed $\Theta$, for very
large $N_c$ the sampling temperature eventually has to be lowered, for the same reasons discussed in Sec.~\ref{sec:entropy2}. Our criterion for optimizing
parameters by minimizing $\langle \chi^2\rangle$ at fixed $\Theta$ then cannot be applied to $N_c$.

In the case of the peak-central constraint, we expect (for the reasons as in Sec.~\ref{sec:entropy1}) that the relevant entropies should eventually
scale as $N_p$ and $N_c$ for the peak and continuum, respectively, and then only the ratio $N_p/N_c$ will have to be optimized when
$N_c$ is sufficiently large. In contrast, with the low-edge constraint the lone edge-$\delta$ is pushed down by an entropy scaling as $N_c$
without any compensating upward entropic pressure, and one can therefore expect problems for large $N_c$ in this case.

The extent to which the final spectrum depends on $N_c$ is indeed highly dependent on which of the three constraints in Fig.~\ref{peakfig} is used. In our tests,
with the above synthetic spectrum and many others, we find that all three constraints produce reasonable results when $N_c$ is rather small, say $N_c=500$.
When increasing $N_c$, both the high- and low-edge constraints produce distortions of the peak, and there is no good agreement with the true spectrum for any
($A_0,N_p$). However, the peak-center constraint, Fig.~\ref{peakfig}(c), works well also when $N_c$ becomes much larger.

The different behaviors of the three parametrizations when $N_c$ is relatively large, $N_c=5000$, are demonstrated in Fig.~\ref{pcompare}, using the same
synthetic spectrum as before. In this case only the peak-center constraint produces good results, and we therefore fix $N_p$ at the optimal value, $N_p=30$,
for said constraint. This $N_p$ value is also close to optimal in the other cases. The $\langle \chi^2\rangle$ minimum for all three constraints is
located at $A_0\approx 0.23$, and the results shown are for the respective optimal $A_0$, with Figs.~\ref{pcompare}(a)-(c) corresponding to the constraints
depicted in parts (a)-(c) of Fig.~\ref{peakfig}.

\begin{figure*}[t]
\includegraphics[width=130mm]{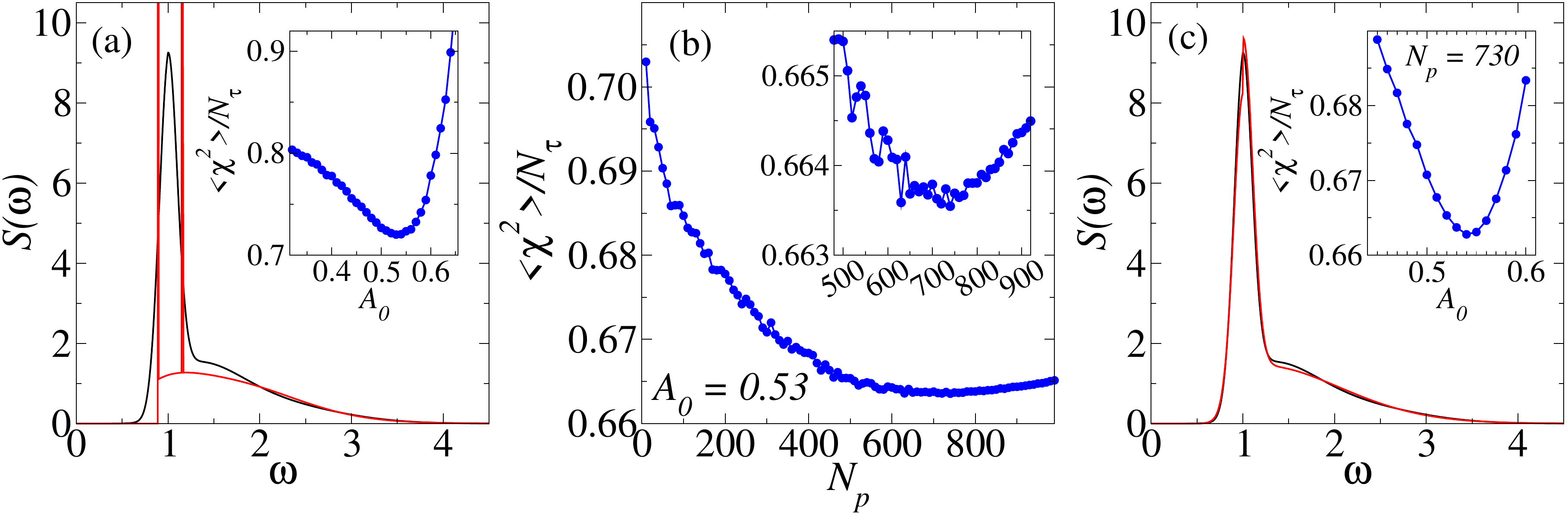}
\caption{Illustration of the line-scan procedure to determine the optimal $A_0$ and $N_p$, using the same $G(\tau)$ data at error level $\sigma=10^{-5}$
as in Figs.~\ref{syntcomp}(a) and \ref{syntcomp}(b); the synthetic spectrum is shown as the black curve in (a). In all cases, $N_c=2\times 10^4$ and
the sampling temperature was $\Theta=0.05$. A scan over $A_0$ with the low-edge constraint and $N_p=2$ fixed produced the goodness-of fit results shown
in the inset of (a). The optimal two-spike spectrum with $A_0=0.53$ is shown as the red curve in (a). In (b), $A_0$ was fixed at the optimal value from
the scan in (a) and a scan over $N_p$ in steps of $10$ was carried out with the peak-central constraint. The optimal $N_p=730$ (extracted approximately
from a fit to the data close to the optimum) was used to produce the spectrum shown in red in (c), where the true spectrum is shown in black. The inset
of (c) shows a final scan over $A_0$ with $N_p=730$, giving an improved optimal peak weight $A_0=0.54$. The spectrum obtained here (not shown) is almost
identical to the one for $A_0=0.53$.} 
\label{broadsynt}
\end{figure*}

As seen in Fig.~\ref{pcompare}(c), with the peak-central constraint the result again agrees very well with the true spectrum, the SAC result being very
similar to that obtained with $N_c=1000$ and $N_p=25$ in Fig.~\ref{chicolor1}(a). With the low-edge constraint used in Fig.~\ref{pcompare}(a) the peak is
too narrow, with a very sharp edge on its right side for the entropic reasons already discussed. With the low-edge constraint, Fig.~\ref{pcompare}(b), the peak
has split, with a very sharp small peak below the true peak (corresponding to a single $\delta$-function of the peak group) and the main part of the peak
shifted up relative to the true peak. These behaviors indicate that the entropic pressures exerted by the continuum become too severe for large $N_c$, except
in the case of the peak-center constraint. The latter does not influence the peak width, only keeps the center down in frequency. This pressure is well
controlled because the entire peak located too far down in frequency will lead to a large penalizing $\chi^2$ value. Thus, we conclude that, while the
low- and high-edge constraints can often produce acceptable results when $N_c$ is relatively small (more so with the low-edge constraint), the peak-center
constraint is much less sensitive to the number $N_c$ of continuum elements and invariably seems to be the best option among the three.

Figures \ref{pcompare}(d)-\ref{pcompare}(f) show individually the peak and continuum contributions to the full average spectra in
Figs.~\ref{pcompare}(a)-\ref{pcompare}(c). Here the ways in which the different constraints affect the continuum are seen clearly. The
edge of the continuum is always sharp when $N_c$ is large and becomes sharper with increasing $N_c$ (eventually forming a step function).
In the case of our favored peak-center constraint, Fig.~\ref{pcompare}(f), the step feature causes a small asymmetry that is barely visible at
the tip of the peak in Fig.~\ref{pcompare}(c). This distorting feature is more serious when the peak is broad (as we will se explicitly below).

Even with the peak-center constraint, there is a detectable difference in the results obtained with $N_c=1000$ in Fig.~\ref{chicolor1}(a) and those with
$N_c=5000$ in Fig.~\ref{pcompare}(c): The continuum between $\omega \approx 1.1$ and $1.3$ falls slightly more below the correct profile in the
latter case, and this distortion can be traced to a slightly larger $A_0$ when $N_c$ is increased. It is possible that the peak asymmetry
associated with the peak-center constraint is at least partially responsible for the slight suppression of weight in the continuum close to the
peak, an effect compensating for the larger weight on the upper half of the peak. This effect is less obvious for smaller $N_c$, when the peak
center fluctuates more. We will present additional evidence for this distortion mechanism further below.

\begin{figure*}[t]
\centering
\includegraphics[width=130mm]{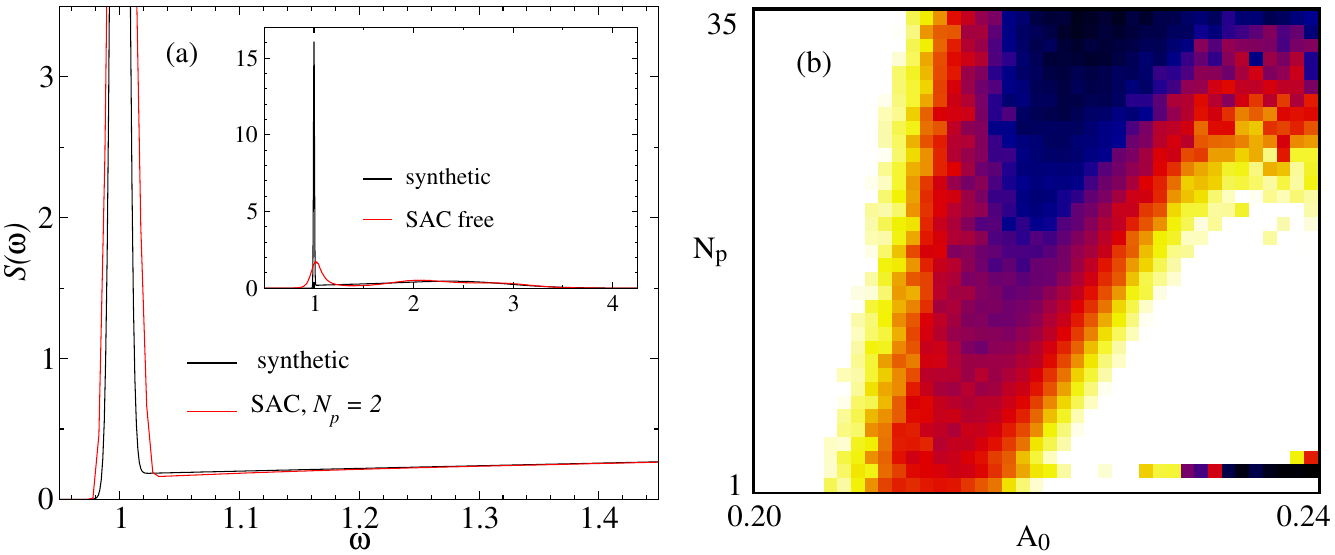}
  \caption{Test of the peak-center constraint, Fig.~\ref{peakfig}(c), for a spectrum with a very sharp quasi-particle peak, shown at full
    height in the inset of (a) as the black curve along with the results of unrestricted SAC shown as the red curve (obtained with both
    frequency and amplitude updates). The error level of the synthetic data was $10^{-6}$. The main figure shows the SAC result for $N_p=2$ (red curve)
    compared with the true spectrum (black curve). The 2D plot in (b) shows a mapping of $\langle \chi^2\rangle$ as in Fig.~\ref{chicolor1},
    but here there is no isolated minimum within the window of reasonable values of $A_0$ and $N_p$, due to the inability of the method to reproduce
    the sharp peak at the present error level. This test was done with $N_c=5000$, sampled at $\Theta=0.5$.}
 \label{chicolor2}
\end{figure*}

We have not comprehensively studied the behavior versus $N_c$ and how to choose an optimal $N_c$. However, all tests that we have carried out
suggest that $N_c$ should not be much larger than needed in order to find a clear $\langle \chi^2\rangle$ minimum in the space $(A_0,N_p)$---the minimum becomes
more pronounced with increasing $N_c$, but at some point the entropic pressures by the continuum can become too large and the results deteriorate. This behavior
is very apparent with the low- and high-edge constraints but less obvious with the peak-center constraint, which we strongly prefer in most cases.

When the quasi-particle peak is broad $N_p$ has to be larger, which implies that also $N_c$ may have to be large. We next consider the same spectrum
with a broad peak that we tested extensively with unrestricted sampling in Fig.~\ref{syntcomp}. To test a possible advantage of the multi-$\delta$
peak parametrization also for such a broad peak, we use the $\bar G(\tau)$ data at error level $\sigma = 10^{-5}$, the same as in
Figs.~\ref{syntcomp}(c) and \ref{syntcomp}(d). In this case $N_c=1000$ gives an optimal $N_p \approx 50$, which is not enough to produce
a smooth lower peak edge. The results are good for $N_c=5000$ and above, and in Fig.~\ref{broadsynt} we present results for $N_c=2\times 10^4$.

Here we also demonstrate a systematic way to efficiently locate the minimum in the $(A_0,N_p)$ space. While we have argued that the low-edge constraint is not
appropriate when $N_c$ is large, we here employ this constraint in another way, illustrated in Fig.~\ref{broadsynt}(a). Applying the low-edge constraint and
sampling with $N_p=2$ produces two sharp spikes, one on either side of the true peak. In general, we have found that the optimal $N_p=2$ peak weight $A_0$ is very
close to that of the optimum in the full space $(A_0,N_p)$, slightly larger than the best $A_0$ obtained for small $N_p$ with the peak-center constraint. The
reason for this improvement is likely that the first $\delta$-function is pushed down more with the low-edge constraint, thus the two $\delta$-functions
can bracket the peak in a more optimal way. After performing a scan over $A_0$ with $N_p=2$ and the low-edge constraint, the so obtained optimal $A_0$ is
used in a subsequent scan over $N_p$ with $A_0$ held fixed,
now using the peak-central constraint. In the present case, with $A_0=0.53$ as illustrated in Fig.~\ref{broadsynt}(b), the optimum is for $N_p\approx 730$.
The resulting spectral function is shown in Fig.~\ref{broadsynt}(c).

Overall, the result is better than those of free sampling in Figs.~\ref{syntcomp}(c) \ref{syntcomp}(d), though in this case the asymmetric tip of the peak,
stemming from the peak-center constraint as discussed above, is seen clearly. To further check the optimum, another scan over $A_0$ can be carried out with
$N_p$ fixed, which in the present case gives $A_0=0.54$ [inset of Fig.~\ref{broadsynt}(c)], only slightly different from the original $A_0=0.53$ from the
$N_p=2$ scan. The resulting spectral function at $A_0=0.54$ and $N_p=730$ essentially overlaps to within the curve thickness with the profile for $A_0=0.53$
in Fig.~\ref{broadsynt}(c). Additional scans, not shown, confirm the location of the minimum. This method of line scans also works with narrow peaks,
e.g., we can reproduce the results in Fig.~\ref{chicolor1}(a) at much less effort.

In Fig.~\ref{broadsynt}(c) it is more obvious than before that the peak asymmetry associated with the peak-center constraint may be responsible
for the mild suppression of the continuum for $\omega \approx 1.5$, to compensate for the slight excess weight on the right side of the peak. This 
excess weight is a consequence of the continuum contributions extending significantly only into the right half of the peak, while the peak contributions
themselves must be sufficiently large to account completely for the left side of the peak. In principle, the constraint could be modified to allow the
continuum to extend also slightly below the peak center (but not all the way to the lower edge to avoid the peak splitting associated with the low-edge
constraint). The asymmetry would then be alleviated. Alternatively, a soft potential could be introduced instead of the
hard constraint to achieve a similar effect. However, in general we favor less adjustable features, and the peak-center constraint should already
be sufficiently good in most cases of moderately narrow quasi-particle peaks.

\subsubsection{Resolution and self-consistent test}

There must of course be some limit to the ability of this method to resolve a very narrow quasi-particle peak. As an example of excessive broadening, we next
consider a synthetic spectrum with a much sharper peak than the previous examples; see the inset of Fig.~\ref{chicolor2}(a), where the unsatisfying result of
unrestricted SAC is also shown for comparison. The color map in Fig.~\ref{chicolor2}(b) reveals no isolated minimum, instead the minimum value of
$\langle \chi^2(A_0,N_p)\rangle$ versus $A_0$ decreases slowly as $N_p$ is increased while the corresponding location $A_0$ of the minimum drifts toward larger
values. The spectrum obtained for $N_p=2$ is shown in Fig.~\ref{chicolor2}(a) and already has a peak broader than that of the true spectrum. The peak width grows
with $N_p$, and at the upper edge of the color map in Fig.~\ref{chicolor2}(a), $N_p=35$, the peak of the best-$A_0$ spectrum (not shown) is already wider
than the peak of the previous synthetic spectrum in Fig.~\ref{chicolor2}(a).

The failure of the method to produce a minimum in the space ($A_0,N_p$) corresponding to close agreement with the true spectrum can be understood as a
sign that the peak is too sharp to resolve at the given error level of the imaginary-time data. Because the spectrum broadens further as $N_p$ is increased
above $2$, there is no $N_p>2$ for which there is a local optimum in the match between the true and sampled peak widths, which, correspondingly, is
reflected in the lack of local $\langle \chi^2\rangle$ minimum in the space ($A_0,N_p$). Thus, when a well-defined minimum cannot be found,
a reasonable assumption is that the spectrum sampled with $N_p=2$ represents a resolution-limited result, where the peak produced is broader than the
true peak. Conversely, in cases where a $\langle \chi^2\rangle$ minimum is present, we assert that the quasi-particle peak should be well resolved,
without artificial broadening (though there can still of course be some distortions).

As already discussed in the context of the second $\langle \chi^2\rangle$ minimum in Figs.~\ref{x2np} and \ref{chicolor1}, $\langle \chi^2\rangle$ can also
exhibit a ``false'' minimum locally when $A_0$ is larger than the true peak weight. The same mechanism should be at play here as well in causing a false
optimum located at increasing $A_0$ with increasing $N_p$ as observed in Fig.~\ref{chicolor2}(b). A second minimum is also apparent for $N_p=2$ in
Fig.~\ref{chicolor2}(b) (and also exist for larger $N_p$ outside the edge of the graph)---the spectrum at that minimum has two well separated spikes
instead of a single peak.

It is actually possible to self-consistently test whether a sampled spectrum has a peak of width larger than the resolution limit set by the
imaginary-time data. To see this, we note that the sampled spectrum in the multi-$\delta$ peak parametrization used here has two contributions
that can be collected separately during the sampling process: the peak and the continuum. Thus, we can write the result for the sampled
normalized spectrum $A(\omega)$ as
\begin{equation}
A(\omega)=A_0A_p(\omega)+(1-A_0)A_c(\omega).
\end{equation}
Now consider a hypothetical spectrum obtained from $A(\omega)$ by removing the peak part, whose first moment we call $\omega_\delta$, and replacing it by
a macroscopic $\delta$-function with the same weight at $\omega_\delta$. Thus, we define
\begin{equation}
A_\delta(\omega)=A_0\delta(\omega-\omega_\delta)+(1-A_0)A_c(\omega)
\end{equation}  
and similarly the final output spectrum $S_\delta(\omega)$ according Eq.~(\ref{barelation}) as usual.
This new spectral function corresponds to a modified QMC imaginary-time function according to Eq.~(\ref{contrel2}):
\begin{equation}
\bar G_\delta(\tau_i)=\bar G(\tau)+A_0\int_0^\infty d\omega [\delta(\omega-\omega_\delta)-A_p(\omega)]\bar K(\tau_i,\omega),
\label{gdelta}
\end{equation}
which in the basis corresponding to the transformation according to Eq.~(\ref{basistransf}) has the same form with
$\bar G_\delta(\tau_i)  \to \bar G_\delta(i)$ and $\bar K(\tau_i,\omega)  \to \bar K(i,\omega)$, where the index $i$ corresponds to
the eigenvalue $\epsilon_i$ of the covariance matrix (Sec.~\ref{sec:qmcdata}). We can now process this imaginary-time function,
using the same value of $A_0$ (and other parameters of the sampling) as before but with different values of $N_p$, to find a new optimal $N_p$ and the
corresponding approximation to $A_\delta(\omega)$ (which cannot have an exact $\delta$-function because of the limitations of the method). If the original
SAC spectrum has a peak of width resolvable by the method, the new peak (representing the $\delta$-function) should be sharper, and, if so, indicates the
degree to which the method can resolve a $\delta$-function when the continuum is that of the original spectrum. Thus, the procedure will deliver the
``instrumental resolution'' of the method (given the noise level in the input data).

The new $\delta$-peak should also be associated with statistical noise, and it is necessary to introduce additional normal-distributed perturbations
to $\bar G_\delta$ beyond the transformation in Eq.~(\ref{gdelta}). Even though the statistical errors associated with the replaced broadened peak are
still present in the imaginary-time data and cannot be removed, just adding the $\delta$-function without new noise would not correspond to a realistic
situation (and indeed, in our tests the $\delta$-function is then resolved to an unrealistically good degree). Thus, we add Gaussian noise of standard
deviation $\propto \sqrt{\epsilon_i}$ to the imaginary-time QMC data $\bar G_\delta$ expressed in the eigenbasis of the covariance matrix,
keeping the covariance matrix itself unchanged.

To fix an appropriate constant of proportionality of the noise, we note that the contribution to $\bar G_\delta(i)$ from the macroscopic $\delta$-function
is $A_0 \bar K(i,\omega_\delta)$. Lacking detailed knowledge of how the statistical errors should be associated with the different parts of the spectral
function, a reasonable assumption is that the $\delta$-peak should be given synthetic noise in proportion to its relative weight in $\bar G_\delta$. Thus, we
add normal-distributed random numbers with the following standard deviation to $\bar G_\delta(i)$:
\begin{equation}
\sigma_\delta(i) = A_0 \frac{\bar K(i,\omega_\delta)}{\bar G_\delta(i)}\sqrt{\epsilon_i}.
\end{equation}
This additional noise (beyond the noise from the QMC simulation) implies that $\langle \chi^2\rangle$ should be somewhat higher than in the
original SAC processing of $\bar G(i)$.

\begin{figure}[t]
\centering
  \includegraphics[width=70mm]{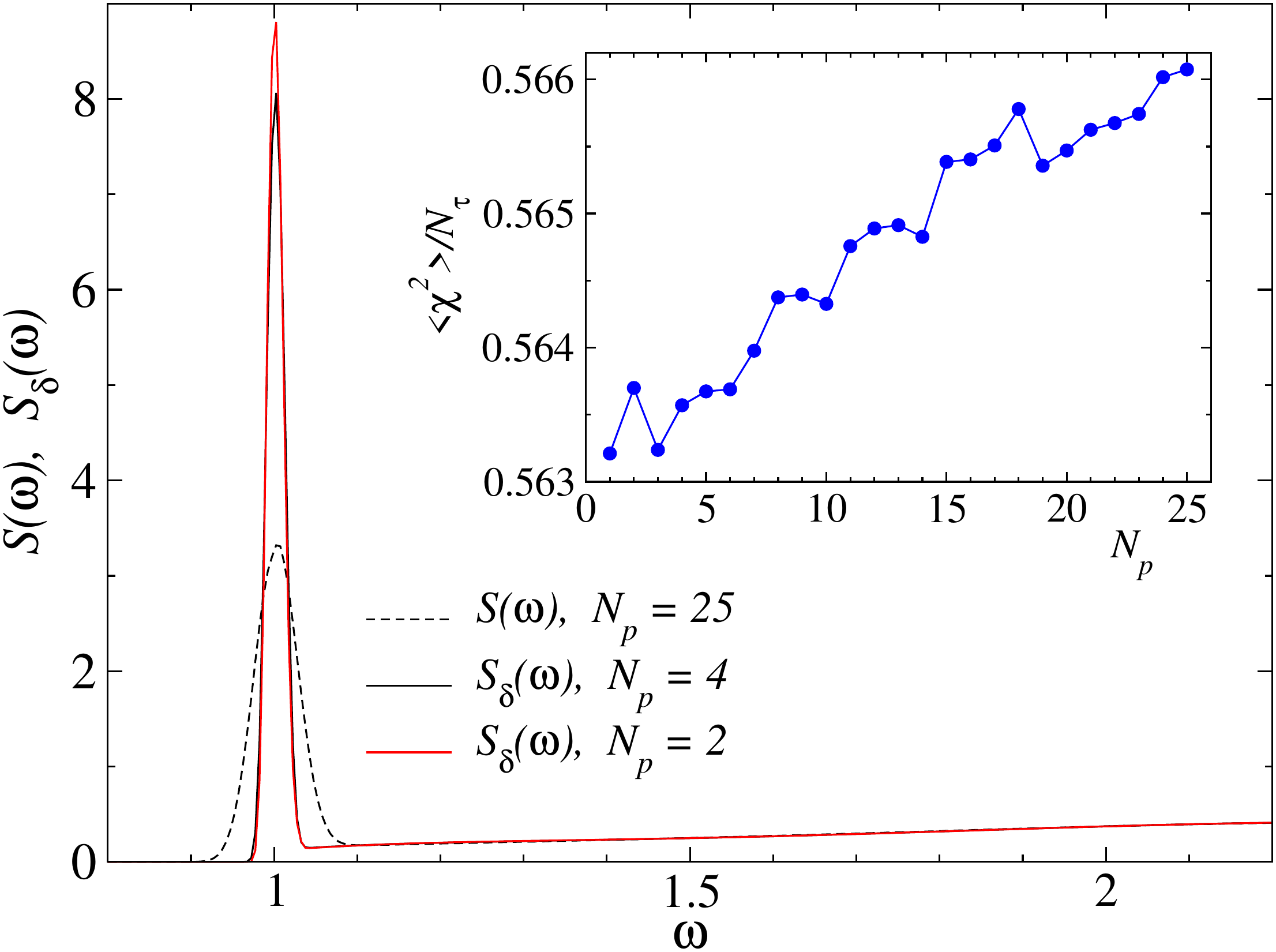}
  \caption{Self-consistent resolution test of the SAC spectrum in Fig.~\ref{chicolor1}(a). The original optimal spectrum with $N_p=25$
    is shown with the black dashed
   curve, and those obtained from the peak-replaced imaginary-time data, Eq.~(\ref{gdelta}), with $N_p=2$ and $N_p=4$ are shown with the red and black
   solid curves, respectively. The inset shows the goodness of the fit in sampling runs where all other parameters are kept the same as in the original
   procedure but $N_p$ is varied. The best goodness-of-fit values at very small $N_p$, and the corresponding narrow peaks in $S_\delta(\omega)$ show that the
   method is capable of properly resolving the original peak width.}
 \label{swdelt}
\end{figure}

In Fig.~\ref{swdelt} we present results of this type of resolution test, using the data underlying the successfully reproduced spectrum in Fig.~\ref{chicolor1}.
The original optimal spectrum was sampled with $N_p=25$, and, with the peak replaced as described above, we keep all parameters the same except for $N_p$
that is scanned over for $N_ p \in [1,25]$. The inset of Fig.~\ref{swdelt} shows that the goodness of the fit systematically decreases as $N_p$ is
reduced, and there is no discernible local minimum beyond the statistical errors (judging from the scatter of the data points). The best estimate
of the intrinsic resolution of the method is then obtained with $N_p=2$. In Fig.~\ref{swdelt} we show results with $N_p=2$ and $N_p=4$, demonstrating
the very weak dependence on the peak width for these small $N_p$ values. These peaks are clearly much sharper than the original peak, thus independently
confirming that the original spectrum is reliable as to the its peak width, i.e., the peak is wider than the resolution limit of the method with
the data quality at hand.

The tests presented above demonstrate that even rather narrow quasi-particle peaks can be resolved to a much higher degree than what may have been expected. We
also again have seen how also the high-energy continuum can be much better reproduced once a low-energy feature has been treated better than what is possible
with either unrestricted sampling in SAC or with the ME method. This aspect of the improved resolution is very striking when comparing the spectum in
Fig.~\ref{chicolor1} with the result of unrestricted sampling in the inset of Fig.~\ref{x2np}(a), where ringing behavior is present far above the much
too broad peak. We coinclude that ``hidden information'' in $\bar G(\tau)$ is able to reproduce the true spectrum once the primary edge distortions
have been removed. 

While there is still room to improve the parametrization in Fig.~\ref{peakfig}, specifically to allow some migration of the continuum $\delta$-functions
also below the peak center, the method as presented here should already be adequate for many applications. In the context of models discussed in this paper,
it could be used to investigate a possible quasi-particle broadening in the 2D Heisenberg model, where so far (Ref.~\cite{shao17} and Fig.~\ref{2dsw})
we have only used the sharp $\delta$-edge.

\section{Edge singularities I}
\label{sec:contedge1}

In some of the most interesting quantum many-body systems, conventional quasi-particles fractionalize into two or more objects. These fractionalized
objects propagate essentially independently, and, as a consequence, the spectral functions of relevance to experiments reflect the collective contributions
from two (in the simplest case) particles. At a fixed total momentum ${\bf q}$ of a state excited in, e.g., inelastic neutron scattering, combinations
of the individual quasi-particle momenta ${\bf q}_1$ and ${\bf q}_2$ with ${\bf q}_1+{\bf q}_2={\bf q}$ (often with a ``semion'' condition on the individual
momenta) produce $S({\bf q},\omega)$ with a contunuum of spectral weight in the range of available values of $\omega=\omega_1 + \omega_2$. Typically, because
of density of states and matrix element effects, the spectral functions of interest exhibit power-law singularities at the lower edge of the continuum.

The perhaps best known and understood example of such an edge-singular spectral function is the dynamic spin structure factor $S(q,\omega)$ corresponding to
spin-$1$ excitations of the $S=1/2$ Heisenberg chain at $T=0$, accessed with the operator $O=S^z_{\bf q}$ in Eq.~(\ref{somegasum}). The predominant contributions to
$S(q,\omega)$ arise from pairs of spinons (each carrying spin $1/2$) \cite{karbach97} and there are contributions also from four, six, etc.~spinons within the BA
framework \cite{caux05a,caux05b,pereira06}. An example of this spectral function was already shown in Fig.~\ref{sw2}, where the difficulties in resolving the 
sharp edge and associated power-law singularity with the unrestricted SAC method, or any other conventional analytic continuation method, are apparent.

In order for analytic continuation to reproduce a sharp edge, this feature has to be imposed in some way from the outset, by
which we mean that the parametrization should be robust in the way this aspect of the spectrum is sampled and not suppressed by entropic
pressures. An extreme example is the macroscopic edge $\delta$-function discussed in Sec.~\ref{sec:deltapeak}. For a generic edge at $\omega=\omega_0$, with
continuously divergent or non-divergent behavior when $\omega \to \omega_0$ from above, we would like to use a parametrization that only imposes a cut-off at
$\omega=\omega_0$. The value of $\omega_0$ is not necessary known in advance but should be determined by the process, which ideally should also not in any
other way dictate the shape of the spectrum. More realistically, the parametrization should only have a weak bias that can largely be counteracted
by imaginary-time QMC data of achievable quality.

The simplest way to impose an edge is simply to optimize the lower bound $\omega_0$, which we already discussed an example of in Sec.~\ref{sec:example2}
(Fig.~\ref{w0fix}) in the continuous-frequency representations (and in the fixed-grid representation in Ref.~\cite{sandvik16}; see also
Ref.~\cite{shao17}). While this approach produces much better results than unrestricted sampling for sharp-edge spectra, there is still rounding of
a divergent peak and poor fidelity at higher frequencies.

A method for reproducing a sharp peak within the fixed-grid SAC representation was to impose a single-peak constraint at the level of the sampling, i.e.,
the amplitudes $A_i$ have a single maximum versus $\omega_i$ in each valid configuration \cite{sandvik16}. The dynamic structure factor of the
Heisenberg chain was then reproduced at unprecedented fidelity. A drawback of this method was that the upper and lower bounds also had to be optimized
(using the generic scanning approach illustrated in Fig.~\ref{fig:optim}), which was time consuming.

The motivation for using the parametrization depicted in Fig.~\ref{fig:spec}(e) to model an edge singularity (a divergence in the case shown) is that the
increasing distance between the $\delta$-functions corresponds to the spectral weight density decreasing monotonically after the edge. Though a monotonically
decaying continuum is of course not completely general, such spectra are, in fact, quite common, and it is worth exploring this case before moving on to the
more general spectra with sharp edges.

A monotonic behavior of the averaged spectrum is not completely guaranteed with the parametrization in Fig.~\ref{fig:spec}(e), as in principle there
could be fluctuations that cause multiple peaks. In practice, however, the fluctuations are
strongly suppressed and we have found a monotonic decays in all cases tested. With this parametrization, the upper and lower frequency bounds do not
have to be optimized---there are no significant entropic pressures to broaden the spectrum when reasonably good $\bar G(\tau)$ data are used (in fact, as we
will show, the entropic tendency is often rather to slightly narrow the spectrum) and the sampling converges to the correct bounds to a remarkable
degree. This approach is therefore much more efficient than the scheme in Ref.~\cite{sandvik16}, and also the results are better. We could also in this
case instead consider a peak at some arbitrary location in the spectrum (or even consider several peaks, which we will not do here but will discuss
qualitatively in Sec.~\ref{sec:pdm}), by modifying the constraint in Fig.~\ref{fig:spec}(e) to monotonically increasing distances on either side of a
minimum distance.

To describe a non-monotonic continuum, we can mix the parametrizations of Figs.~\ref{fig:spec}(e) and \ref{fig:spec}(b), at a ratio that is again optimized
as a generic constraint. In this section we will develop the machinery for a monotonic decay and consider the general case of unrestricted continuum
in Sec.~\ref{sec:contedge2}. Here, in Sec.~\ref{sec:hbergedge}, we again use the dynamic structure factor of the Heisenberg
chain as an example to illustrate the sampling technique and the kind of results that are produced with just the basic monotonicity condition and
no further optimization. In Sec.~\ref{sec:triangle} we study a synthetic sharp-edge spectrum with a non-divergent continuum, where a parameter
regulating the edge sharpness is optimized.

\subsection{Sampling with monotonicity constraint}
\label{sec:hbergedge}

We start the sampling process from a configuration with monotonically increasing spacing between the equal-amplitude  $\delta$-functions, of the kind
illustrated in Fig.~\ref{fig:spec}(e). Typically we initialize the frequencies $\omega_i$ ($i=1,\cdots,N_\omega$) with quadratically increasing distance,
$d_i=\omega_{i+1}-\omega_{i}$, within reasonable lower and upper bounds (e.g., based on initial free-sampling runs such as those in Fig.~\ref{sw1}) or
aided by sum rules. We stress that we do not optimize any parameter in the most basic application of this parametrization, and the constraint is just
the monotonicity condition that is maintained during every step of the sampling. This approach is suitable when the continuum diverges at the edge,
and non-divergent edges will require a modification of the monotonicity constraint with an optimized parameter, which we will demonstrate
in Sec.~\ref{sec:triangle} (and later in a different way in Sec.~\ref{sec:contedge2}).

\subsubsection{Single-$\delta$ update}

\begin{figure}[t]
\centering
\includegraphics[width=75mm]{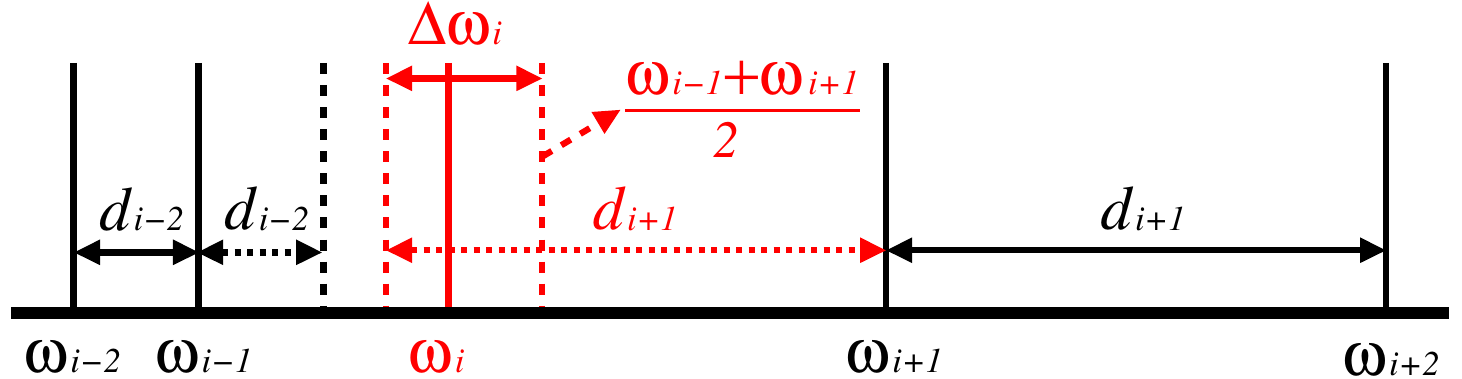}
\caption{Illustration of the allowed window $\Delta\omega_i$ for moving the $i$th $\delta$-function (here with $2 < i < N_\omega$) from its current
frequency $\omega_i$ with the space-increasing constraint maintained. The right boundary is $(\omega_{i-1}+\omega_{i+1})/2$ and the left boundary is
${\text{max}}(\omega_{i-1}+d_{i-2},\omega_{i+1}-d_{i+1})$, where, as marked in the drawing,  $d_{i-2}=\omega_{i-1}-\omega_{i-2}$. For
$i \in \{1,2,N_\omega\}$ slightly different rules apply, as described in the text.}
\label{fig:edge-1}
\end{figure}

\begin{figure*}[t]
\centering
\includegraphics[width=90mm]{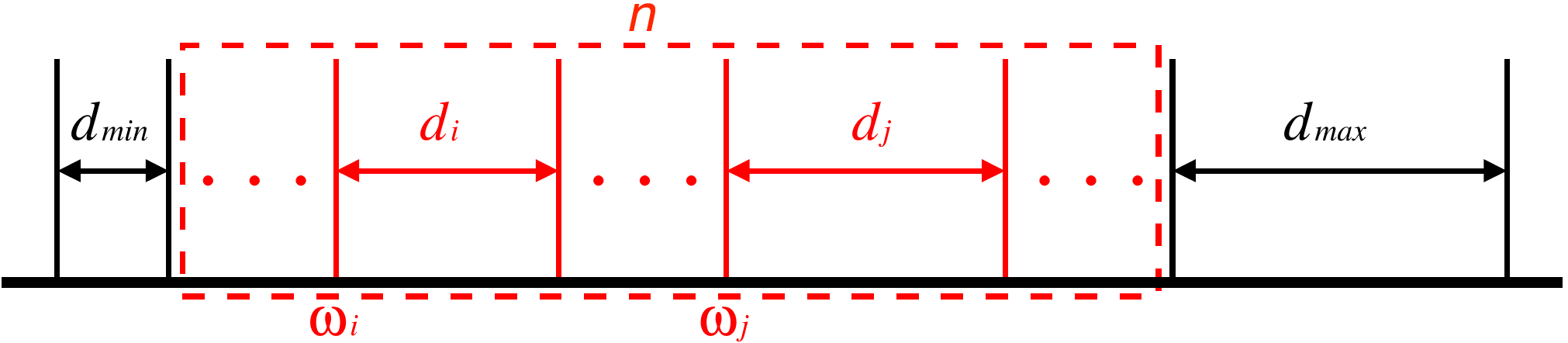}
\caption{Principles of collective updating of a group of $n$ $\delta$-functions with the distance-monotonic constraint maintained. A list
of distances $d_i=\omega_{i+1}-\omega_i$ is used.  By randomly selecting $i$ and $j$, and with $d=d_i+d_j$ kept fixed
(to maintain the sum of all distances, indicated with the dashed line below the horizontal axis), a new value $d'_i$ is chosen randomly between
${\text{max}}(d_{min},d-d_{max})$ and ${\text{min}}(d_{max},d-d_{min})$, and then $d'_j=d-d'_i$. This process is repeated $n/2$ times with
different pairs $i,~j$, after which the updated distances are sorted for monotonicity before converted to new frequency values.
The update is accepted or rejected in the standard (Metropolis) way based on the change in $\chi^2$.}
\label{edge-2}
\end{figure*}

A straight-forward sampling algorithm amounts to randomly choosing one of the $\delta$-functions and moving it anywhere within the window
$\Delta_{\omega_i}$ determined from the monotonically space-increasing constraint, as illustrated in Fig.~\ref{fig:edge-1}. The cases
$i=1$, $2$, and $N_\omega$ require special treatment, and we first consider the generic cases $i \in \{3,\ldots,N_\omega-1\}$. The window
$\Delta\omega_i$ within which $\omega_i$ is chosen is situated as follows: The upper boundary is $(\omega_{i-1}+\omega_{i+1})/2$
in order for the condition $d'_{i-1} \le d'_i$ on the new distances to hold true after the move. To satisfy $d_{i-2} \le d'_{i-1}$ and  $d'_i \le d_{i+1}$,
the left boundary must be the larger of $\omega_{i-1}+d_{i-2}$ and $\omega_{i+1}-d_{i+1}$. For moves within the allowed window, the Metropolis acceptance
probability is applied. In the case of $\omega_2$, there is no constraining distance
$d_{i-2}$ and only the second of the left-boundary conditions apply. As for $i=1$ and $i=N_\omega$, fixed windows centered at $\omega_i$ can be used,
with their widths adjusted in the standard way to achieve an acceptance rate close to $0.5$. Any move violating the constraint is of course rejected
outright.

In practice the allowed windows $\Delta\omega_i$ are small and of course scale approximately as $1/N_\omega$.
The change in $\chi^2$ in a tentative move is then also small, thus leading to a large acceptance rate---over $90\%$ if $N_\omega$ is $80$ or larger in
the example that we will consider below. With the acceptance rate being large, one might think that the configuration should change significantly
after one updating sweep of $N_\omega$ moves. However, $\Delta\omega_i$ is typically much smaller than the distance $d_i$, and in practice
each $\delta$-function tends to fluctuate around a mean location. The edge frequency $\omega_1$ also moves very little once equilibrium has been
reached, while $\omega_{N_{\omega}}$ has the largest (though still typically small) fluctuations.

The limited fluctuations do not just stem from inefficiencies in the sampling algorithm but reflect the way the space-increasing constraint suppresses
the configurational entropy. As we will see below, the low entropy is at the heart of the benefit of the parametrization in its ability to resolve details
of spectrum with an edge followed by a monotonically decaying continuum. The restricted fluctuations also cause some problems, as we will discuss further below.

\subsubsection{Multi-$\delta$ update}

A more efficient algorithm for updating the frequencies $\omega_2,\cdots,\omega_{N_\omega-1}$ is to collectively move several consecutive
$\delta$-functions. There are many ways to accomplish this, and here we only discuss what is perhaps the simplest (and likely most efficient) method.

For a group of $n$ $\delta$-functions at current locations $\omega_k,\ldots,\omega_{k+n-1}$, we work with a corresponding list of distances $d_i$
as is illustrated in Fig.~\ref{edge-2}. In a series of moves performed before applying the Metropolis probability, we randomly select $d_i$ and $d_j$,
to be updated with $d=d_i+d_j$ conserved. The minimum and maximum new distances are dictated by the distances to the left
and right of the group, $d_{\rm min}=d_{k-1}$ and $d_{\rm max}=d_{k+n}$, respectively.
The goal of this update is to keep the distance between $\omega_{k-1}$ and $\omega_{k+n}$ unchanged, and a complete update consists
of several such random pair moves (so that most of the $n$ distances change).
Since both the updated distances $d'_i$ and $d'_j$ have to be in $[d_{\rm min},d_{\rm max}]$, the updated distance
$d'_i$ is chosen randomly between $[{\text{max}}(d_{min},d-d_{max})]$ and $[{\text{min}}(d_{max},d-d_{min})]$, and then $d'_j=d-d'_i$. After repeating
this procedure $n/2$ times, the updated distances are sorted into the required monotonically increasing order, which are then converted into
updated frequencies $\omega_i$ in the group. The change in $\chi^2$ is computed for the acceptance probability.

It can easily be verified that the above algorithm satisfies detailed balance, as it redistributes in an unbiased way how a frequency segment is cut up into pieces.
With $i$ and $j$ chosen at random among all the pieces, the ordering aspect plays no role in the selection process. We define one sweep of multi-$\delta$
updates as consisting of $N_\omega/n$ attempts. The acceptance rate decreases with the set size $n$, and we choose this size in the standard way so that
the acceptance rate is approximately $0.5$. For given $n$, the acceptance rate depends strongly on the location of the group (its first index
$i$) of $\delta$-functions within the spectrum, and it is therefore useful for overall efficiency to adapt the group size to the location,
using the $i$-dependent acceptance rate.

\subsubsection{Example and practical considerations}
\label{sec:heismono}

\begin{figure*}[t]
\centering
\includegraphics[width=110mm]{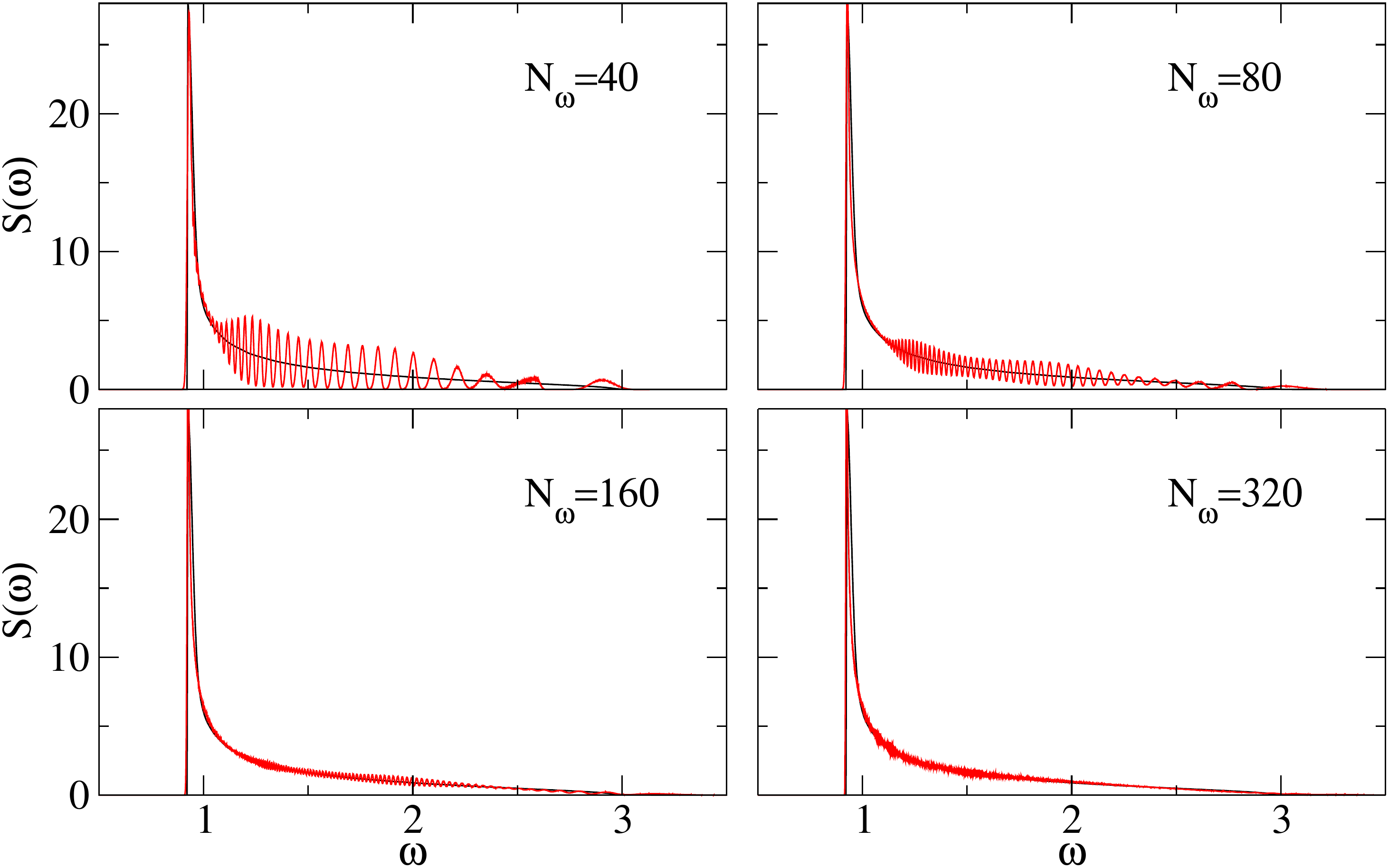}
\caption{SAC spectral function $S(4\pi/5,\omega)$ for the $L=500$ Heisenberg chain (read curves) obtained with the distance-monotonicity constraint in
  Fig.~\ref{fig:spec}(e). Different numbers $N_\omega$ of $\delta$-functions were used, as indicated in the respective panels, and the average spectral
  density was collected in histograms with the same bin size in all cases. The results are compared with the BA calculation for the same system size
  (black curves). The oscillations reflect the tendency of the individual $\delta$-functions to fluctuate narrowly around mean positions. With increasing
  $N_\omega$, the fluctuations of the frequencies relative to their mean separation become large enough to produce smooth average spectra.
  The noise in the $N_\omega = 320$ spectrum is partially due to the very small histogram bins used in order to show the oscillations present in
  the other cases.}
\label{nobroaden}
\end{figure*}

While the multi-$\delta$ updates considerably speed up the sampling, there are still some details that have to be addressed in order to obtain good
results. In discussing a number of issues below, we will point to Fig.~\ref{nobroaden}, which shows $S(q,\omega)$ of the $L=500$ Heisenberg chain at $q=4\pi/5$,
i.e., the same case as the results in Fig.~\ref{sw2} for unrestricted sampling. The spectra presented here were obtained using multi-$\delta$ updates for
$N_\omega=40,80,160$, and $320$, and are graphed together with the BA result. We stress again here that the sampling with the constraint is
by itself sufficient to locate the edge of the spectrum reasonably precisely (to within less than $1\%$ in Fig.~\ref{nobroaden}), and there is no other input
to the process beyond the $\bar G(\tau)$ data and the use of this specific parametrization. Once the spectrum has converged the edge fluctuations
are also very small (thus resulting in a histogram with a sharp peak) at the sampling temperature $\Theta$ satisfying the same criterion,
Eq.~(\ref{eq:chi2}), as used previously in Sec.~\ref{sec:theta}.

When judging the results, it should again be kept in mind that the BA computed $S(q,\omega)$ is not complete but contains about $98\%$ of the total
spectral weight in the case of $q=4\pi/5$. Moreover, even though a chain of $500$ spins may seem quite large, the exact $T=0$ spectrum, Eq.~(\ref{somegasum}),
likely still consists of less than 100 distinct $\delta$-functions with significant spectral weight. Results obtained with matrix product states indicate
only about $10$ $\delta$-functions for $L=64$, $q=4\pi/5$ \cite{wang19} and for $L=100$, $q=\pi$ there is a similar number of sharp peaks \cite{xie18}.
The BA results have been broadened to produce a continuum \cite{caux05a,cauxdata}. Given these caveats, the agreement is very good for the entire spectrum
for $N_\omega=320$. For smaller $N_\omega$, oscillations in the tail part of the spectrum are observed that will be explained below.

To apply the criterion for the sampling temperature $\Theta$, Eq.~(\ref{eq:chi2}), the minimum goodness-of-fit $\chi^2_{\rm min}$ is again determined using
simulated annealing. The ultimate form of the spectrum with a small number of sharp peaks in the unrestricted sampling (discussed in detail in
\ref{app:lowtheta}) cannot be realized with the monotonicity constrained parametrization, and therefore $\chi^2_{\rm min}$ will be slightly larger in this case.
The goodness-of-fit should still be statistically acceptable if the parametrization is appropriate, i.e., if the actual spectrum can be well described as a
sharply divergent edge and monotonically decaying tail, which the parametrization mimics for large $N_\omega$.

Since the distance-monotonicity represents a very strong constraint on the fluctuations of the spectrum, the changes with $\Theta$
are much smaller than with unrestricted sampling. The same criterion for $\Theta$ as before still leads to optimal results, with only weak sensitivity
to the exact value of the factor $a\lesssim 1$ used in Eq.~(\ref{eq:chi2}).
We will show examples of the $\Theta$ dependence in Sec.~\ref{sec:monop}, where we apply the method to a
synthetic spectrum for which the SAC results can be evaluated without any uncertainties on the true shape of the spectrum.

The first problem encountered due to the rather inefficient constrained sampling is that it can be difficult to reach good $\langle \chi^2\rangle$ values,
especially $\chi^2_{\rm min}$, when $N_\omega$ is large when starting from a typically large-$\chi^2$ initial configuration. This problem stems primarily from
the difficulty of the edge to migrate to its correct position when starting the sampling with $\omega_1$ far from the true edge. Therefore, we have found it
helpful to start with a small number $N_\omega$ of $\delta$-functions ($N_\omega=10$, say), in which case the edge converges relatively rapidly. Having stored the
configuration with the smallest $\chi^2$ value, at the next stage we double $N_\omega$ by inserting one $\delta$-function between every two in the old
configuration (and one more at the end), setting their amplitudes to half the previous value. This doubled configuration becomes the starting point for a new
sampling run. The doubling is repeated until the desired $N_\omega$ is reached. This procedure can be carried out at a rather high value
of $\Theta$, before the simulated annealing procedure to fix $\Theta$ is carried out. Another option is to carry out an initial 2D scanning procedure over
the lower and upper bounds of a spectrum without sampling, saving the configuration with the best $\chi^2$ value and subsequently using it to start the
annealing procedure.

Another issue, which can be observed for the small-$N_\omega$ cases in Fig.~\ref{nobroaden}, is that the collected spectral function $S(q,\omega)$ exhibits
a complicated unrealistic structure of the tail. The oscillations correspond to the individual $\delta$-functions fluctuating narrowly around their mean
frequencies $\langle \omega_i \rangle$. When $N_\omega$ is large enough, the fluctuations of $\omega_i$ eventually do become larger than the mean separation
$\langle d_i \rangle = \langle \omega_{i+1}-\omega_i \rangle$ for all $i$, and a smooth spectrum is then obtained if sampled sufficiently. The oscillating
behavior for a range of frequencies is observed clearly in Fig.~\ref{nobroaden} for $N_\omega=40$ and $80$. Though the peaks broaden somewhat with
increasing sampling time, the peak structure is real and does not vanish until $N_\omega$ becomes larger; there is still some hint of oscillations
even for $N_\omega=160$ in Fig.~\ref{nobroaden}, while for $N_\omega=320$ the spectrum appears rather noisy, due to the small histogram bins used,
without any obvious oscillations.

A simple way to circumvent the oscillations is to broaden the $\delta$-functions at the stage of accumulating the histogram. Since the spacing between
the $\delta$-functions, for appropriate choices of $N_\omega$, is small on the scale of the true features of the spectrum, such a procedure is a reasonable
way to obtain a smooth continuum. A convenient way to impose appropriate broadening automatically adapted to the scale of details of the spectrum is to
regard the $\delta$-function at $\omega_i$ as a uniform distribution of the weight $1/N_\omega$ between $\omega_{i}$ and $\omega_{i+1}$, which we did for the
remaining results presented in this section. In principle, the spectrum could also mathematically [in the integrations $S(\omega)$ giving $G(\tau)$ for
the $\chi^2$ calculation], be regarded as a sum over such finite-width boxes. This would be much more time consuming, however, and for reasonably large
$N_\omega$ the differences would be negligible. In Sec.~\ref{sec:contedge2}, we will introduce a way of directly relating the mean spacing
$\langle d_i\rangle =\langle \omega_{i+1}-\omega_i\rangle$ to a local spectral density.

\begin{figure}[t]
\centering
\includegraphics[width=70mm]{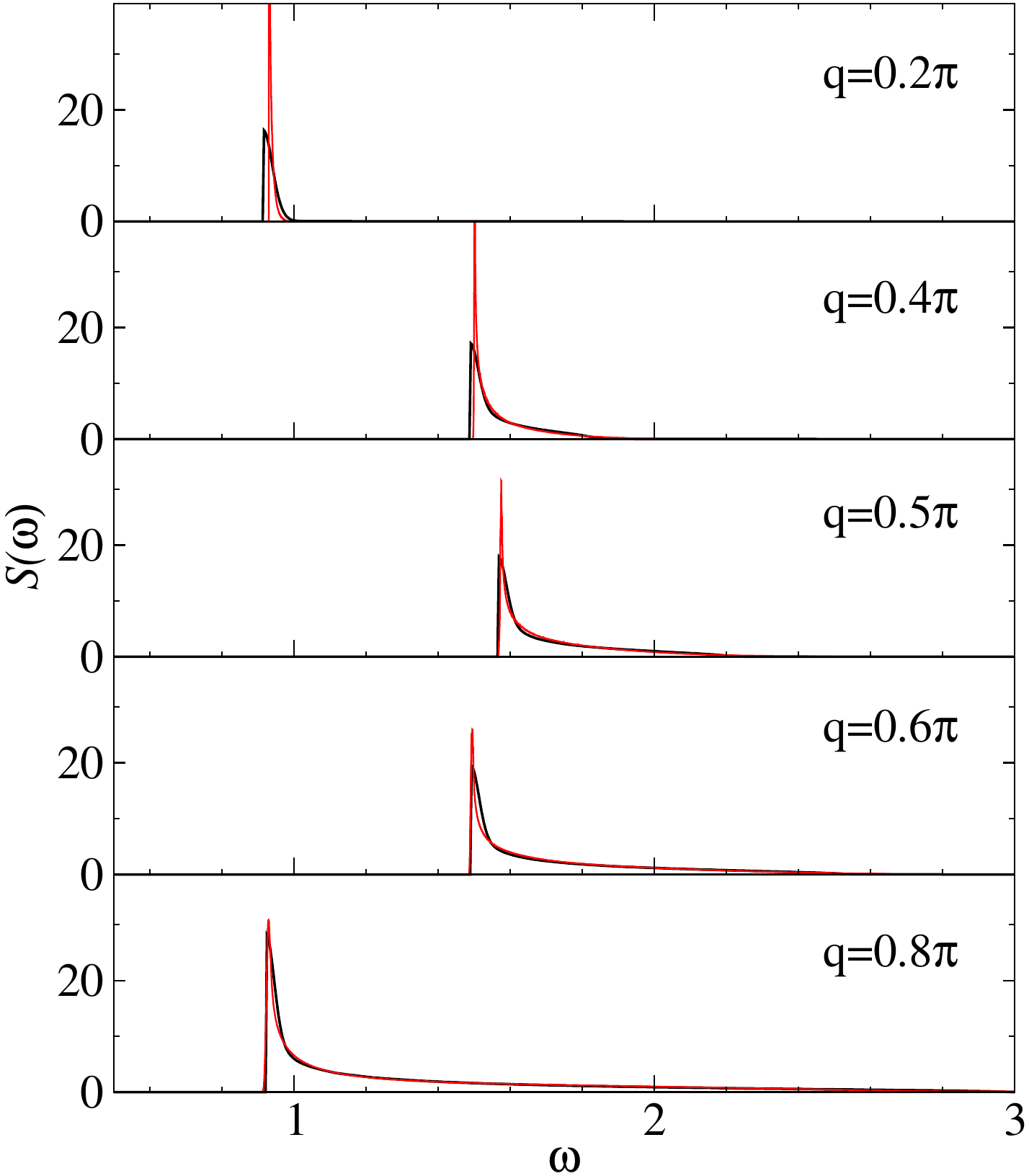}
\caption{SAC spectral functions $S(q,\omega)$ (read curves) for selected momenta $q$ of the Heisenberg chain sampled
with $N_\omega=80$ $\delta$ functions (red curves). The broadening strategy explained in the text was used to smoothen out the
oscillations seen in the $q=4\pi/5$ results in in Fig.~\ref{nobroaden} for small $N_\omega$. The results are are compared with the
corresponding BA results (black curves), which have also been subject to smoothing \cite{cauxdata} (even more so than the SAC results,
resulting in more rounded peaks).}
\label{sw-1d}
\end{figure}

Figure \ref{sw-1d} shows results for several momenta, again comparing with BA results for the same Heisenberg chain of size $L=500$. In all cases, we used
only a modest number of $\delta$-functions, $N_\omega=80$, and the $\delta$-broadening strategy explained above was employed. With the exception of the case of
the smallest momentum, $q=\pi/5$, both the edge and the tail are reproduced very well. For $q=\pi/5$ the SAC result is somewhat too narrow though the result
is still much better than what normally would be expected with numerical analytic continuation. Here it should again be noted that the BA spectra are broadened
and the edges are therefore not divergent. The internal broadening applied in the SAC calculations at the stage of accumulating the spectrum
is very small close to the edge, and the peaks are therefore much sharper. Results such as these can simply not be obtained with conventional
SAC or ME methods.

The total computation time (single core) used for each of the spectral functions in Fig.~\ref{sw-1d}
was only of the order of a few minutes. We also stress again
that there was no other input to these calculations beyond the QMC imaginary-time data and the choice of parametrization of the spectrum; the method of
successive doubling of $N_\omega$ from $10$ to $80$ automatically adapted the edge location and the value of $\langle \chi^2\rangle/N_\tau$ [again
fixed according to Eq.~(\ref{eq:chi2}) after simulated annealing to find $\chi^2_{\rm min}$] was statistically good ($<1$) in all cases.

It is clear that the parametrization used here favors an edge with a sharp peak, and the same entropic effect also some times tends to compress the overall
width of the spectrum, as seen clearly in the $q=\pi/5$ result in Fig.~\ref{sw-1d}. In Sec.~\ref{sec:monoentropy} we will demonstrate explicitly that the
entropic pressures produce an inverse square-root singularity natively, when a spectrum is sampled within fixed bounds without imaginary-time data input.
This form is close to the singularity of $S(q,\omega)$ for $0 < q < \pi$ of the Heisenberg chain in the thermodynamic limit, but in the latter
there is also a multiplicative
logarithmic correction to the power law \cite{bougourzi96,karbach97}. In the presence of the QMC data, the shape of the peak is sufficiently pulled away from the
native form to match the BA data remarkably well in Figs.~\ref{nobroaden} and \ref{sw-1d}. Moreover, the overall shape of the spectrum, the tail of which
depends strongly on $q$, is highly non-trivial, and it was not a priori clear that any analytic continuation with QMC data of typical quality would be able
to resolve it as well as we have demonstrated here. Results almost as good were obtained in Ref.~\cite{sandvik16}, but at much higher cost from
explicitly optimizing the lower and upper spectral bounds on a frequency grid [the parametrization in Fig.~\ref{fig:spec}(a)].

\subsection{Edge optimization}
\label{sec:triangle}

Given that the parametrization used here favors a divergent edge, a relevant question is whether a sharp non-divergent edge also
can be reproduced in this way. The answer is yes, but the entropic pressure favoring the divergence has to be suppressed. To do this, we here introduce a simple
parameter that acts to regularize the edge, namely, a minimum distance $\Delta\omega$ between the first two $\delta$-functions,
\begin{equation}
d_1 \equiv \omega_2-\omega_1 \ge \Delta\omega.
\label{dconstraint}  
\end{equation}
The parameter $\Delta\omega$ is optimized by scanning as in Fig.~\ref{fig:optim} and held fixed during the sampling,
with the increasing spacing between all $\delta$-functions maintained as before; $d_{i+1} \ge d_i$ for all $i \in \{1,\ldots,N_\omega-1\}$. The
modifications of the sampling algorithm described in the previous section are trivial.

\begin{figure}[t]
\centering
\includegraphics[width=70mm]{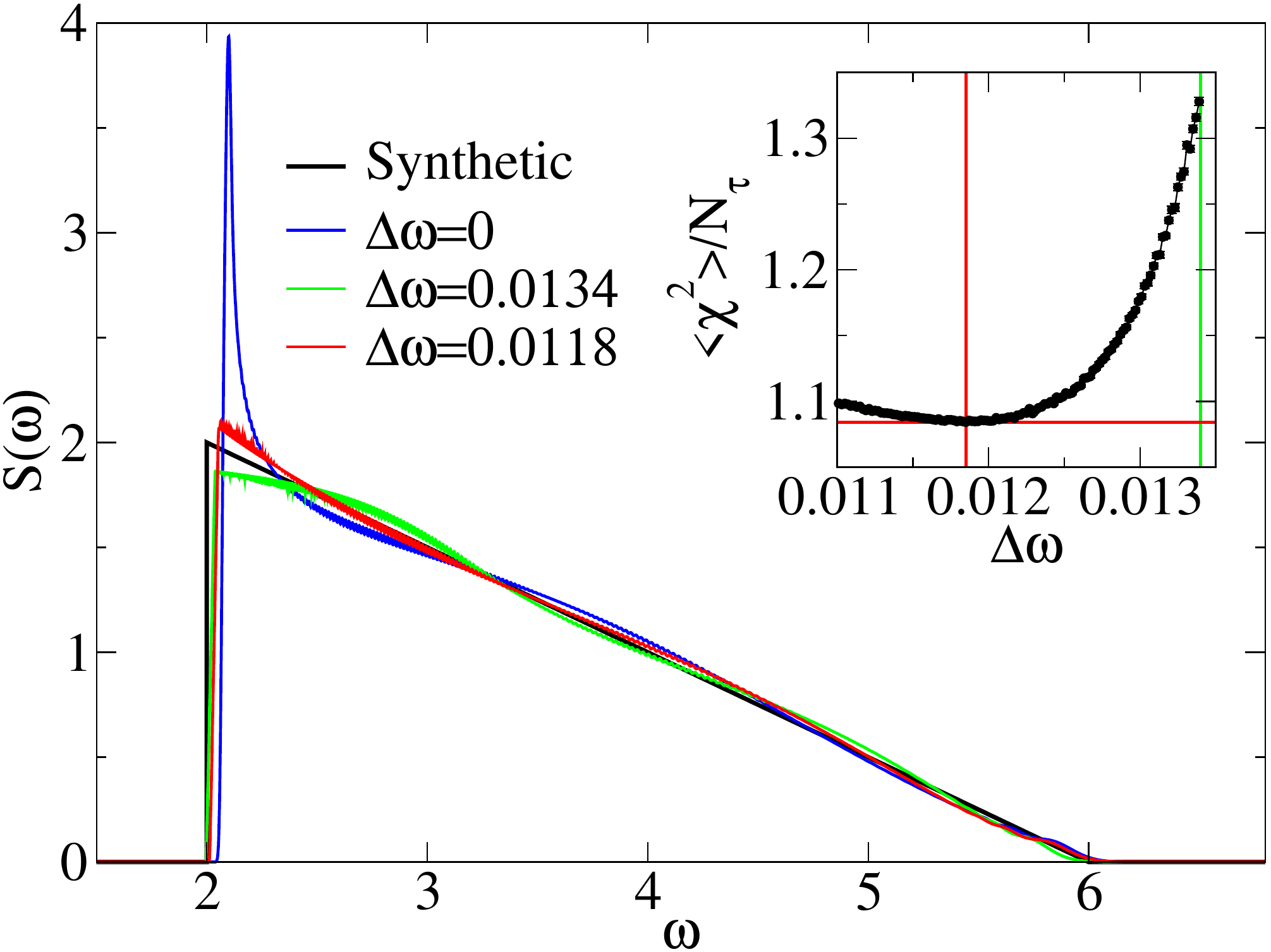}
\caption{Test with synthetic data where the correct spectral function has a triangular shape (shown in black). The result of SAC sampling with only
  the distance-increasing constraint is shown in blue. The goodness-of-fit scan over the minimum edge distance $\Delta\omega$ of the $\delta$-functions
  is shown in the inset, with the optimal value indicated by the red lines. The red and green curves in the main graph show
  spectra sampled with the optimal value of the parameter and at a slightly ($10\%$) higher value (the green
  line in the inset), respectively. In all cases, $N_\omega=160$ $\delta$-functions were used, and when collecting the histogram the spectral
  weight at $\omega_i$ was distributed uniformly within the range $[\omega_{i},\omega_{i+1}]$.}
\label{tri}
\end{figure}

Here we use a synthetic test spectrum that may appear contrived from the physics standpoint but illustrates very well the power of the method---a triangular
profile illustrated in Fig.~\ref{tri}. We constructed $G(\tau)$ from this spectrum and added correlated noise at the level $\sigma=10^{-6}$. If we sample
with only the distance-increasing constraint as described in the previous section, a very sharp edge peak is obtained, as shown with the blue curve in
Fig.~\ref{tri}, and after this peak some ringing can be observed. The inability to correctly resolve the shape of the edge reflects the fact that the parametrization
has an inherent entropic pressure favoring a sharp edge peak, which we will discuss further in Sec.~\ref{sec:monoentropy}.

We next optimize the edge parameter $\Delta\omega$ by scanning and identify a minimum in $\langle \chi^2\rangle$, as shown in the inset of
Fig.~\ref{tri}. Here the minimum appears to be rather shallow, but note that the figure zooms in on a narrow range of $\Delta\omega$ values.
The red and green curves in Fig.~\ref{tri} show the spectrum sampled at the $\Delta\omega$ value corresponding to the $\langle \chi^2\rangle$ minimum
and a slightly larger value, respectively. At and close to the optimum, the entire triangle is resolved well, including the almost linear
decay terminating at the correct upper bound.

For this non-divergent spectral function, the edge optimization within the otherwise unrestricted SAC, illustrated in Fig.~\ref{w0fix}, can also be used
with very good results. Here we used the distance-monotonic parametrization to demonstrate that it can also reproduce a non-divergent edge (of an otherwise
monotonically decaying spectrum). The non-divergent edge can be essentially of any shape with this approach, as long as it is monotonic.

Applying the same edge optimization to $S(q,\omega)$ of the Heisenberg chain, we also find a minimum in $\langle \chi^2 \rangle$, but in this case the
optimal value of $\Delta\omega$ is very close to $0$. The spectrum (not shown here) at the optimal point is almost indistinguishable from the results
in Figs.~\ref{nobroaden} and \ref{sw-1d}. This example as well illustrates the success of the constraint optimization method.

\section{Edge singularities II}
\label{sec:contedge2}

Here we present further insights into and generalizations of the SAC method for edge singularities. In Sec.~\ref{sec:density} we first
discuss an alternative way of defining the average spectrum as a local density of $\delta$-functions which allows more straight-forward access to the
exponent governing a power-law singularity. In Sec.~\ref{sec:monoentropy} we demonstrate that the native entropy driven form of a spectrum
parametrized with monotonically increasing distances produces a singularity of the form $S(\omega) \to (\omega - \omega_0)^{p}$ with $p=-1/2$
when $\omega \to \omega_0$. In order to model a different asymptotic form of the singularity, in Sec.~\ref{sec:monop} we modify the parametrization in
Fig.~\ref{fig:spec}(e) by making the formerly constant amplitudes $A_i$ dependent on $i$ in a way that can produce a native power-law form with any
exponent $p$ (positive or negative). Finally, in Sec.~\ref{sec:monomixed} we discuss how to mix the parametrizations in Figs.~\ref{fig:spec}(b)
and \ref{fig:spec}(e) to model a power-law singularity followed by an arbitrary (not necessarily monotonically decaying) continuum.

\subsection{Spectral density and asymptotic form}
\label{sec:density}

As we observed in Sec.~\ref{sec:heismono} (Fig.~\ref{nobroaden}), with the distance-monotonicity imposed, the locations $\omega_{i}$ of the individual
$\delta$-functions fluctuate only weakly around well defined mean values $\langle \omega_{i}\rangle$. In contrast, in a continuum sampled without restrictions,
even if the frequencies are ordered so that $\omega_i \ge \omega_{i-1}$ at the stage when expectation values are accumulated, the fluctuations in $\omega_i$
are large compared to the mean separation $\langle \omega_i-\omega_{i-1}\rangle$.

\begin{figure}[t]
\centering
\includegraphics[width=75mm]{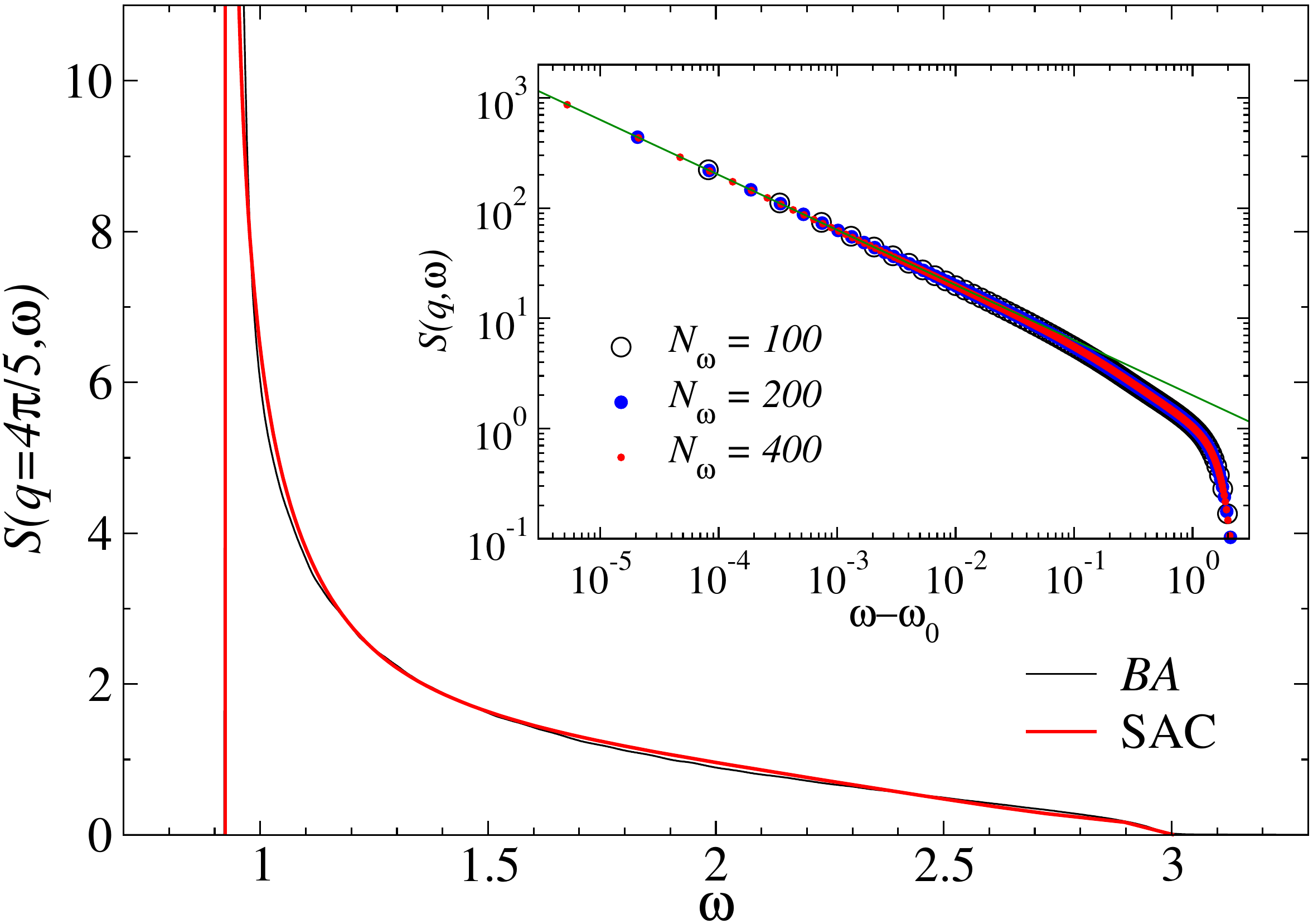}
\caption{Constrained SAC results (red curve) for the dynamic structure factor at $q=4\pi/5$ of the Heisenberg chain with $L=500$ spins, sampled with
$N_\omega = 400$ distance-monotonic $\delta$-functions and using the self-generated grid definition Eq.~(\ref{sdensity}) for the spectral weigh density.
The corresponding BA result \cite{caux05a,caux05b,cauxdata} is shown as the black curve. The inset shows results for different numbers $N_\omega$ of sampled
$\delta$-functions on a log-log scale after subtraction of the sampled mean edge position $\omega_0=\langle\omega_1\rangle$ from $\omega$. 
The blue line shows the form $(\omega-\omega_0)^{-1/2}$. The sampling temperature was $\Theta=1$ for $N_\omega=100$ and slightly lower for $N_\omega=200$ and $400$
[determined according to Eq.~(\ref{eq:chi2}) with $a=0.5$].}
\label{fig:hbergsdens}
\end{figure}

Because of the small frequency fluctuations in the monotonicity constrained SAC, it is appropriate to convert the distance between successive mean
frequencies into a local spectral density by
\begin{equation}
S(\omega_{i+1/2})=\frac{(A_{i}+A_{i+1})/2}{\langle \omega_{i+1}\rangle-\langle \omega_{i}\rangle},
\label{sdensity}
\end{equation}
where we have distributed half of the spectral weight at $\omega_i$ and $\omega_{i+1}$ in the window between these frequencies and assign this
weight to the mid point,
\begin{equation}
\omega_{i+1/2} = \frac{\langle \omega_{i+1}\rangle+\langle \omega_{i}\rangle}{2}.
\label{selfomega}
\end{equation}
We will later generalize the equal-amplitude $\delta$-functions, $A_i=1/N_\omega$, to $i$ dependent amplitudes and therefore include these in the
definition Eq.~(\ref{sdensity}) of the spectral density.
We also note that the spectrum should finally be multiplied by the pre-normalization value of $\pi\bar G(\tau=0)$ to correct for the
normalization used in the sampling, as discussed in Sec.~\ref{sec:qmcdata}.
The mean frequencies defined in Eq.~(\ref{selfomega}) form a self-generated nonlinear grid which automatically adjusts to the singularity by forming an increasing
density of points at the same rate as $S(\omega)$ diverges. We have confirmed that this manner of defining the average spectrum agrees extremely well with the
histogram method where they can be compared (i.e., when the histogram bins are sufficiently small on the scale of the relevant variations in the spectral function).

Fig.~\ref{fig:hbergsdens} shows results for the Heisenberg chain with 500 spins at $q=4\pi/5$. Here the QMC data $\bar G(\tau_i)$ were calculated on the
non-linear grid shown in Fig.~\ref{fig:gtau} (the data used are exactly those in the figure) at temperature $T=10^{-3}$. $N_\omega=400$ $\delta$-functions
were used in the SAC averaging and the definition Eq.~(\ref{sdensity}) was used for the final spectrum displayed in the figure. The lower bound is
within $0.5\%$ of the BA result and the upper bound as well is very closely reproduced. Here it should be noted that the edge is by definition completely
sharp when the spectrum is graphed based on the density definition in Eq.~(\ref{sdensity}). However, the lowest frequency $\omega_1$
fluctuates very little, and the edge is sharp (and the peak very tall) also when the spectral density is collected in a histogram in the
standard way (as we did in Fig.~\ref{nobroaden}). With the histogram, it is more difficult to investigate the asymptotic form of the spectrum,
since a small bin size is needed close to the edge and the results are more affected by noise.

With the self-generated frequency points becoming very dense as the edge is approached, it is indeed possible to use Eq.~(\ref{sdensity}) to investigate
the asymptotic form of the divergence. As shown in the inset of Fig.~\ref{fig:hbergsdens}, the SAC perfectly captures the known power law form
$(\omega-\omega_0)^{-1/2}$ when $\omega -\omega_0 \lesssim 0.1$. However, there should also be a multiplicative logarithmic correction to this
form in the Heisenberg chain \cite{bougourzi96,karbach97}, which is not apparent in the SAC result. We will address this issue further in
Sec.~\ref{sec:monop}.

\subsection{Entropic pressure}
\label{sec:monoentropy}

It is clear that the $\delta$-functions constrained by the monotonically increasing spacing will favor a spectrum with a sharp edge at some
frequency $\omega_0$, with only small fluctuations when the $\bar G(\tau)$ data are good. It is not a priory clear, however, exactly what kind of edge
forms asymptotically for $\omega \to \omega_0$. If the frequencies are sampled in the absence of data without any bounds, they will clearly
diffuse out and cover the infinite frequency range, and to test for a native shape of the average spectrum in the absence of $\bar G(\tau)$
data we have to impose bounds.

\begin{figure}[t]
\centering  
\includegraphics[width=77mm]{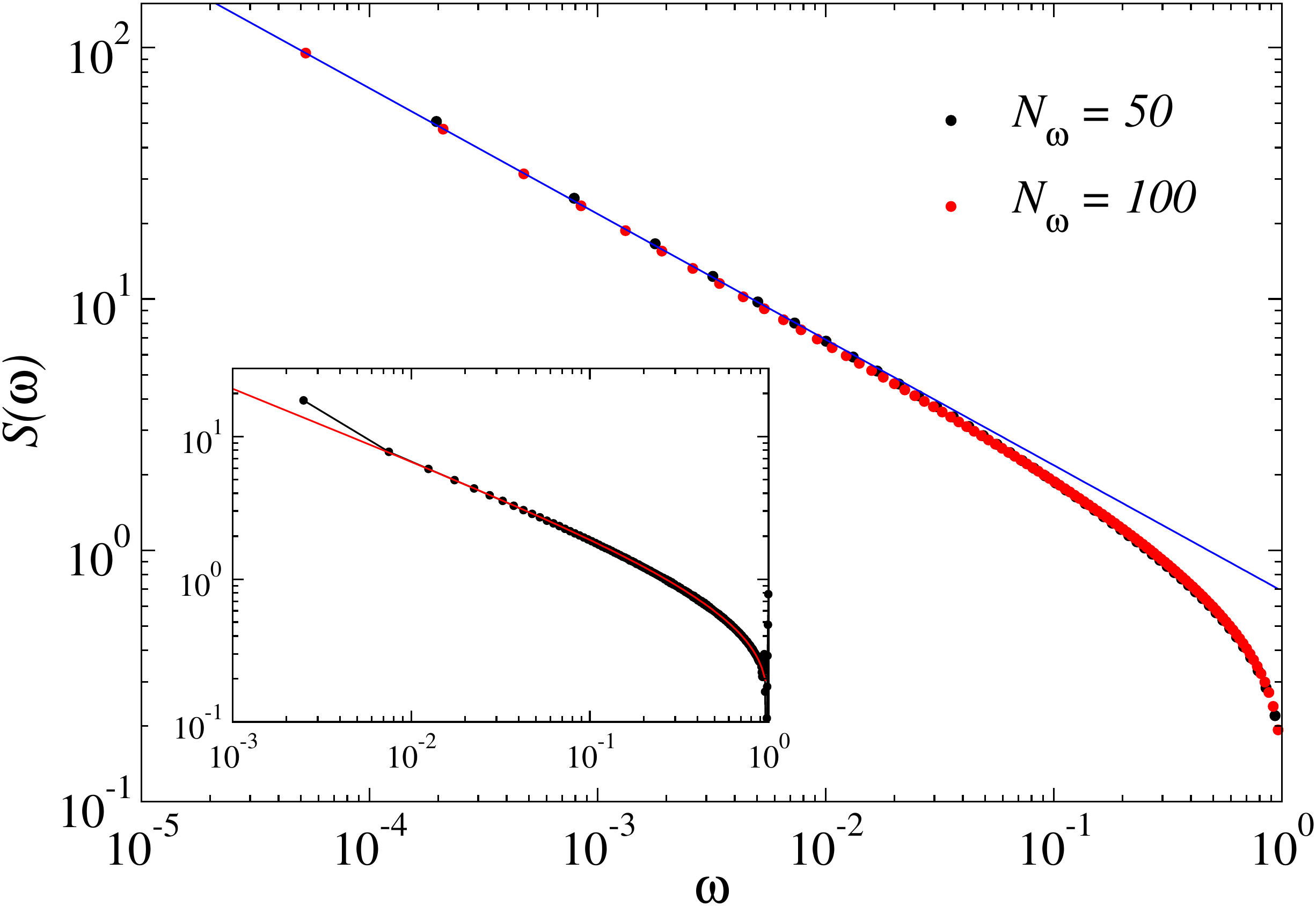}
\caption{Spectral functions obtained by purely entropic sampling of $N_\omega=50$ (black dots) and $N_\omega=100$ (red dots)
$\delta$-functions with the monotonic distance constraint
[Fig.~\ref{fig:spec}(e)] within the bounds $\omega_1=0$ and $\omega_{N_\omega}=1$. The spectral density, graphed on a log-log scale, was extracted from the
mean frequencies $\langle \omega_i\rangle$ according to Eq.~(\ref{sdensity}) on the grid defined by the same data points according to
Eq.~(\ref{selfomega}). The blue line has slope $-1/2$, demonstrating that $S(\omega)$ takes the asymptotic form $\omega^{-1/2}$ for $\omega \to 0$.
The inset shows the same $N=100$ result (for clarity using line segments connecting the data points in the main graph) compared with $S(\omega)$
collected in a histogram with bin width $\Delta=0.005$ (with the points located at the center of the histogram bins).}
\label{n50n100}
\end{figure}

In the case of unconstrained sampling, the mean spectral density will trivially be uniform between any applied fixed bounds,
effectively providing a flat default model between those bounds.
To investigate a corresponding native form when the monotonicity constraint is imposed, we sample this parametrization within the bounds $\omega \in [0,1]$
for different numbers $N_\omega$ of $\delta$-functions and define the spectral density from the resulting mean frequencies according to Eq.~(\ref{sdensity}).
Results are shown in Fig.~\ref{n50n100} for $N_\omega=50$ and $N_\omega=100$, for which the profile is already well converged except very close to the
lower bound. We observe clearly an $\omega^{-1/2}$ behavior for small $\omega$. The inset shows a comparison of the $N_\omega=100$ results with the standard way
of collecting the density in a histogram. The two representations of $S(\omega)$ agree fully, except for the lowest-$\omega$ histogram bin for which the
total weight in that bin can, naturally, not capture the correct divergent form. There are also deviations at the highest frequencies, where the histogram
weights show large oscillations stemming from the fact that the last few $\delta$-functions do not fluctuate much on the scale of their typical spacing
(as in Fig.~\ref{nobroaden}, which is not easy to see on the log scale in Fig.~\ref{n50n100} because these point are piled up at the upper bound).
The density definition Eq.~(\ref{sdensity}) circumvents this issue by construction.

These results show that, within a fixed frequency window, the entropic pressures of the monotonicity constraint induce a singularity of the form
$(\omega-\omega_1)^{-1/2}$, i.e., the same form found in Fig.~\ref{fig:hbergsdens} for the Heisenberg chain.  Thus, the correct edge type (apart
from the expected logarithmic correction) had been implicitly supplied to the process, along with the monotonic decay above the edge. Note,
however that the $(\omega-\omega_1)^{-1/2}$ form is exhibited only very close to the edge (see the inset of Fig.~\ref{fig:hbergsdens}), with
deviations starting already at $\omega-\omega_0 \approx 0.1$ and good agreement with the BA spectrum is observed over the full profile (except very
close to the edge, where the BA result has been broadened).

The native shape obtained with this parametrization, when sampled within a fixed frequency window, also exhibits the asymptotic form only very close
to the lower edge, as seen in Fig.~\ref{n50n100}. While in both cases the decay is faster than $(\omega-\omega_0)^{-1/2}$ away from the edge, the shapes of
the tails are clearly different in Figs.~\ref{fig:hbergsdens} and \ref{n50n100}. Thus, the $\bar G(\tau)$ data, when the statistical errors are small
enough, can pull the edge away from its native entropic form toward the correct form except very close to the edge, where the entropy imposes the native
form. It is again noteworthy how even the upper bound of the spectrum is very well reproduced in Fig.~\ref{fig:hbergsdens}.

\subsection{Singularity with arbitrary exponent}
\label{sec:monop}

An important question now is whether the method introduced here is only applicable to singular edges of the form $(\omega-\omega_1)^{-1/2}$. We already
saw in Sec.~\ref{sec:triangle} that the divergence can be quenched by constraining the first spacing $d_1 \equiv \omega_2-\omega_1$ according
to Eq.~(\ref{dconstraint}), but here we are interested in modeling more general divergent forms, and also non-divergent power-law singularities. 

It should be noted that the asymptotic form of the peak is not always of critical importance as it typically applies only very close to the
edge (as seen in Fig.~\ref{fig:hbergsdens}) and it may be more important to accurately determine the location of the edge and the profile away
from the very close proximity of the peak. A formally incorrect exponent for the divergence is also much preferable to the drastic rounding of 
the peak and associated distortions at higher frequencies introduced with other methods (as in the unrestricted sampling results in Fig.~\ref{sw2}, and
even with the correct lower bound imposed in Fig.~\ref{w0fix}). Nevertheless, it would be additional icing (or whipped cream) on the cake if asymptotic
power laws could also be reproduced with the correct exponent. We here show that this is actually possible with a further generalization of the
monotonicity constraint, by introducing an amplitude profile with an adjustable parameter to accommodate generic asymptotic power-law edges
with positive or negative exponents.

\begin{figure*}[t]
\centering
\includegraphics[width=140mm]{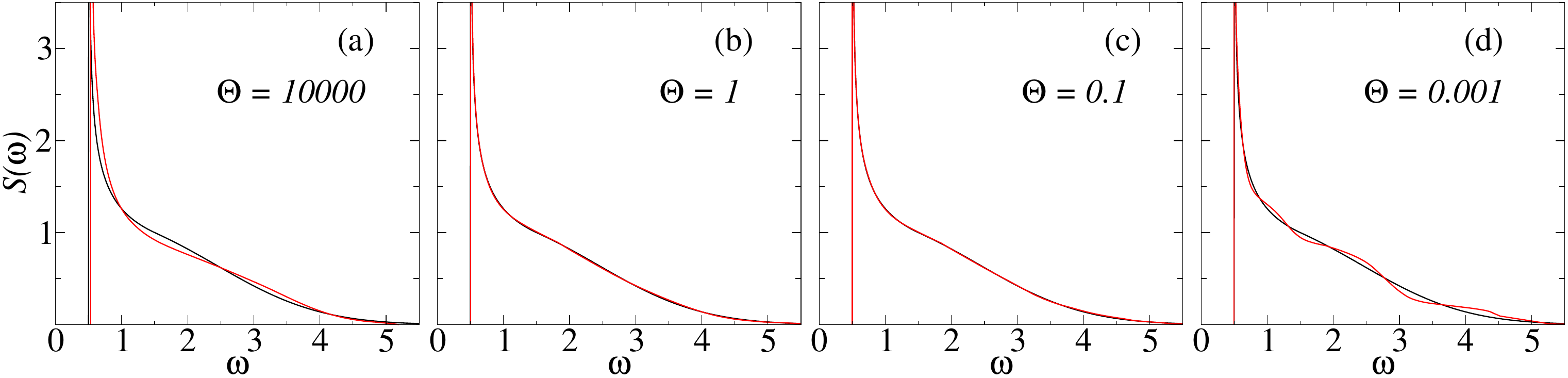}
\caption{Synthetic $T=0$ spectrum with edge divergence exponent $p=-1/3$ (black curves) reproduced with SAC (red curves) using $N_\omega=400$
$\delta$-functions with  monotonicity constraint and the dynamic form Eq.~(\ref{lnamp}) of the amplitudes, with $c=2p+1=1/3$, $\epsilon=0$, and
$n_0$ importance sampled along with the frequencies. Results are shown at four sampling temperatures:
(a) $\Theta=10^4$ ($\langle \chi^2\rangle/N_\tau \approx 2.5\times 10^4$),
(b) $\Theta=1$ ($\langle \chi^2\rangle/N_\tau \approx 0.99$), (c) $\Theta=10^{-1}$ ($\langle \chi^2\rangle/N_\tau \approx 0.86$),
(d) $\Theta=10^{-3}$ ($\langle \chi^2\rangle/N_\tau \approx 0.83$). The noise level in $G(\tau)$ was $4\times 10^{-6}$ and 47 $\tau$ values
on a roughly quadratic grid between $\tau=0.05$ and $\tau=16.2$ were used.}
\label{sw33}
\end{figure*}

Consider an arbitrary power-law edge; $S(\omega \to \omega_0^+) \propto (\omega-\omega_0)^{p}$. We first make use of the fact that the native entropic
pressure on the $\delta$-functions constrained as in Fig.~\ref{fig:spec}(e) corresponds to $p=-1/2$. For large $N_\omega$, we can consider the denominator
of Eq.~(\ref{sdensity}) as the derivative of $\omega$ with respect to $i$. Setting $\omega_0=0$ for simplicity, we then have
$S(\omega) \propto (d\omega/di)^{-1}$. Assuming $\omega_{i+1}-\omega_i = d\omega/di \propto i^a$, we have $\omega \propto i^{a+1}$ and
$d\omega/di \propto S^{-1}(\omega) \propto \omega^{a/(a+1)}$. Given that the actual exponent here should be $a/(a+1)=1/2$, we have $a=1$,
i.e., for constant $A_i$ we will obtain $\langle\omega_{i+1}-\omega_i\rangle \propto i$ and $\omega \propto i^2$ for $i \ll N_\omega$ and large
$N_\omega$. If we now let the the amplitudes vary with the index $i$ as
\begin{equation}
A_i \propto i^c,
\label{anform}
\end{equation}
and using $i \propto \omega^{1/2}$, the spectral density as defined in Eq.~(\ref{sdensity}) gives
\begin{equation}
S(\omega) \propto (\omega-\omega_0)^{(c-1)/2},
\label{scform}
\end{equation}
by entropy alone for $\omega$ close to $\omega_0$.
Thus, if we want a specific exponent $p$, we should choose $c=2p+1$ in Eq.~(\ref{anform}).
This entropy favored form should prevail also when sampling without explicit frequency bounds in
the presence of imaginary-time data with the probability distribution Eq.~(\ref{psg}), as it did in the case of the Heisenberg chain with constant amplitudes
(Fig.~\ref{fig:hbergsdens}), because the lower edge is always eventually, for sufficiently large $N_\omega$, entropy dominated. 

The original constrained parametrization with constant $A_i$ has the advantage that it can easily reproduce an arbitrary monotonic decay. To preserve this
property for arbitrary $p$, we let the modified amplitudes $A_i \propto i^c$ cross over into the constant form, which we implement by 
\begin{equation}
\ln(A_i)=b+cx_i/2 \pm \sqrt{(cx_i/2)^2+\epsilon^2},~~~~x_i \equiv \ln ({i}/{n_0}).
\label{lnamp}
\end{equation}
where $b$ is the constant normalizing the sum of amplitudes and  $+$ and $-$ correspond, respectively, to $c<0$ and $c>0$. The parameter $n_0$ in the
definition of $x_i$ is real-valued in the range $n_0 \in \{1,N_\omega\}$ and sets the point of cross-over to the constant amplitudes. Where this cross-over
takes place should be dictated by the QMC data, and therefore $n_0$ is sampled along with the other Monte Carlo updates of the spectrum (note that all
amplitudes have to be re-normalized when $n_0$ is changed). The parameter $\epsilon$ in Eq.~(\ref{lnamp}) imposes a rounding of the cross-over, but we have
found that such a parameter is mostly not needed, as the fluctuations in the sampling of $n_0$ impose a de facto natural (data-imposed) small rounding as
well; thus we use $\epsilon=0$ henceforth though in some cases it may still be better to use some small non-zero value.

The eventual goal of this modified parametrization will be to determine the best value of the exponent $p$ according to the generic scheme
in Fig.~\ref{fig:optim}. We will show that such optimization indeed can be carried out and produces surprisingly good results. We first present some
results where the correct exponent is imposed from the outset, with the aim of investigating how well the edge location and the overall profile can
be determined when reproducing a synthetic, non-trivial spectrum.

In Fig.~\ref{sw33} we present tests for a synthetic spectrum with edge exponent $p=-1/3$, with the power law cut off at high frequency by an exponential
tail. To make the profile more complicated we also added a broad Gaussian, cut off at the edge and with other parameters chosen such that the spectrum
is still monotonically decaying above the edge at $\omega_0=1/2$. We first present results at four different sampling temperatures $\Theta$, to
investigate the role of the optimal-$\Theta$ criterion with this parametrization (which we postponed in Sec.~\ref{sec:contedge2}).

\begin{figure*}[t]
\begin{center}
 \includegraphics[width=110mm]{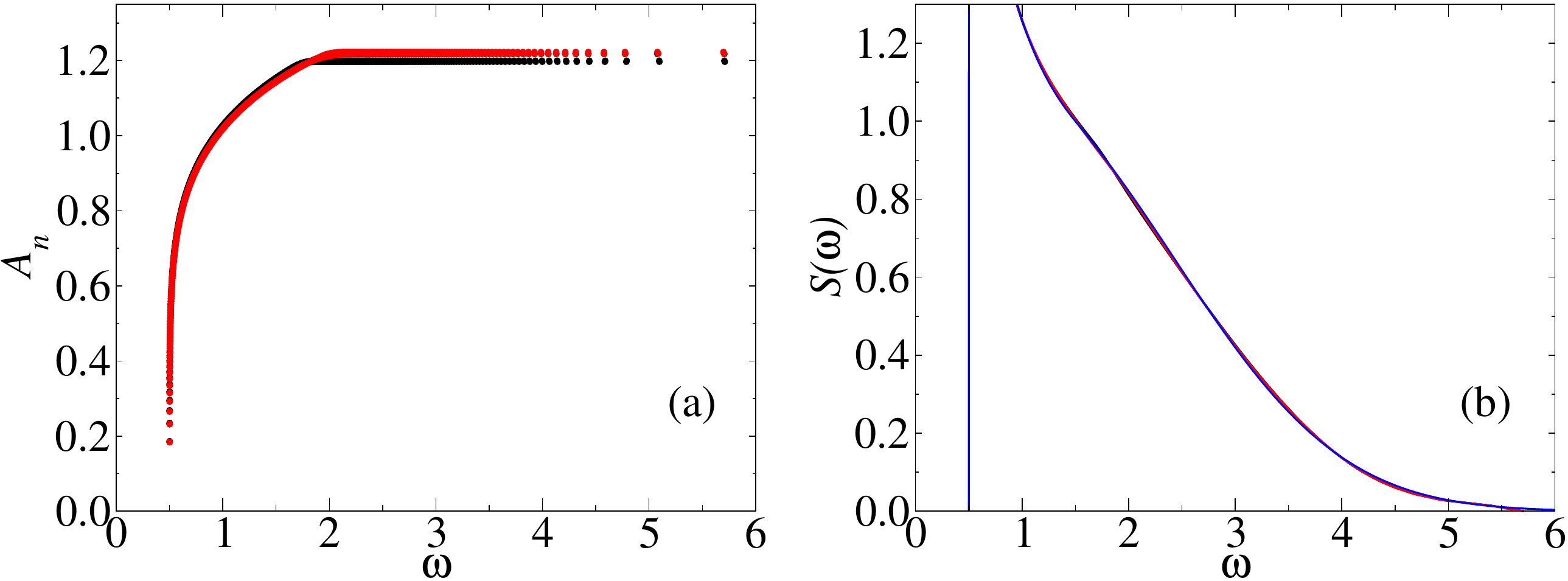}
\caption{Results of five independent runs for the same synthetic spectrum and $\bar G(\tau)$ data as in Fig.~\ref{sw33}, all at $\Theta=1$. The mean amplitudes
(here multiplied by $N_\omega$, so that the average is $1$)
resulting from the sampling of $n_0$ according to Eq.~(\ref{scform}) with $\epsilon=0$ and $c=2p+1=1/3$ are shown in (a) vs the corresponding mean
frequencies $\omega=\langle \omega_i\rangle$. The five runs have converged to two different groups; the amplitudes in the first group (three of the runs)
are shown in black and those of the second group (two runs) are shown in red. Within these groups, the data points are indistinguishable on the scale shown here.
The corresponding spectral functions are shown using the same color coding in (b), with the exact spectrum drawn in blue. All the curves fall very
close to each other and are hard to distinguish.}
\label{n0runs}
\end{center}
\end{figure*}

At $\Theta=10^4$,
Fig.~\ref{sw33}(a), where $\langle \chi^2\rangle$ is very far from acceptable, the spectral weight is still roughly in the correct window but the edge
is a bit too high and there are clear deviations from the exact spectrum everywhere. In contrast, at $\Theta=1$ and $\Theta=0.1$, Figs.~\ref{sw33}(b)
and \ref{sw33}(c), the $\langle \chi^2 \rangle$ values are good and no significant deviations from the exact spectrum can be seen on the scale of
the figures (the edge in both cases is less than $0.5\%$ above the correct frequency $\omega_0$). Close to the $\Theta \to 0$ limit, in
Fig.~\ref{sw33}(d) the spectrum for $\Theta=10^{-3}$ again has deteriorated, but the edge is still very close to its correct location. Unlike the case
of free sampling (e.g., in Fig.~\ref{sw2}), with the monotonicity constraint no multi-peak structure can appear as $\Theta \to 0$, and instead a
series of still monotonically decaying shoulders have formed. The main messages here are that (i) $\Theta > 0$ is still required for the best fidelity
and (ii) the sampled spectrum is very insensitive to the exact value of $\Theta$ in the range of reasonable values of $a \le 1$ in the
$\langle \chi^2\rangle$ Eq.~(\ref{eq:chi2}), here corresponding to $\Theta \lesssim 1$.

We next demonstrate in more detail how the cross-over between the imposed asymptotic power-law, Eq.~(\ref{anform}), and the constant amplitudes is realized
when $n_0$ in Eq.~(\ref{lnamp}) is sampled together with the other degrees of freedom of the spectrum (as was also done for Fig.~\ref{sw33}). Fig.~\ref{n0runs}(a)
shows results for the mean values of amplitudes from five independent $\Theta=1$ runs with the same $\bar G(\tau)$ data as in Fig.~\ref{sw33}, graphed
versus the corresponding mean frequencies. The cross-over behavior is rather sharp, but no apparent anomalies are seen (on the scale of the plot) at the
corresponding cross-over frequencies in the spectral functions in Fig.~\ref{n0runs}(b). The spectral function is less sensitive to $n_0$ than the
amplitudes (i.e., the sampling can compensate for a sub-optimal $A_i$ form), which is illustrated by the fact that the amplitudes of the five runs converged
to two groups of slightly different mean values but no such obvious grouping can be observed in the spectral functions. In all cases, the broad tail of
the synthetic spectrum is reproduced essentially perfectly, illustrating that the mean sampled frequencies adapt correctly to the amplitudes. If the
simulated annealing process is sufficiently slow, and the final spectrum is sampled long enough, the amplitudes (i.e., the cross-over point $n_0$)
should also converge consistently. The runs here were rather short (less than a minute for each of 150 annealing steps and about 30 minutes spent at $\Theta=1$).

\begin{figure}[t]
\begin{center}
\includegraphics[width=80mm]{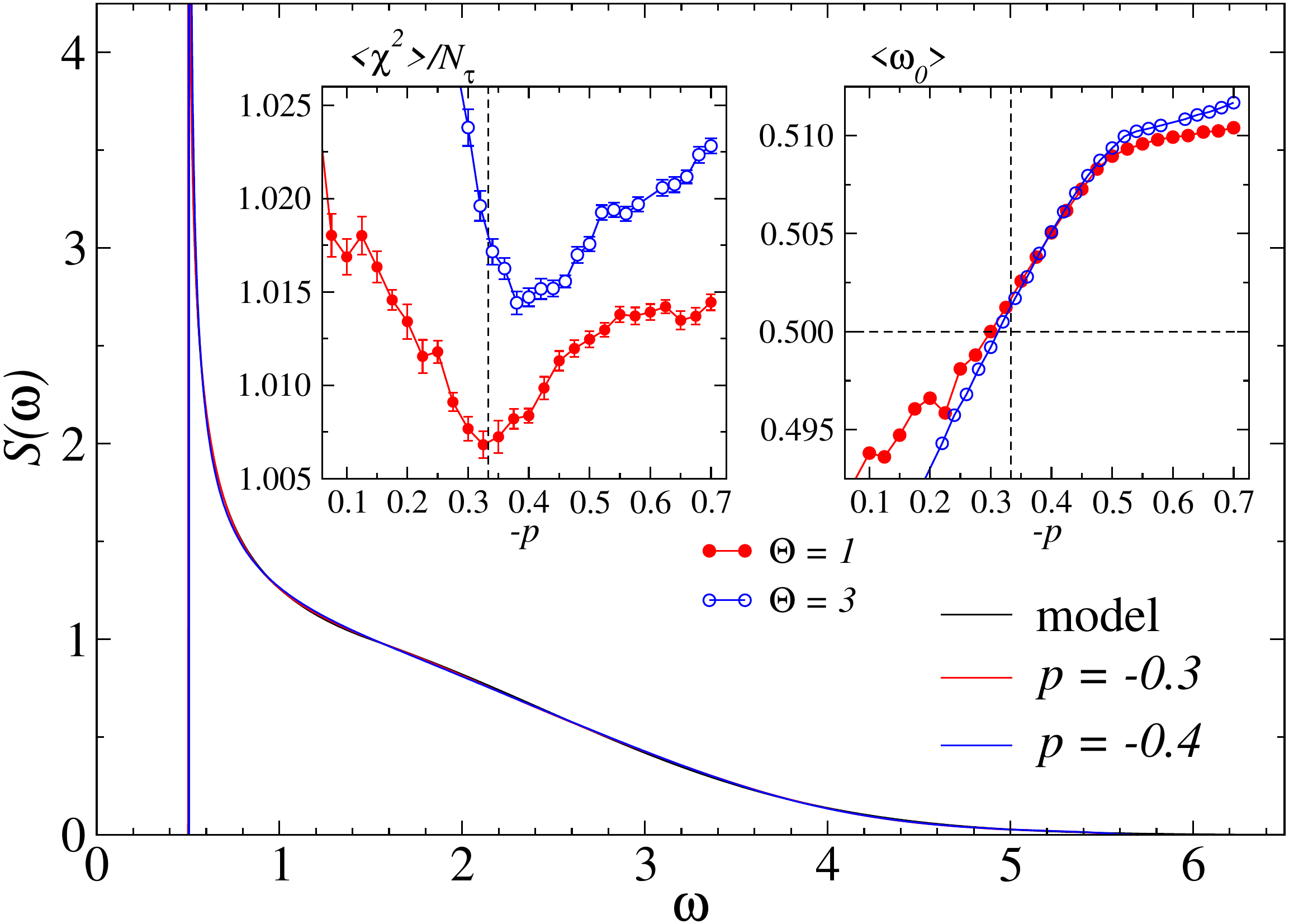}
\caption{Synthetic spectrum with edge exponent $p=-1/3$ (black curve) and almost indistinguishable results from SAC with $p=-0.3$ and $-0.4$
(blue and red curves), sampled with $N_\omega=400$ at $\Theta=1$. The left inset shows the mean goodness of the fit obtained by sampling with different exponents
$p$ at $\Theta=1$ (red circles) and $\Theta=3$ (blue circles, where $0.26$ has been subtracted from the results); the vertical dashed line indicates the
correct value $p=-1/3$. The right inset shows the location of the edge versus the exponent for the same $\Theta$ values; the correct value $1/2$ is
indicated with the horizontal dashed line.}
\label{p33scan}
\end{center}
\end{figure}

If the exponent $p$ is not known, it can be determined at least approximatively by monitoring $\langle \chi^2\rangle$ at fixed $\Theta$, as illustrated in
Fig.~\ref{p33scan} for runs using the same $p=-1/3$ data as above. As shown in the left inset of the figure, a minimum forms in $\langle \chi^2(p)\rangle$ 
very close to the correct value of the exponent when sampling at $\Theta=1$. At a higher temperature, $\Theta=3$ (where we have subtracted
a constant from $\langle \chi^2\rangle/N_\tau$ in order to better visualize the two data sets in the same graph), the minimum is shifted to 
$p \approx -0.38$. Thus, we see a similar trend as in the tests with $\delta$-function edge in Sec.~\ref{sec:delta2}, where the correct
value of the constraint is approached when the sampling temperature is sufficiently low (if the data quality is also sufficiently good), though
the optimum is also harder to discern for lower $\Theta$. The location of the edge, shown versus $p$ in the right inset of Fig.~\ref{p33scan}, also
takes its best value (within $0.5\%$ of the correct value when $\Theta=1$) when the exponent $p$ is in the close neighborhood of its correct value.

While the input value of $c=2p+1$ in the amplitude form, Eq.~(\ref{anform}), dictates the asymptotic form of the edge singularity, the shape of the spectrum
away from the very close vicinity of the peak is not very sensitive to the exact value of $c(p)$ used. The main part of Fig.~\ref{p33scan} shows spectra for
both $p=-0.3$ and $p=-0.4$, but these curves are too close to each other and the correct profile to be distinguishable in the plot.

The fact that $p$ at the $\langle \chi^2\rangle$ minimum (left inset of Fig.~\ref{p33scan}) is considerably better at $\Theta=1$ than at $\Theta=3$ indicates that
the data quality in this case is barely sufficient to extract the correct asymptotic exponent, though the resolution of the overall profile, except for the
very tip of the divergent peak, is already very stable, with almost indistinguishable results obtained at $\Theta=1$ and $\Theta=3$ (not shown). We stress again
that the error level is realistic for actual (long) QMC simulations; $\sigma = 4\times 10^{-6}$, which is close to the noise level in the QMC data for
the Heisenberg chain used previously and graphed in Fig.~\ref{fig:gtau}. 

Here it should be noted that very long sampling runs are required in order for $\langle \chi^2\rangle$ to be determined precisely enough to detect the
$\langle \chi^2\rangle$ minimum---about 30 CPU hours was used for each $p$ value (in parallel runs)
in the case shown in Fig.~\ref{p33scan}, in order to reach sufficiently
small error bars. It is nevertheless remarkable that information on the value of $p$ is contained in the $\bar G(\tau)$ data. With still lower
noise level, the optimal exponent becomes easier to identify. We also point out the well known fact that the effort of generating good imaginary-time
data with QMC simulations normally far exceeds what is required even in slow SAC procedures such as the above.

We stress that the results shown here are typical in how edge exponents can be resolved with data of similar noise levels and with monotonically
decaying continua of various shapes. With decreasing (more negative) exponent $p$ (sharper divergence), the sampling of the spectrum becomes less efficient,
however, and the slower convergence of $\langle \chi^2\rangle$ (larger error bars for given sampling time) can pose an obstacle for determining the exponent
to good precision when $p \lesssim -0.8$. However, as also exemplified above, even if the exponent is formally not quite correct, the overall profile of
the spectrum, including the location of the edge, is invariably well reproduced.

We have also carried out scans similar to those in Fig.~\ref{p33scan} with the Heisenberg simulation data (same as in Fig.~\ref{fig:hbergsdens}). We
find that the optimal exponent is $p\approx -0.7$ in this case. The more negative value than the naively expected $p=-0.5$ can be understood on account
on the fact that there is a multiplicative correction to the power-law edge \cite{bougourzi96,karbach97}. Apart from minor changes in the peak
form very close to $\omega_q$, the spectrum overall is essentially unchanged (and therefore not shown here) from the results in Fig.~\ref{fig:hbergsdens}.

We elaborate a bit more on the asymptotic divergence of $S(q,\omega)$ in the case of the Heisenberg chain. It is well known that logarithmic corrections
in general (e.g., in the distance dependence of static correlation functions) can lead to ``effective'' power laws with slightly different exponents when
analyzed within a finite range of the argument (see, e.g., results for the correlation function of the Heisenberg chain in Ref.~\cite{sandvik10}). It is
therefore not surprising that analytic continuation based on a small imaginary-time data set can produce a similar, larger (in magnitude) effective exponent
in the dynamic structure factor, in the above example $p \approx -0.7$ instead of $p=-0.5$, to account for the substantial correction from the also
divergent log factor. It should also again be noted that, even for $L=500$,
the $T=0$ spectrum for the Heisenberg chain is a sum of a not very large number of $\delta$-functions \cite{wang19}, and this deviation from a strictly
continuous spectrum may also have some impact on the analytic continuation with a parametrization that mimics a true continuum to a higher degree
when $N_\omega$ is large.

\begin{figure}[t]
\centering
\includegraphics[width=81mm]{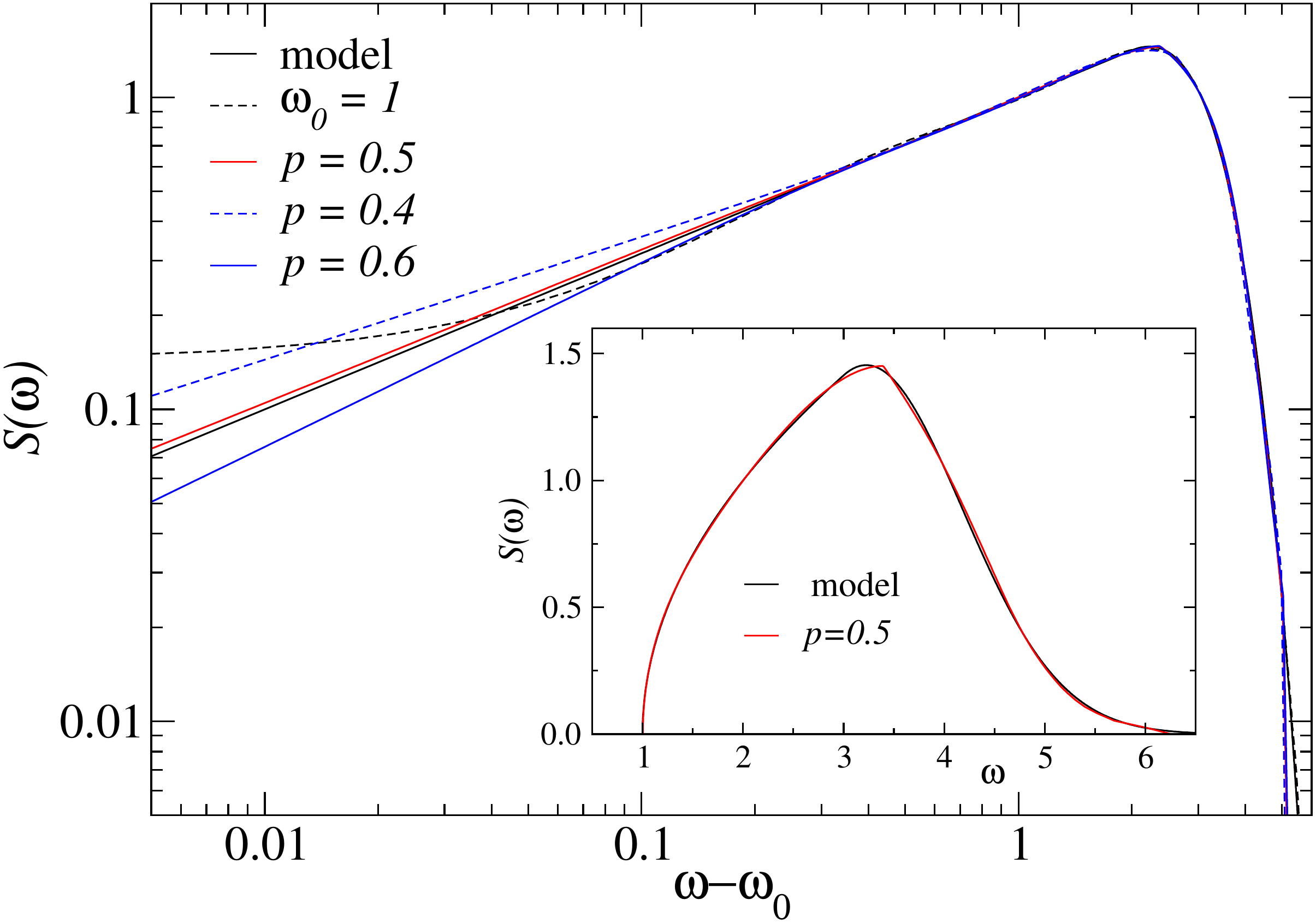}
\caption{Synthetic spectrum with edge of the form $(\omega-\omega_0)^{1/2}$ (black curve) together with SAC results with the correct edge exponent $p=0.5$
  (red curve), as well as $p=0.4$ (blue dashed curve) and $p=0.6$ (blue solid curve). In order to discern asymptotic power-law behaviors in this log-log plot,
the results are graphed vs $\omega-\omega_0$, where the edge location $\omega_0$ is determined from each individual SAC spectrum ($\omega_0 \approx 1.015$ for
$p=0.4$, $\omega_0 \approx 1.003$ for $p=0.5$, and $\omega_0 \approx 0.992$ for $p=0.6$). A result obtained with unrestricted frequency updates above the imposed
correct edge location $\omega_0=1$ [shown on a linear scale in Fig.~\ref{scanm50}(b)] is also included (black dashed curve). The inset shows the exact synthetic
and $p=0.5$ SAC spectra vs $\omega$ on a linear scale. The error level of the QMC data was $\sigma = 10^{-6}$ and $N_\omega=800$.}
\label{m50}
\end{figure}

We discuss one more example, where the edge is not divergent but is controlled by a power law with positive exponent. We test the form $(\omega -1)^{1/2}$,
which we further multiply by an exponential function ${\rm e}^{-(\omega-3)}$ for $\omega > 3$; this synthetic spectrum is shown in the inset of
Fig.~\ref{m50}. Here the SAC sampling was first run with the correct input exponent $c=2$ corresponding to $p=0.5$ in the amplitude form Eq.~(\ref{anform}).
The resulting spectrum agrees very well with the exact profile. The cross-over between the power-law increase and the exponential decay is rather sharp in
the synthetic spectrum, and the error in the SAC spectrum is also the largest at the peak (which is seen more clearly on the linear scale used in the inset
of Fig.~\ref{m50}), with a rather sharp kink corresponding to only small fluctuations in the cross-over parameter $n_0$ in Eq.~(\ref{lnamp}). It is still
noteworthy that the sampling of $n_0$ converges the cross-over to the correct frequency region. Using a small non-zero value of $\epsilon$ in the cross-over
form leads to a more rounded maximum, but here we want to avoid additional parameters and consistently show results for $\epsilon=0$. The far tail of the exponential
decay is somewhat suppressed, though this is not seen very clearly on the scales used in the plots. This minor deficit is due to the fact that
$\langle n_0\rangle \approx 660$ is close to $N_\omega=800$, and the number of $\delta$ functions available for the tail part is therefore rather small.
For $N_\omega=1600$ (not shown here), the tail resolution is further improved but the sharp kink at the peak still persists.

To test the sensitivity of the spectrum overall to the amplitude exponent $c$ used, we also carried out runs with $c$ corresponding to other edge exponents $p$.
Results for $p=0.4$ and $p=0.6$ are shown along with the $p=0.5$ spectrum in the main log-log plot in Fig.~\ref{m50}. The location of the edge $\omega_0$
is determined to better than $0.5\%$ when the correct value $p=0.5$ of the exponent is used, while using $p=0.4$ and $p=0.6$ delivers $\omega_0$ about $1.5\%$
too high and $1\%$ too low, respectively. To observe the asymptotic power law behaviors at the edge, the spectra are graphed versus $\omega-\omega_0$ in
Fig.~\ref{m50}, with the edge location $\omega_0(p)$ defined as $\omega_0=\langle \omega_1\rangle$ for given exponent $p$ [i.e., slightly below the first
frequency in the grid definition Eq.~(\ref{selfomega})]. For small $\omega-\omega_0$, the different power-law behaviors dictated by the imposed exponent
$c$ of the amplitudes in Eq.~(\ref{anform}), but for $\omega-\omega_0 \gtrsim 0.2$ even the $p=0.4$ and $p=0.6$ spectra cross-over into the correct $p=0.5$
form until the peak of the spectrum at $\omega=3$.

\begin{figure*}[t]
\centerline{\includegraphics[width=160mm, clip]{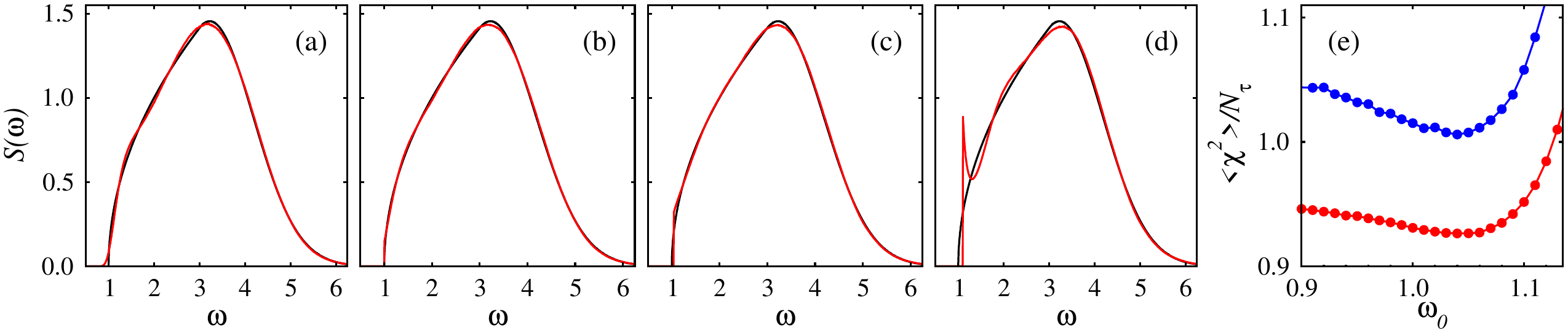}}
\vskip-1mm
\caption{Results of SAC (red curves) with frequency-only updates [the parametrization in Fig.~\ref{fig:spec}(b)] for the
  same synthetic spectrum as in Fig.~\ref{m50} (black curves). The spectrum obtained with
  unrestricted sampling is shown in (a), while a lower bound was imposed in the other cases; $\omega_0=1$ (the correct bound) in (b), $\omega_0=1.04$
  (where $\langle \chi^2\rangle$ is minimized) in (c), and $\omega_0=1.1$ in (d). Panel (e) shows the goodness of fit vs the lower bound at two sampling
  temperatures; $\Theta=1$ (red circles) and $\Theta=2$ (blue circles). The spectral functions in (a)--(d) were obtained with $\Theta=1$.}
\label{scanm50}
\vskip-2mm
\end{figure*}

The results in Fig.~\ref{m50} were obtained by sampling at $\Theta=1$, resulting in $\langle \chi^2\rangle/N_\tau \approx 0.92$ for $p=0.5$.
In a scan over many values of $p$, the $\langle \chi^2\rangle$ value does not change appreciable until $p < 0.15$ or $p > 0.60$. Thus, in this case, it is
much more difficult (than in the case of a divergent edge) to determine the exponent if it is not known; especially to determine a lower bound of $p$.
This difficulty can be understood because the
spectral weight at the edge is very small when $p > 0$, and a change in $p$ from the correct value can then be largely compensated for (to fit the
data) by a shift in the edge location, unless the noise level of $\bar G(\tau)$ is very low (much lower than $\sigma = 10^{-6}$ used here).

Despite the difficulties of independently determining $p$ in this case, the level of fidelity at the edge when $p$ is at or close to the correct value
would be impossible to obtain with standard analytic continuation methods. To further illustrate this point, Fig.~\ref{scanm50}(a) shows results for the same
synthetic spectrum obtained with unrestricted SAC, using equal-amplitude $\delta$-functions. While the agreement with the correct form is actually
quite good, even somewhat better than the results in the inset of Fig.~\ref{m50} at the peak and high-frequency tail, there is no clear asymptotic
power-law behavior. The edge is rounded and there is associated mild ringing at higher frequencies. Though overall the edge distortion is not very
dramatic, it would still be difficult to extract an exponent.

The method of just constraining the sampling with a lower bound $\omega_0$ also does not work very well. If the exact lower bound is used, as in
Fig.~\ref{scanm50}(b), the agreement improves over Fig.~\ref{scanm50}(a), but now there is a small jump at the edge and noticable (though milder)
ringing behavior. The same spectrum is also shown in the log-log plot in Fig.~\ref{m50} (dashed black curve), where the deviations from the asymptotic
power-law behavior can be observed more clearly. Like the other results, the correct form is still well reproduced above $\omega = 0.2$. Independently
determining the edge location is difficult in this case. Results with $\omega_0=1.04$ and $1.1$ are shown in Figs.~\ref{scanm50}(c) and \ref{scanm50}(d),
respectively. The former case actually corresponds closely to the $\langle \chi^2\rangle$ minimum, as shown in Fig.~\ref{scanm50}(e) for two different
sampling temperatures $\Theta$, while the latter is clearly too high according to $\langle \chi^2\rangle$.

The reason why this simple lower-bound method fails here is again that the spectral weight close to the edge is very small. In some other cases we have
tested, the small sharp spike at the edge in Fig.~\ref{scanm50}(d) appears even before the $\langle \chi^2\rangle$ minimum. The example again shows
how entropic pressures can lead to distortions when insufficient constraints are imposed, as we saw also for a case of a divergent edge sampled
with the simplest type of constraint in Fig.~\ref{w0fix}.
In this particular case illustrated in Fig.~\ref{scanm50}, the unrestricted sampling above the edge cannot reproduce the power-law
form well enough for the simple optimization of the lower bound to work well, even at a low noise level of the $\bar G(\tau)$ data.

In general, for any spectrum that is expected to host an edge, it is useful to first carry out optimization of just the lower bound (after the very first
step of carrying out unrestricted SAC). Even if the results may
still not be satisfactory, they can give hints as to what the correct edge shape may be, e.g., a sharp peak indicating a divergent edge, as in Fig.~\ref{w0fix},
or a jump (in some cases even a small spike), as Fig.~\ref{scanm50}(b), suggesting
the possibility of a power law with a positive exponent. Tests with better parametrizations can then follow.

\subsection{Generic continuum}
\label{sec:monomixed}

\begin{figure}[t]
\centering
\includegraphics[width=75mm]{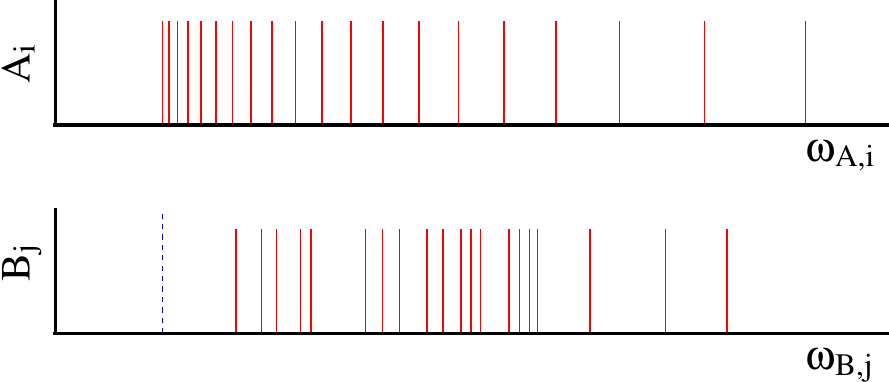}
\caption{Parametrization of the spectrum as two sets of $\delta$-functions. The first set A, with amplitudes $A_i=W_A/N_A$ is sampled under 
the monotonicity constraint $\omega_{A,i+1}-\omega_{A,i} > \omega_{A,i}-\omega_{A,i-1}$ (here illustrated with all equal amplitudes, corresponding to the
edge exponent $p=-1/2$), while the second set with weights $B_j=W_B/N_B$ is only constrained by ${\rm min}\{\omega_{B,j}\} > \omega_{A,1}$ 
(the vertical dashed line). The limiting cases $W_A=0$ and $W_B=0$ correspond to free sampling and monotonically decaying continuum, respectively.}
\label{fig:mixed}
\end{figure}

For spectra that are not monotonically decaying above the edge, we can mix the parametrizations with the monotonicity  constraint (part A with $N_A$
$\delta$-functions) and the equal-amplitude unconstrained $\delta$-functions (part B with $N_B$ $\delta$-functions); see Fig.~\ref{fig:mixed}. In
the simplest case, the A part corresponds to a divergent edge with exponent $p=-1/2$, in which case the amplitudes are $A_i=W_A/N_A$ for part A and $B_j=W_B/N_B$,
where $W_A+W_B=1$ and the A-part weight $W_A \in (0,1)$ should be determined as an optimized constraint. When sampling, the B part is not completely
unrestricted, as the lower edge of the A part serves as a lower bound for the B $\delta$-functions. Other edge exponents can also be imposed or
optimized, though in the latter case a 2D scan over both the exponent $p$ and the mixing parameter $W_A$ would be required.

To still take advantage of the definition of the spectral weight density in Eq.~(\ref{sdensity}), we combine this form for the A part
of the spectrum with the B part recast into a histogram based on the self-generated grid points in Eq.~(\ref{selfomega}). This merger of the two
different representations of the spectral density only has to be performed at the output stage. The working histogram for the B part during the
sampling process is defined with a very small frequency spacing, the same $\delta_\omega$ as used for the micro grid on which the precomputed kernel and
its derivative are stored when the continuous (double precision) frequencies $\omega_i$ are used, as detailed in Sec.~\ref{sec:contsamp}. For any
part of the B continuum extending above the highest frequency on the self-generated grid, a regular histogram with equal-size bins is used.

\begin{figure}[t]
\centering
\includegraphics[width=75mm]{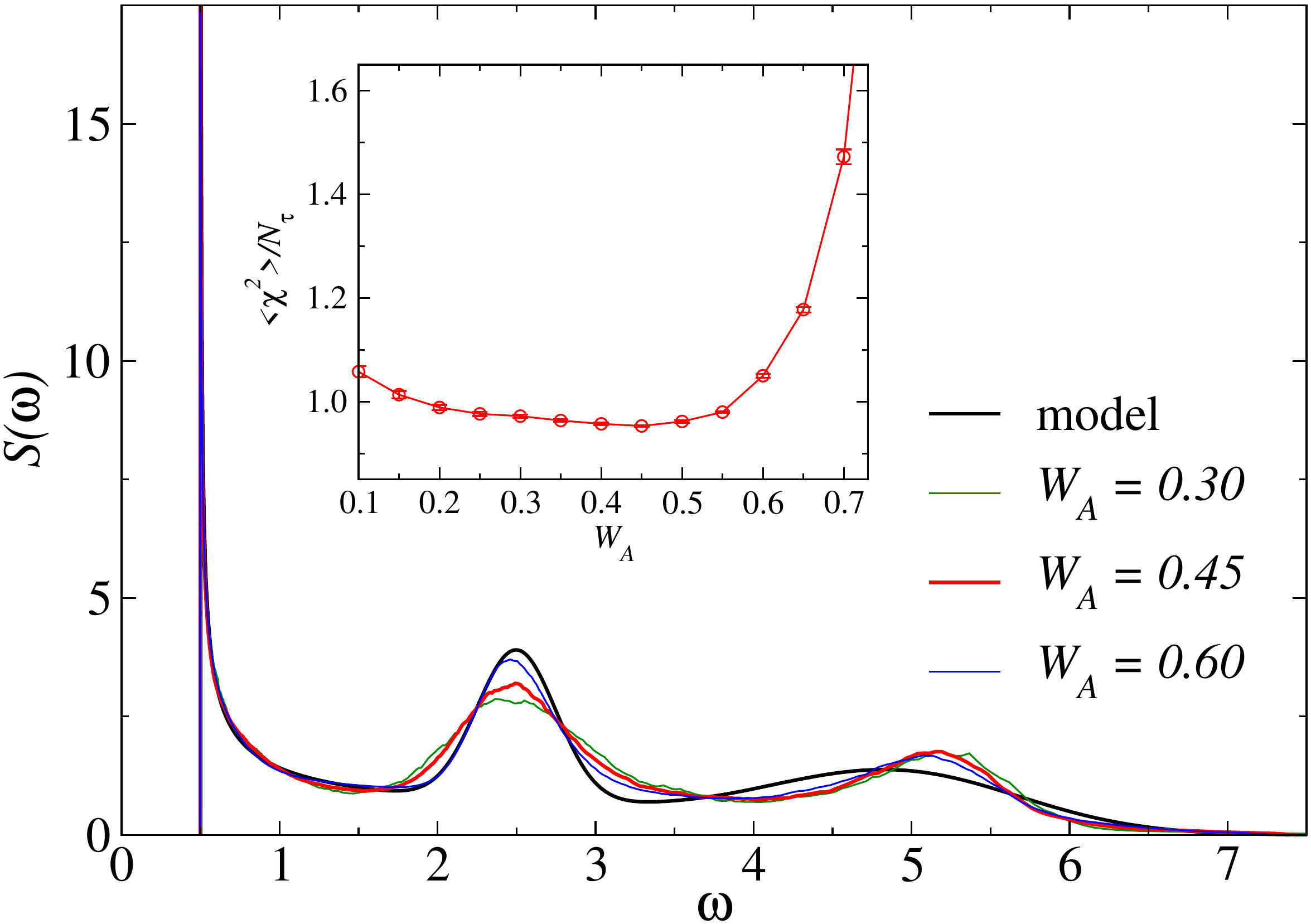}
\caption{Synthetic spectrum with edge singularity of the form $(\omega-\omega_0)^{-1/2}$ followed by two broad maxima (black curve) reproduced using
the SAC parametrization in Fig.~\ref{fig:mixed}. The error level of the $\bar G(\tau)$ data was $\sigma = 10^{-6}$. The inset shows the goodness of
the fit vs the relative weight $W_A$ of the $A$ (singular) part. The optimum is at $W_A \approx 0.45$. In the main figure the spectrum is shown at
$W_A=0.30$, $0.45$, and $0.60$. All the results were obtained with $N_A=400$, $N_B=200$, and with the sampling temperature set to $\Theta=1$.}
\label{twobump}
\end{figure}

Fig.~\ref{twobump} shows an example of a challenging synthetic spectrum where there are two maxima following the edge. Here the edge exponent is
$p=-1/2$, and we provide this as input, i.e., using equal $\delta$-function amplitudes as in Fig.~\ref{fig:mixed}(a). An important aspect of this test is that the
spectrum is not very sensitive to the exact value of $W_A$ as long as the goodness of the fit is reasonable and not too far from the optimal value.
The edge location is less than $0.7\%$ too low at the optimal (within the resolution of the test) $W_A=0.45$, slightly worse at $W_A=0.3$, and slightly
better at $W_A=0.6$. The first maximum is also better reproduced at $W_A=0.6$, suggesting that it may be better to use $W_A$ slightly higher than
indicated by the best goodness of fit. However, we have not been able so far to define a universal concrete criterion beyond the $\langle \chi^2\rangle$
minimum. Even with the imperfect mixing parameter, the results still capture the edge well and reproduce the two higher maxima reasonably well.

In general, we find that the results of this parametrization are also not very sensitive to the ratio $N_B/N_A$ of the number of $\delta$-functions
in the two different sets, as long as both are sufficiently large. In Fig.~\ref{twobump} we used $N_A=400$ and $N_B=200$. As in the case of the
unrestricted parametrizations, the sampling efficiency is better with larger $N_B$.

In principle, a non-divergent edge singularity ($p>0$) can also be mixed with the generic continuum. However, the $\delta$-functions
of the continuum will then have an outsize influence when they appear at the edge, causing a discontinuous jump instead of a smooth power-law decay to zero.
In contrast, for $p<0$ the divergent spectral weight far outweighs any small contributions from the B-set close to the edge.

One promising approach for edges with $p>0$, that we have not yet explored fully, is to consider ordered frequencies $\omega_{j+1} > \omega_j$
for the continuum part and impose monotonically increasing weights, $B_{j+1} > B_j$ with $B_j \to 0$ for $j \to 1$ (e.g., linearly
increasing, $B_j \propto j$). If the order is maintained during the sampling of the frequencies, the jump at the edge is avoided. To remedy
potentially excessively large (relative) amplitudes for $j \to N_B$, a cross-over form similar to Eq.~(\ref{lnamp}) can be implemented.

\section{Heisenberg ladders}
\label{sec:ladders}

Here we present applications of constrained SAC methods to 2- and 3-leg Heisenberg spin ladders. These systems have been studied for a long time
as examples of even-odd effects in the ``dimensional cross-over'' from 1D to 2D antiferromagnetism \cite{barnes93,dagotto96}. For ladders with an even
number $n$ of legs (chains) the ground state is gapped, with the gap decreasing exponentially with increasing $n$ (similar to the Haldane gap of integer-spin
chains \cite{khveshchenko94}), while for odd $n$ the spectrum is gapless.

An $n$-leg ladder consists of $n$ chains of length $L$ with intra- and inter-chain couplings $J_\parallel$ and $J_\perp$, respectively, with Hamiltonian 
\begin{equation}
  H = J_\parallel\sum_{c=1}^n \sum_{i=1}^L {\bf S}_{c,i} \cdot {\bf S}_{c,i+1} + J_\perp \sum_{c=1}^{n-1} \sum_{i=1}^L{\bf S}_{c,i} \cdot {\bf S}_{c+1,i},
\label{hhladder}
\end{equation}
where ${\bf S}_{c,i}$ is a spin-1/2 operator on site $i$ of chain $c$. Here we only consider the case $J_\parallel=J_\perp=1$ and apply periodic
boundary conditions on the chains (but not for the interactions between the chains).

Although the basic mechanism of the gap formation for even $n$ and the gapless spinon spectrum for odd $n$ (which is similar to that of the Heisenberg chain,
$n=1$) is well understood \cite{barnes93,white94}, the dynamic structure factor contains contributions from different low-energy processes and is expected to be
rather complicated. It is very difficult to analytically predict the full line shape for fixed longitudinal momentum $q$ (and even or odd parity sector, which
we will define further below), though some results are available for band edges in the 2-leg ladder versus $q$ \cite{sushkov98,knetter01}. Numerical calculations
of excitations have largely focused on the 2-leg case \cite{barnes93,yang98,schmidiger12}. A dominant isolated $\delta$ function has been identified at the $q$
dependent gap for $q$ at and close to the gap minimum at $q=\pi$.
Much smaller spectral weight has been detected at higher energies, far above the gap when $q\approx \pi$, but the profile
above the gap has not been resolved in detail. For the $3$-leg ladder, we are not aware of any previous numerical results for the dynamic structure factor.

We only present a limited number of preliminary calculations here, for both $n=2$ and $n=3$, as illustrations of the power of constrained SAC. We leave
further results for a future publication.

\subsection{$S(q,\omega)$ of the 2-leg ladder}
\label{sec:hladd2}

The 2-leg ladder hosts a reflection symmetry; the Hamiltonian Eq.~(\ref{hhladder}) for $n=2$ is invariant under permutation of the chain index,
$c \in \{1,2\} \to \{2,1\}$. Thus, there are even and odd excitations of the even ground state. The system is gapped and the lowest excitation for given
parity sector and momentum $q$ can be understood as a propagating rung triplet (often referred to as a triplon), out of which composite excitations at higher
energy can also be formed \cite{barnes93}. Here we focus on the odd triplet channel, i.e., with the operator
\begin{equation}
O_q=S^z_{1,q}-S^z_{2,q}
\end{equation}
in Eq.~(\ref{gtaudef1}). We consider only $q=\pi$, where the lowest triplet excitation defines the gap $\Delta$ of the system.

Early studies extracted the gap $\Delta \approx 0.50$ using Lanczos exact diagonalization \cite{barnes93}, which in this case converges rapidly as a function
of the ladder length $L$ because the correlation length is only a few lattice spacings. Extrapolations using $L \le 14$ in Ref.~\cite{yang98} gave
$\Delta=0.502$. Higher excitations are formed at $q\approx \pi$ from an odd number of the elementary triplons. While bound states have been predicted in
the two-triplon sector \cite{sushkov98}, they appear in parts of the Brillouin zone that are not relevant to the lowest three-triplon excitations at $q=\pi$.
Thus, the odd-sector dynamic spin structure factor at $q=\pi$ should consist of an isolated $\delta$-function at the gap $\Delta$, and there should be a
second gap $3\Delta$ above which a continuum of three-triplon states form.

It is well known \cite{barnes93} that the first $\delta$-function completely dominates the spectral weight at $q=\pi$, containing $96.7\%$ of the total
weight in a chain of length $L=12$ \cite{yang98}. In the exact diagonalization results  \cite{barnes93,yang98} spectral weight above the gap starts
at $\omega \approx 2.5 \approx 5\Delta$ and extends up to $\omega \approx 5$. The lack of spectral weight close to the three-triplon bound $3\Delta$
can likely be explained by strong finite-size effects in the form of repulsive interactions between triplons that push the energy of the lowest three-triplon
state to higher energy. A more recent DMRG study \cite{schmidiger12} for an experimentally relevant case of $J_\parallel < J_\perp$ showed the presence of
spectral weight far above the $\delta$-function at $\omega=\Delta$. However, also in this case there is no spectral weight close to $3\Delta$, even though
$L$ was as large as $200$. The open boundary conditions used in DMRG studies may play some role in pushing the three-triplon states to higher energies,
unless $L$ is much larger.

\begin{figure*}[t]
\centering
\includegraphics[width=105mm]{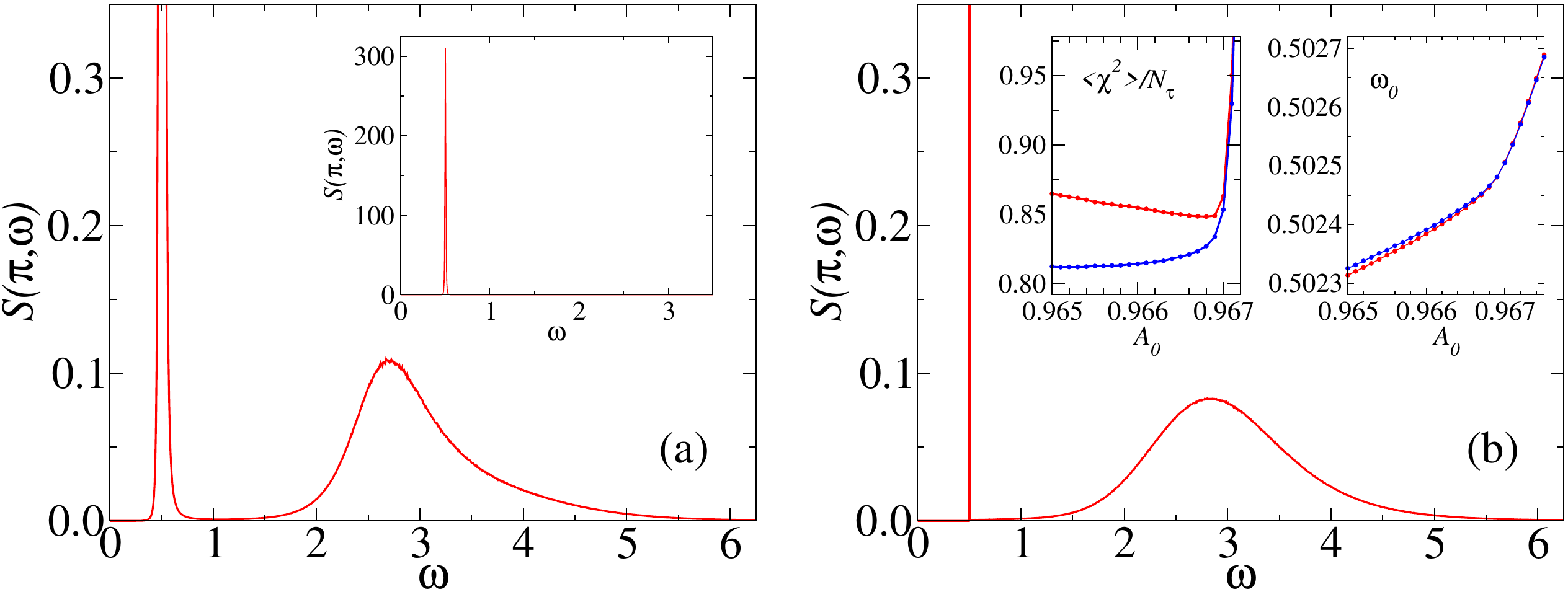}
\caption{SAC results for the odd-mode $q=\pi$ dynamic structure factor of the 2-leg Heisenberg ladder of length $L=256$. Results of unrestricted sampling
  with both frequency and amplitude updates [the parametrization in Fig.~\ref{fig:spec}(c)] are shown in (a), with the main panel focusing on the small spectral
  weight above the dominant peak and the inset showing the entire peak on a scale where the rest of the spectrum is invisible. The spectrum obtained when
  imposing a macroscopic edge $\delta$-function [Fig.~\ref{fig:spec}(d)] is shown in (b), at $A_0=0.9668$, corresponding to the location of the minimum in
  $\langle \chi^2\rangle/N_\tau$ at sampling temperature $\Theta \approx 0.07$, shown with red circles in the left inset; the blue circles show results at
  $\Theta \approx 0.036$, where there is no minimum in this range of $A_0$. The right inset shows the corresponding locations of the $\delta$-function
  (the gap).}
\label{hladd2a}
\end{figure*}

We here present results obtained for a ladder of length $L=256$ with periodic boundary conditions at inverse temperature $\beta=128$, which for all practical purposes
puts the system in the ground state since $T/\Delta \approx 0.016$. For the imaginary-time correlation functions, we used a linear grid of $\tau$ points at
spacing $\Delta_\tau = 0.06$ for $\tau<1$, followed by a roughly quadratically increasing $\tau$ spacing for a total of 28 points up to $\tau \approx 17.6$.
The error level of $\bar G(\tau)$ was $\sigma \approx 5\times 10^{-6}$.

Figure \ref{hladd2a}(a) shows SAC results based on unrestricted sampling with $N_\omega=2 \times 10^4$ at $\Theta \approx 7 \times 10^{-4}$, which satisfies the
criterion Eq.~(\ref{eq:chi2}) with the standard value of the coefficient, $a\approx 0.5$ (the minimum goodness-of-fit being $\chi^2_{\rm min}/N_\tau \approx 0.73$).
We used both frequency and amplitude updates, which is preferred especially when the spectrum has two almost separated features and it is
more difficult for frequency-only sampling to converge to the optimal spectral weight distribution. As we have seen in many of our tests in the preceding
sections, amplitude updates also lead to sharper peaks, and this is also the case with the leading peak here. We indeed observe a very sharp peak at
$\omega \approx 0.50$ as expected, as well as the likewise expected (based on the previous Lanczos calculations \cite{barnes93,yang98}) much smaller and
broadly distributed spectral weight up to $\omega \approx 5$. With only frequency updates (results not shown), the main peak is much broader and the
continuum exhibits two peaks, likely as a result of distortions induced by the more significant broadening of the main peak.

Given the expected perfect $\delta$-function at the gap, we next use the parametrization in Fig.~\ref{fig:spec}(d) and the method of optimizing the edge-$\delta$
weight $A_0$, described in Sec.~\ref{sec:deltapeak}. In this case we sampled only with frequency updates, as we normally do when a constraint is imposed,
with $N_\omega=1000$. Since the optimal amplitude $A_0$ is large, $\chi^2$ is very sensitive to the location $\omega_0$, which therefore fluctuates
only very little about its optimum. It is then important to use a sufficiently small spacing $\delta_\omega$ in the micro grid representing the continuous
frequency space (discussed in Sec.~\ref{sec:contsamp}). In this case we used $\delta_\omega = 10^{-6}$.

Results are shown in Fig.~\ref{hladd2a}(b). In the left inset we show the goodness of
fit versus $A_0$ at two sampling temperatures $\Theta$. The spectral function in the main graph is the result obtained at the
$\langle \chi^2\rangle$ minimum, $A_0=0.9668$, at the higher $\Theta$ value, which corresponds closely to the optimal-$\Theta$ criterion with $a = 0.5$.
The shape of the continuum above the $\delta$-function is slightly different from that obtained with unrestricted sampling in Fig.~\ref{hladd2a}(a), which is
expected because the broadening of the $\delta$-peak, caused by the entropy effects in unrestricted sampling, also induce secondary distortions.

While a well pronounced minimum is seen at the higher $\Theta$ value in Fig.~\ref{hladd2a}(b), at the lower value
the minimum is significantly shifted to the left and is not seen in the graph. The rapid shift of the minimum when $\Theta$ is lowered is reminiscent of
the tests in Fig.~\ref{broadened-1}, where the minimum eventually vanishes when the $\delta$-function parametrization is ill suited; in that case because of
the large width of the quasi-particle peak of the synthetic spectral function studied. In the present case we have no reason to expect a quasi-particle peak
of finite width, but there is another reason for the constraint applied here to be insufficient: there should be a second gap at $3\Delta$. Without imposing
the second gap, some spectral weight leaks out into the region between the two gaps, which clearly is the case in Fig.~\ref{hladd2a}(b).
We know that such a distortion also propagates to ``compensating'' (in the sense of fitting) re-distribution of weight in other parts of the spectrum.
Thus, we are motivated to apply the constraint on the continuum in a different way from before: instead of the microscopic $\delta$-functions having
a lower bound $\omega_0$, they should now not be allowed to fall below $3\omega_0$.

\begin{figure}[t]
\centering
\includegraphics[width=75mm]{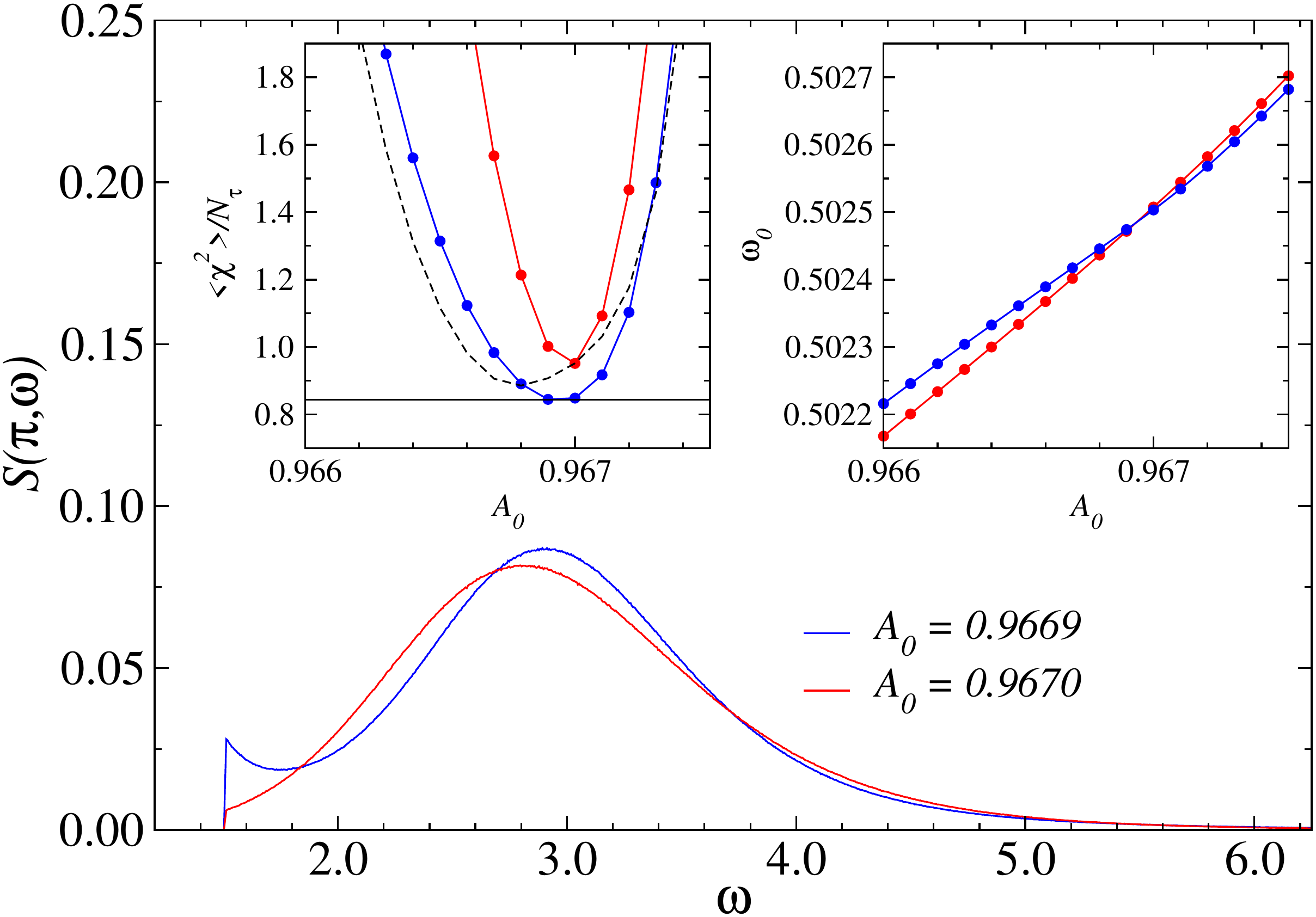}
\caption{Results based on the same imaginary-time data as in Fig.~\ref{hladd2a}, but with the second gap $3\Delta$ implemented in SAC as a modified constraint
  on the microscopic $\delta$-functions (here with $N_\omega=1000$) where, for given sampled location $\omega_0$ of the macroscopic $\delta$-function,
  the lower bound is at $3\omega_0$. Only the continuum above the second gap is graphed here, for two values of $A_0$ as indicated in the legends.
  These $A_0$ values correspond to the two minima in the left inset, where the sampling temperatures were $\Theta=0.30$ (red circles) and $0.017$
  (blue circles). The horizontal solid line corresponds to the standard optimal-$\Theta$ criterion, while the dashed curve indicates the refined criterion,
  Eq.~(\ref{chi2a0}). The right inset shows the corresponding $A_0$ dependence of the mean location of the $\delta$-peak.}
\label{hladd2b}
\end{figure}

With this very simple modification of the $\delta$-constraint, we obtain the results for the continuum weight shown in Fig.~\ref{hladd2b}. Now a sharp edge has
emerged above the second gap, but the exact shape of the edge is strongly dependent on $A_0$, as illustrated with the two different cases shown. The very sharp
$\langle \chi^2\rangle$ minimum shifts very slightly to lower $A_0$ as $\Theta$ decreases, as shown in the left inset of the figure, but the effect on the
distribution of the small amount of spectral weight in the continuum ($\approx 3\%$ of the total) is significant.

In light of this sensitivity of the continuum, we aply a more sophisticated version of the optimal-$\Theta$ criterion, Eq.~(\ref{eq:chi2}). While normally
we take $\chi^2_{\rm min}$ to be the lowest possible goodness-of-fit value with any parametrization, we can also use the minimum value $\chi^2_{\rm min}(A_0)$ that
is attainable for a given $A_0$, which we again obtain by simulated annealing to very low $\Theta$. Thus the target value is
\begin{equation}
\langle \chi^2(A_0)\rangle = \chi^2_{\rm min}(A_0) + a\sqrt{2\chi^2_{\rm min}(A_0)}.
\label{chi2a0}
\end{equation}
In all the test cases studied previously, the sensitivity of  $\chi^2_{\rm min}(A_0)$ to the value of $A_0$ when it is close to
optimal was not as significant and we did not consider fine details of the spectrum depending so strongly on $A_0$. The sensitivity in the present
case is clearly caused by the very large isolated $\delta$-peak.

The dashed curve in the left inset of Fig.~\ref{hladd2b} shows the refined criterion for $\langle \chi^2(A_0)\rangle$ with $a=0.5$ in Eq.~(\ref{chi2a0}).
In the simulated annealing
processes for several values of $A_0$ we also saved the mean values $\langle \chi^2(A_0,\Theta)\rangle$ (and also the spectral functions at all $A_0$ values
studied), thus allowing us to find $\Theta$ such that the computed and target values coincide at a point that is also a $\langle \chi^2(A_0)\rangle$ minimum.
The red points in left inset of Fig.~\ref{hladd2b} show results at $\Theta \approx 0.30$, where the conditions are fulfilled. Moreover,
$\langle \chi^2(A_0)\rangle/N_\tau$ is safely below $1$ at this point, thus indicating statistical soundness of the procedure. For reference, we also 
show results at $\Theta=0.017$ (blue points), where the $\langle \chi^2(A_0)\rangle$ minimum (to within the resolution $\Delta_\Theta=0.0001$ used here)
is at the value corresponding to $\chi^2_{\rm min}/N_\tau \approx 0.73$ for free sampling.

As regards the spectral functions in Fig.~\ref{hladd2b}, at the higher $\Theta$ (red curve) there is only a small step at the edge, while a sharp peak has formed
at the lower value (blue curve). Most likely, this spectrum is already showing signs of overfitting, given that the fluctuations relative to $\chi^2_{\rm min}$
are too restricted, as seen in the left inset of Fig.~\ref{hladd2b}. The distance between the dashed curve and the point at $A_0 = 0.9669$ corresponds to the
$\Theta$-fixing criterion Eq.~(\ref{chi2a0}) with $a \approx 0.2$, well below our standard value $0.5$. The high sensitivity of the continuum to the value of $a$
of course still implies that the data quality is not quite good enough for reliably studying the continuum, though it is clear that our results are consistent
with the continuum starting at $\omega = 3\Delta$.

With the present data quality we cannot completely exclude a spectrum with an edge peak at $\omega=3\Delta$, but we have also found that the trend with
improving data quality is for $A_0$ to increase somewhat (though unlikely going significantly above the present range of values),  which makes a peak even
less likely. It is possible that the small step (as in the $A_0=0.9670$ curve in Fig.~\ref{hladd2b}) eventually also vanishes (in case of a further small
increase in the optimal $A_0$ value), though we expect spectral weight to extend all the way to $\omega=3\Delta$

When the edge peak is sharp in Fig.~\ref{hladd2b}, for $A_0=0.9669$, it looks vaguely similar to the spinon edge in the Heisenberg chain when only the
lower bound is imposed in Fig.~\ref{w0fix}. In the ladder, it is the momenta of three triplons (instead of two spinons) that have to add up to the total momentum,
here $q=\pi$. It should be noted that, while bound states of two triplons have been predicted \cite{sushkov98,knetter01}, these states do not appear when the total
momentum is close to $0$ (i.e., both triplons have individual $q$ close to $\pi$), and all calculations agree that the two-triplon states at $q=0$ have a lower
edge starting exactly at $\omega = 2\Delta$. Similarly, we do not expect any bound states of the lowest three-triplon $q=\pi$ states; thus, the lower edge should
indeed be exactly at the imposed energy $\omega=3\Delta$ (for sufficiently large $L$, so that any effects of weakly repulsive triplon-triplon interactions are small).

For $q$ close to $\pi$, the triplon is known to be only weakly dispersing \cite{sushkov98,knetter01}, and therefore there is a high density of states
right at the $3\Delta$ edge. However, matrix elements play a crucial role in the shape of the edge, and most likely, as we have argued above, there is
no divergence at the edge, which instead could have a non-divergent power law form (if the step feature in Fig.~\ref{hladd2b} eventually goes away).
The broad maximum centered at $\omega \approx 3$ in Fig.~\ref{hladd2b}
likely arises from a bound triplon pair together with a third triplon. Five (and higher) triplon contributions should have extremely small spectral weight. 

The first gap is determined very precisely, as seen in the right inset of Fig.~\ref{hladd2b}, with realistic values of $\Delta$ between $0.50245$ and $0.50255$ based
on a conservatively acceptable range of $A_0$ values. While this result and the relative spectral weight agree with previous results
(but are more precise), there are important differences in the continuum. Previous
studies have not resolved the second gap $3\Delta$, as we discussed above. While the Lanczos results can be explained by finite-size effects, the DMRG results are
harder to interpret. Fig.~1(a) of Ref.~\cite{schmidiger12}, shows spectral weight above the gap, starting at $\approx 5\Delta$ for $q=\pi$ (where $\Delta$
is smaller than in our case because of the different coupling ratio). We also observe significant weight in this regime---the larger broad peak in
Fig.~\ref{hladd2b}. Given that we have only enforced known aspects of the spectral function, the weight extending all the way to $3\Delta$
should correctly represent the three-triplon excitations, though with details of the edge that cannot be conclusively determined
based on the current $\bar G(\tau)$ data.

Reference \cite{schmidiger12} did not offer any physical interpretation of the weight observed far above the expected second gap $3\Delta$ close to
$q=\pi$, and it remains a mystery why the results differ so much from ours in this regard. We are not aware of any other reliable studies of the dynamic structure
factor of the ladder, and future studies, with DMRG and other techniques, are called for to clarify this issue. We note again that the system size we have used
here, $L=256$, is about 100 correlation lengths \cite{sandvik10}, and it is unlikely that there are significant finite size effects left, considering also
the periodic boundary conditions (in contrast to the open boundaries used in the DMRG calculations). The SSE QMC calculations do not introduce any systematical
errors in $\bar G(\tau)$. We plan to further improve the imaginary-time data and carry out studies for other values of $q$ and well as of the even rung mode.

\subsection{$S(q,\omega)$ of the 3-leg ladder}
\label{sec:hladd3}

\begin{figure*}[t]
\centering
\includegraphics[width=105mm]{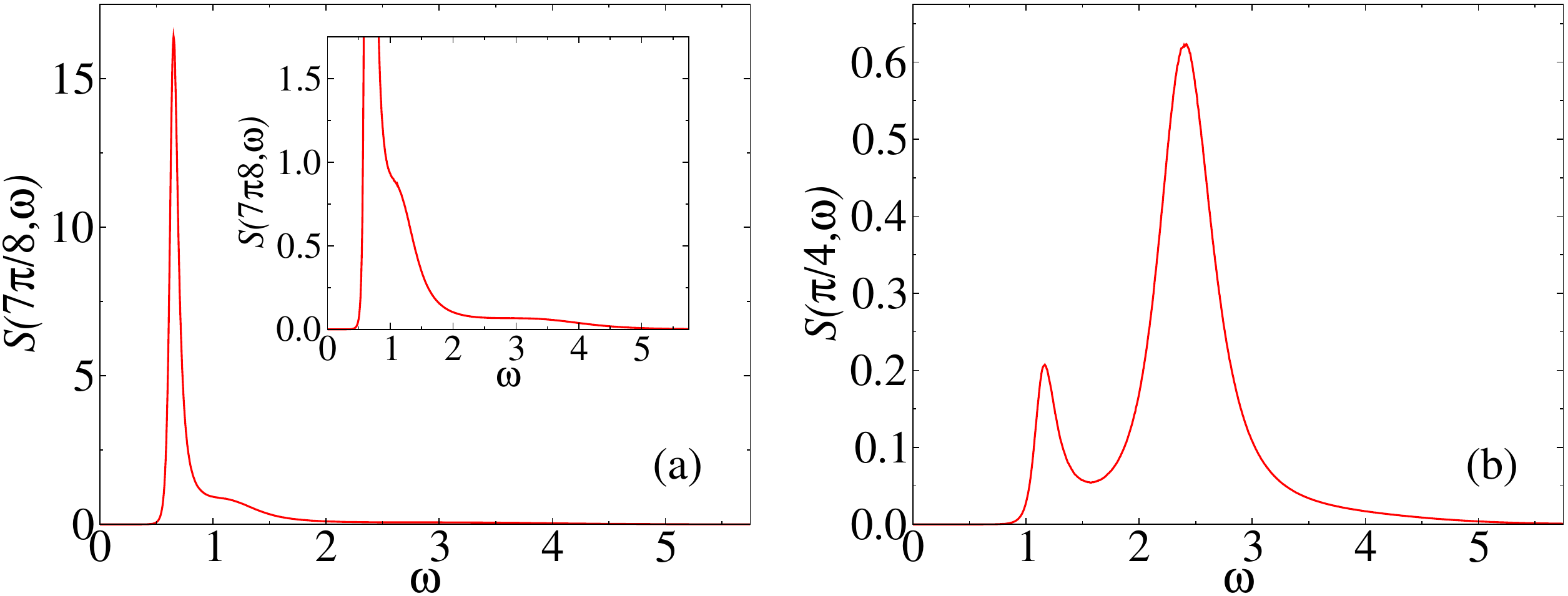}
\caption{SAC results for the dynamic structure factor in the even parity channel of the 3-leg ladder at $q=7\pi/8$ in (a) and $\pi/4$ in (b). In both cases,
unrestricted sampling with $N_\omega=2000$ $\delta$-functions was carried out, with both frequency and amplitude updates included. The sampling temperature
$\Theta$ was adjusted according to the standard criterion, Eq.~(\ref{eq:chi2}) with $a=0.5$, corresponding to $\langle \chi^2\rangle /N_\tau = 0.91$ and $0.84$
in (a) and (b), respectively. The inset of (a) shows a more detailed view of the tail of the $q=7\pi/8$ spectrum.}
\label{hladd3a}
\end{figure*}

For the 3-leg ladder, $n=3$ in the Hamiltonian Eq.~(\ref{hhladder}),
we are not aware of specific numerical results for $S(q,\omega)$, while there are
many prior results for static correlations and thermodynamic properties, e.g., Refs.~\cite{frischmuth96,sandvik10}. Thus, in this case we do not have any previous
benchmarks for comparison. Though qualitatively it is clear what type of low-energy elementary excitations to expect \cite{dagotto96,white94}, the contributions
to the dynamic structure factor from composite excitations at higher energy have not been studied to our knowledge.

Given that each rung in isolation has a two-fold degenerate ground state, the low-energy excitations will map onto those of the single $S=1/2$ Heisenberg chain.
As in the 2-leg ladder, there are again two distinct modes corresponding to even and odd parity, defined as reflection about the central chain in the Hamiltonian
Eq.~(\ref{hhladder}), i.e., permuting the chain index, $c \in \{1,2,3\} \to \{3,2,1\}$. For the mapping, which we will not carry out formally here, it is useful to
first consider the gapless points, $q=0$ and $q=\pi$, of the Heisenberg chain. In the 3-leg ladder, the very lowest excitation should correspond to fully
antiferromagnetic fluctuations, which would be most naturally accessed with the staggered rung $i$ operator $S^z_{1,i}-S^z_{2,i}+S^z_{3,i}$. The uniform long-wavelength
fluctuations, $q\to 0$, should also be gapless, and here the in-phase rung operator $S^z_{1,i}+S^z_{2,i}+S^z_{3,i}$ may seem like the best option. However, both
the staggered and uniform operators above are even under reflection, and, thus, we can select one of them to study all $q \in (0,\pi)$ between the gapless points
(and we could also in principle just consider the central-chain spins $S^z_{i,2}$).

For any $q$ value, the spectral function in the even sector should have the same asymptotic $\omega \to \omega_q$
edge shape as that of the Heisenberg chain. Given that the dominant fluctuations are antiferromagnetic, we expect the largest total spectral weight with
the staggered rung operator (which is also borne out by our results for the total spectral weight; the static structure factor), which we therefore
use here. After Fourier transformation, 
\begin{equation}
O_q=S^z_{q,1}-S^z_{q,2}+S^z_{q,3}.
\end{equation}
is the operator used in the imaginary-time correlation function in Eq.~(\ref{gtaudef1}).

We here report results for the  dynamic spin structure factor based on $\bar G_q(\tau)$ calculated with the SSE method on a 3-leg ladder system of length
$L=512$ at inverse temperature $\beta=8192=16L$. The very low temperature is chosen in order to effectively obtain $T=0$ results even for
$q$ very close to the gapless points. Here we will not consider extreme cases, however, and study only $q=\pi/4$ and $q=7\pi/8$ as two characteristic
examples with different features in the dynamic structure factor. In both cases, the error level of the imaginary-time data is $\sigma \approx 10^{-5}$.
We used a linear $\tau$ grid with $\Delta_\tau=0.05$ up to $\tau=1$ and a roughly
quadratic grid thereafter, for $\tau$ up about $6.5$ ($40$ $\tau$ points in total) and $14$ ($52$ points) for $q=\pi/4$ and $q=7\pi/8$, respectively
[with the $\tau$ cut-off set at relative error of $\approx 20\%$ in $\bar G_q(\tau)$].

\begin{figure*}[t]
\centering
\includegraphics[width=105mm]{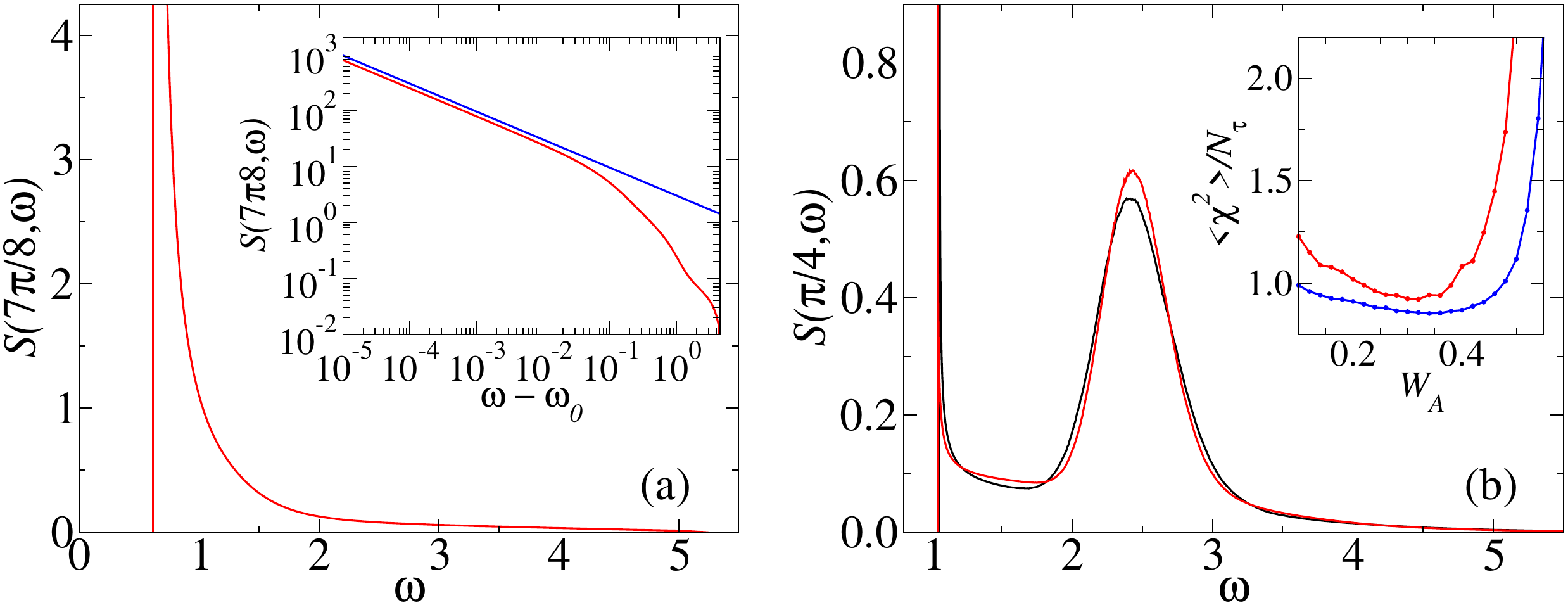}
\caption{Constrained SAC results for the same 3-leg ladder structure factors
  as in Fig.~\ref{hladd3a}. In (a), for $q=7\pi/8$ the distance-monotonic parametrization with all equal
  amplitudes [Fig.~\ref{fig:spec}(e)] was used with $N_\omega=400$. The main graph focuses on the tail of the spectrum and the inset shows the entire
  spectrum (red curve) in a log-log plot [with the blue line showing the asymptotic $(\omega-\omega_q)^{-1/2}$ form for reference]. In (b), for $q=\pi/4$ 
  a generic continuum had to be included, using the mixed parametrization in Fig.~\ref{fig:mixed} with a total
  of $N_\omega=800$ $\delta$-functions, whereof $400$ in the B (continuum) part. The goodness of the fit vs the weight in the A (edge) part is shown in the
  inset for two sampling temperatures; $\Theta=0.3$ (blue circles) and $\Theta=0.1$ (red circles). The two spectra in the main graph were sampled with
  $W_A=0.34$ (black curve) and  $W_A=0.40$ (red curve), both at $\Theta=0.1$.}
\label{hladd3b}
\end{figure*}

We again start with unrestricted SAC sampling, using both amplitude and frequency updates. Results are shown in Fig.~\ref{hladd3a}. As expected,
the overall spectral weight is much larger for $q=7\pi/8$, Fig.~\ref{hladd3a}(a), where there is a single sharp peak followed by two shoulder-like features.
This spectrum can be compared to that obtained with unrestricted SAC for the Heisenberg chain at the only slightly different momentum $q=4\pi/5$ in
Fig.~\ref{sw2}, where there is also a sharp peak but in that case followed by a second distinct peak. We know that the second peak is an artifact of the
unrestricted sampling, and the shoulders seen in Fig.~\ref{hladd3a}(a) are likely also produced by entropic distortions of the expected sharp spectral
edge.

In Fig.~\ref{hladd3a}(b), the spectrum for $q=\pi/4$ looks very different, with a small low-frequency peak followed by a larger broad maximum.
Here it is unlikely that distortions of the edge would artificially
induce such a large second maximum; thus, both features should be real though distorted by entropic
effects. In the Heisenberg chain, the correct spectral function never has two peaks; for any $q$ there is a sharp edge followed by a monotonic decay of
spectral weight (Fig.~\ref{sw-1d}). Thus, the result of unrestricted sampling already suggests a more complex continuum for the 3-leg ladder in some
parts of the Brillouin zone. 

Moving on to constrained SAC, here it is of course natural to test the edge parametrization, Fig.~\ref{fig:spec}(e), which we explored extensively
in Secs.~\ref{sec:contedge1} and \ref{sec:contedge2}. We keep the edge exponent at $p=-0.5$, which we know works well for the Heisenberg chain even
though the logarithmic correction to the power-law form is strictly not captured correctly very close to the edge. For $q=7\pi/8$, this parametrization
indeed works very well, with $\chi^2_{\rm min}/N_\tau$ obtained by simulated annealing being well below $1$.
Applying the standard $\Theta$ criterion for the final sampling
temperature $\Theta$, we obtained the spectrum shown in Fig.~\ref{hladd3b}(a). Comparing with Fig.~\ref{hladd3a}(a), there are no shoulders left, only a smooth
decay away from the edge.

For $q=\pi/4$, the same parametrization does not work, which is expected because of the double-peak structure with dominant second maximum in
Fig.~\ref{hladd3a}(b). Instead, we have to use the mixed parametrization illustrated in Fig.~\ref{fig:mixed}. The inset of Fig.~\ref{hladd3b}(b) shows the
goodness of the fit versus the weight in the A (edge) part of the spectrum, obtained at two different $\Theta$ values. There is a clear minimum in both cases,
at marginally higher $W_A$ at the lower $\Theta$. The main graph in Fig.~\ref{hladd3b}(b) shows spectra obtained at $W_A=0.34$ (corresponding to the optimum
at the lower $\Theta$ value) and at $W_A=0.40$ (slightly above both of the optima). The two-peak structure revealed already in Fig.~\ref{hladd3a}(b) persists,
but now of course the edge is sharp, while the second maximum is only slightly different from the unrestricted SAC result. There is also a flat portion
of spectral weight between the two peaks.

We have also carried out SAC runs including the mixed parametrization at $q=7\pi/8$, where, as mentioned above, $\langle \chi^2\rangle$ is statistically
very good even with only the A part of the spectrum. When including a B part at just $0.1\%$ of weight ($W_A=0.999$), $\langle \chi^2\rangle$  already
increases significantly (though the fit is still acceptable), thus indicating that there is no statistical advantage of the B part here. The
resulting spectrum in this case is also still monotonically decaying above the edge. Thus, the monotonic form here is completely sufficient and
the profile in Fig.~\ref{hladd3b}(a) should be very close to correct (similar to our good results for the Heisenberg chain).

These results demonstrate consistency with the expected edge in the spectral function, which arises from deconfined spinons according to a single-chain
description of the $S=1/2$ rung degrees of freedom. They also illustrate the contributions from other excitations not present in the Heisenberg chain.
The fact that the tail of the continuum at $q=7\pi/8$ is much thinner than in the Heisenberg chain can likely be understood as the initial effects of the
dimensional cross-over into the 2D Heisenberg model \cite{dagotto96}. In the 2D case, there is a sharp magnon quasi-particle followed by a broad continuum
\cite{shao17}, and the thinning of the edge peak in the 3-leg ladder can perhaps be understood as an evolution with $n$ (odd) of the peak to a a very narrow
magnon quasi-particle peak (essentially a $\delta$-function \cite{shao17,chernyshev09}). The second maximum for $q=\pi/4$ originates from rung degrees of freedom
not present in the single chain. As in the many test cases studied in Secs.~\ref{sec:contedge1} and \ref{sec:contedge2}, we expect that the edge frequency
$\omega_q$ is very close the actual value. It will be very interesting to study the dispersion relation as well as the systematic emergence and evolution
of the second maximum. We plan to study these features in detail in the future, along with the odd-mode spectral functions.

\section{Maximum entropy methods}
\label{sec:maxent}

Here we discuss the relationships between SAC and the ME method and also present SAC-inspired potential improvements of the latter.
We gave a very brief review of the ME method in Sec.~\ref{sec:methods_b} and here first elaborate further on some key facts.

In terms of the spectral function $A(\omega)$, which is defined only for $\omega \ge 0$ in Eq.~(\ref{barelation}),
the postulated probability distribution given the QMC data $\bar G$ is
\begin{equation}
P(A|\bar G) \propto {\rm e}^{-\chi^2(A)/2+\alpha E(A)},
\label{pmaxent2}
\end{equation}
where $A(\omega)$ depends on $\bar G$ according to Eq.~(\ref{eir}) and $E(A)$ is the Shannon information entropy with respect to a default model $D(\omega)$,
Eq.~(\ref{esdef}).

Normally the entropy is expressed in terms of the original spectral function $S(\omega)$, which is defined on the entire frequency axis
$\omega \in (-\infty,\infty)$ and is related to $A(\omega)$ with $\omega \ge 0$ according to Eq.~(\ref{barelation}). The default model should also satisfy
the same bosonic relationship as $S(\omega)$, $D(-\omega)={\rm e}^{-\beta\omega}D(\omega)$, but even then the entropy computed according to Eq.~(\ref{esdef})
with $S$ replaced by $A$ is not exactly the same as the original definition with $S$, except in the limit $T \to 0$ (and with very small differences for a
gapped spectrum if $T$ is below the gap).

With SAC, we normally parametrize and sample $A(\omega)$ [though in principle $S(\omega)$ can also be used]. In order to be able to compare results obtained
with the ME method, we also use $A(\omega)$ for the latter. We always use a flat default model, which can be neglected (i.e., $D=1$) in the entropy expression,
and in the discrete computer implementation with $N_\omega$ histogram bins [here strictly speaking with $\delta$-functions at frequencies
$\omega_i=(i+1/2)\Delta_\omega$] we use the Shannon entropy in the form
\begin{equation}
E_{\rm SH} = - \sum_{i=1}^{N_\omega} A_i \ln (A_i), \label{esh2}
\end{equation}
where the amplitudes are normalized as $\sum_i A_i = 1$.

The entropic weighting factor $\alpha$ in Eq.~(\ref{pmaxent2}) can be adjusted  according to one of several proposed
criteria \cite{gull84,silver90,gubernatis91,jarrell96} and the spectrum $A(\omega)$ is normally  sought that maximizes the probability at that $\alpha$.
Alternatively, in Bryan's method \cite{bryan90} the probability is augmented by a prior distribution of $\alpha$ values; typically
$P(\alpha) \propto 1/\alpha$;
\begin{equation}
P(A,\alpha|\bar G)= P(A|\bar G)P(\alpha).
\label{pmaxenta2}
\end{equation}
Then either $A(\omega)$ and $\alpha$ are optimized together for maximum likelihood or $A_\alpha(\omega)$, which maximizes the probability for given
$\alpha$, is integrated over $\alpha$.

For a fixed value of $\alpha$, obtaining the maximum probability spectrum corresponds to minimizing
the functional
\begin{equation}
F(A)=\chi^2(A)/2 - \alpha E(A),
\label{functional}
\end{equation}
which can be done either by some stochastic approach, a Newton-type optimizer \cite{bergeron16}, or a more sophisticated non-negative least squares
method \cite{Koch18,Ghanemthesis}. In a slightly different approach \cite{boninsegni96,kora18}, instead of aiming for the maximum-probability solution,
the amplitudes (and optionally also $\alpha$) are sampled for the average spectrum, as in SAC, using one of the above probability distributions,
either Eq.~(\ref{pmaxent2}) or Eq.~(\ref{pmaxenta2}). 

Here, in Sec.~\ref{sec:maxent1} we propose a new simple way to determine an appropriate $\alpha$ value, inspired by our criterion for $\Theta$
in SAC. In Sec.~\ref{sec:sacme} we present new insights into the relationship between SAC and ME methods based on the different forms of the
entropy corresponding to different SAC parametrizations (continuing the discussion started already in Sec.~\ref{sec:entropy1}). In Sec.~\ref{sec:maxent2}
we address the issue of maximizing the ME probability versus sampling $A(\omega)$ over the full distribution.

\subsection{Criterion for the factor $\alpha$}
\label{sec:maxent1}

Besides the standard arguments for optimizing $\alpha$ \cite{jarrell96}, it has also been proposed to fix $\alpha$ at the value maximizing
$d\ln \chi^2(\alpha)/d\ln(\alpha)$ \cite{bergeron16}, in analogy with the proposal by Beach \cite{beach04} to fix the sampling temperature $\Theta$ in SAC
at the value maximizing the logarithmic derivative of $\langle \chi^2(\Theta)\rangle$. However, as mentioned in Sec.~\ref{sec:theta}, in SAC a broad maximum
is often located at $\Theta$ high enough to cause a suboptimal fit if sampling there. We will see below that the logarithmic derivative of $\chi^2(\alpha)$ has
a qualitatively similar form, and, therefore, this way of determining $\alpha$ cannot in general be the best course of action.

Our method of fixing the sampling temperature using the $\langle \chi^2\rangle$ criterion in Eq.~(\ref{eq:chi2}) can also be directly taken over into the
ME method, by minimizing the functional in Eq.~(\ref{functional}) for a range of decreasing $\alpha$ values and monitoring $\chi^2$ as it convergences
close to its minimum value $\chi^2_{\rm min}$. The minimum value $\chi(\alpha \to 0) \to \chi^2_{\rm min}$ should of course be the same (in practice very
close to the same) as obtained in the SAC simulated annealing procedure, where $\langle \chi(\Theta \to 0)\rangle \to \chi^2_{\rm min}$.
Using the same arguments that we discussed at length in in Sec.~\ref{sec:theta}, the value of $\alpha$ corresponding to an optimal balance
between entropy maximization and data fitting is postulated as:
\begin{equation}
\chi^2(\alpha) = \chi^2_{\min} + a\sqrt{2\chi^2_{\min}}.
\label{eq:chi2me}
\end{equation}
This criterion of course guarantees a statistically acceptable fit from the outset with the factor $a \lesssim 1$.

We here test the $\alpha$-fixing scheme using the same $L=16$ and $L=500$ Heisenberg chain data that we used previously in the tests
of unrestricted SAC sampling in Sec.~\ref{sec:theta}. To find the spectrum minimizing the functional $F(A)$, Eq.~(\ref{functional}), we use a uniform
frequency grid in the range from $\omega=0$ to $\omega=5$; the same cut-off value as used with the SAC method in Figs.~\ref{sw1}(a) and \ref{sw2}(a).

To optimize the spectrum, we employ the same Monte Carlo procedures with 2- and 3-amplitude updates as for the SAC fixed-grid sampling (described
in Sec.~\ref{sec:gridsamp}), but in this case at $\Theta=0$ (i.e., only updates leading to higher probability are accepted). It can still be beneficial
to start the process by first sampling at $\Theta>0$, with some initial large value of $\alpha$, and annealing to some low $\Theta$ to obtain a good starting
configuration for the subsequent $\Theta=0$ run. After this initial step, $\alpha$ is reduced in a way similar to the simulated annealing process
(dividing by $1.05$ or $1.1$ each time and with a large number of updating sweeps for each $\alpha$ value). Given the ability of the 3-amplitude
moves, in particular, to re-distribute spectral weight among different parts of the spectrum, we do not expect any issues related to potential local
minima of $F(A)$. Indeed, in test runs with different random number seeds, the same spectral functions were obtained versus $\alpha$, as long as
sufficiently many Monte Carlo sweeps were carried out.

\begin{figure}[t]
\centering
\includegraphics[width=75mm]{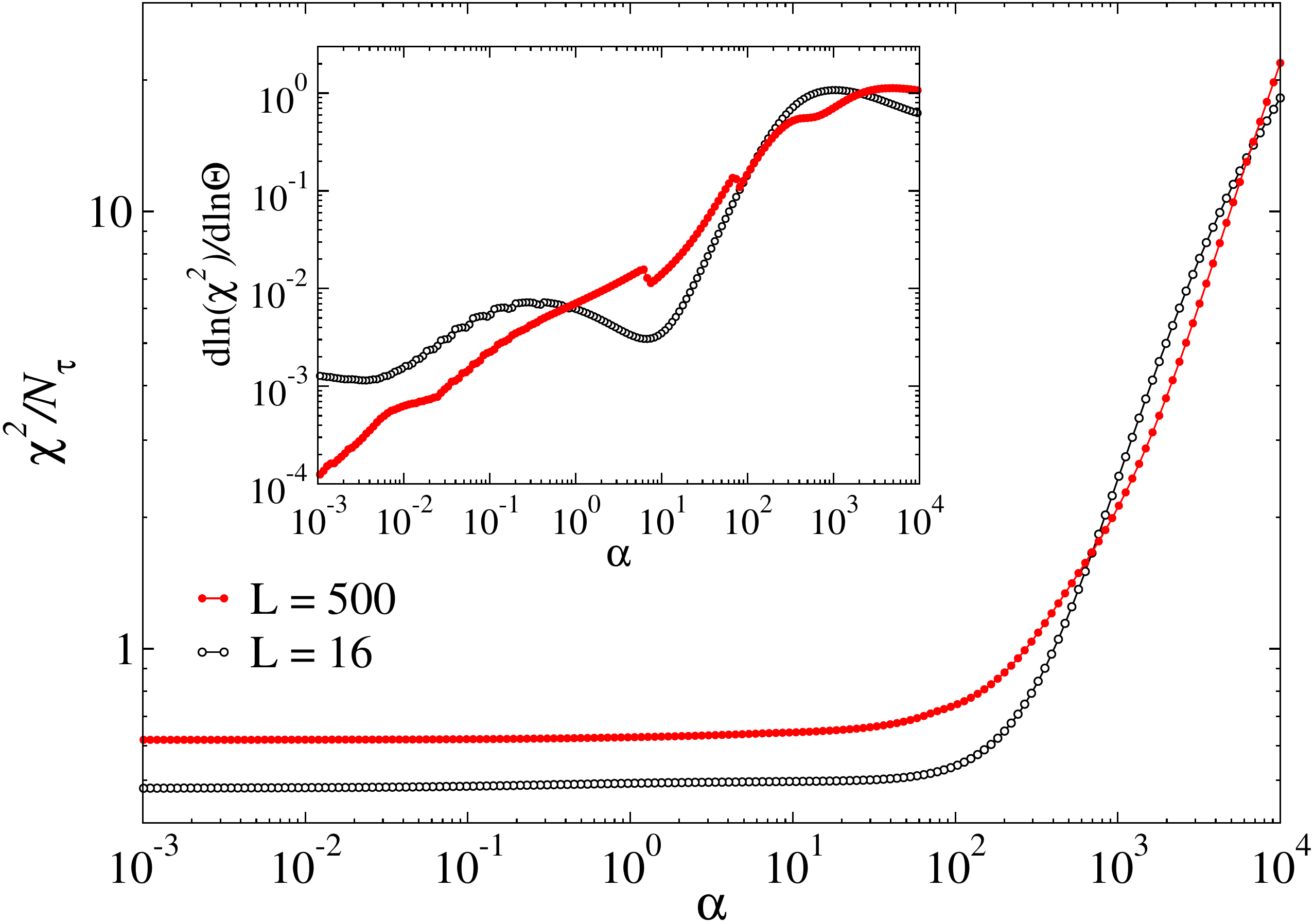}
\caption{Goodness-of-fit vs the entropy weighting parameter obtained by maximizing the probability Eq.~(\ref{pmaxent2}). For the $L=16$ case
($\beta=2$, $q=\pi/2$) the same Heisenberg $\bar G(q,\tau)$ data as in Figs.~\ref{fig:x2} and \ref{sw1} were used, and the $L=500$ ($\beta=500$, $q=4\pi/5$)
results are based on the same data as in Fig.~\ref{sw2}. The inset shows the corresponding numerical log derivatives.}
\label{x2_me}
\end{figure}

Figure \ref{x2_me} shows $\chi^2$ versus $\alpha$ for the Heisenberg chains, with the log derivatives shown in the inset. Similar to the SAC results
in Fig.~\ref{fig:x2}, for $L=16$ the maximum derivative corresponds to a suboptimal fit with $\chi^2/N_\tau \approx 2.5$ and for the $L=500$ system the value
is higher still. Below the global maximum, a second broad maximum can be observed at much lower $\alpha$ in the case of $L=16$ (and going beyond the
left edge of the figure, there is also additional nontrivial structure), while for $L=500$ there are several small but sharp local maxima and shoulders.
The minimum $\chi^2$ values reached at the end of the process are fully compatible with the SAC $\Theta$ annealing results. The ultimate low-$\alpha$ ME
spectra also exhibit the same few-peak structure that we discussed previously in Sec.~\ref{sec:theta} and illustrated in the $L=500$ case in
Fig.~\ref{sw2} (see also \ref{app:lowtheta} for results at still lower $\Theta$). The good agreement between the two different ways of
approaching the pure $\chi^2$ fitting limit also supports the ability of our method to correctly find the maximum-probability ME spectrum versus $\alpha$.

Even though the small-$\alpha$ limit is correct, one might perhaps suspect that the rather sharp features seen in the log derivative for $L=500$ in
Fig.~\ref{x2_me} could be associated with metastability. However, the results are fully reproducible when carrying out the $\alpha$ annealing from different
starting values and with different types of Monte Carlo updates included. Thus, we believe that there are no problems with the stochastic optimization
algorithm. Since the process results in a single spectrum maximizing the probability (as opposed to sampling an average), it is also not too surprising
that there could be certain $\alpha$ values at which rather sudden minor re-organizations of the spectrum take place. These sharp features are only
clearly visible in the derivatives, however, and will not in any way be detrimental when applying our new criterion, Eq.~(\ref{eq:chi2me}), to determine
the ``best'' $\alpha$.

\begin{figure}[t]
\centering
\includegraphics[width=84mm]{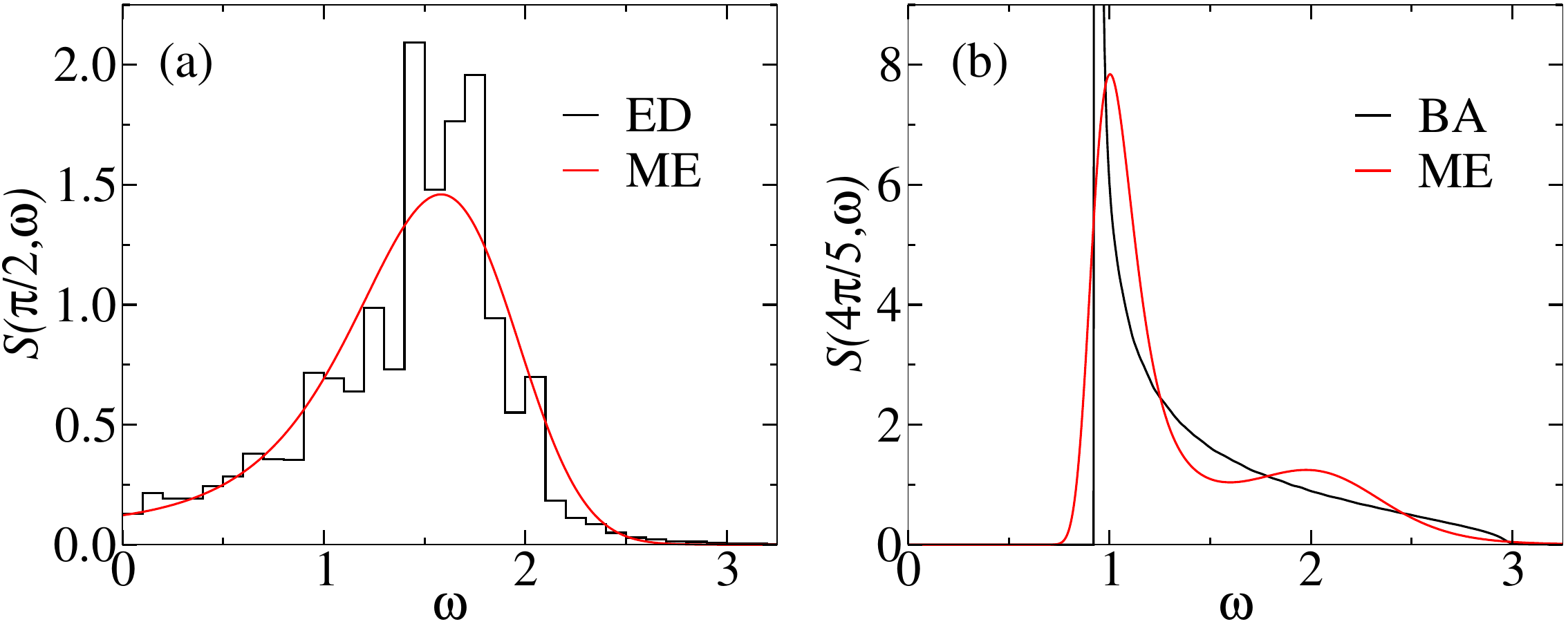}
\caption{ME results for the spectral functions of the same Heisenberg chains as in Figs.~\ref{sw1} and \ref{sw2}, $L=16$ at $\beta=2$ in (a) and
$L=500$ at $\beta=500$ in (b), shown in red. The corresponding exact diagonalization and BA results are shown in black. The criterion in
Eq.~(\ref{eq:chi2me}) was used with $a=0.5$ to fix the value of the parameter $\alpha$, based on the data shown in Fig.~\ref{x2_me}.}
\label{sw_me}
\end{figure}

In Fig.~\ref{sw_me} we show the spectral functions obtained with $a=0.5$ in Eq.~(\ref{eq:chi2me}), for both $L=16$ and $L=500$. Note that here we
show $S(\omega)$, obtained from $A(\omega)$ according to Eq.~(\ref{barelation}). It is striking how similar these results are to those of the
unrestricted SAC in Figs.~\ref{sw1}(b) and \ref{sw2}(b), i.e., those obtained with equal-weight $\delta$-functions sampled in the frequency continuum
in the range of good values of $\langle \chi^2(\Theta)\rangle/N_\tau$. In fact, the curves for matching goodness-of-fit values fall on top of each other,
not only in the cases shown in Fig.~\ref{sw_me} (where we only graph the ME results) but for any pairs of values $(\alpha,\Theta)$ such that
$\chi^2(\alpha) = \langle \chi^2(\Theta)\rangle$. These essentially identical results from the two methods are explained by an {\it exact}
mapping between them, as we explain next.

\subsection{Relationship to SAC}
\label{sec:sacme}

In Sec.~\ref{sec:entropy}, we computed the exact entropy of the SAC spectrum with equal-amplitude $\delta$-functions accumulated in a histogram, resulting
in $E_{\rm EA}$ in Eq.~(\ref{eea}). Except for the factor $N=N_\omega$, this expression is the information entropy $E_{\rm SH}$ used in the ME method with
a flat default model. We also discussed the GK entropy $E_{\rm GK}$, Eq.~(\ref{egk}), pertaining to amplitudes sampled on a fixed grid, and our conjectured
mixed entropy $E_{\rm MX}$, Eq.~(\ref{emx}), for a spectrum with both frequencies and amplitudes sampled.

For the sake of convenience and uniformity of the notation, we here repeat the three entropy expressions but now in the discrete summation form
\begin{subequations}
\begin{eqnarray}
E_{\rm EA} & = & - \sum_{i=1}^{N_\omega} A_i \ln (A_i) ~=~ E_{\rm SH} \label{eea2} \\
E_{\rm GK} & = & + \sum_{i=1}^{N_\omega} \ln (A_i), \label{egk2} \\
E_{\rm MX} & = & - \sum_{i=1}^{N_\omega} (A_i-\gamma/N_\omega) \ln (A_i), \label{emx2}
\end{eqnarray}
\label{allentropies}
\end{subequations}
where again $\sum_i A_i=1$ and we have left out the factor $N_\omega$ in front of the expressions,
for convenience below when we relate the SAC and ME methods to each other.

In the case of SAC, we sample at a temperature $\Theta$ using Eq.~(\ref{psg}), while in the ME probability Eq.~(\ref{pmaxent2}) there is no such
adjustable temperature (i.e., $\Theta=1$). In Sec.~\ref{sec:entropy} we found that the optimal sampling temperature scales with the inverse of the
number of $\delta$-functions, and we therefore introduce a normalized temperature $\theta$ through;
\begin{equation}
\Theta =\theta/N_\omega. 
\end{equation}
The SAC generated spectral weight distribution is always, for any of the parametrizations, collected in a histogram, and we know the forms of
the entropies $N_\omega E_{\rm X}$ [where X is one of EA, GK, or MX in Eqs.~(\ref{allentropies})] of the spectra represented by those histograms.
We can therefore write down probability distributions for those histograms defined by amplitudes $A_i$ in the same way as done in the ME method.
The probability distributions for the ME and SAC spectra are
\begin{subequations}
\begin{eqnarray}
P_{\rm ME,X}(A)  & \propto & {\rm e}^{-[\chi^2(A)/2-\alpha E_{\rm X}(A)]},\label{pme} \\
P_{\rm SAC,X}(A) & \propto & {\rm e}^{-N_\omega[\chi^2(A)/2-\theta E_{\rm X}(A)]/\theta}. \label{psac}
\end{eqnarray}
\end{subequations}
The factor $N_\omega$ multiplying the exponent in Eq.~(\ref{psac}) is very important, originating from the sampling entropies corresponding to the different
parametrizations (Sec.~\ref{sec:entropy}). Its presence here implies that increasing $N_\omega$ in SAC eventually completely enforces the minimization of
$\chi^2/2-\theta E_{\rm X}$. Thus, for large $N_\omega$, sampling the original spectrum in some specific parametrization X leads exactly to the same
spectrum as does minimization of $\chi^2/2-\alpha E_{\rm X}$ in the ME method with $\alpha=\theta$.

The probability distribution in Eq.~(\ref{psac}) looks like that in a conventional statistical mechanics problem (with $\chi^2$ corresponding to the energy
density, $E_{\rm X}$ to the relevant form of the entropy density, $N_\omega$ to the number of particles, and $\theta$ to the temperature $T$), and the claim (mentioned
in passing several times) that fluctuations play no role in Eq.~(\ref{psac}) when $N_\omega \to \infty$ may therefore seem unfounded. However, there are
fundamental differences between the
configuration spaces in SAC and conventional statistical mechanics. In particular, as we discuss in more detail in \ref{app:statmech}, in SAC there is
in practice no analogy to low-energy infrared fluctuations when $N_\omega \to \infty$, because the spectrum has well defined upper and lower bounds,
corresponding to a volume that does not grow with $N_\omega$. Thus, the ''thermodynamic limit'' here is an infinite-density limit, in contrast to the normally
fixed finite density in statistical mechanics. All possible fluctuations when $N_\omega \to \infty$ are then suppressed. This is one of the most important
previously neglected aspects of the problem of relating SAC to the ME method.

Since two of the entropies, $E_{\rm GK}$ and $E_{\rm MX}$, also have constant factors in front that we have neglected (discussed in Sec.~\ref{sec:entropy}),
the exact relationship between the SAC and ME parameters is $\alpha=x\theta$ with some $N_\omega$ independent factor $x$. We can circumvent this
uncertainty by comparing results for which $\langle \chi^2(\Theta)\rangle$ in the SAC case equals $\chi^2(\alpha)$ in the ME case. The mapping between
the two methods is then complete and testable in practice. We next confirm this statement by actually performing comparisons with all three parametrizations
in SAC and the corresponding entropies, Eqs.~(\ref{allentropies}), in the ME method.

\begin{figure*}[t]
\centering
\includegraphics[width=110mm]{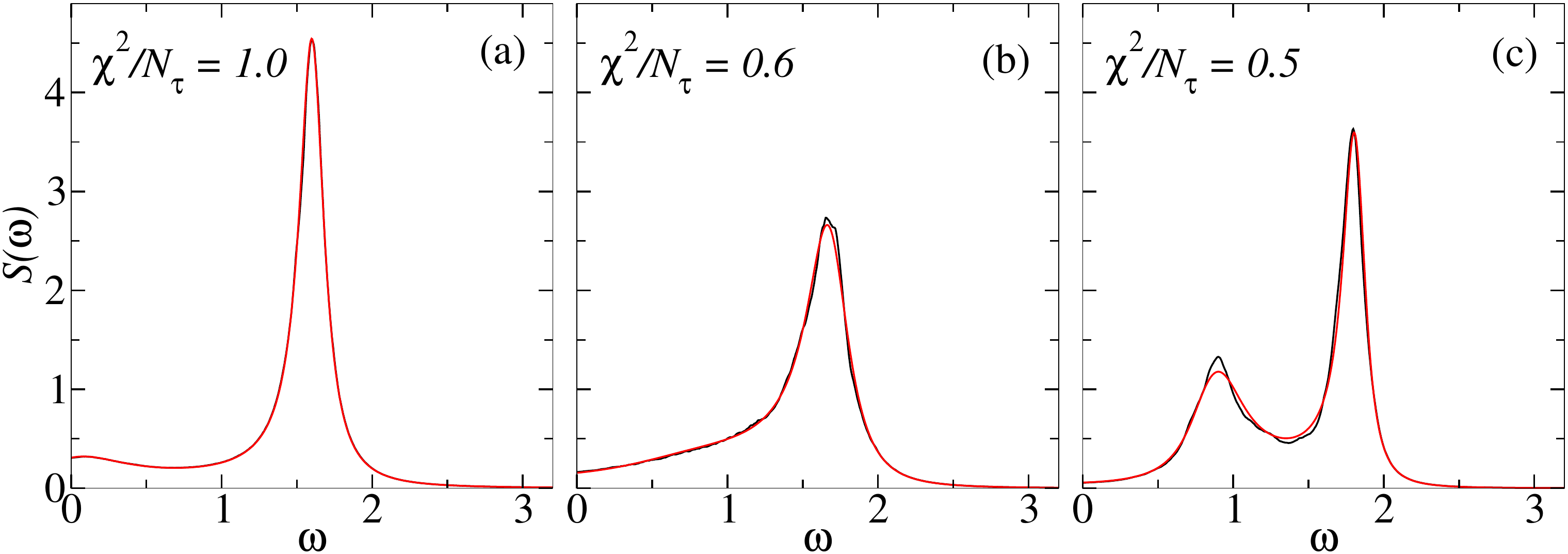}
\caption{Dynamic structure factor for the same Heisenberg chain as in Fig.~\ref{sw1}, obtained by SAC with amplitudes sampled on a fixed grid with 1000
points at spacing $\Delta_\omega=0.005$ (black curves) and with the ME method with the fixed-grid entropy, Eq.~(\ref{egk2}) (red curves). Results are shown
at three values of $\chi^2/N_\tau$ in the ME method, and these values match $\langle \chi^2\rangle /N_\tau$ in the SAC calculations. The values
are $\chi^2/N_\tau=1.0$ in (a), $0.6$ in (b), and $0.5$ in (c). The red ME curve in (a) almost completely covers the SAC result.}
\label{newent}
\end{figure*}

\begin{figure*}[t]
\centering
\includegraphics[width=112mm]{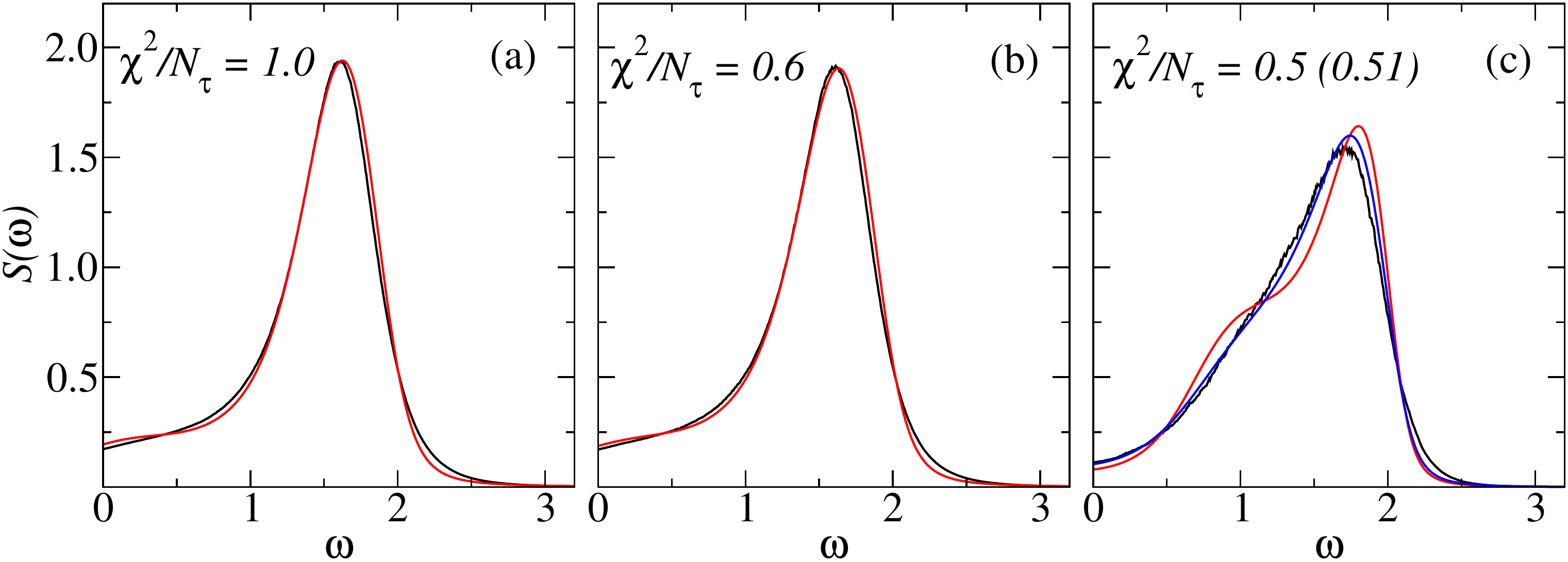}
\caption{Results as in Fig.~\ref{mixent} but for SAC with both amplitude and frequency updates (black curves) and the ME method with the mixed entropy,
Eq.~(\ref{emx2}), with $\gamma=0.44$ (red curves). In (c), results obtained at $\langle \chi^2\rangle/N_\tau=0.51$ are also shown (the blue curve).}
\label{mixent}
\end{figure*}

\subsubsection{Tests with Heisenberg QMC data}

We already concluded in Sec.~\ref{sec:maxent1} that SAC results for the Heisenberg chains based on sampling the equal-amplitude $\delta$-functions
resulted in spectra almost identical to those obtained with the ME method [comparing results in Fig.~\ref{sw1} and Fig.~\ref{sw2} with those in
Fig.~\ref{sw_me}]. The fact that this SAC parametrization indeed generates the same Shannon form of the entropy used in the conventional ME method
explains this remarkable agreement between the two approaches when $\alpha$ and $\Theta$ are chosen consistently so that $\langle \chi^2(\Theta)\rangle
= \chi^2(\alpha)$.

To make SAC different from the conventional ME method, another parametrization has to be used. Indeed, from the results in Figs.~\ref{sw1}
and \ref{sw2}, it is clear that sharper peaks are formed when the amplitudes are also sampled, and the results, e.g., those in Figs.~\ref{sw1}(a)
and \ref{sw1}(c), cannot be reproduced by any choice of $\alpha$ in the conventional ME method. Whether or not the amplitude fluctuations will
have a favorable effect on the average spectrum may not always be clear, though in Sec.~\ref{sec:theta} we presented evidence from several
examples that pointed to better results overall. Sampling amplitudes on a grid consistently appears to be the least preferred option in practice
(with peaks often being too sharp).

To test the fixed-grid case, results with $E_{\rm GK}$ in the ME method are shown in Fig.~\ref{newent} for the $16$-site Heisenberg chain, along
with the previous grid SAC results from Fig.~\ref{sw1}(a). We consider three different sampling temperatures $\Theta$ and corresponding $\alpha$
values, matching the two methods by the goodness of the fit as explained above. Note that we here again present results for the original spectral
function $S(\omega)$ according to Eq.~(\ref{barelation}). For each goodness-of-fit value $\chi^2(\alpha)$ in the ME case, we observe
excellent agreement with the SAC results at the matching value of $\langle \chi^2(\Theta)\rangle$. The small deviations are primarily due to some jaggedness
of the SAC spectra in Figs.~\ref{newent}(b) and \ref{newent}(c), which is a consequence of the rather inefficient sampling on the grid at low $\Theta$.
There may also be some very small effects of the goodness-of-fits not being perfectly matched, and also from the fact that the $N_\omega \to \infty$ limit
is not completely realized in the SAC case (grid sampling being slow, we only obtained well converged results up $N_\omega=1000$).

These results give further credence to the rather involved functional-integral calculations leading to $E_{\rm GK}$ by Ghanem and Koch \cite{ghanem20a}.
The variations among the results for different parametrizations
also illustrate perfectly that details of the SAC stochastic process impact the functional form of the entropy, not
just an unimportant factor. This aspect of SAC was not always recognized in the past, e.g., in Refs.~\cite{beach04,bergeron16}. Beyond analytic
continuation, the Shannon entropy is also not a universal form of the entropy that should always be assigned to a curve---see, e.g.,
Ref.~\cite{balestrino09}, where both the Shannon entropy and a form analogous to the GK entropy appear in a completely different context.

In the case of sampling with both frequency and amplitude updates, with the parametrization in Fig.~\ref{fig:spec}(c),
our conjectured form $E_{\rm MX}$ in Eq.~(\ref{emx2}) contains
the unknown ``mixing factor'' $\gamma$. We also proposed in Sec.~\ref{sec:entropy} that $\gamma^{-1}$ should be approximately the effective width
of the spectrum, though it is not clear exactly how such a width should be defined. In the tests here, we adjusted $\gamma$ so that the
peak height of the spectrum (using the same $L=16$ Heisenberg structure factor as before) is the same in the results of both methods when
$\chi^2(\alpha)/N_\tau=\langle \chi^2(\Theta)\rangle/N_\tau=1$, which gives roughly $\gamma=0.44$. The inverse of this number $\gamma^{-1} = 2.27$ indeed represents
a reasonable effective width of the spectrum.

\begin{figure*}[t]
\centering
\includegraphics[width=110mm]{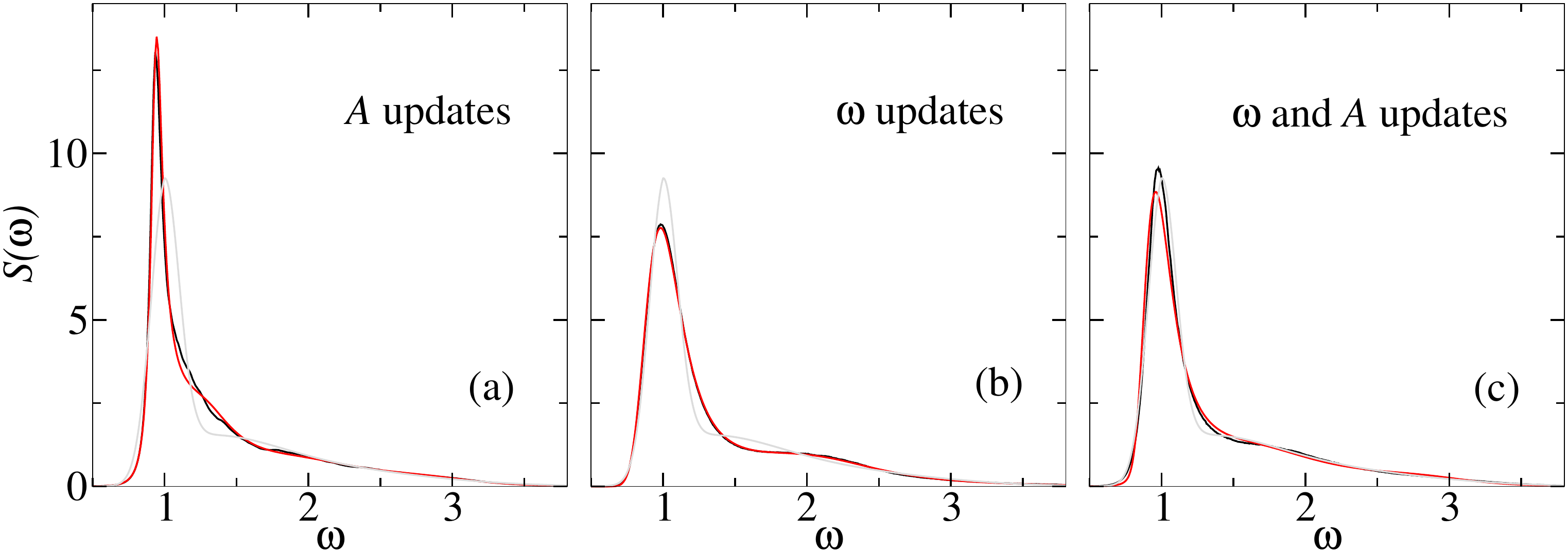}
\caption{Comparison of SAC (black curves) and ME results (red curves) for the synthetic spectrum previously studied in Fig.~\ref{syntcomp},
  here shown with dim gray curves in order to not obscure the SAC and ME results. Three different SAC parametrizations and associated entropies in the ME
  method were used with the data at error level $\sigma=10^{-6}$:
  Amplitudes on a fixed grid and the GK entropy, Eq.~(\ref{egk2}) in (a), equal-weight $\delta$-functions and the Shannon entropy, Eq.~(\ref{eea2}) in (b),
  and both frequency and amplitude sampling in (c), where the mixed entropy, Eq.~(\ref{emx2}) with $\gamma=0.5$ was used. The goodness of fit in all cases is
  $\chi^2_{\rm ME}/N_\tau=\langle \chi^2\rangle_{\rm SAC}/N_\tau=0.97$.}
\label{syntme}
\end{figure*}

As seen in Fig.~\ref{mixent}(a), the entire spectral function matches the SAC result well, though not as perfectly as in Fig.~\ref{newent}. Keeping
the same value of $\gamma$, the result at $\chi^2/N_\tau=0.6$ in Fig.~\ref{mixent}(b) has not changed much and the match between the SAC and ME spectra
is still good. At $\chi^2/N_\tau=0.5$, Fig.~\ref{mixent}(c), the ME result deviates more from the SAC spectrum, with a precursor of the eventual splitting
of the maximum into two peaks already setting in. In the SAC case the splitting starts at lower sampling temperature and is very
pronounced when $\langle \chi^2\rangle/N_\tau=0.48$ in Fig.~\ref{sw1}. In Fig.~\ref{mixent}(c) we also show the ME result
at slightly larger $\alpha$, corresponding to $\chi^2/N_\tau=0.51$, and this result matches quite well the SAC spectrum at
$\langle \chi^2\rangle/N_\tau=0.50$. It should be noted that this particular profile is not realized with any of the other parametrizations
that we have used in SAC. It is also possible to match the spectra better at the same goodness-of-fit values by changing the value of $\gamma$
slightly between the different cases in In Fig.~\ref{mixent}.

Based on these tests we conclude that the mixed entropy with fixed $\gamma$ indeed appears to be realized in SAC when sampling both the frequencies and
the amplitudes. However, since the entropy $E_{\rm MX}$ is not exact, the perfect match between the sampled spectrum and the one maximizing the ME probability
is ruined when $\Theta$ and $\alpha$ are adjusted for $\chi^2_{\rm ME}(\alpha)/N_\tau=\langle \chi^2(\Theta)\rangle_{\rm SAC}/N_\tau$. The closest match is obtained
with some small (at least in the example studied) deviation from this inequality. In other words, for a series of spectra $S_\Theta$ and $S_\alpha$
computed with the two methods, with the same $\bar G(\tau)$ data and with optimal entropy mixing $\gamma$, there should still be some function
$\alpha(\Theta)$ such that $S_\Theta \approx S_\alpha(\Theta)$ holds to a good approximation (but clearly the spectra cannot be exactly the same if the
goodness-of-fit values are not the same). Alternatively, the best $\gamma$ may have a weak dependence on $\alpha$, but we have not systematically
explored how to change $\gamma$ for the best match between the SAC and ME spectra.

The mixed entropic form with the parameter $\gamma$ also offers a broader range of ME methods, with $E_{\rm MX}(\gamma)$ versus $\gamma$ smoothly interpolating
between the entropies exactly corresponding to grid sampling and equal-amplitude frequency sampling. To use this entropy in practice, we need a way to determine
the optimal $\gamma$ value. Heuristically, we can just set $\gamma^{-1}$ equal to the
length of a box (corresponding to a flat default) with the same first and second moments as the spectrum, which can be determined to a good
approximation from $\bar G(\tau)$. The moments correspond to derivatives of $\bar G(\tau)$ at $\tau=0$ \cite{sandvik98b} and are, thus, encoded
in the short-time correlation function. We can also determine the moments from the spectrum itself during the initial stages of optimization procedure,
or from a previous SAC calculation with any parametrization. In the above case, this approach gives $\gamma=0.36$, for which the peak height of the
spectrum at $\chi^2/N_\tau=0.6$ (which is in the acceptable range according to our criterion) only is slightly less than in Fig.~\ref{mixent}
and the result overall matches well the exact diagonalization result. As we will see below in Sec.~\ref{sec:maxent2}, the mixed entropy will
also be relevant in the case of sampling spectra within the ME approach.

\subsubsection{Tests with synthetic data}

For our final test of the three different entropic forms, we consider the synthetic spectrum previously investigated in Sec.~\ref{sec:example3} by SAC
calculations with the two continuous frequency parametrization. We consider the case of imaginary-time data with error level $\sigma=10^{-6}$, for which
SAC results were shown in Fig.~\ref{syntcomp}(e) and \ref{syntcomp}(f), and compare these results with maximum-probability spectra obtained with the ME
method with the corresponding entropic forms Eqs.~(\ref{allentropies}). Here we include also the fixed-grid case that was not studied in Sec.~\ref{sec:example3}.
We match the goodness of fit values in the two methods according to our criteria in Eqs.~(\ref{eq:chi2}) and (\ref{eq:chi2me}), with $a=0.5$ as before, i.e.,
we again have $\chi^2(\alpha)/N_\tau=\langle \chi^2(\Theta)\rangle/N_\tau$ but now focusing only on the statistically optimal spectra.

According to the rigorous mappings between the SAC and ME methods when sampling either equal-amplitude $\delta$-functions or fixed-grid amplitudes, which
lead to entropies given by Eqs.~(\ref{eea2}) and (\ref{egk2}), respectively, the two methods should produce identical results as long as $N_\omega$ is sufficiently
large. In the case of sampling both frequencies and amplitudes, we again expect the mixed entropy, Eq.~(\ref{emx2}), to be a good approximation to its unknown
exact form if $\gamma$ is appropriately chosen. Here we just use the method described above of matching the first and second moments of a flat default
model, which gives $\gamma=0.50$.

Results are shown in Fig.~\ref{syntme}. Indeed, the agreement between the SAC and ME results is almost perfect in the case of the fixed grid and
frequencies-only sampling, Figs.~\ref{syntme}(a) and \ref{syntme}(b). In the case of combined amplitude and frequency sampling, Fig.~\ref{syntme}(c),
the agreement is also quite good, thus providing further support for the mixed entropy form, and also for the simple heuristic way of determining the
mixing coefficient $\gamma$. 

\subsubsection{Conclusions on SAC to ME mapping}

The test results presented above leave little doubt that the unconstrained SAC method delivers exactly the same spectrum as the maximum-probability ME method,
provided that the appropriate form of the entropy is used in the ME method and that $\Theta$ and $\alpha$ are chosen for matching the goodness
of fit of the two spectra. Further, $N_\omega$ must be sufficiently large for convergence to the $N_\omega \to \infty$ limit in the SAC case (and
naturally a fine enough grid has to be used for the ME calculations, though the convergence here appears to be typically faster). Thus,
we have finally established the exact relationship between the two methods. While some of these relationships were anticipated in previous works,
they were never spelled out in a definite, consistent, and systematic manner. Some aspects of the claims are incomplete or incorrect. Before
gathering and further discussing our conclusions on the mapping, we summarize some of the key prior insights and views.

Beach, in his pioneering work on the ME--SAC equivalence \cite{beach04}, used a mean-field approach that did not produce the factor $N_\omega$ of the entropy and
also involved the wrong functional form of the entropy for the parametrization used (both frequencies and amplitudes sampled). The Shannon entropy was derived
without taking into account the details of the stochastic process, which instead should result approximately in the mixed form introduced here. The lack of $N_\omega$
scaling is natural within mean-field theory, and the Shannon entropy was assumed to be generic relative to an arbitrary default model (which also was first
incorporated into SAC by Beach).

Bergeron and Tremblay correctly noted the factor $N_\omega$ in their derivation of the Shannon entropy, but regarded it not as arising from a parametrization
with $\propto N_\omega$ degrees of freedom but as a conceptual device. They also did not discuss the details of the mapping beyond stating that the entropy
derivation  ``suggests a connection'' \cite{bergeron16} between SAC and the ME method. Working with a histogram representation, the form of the entropy was
again incorrect (i.e., not $E_{\rm GK}$), and there was no mention of different entropic forms depending on the SAC parametrization.

Fuchs, Pruschke, and Jarrell \cite{fuchs10} explicitly stated that SAC minimizes a functional including an entropy, as in Eq.~(\ref{functional}). However,
they regarded this entropy as a macroscopic thermodynamic entropy (i.e., a single number generated by the averaging process under given conditions), not one directly
expressed for each spectrum $A$ as $E(A)$ originating from the degrees of freedom of the stochastic process. Their main aim was to fix the sampling temperature
$\Theta$ (or eliminate it by integration) from the problem by using Bayesian inference. The final conclusion was that SAC is better than the ME method because
the exact probability distribution is sampled instead of maximized. Here we have shown that the two approaches, under the prevailing conditions of unrestricted
sampling used in Ref.~\cite{fuchs10}, give exactly the same results (strictly speaking when $N_\omega \to \infty$, where the fluctuations of the average
spectrum vanish).

The entropy $E_{\rm GK}$ \cite{ghanem20a} was extremely useful in our work presented above, but Ghanem and Koch also did not take the final step to
show the exact equivalence between the methods under different conditions, although they realized the non-universality \cite{ghanem20b} of results
obtained in different sampling spaces. The crucial role of the sampling space in producing different entropic pressures was indeed also the motivation
for switching from a fixed frequency grid to $\delta$-functions in continuous frequency space in earlier work ny one of us and collaborators \cite{qin17},
but it was not realized
that the fixed-apmplitude case maps to the conventional ME method and that it is actually the fixed frequency grid that leads to a new form, Eq.~(\ref{egk}),
of the entropy \cite{ghanem20a}.

The mixed entropy, Eq.~(\ref{emxw}), that we have proposed here is not based on a formal derivation and is likely not exact in general. It still
offers an interesting perspective on the case of fluctuating amplitudes and frequencies, including its use in generalized ME methods. It would be
interesting to further analyze the entropy in this space using analytical approaches. While a functional-integral representation may not formally exist
in this case, as pointed out by Ghanem and Koch \cite{ghanem20a}, the stochastic process of SAC is always well defined. There is no dependence on the
discretization as long as the histogram used to collect the spectrum has sufficiently small bins to capture all details of the average. We have
computed the entropy of the spectrum exactly in the case of all-equal amplitudes [Sec.~\ref{sec:entropy1}] by using a ``particles on a line'' method
but have not yet been able to incorporate the fluctuating amplitudes analytically within this approach.

The entropic forms also formally require the limit $N_\omega \to \infty$, which in practice is realized for values of $N_\omega$ that can be easily reached
at least with the parametrizations in the frequency continuum (with larger $N_\omega$ required as the error level of the imaginary-time data decreases).
We here take the point of view that the unrestricted (without any constraints)
SAC methods should be considered in the large-$N_\omega$ limit, which is also conceptually different from
the approach of Ghanem and Koch \cite{ghanem20b}, who focused on sampling at $\Theta=1$ in a range of $N_\omega$ for which the fit is still good (similar
in this regard to Ref.~\cite{sandvik16}). Both points of view are valid, and it is possible that additional spectral structure can be resolved
with an optimal $N_\omega$ that is not yet in the ME limit, so that there are still fluctuations about the maximum-probability ME spectrum. In our
$\Theta=1$ tests here, presented in Fig.~\ref{ndep}, we did not observe any advantage of taking $N_\omega$ away from the large-$N_\omega$ limit
(with $\Theta$ adjusted), but any gain in resolution may depend on the problem studied. The use of a default model may also play some role.

As a final remark on the three different representations and associated entropies, we note again that the underlying sampling of both frequencies and
amplitudes appears to generate the best balance between resolution (sharper peaks than with only frequency sampling) while avoiding overly sharp features.
We have seen several examples of this behavior in our tests in the preceding sections. These apparently generic behaviors are also well represented in
Fig.~\ref{syntme}, where clearly the amplitudes and frequency sampling (mixed entropy) gives the best match with the exact spectrum. The frequencies-only
sampling (conventional Shannon entropy) results in the broadest spectrum, while the fixed-grid (GK entropy) results in a much too sharp peak.

Here it is also worth emphasizing that the error level in the tests in Fig.~\ref{syntme}, $\sigma = 10^{-6}$, is achievable in QMC simulations, but normally
only at great effort. The fact that the three different parametrizations and corresponding entropies still produce such different results then
clearly demonstrates that it is important to use the best possible SAC parametrization (or entropic form in the ME method). For work with unrestricted
sampling, all indications are that the combined amplitude and frequency sampling should
be the best option. Similarly, in the ME method the mixed entropy should be optimal. To fix the value of $\gamma$, our simple spectrum-width approach
with a flat default model works well. Using the mixed entropy with this $\gamma$ is likely the best ME method with a flat default. Other default models
could also be used with the more general form, Eq.~(\ref{emxw}), of the mixed entropy. We leave further exploration of the ME method with the mixed entropy
for the future.

With constrained SAC, we typically sample only the frequencies, since it appears advantageous to minimize the remaining entropic pressures once the
constraint has been optimized, as exemplified in Fig.~\ref{w0fix}. It is still possible that continua with more spectral details would benefit from also
including amplitude updates. We also note that some of the SAC constraints and optimization methods that we have developed can also be adapted to
ME method. The above considerations will then also apply as to what form of the entropy to use in that case.

\subsection{Maximum probability versus sampling}
\label{sec:maxent2}

As an alternative to finding the spectrum minimizing the functional, Eq.~(\ref{functional}), as typically done in the ME method (some times with further
integration over $\alpha$), the spectrum can also be averaged by sampling as in SAC. Then either the probability distribution Eq.~(\ref{pme}) for a fixed
value of $\alpha$ (optimized in some way) can be used, or $\alpha$ can also be sampled using a version of Eq.~(\ref{pmaxenta2}) (with the prior distribution
normally taken $\propto \alpha^{-1}$, as  discussed, e.g., in Ref.~\cite{jarrell96}). The sampling approach may from the outset seem like the better
option, because the use of the optimal spectrum appears to presuppose a distribution with a sharp maximum. In practice, the maximum may not not be very sharp,
and including the contributions from near-optimal solutions will then significantly affect the spectrum. The ME sampling approach has been used, in particular,
in the context of the dynamic structure factor of $^4$He \cite{boninsegni96,kora18}.

It is interesting to compare and contrast the sampling of the spectrum within the ME framework and the SAC method. When sampling, there is a native
configurational entropy of the spectrum that depends on the parametrization used, as we discussed exhaustively in the previous subsection. The fact that the
entropy is always extensive implies that it will ultimately (for a large number $N_\omega$ of sampled degrees of freedom) drive the SAC spectrum to a bad
fit, unless the sampling temperature $\Theta$ is lowered at the rate $1/N_\omega$. When sampling the ME probability, the Shannon entropy is explicitly used
in weighting the configurations according to Eq.~(\ref{pme}), but there is also configurational entropy of the sampling space, exactly as in the SAC method.
Thus, with no $\Theta$ to adjust, the sampled ME spectrum should eventually, when a large number $N_\omega$ of parameters is sampled, flow to poor solution,
in this case to the default model (which typically would correspond to a very large $\langle \chi^2\rangle$ value),

In the case of sampling amplitudes on a grid, which as far as we are aware is the only parametrization that has been considered with the ME approach, the sampling
generates the  GK entropy $E_{\rm GK}$, Eq.~(\ref{egk2}), which implicitly combines with the Shannon entropy used for the prior weight. Thus, it can be expected
that the total entropy of the spectrum in this case is exactly the mixed entropy $E_{\rm MX}$ defined in Eq.~(\ref{emx2}), again for some value of $\gamma$. Here
we will not be concerned with the exact form of the combined entropy, though it would also be interesting to further explore the mixed entropy in this context
as well.

We will focus on the fact that the sampling part of the entropy is extensive, which has been ignored so far in this context. There is no reason
to expect that $\Theta=1$ sampling according to the distribution in Eq.~(\ref{pmaxent2}) would avoid the entropic catastrophe when the number of
histogram bins is large. As an example demonstrating this potential problem, we consider the $16$-site Heisenberg chain as before. We used the
distribution Eq.~(\ref{pme}) to sample a histogram in the range $\omega \in [0,5]$ (effectively with a flat default model in this range), using
different numbers $N_\omega$ of bins.

\begin{figure}[t]
\centering
\includegraphics[width=80mm]{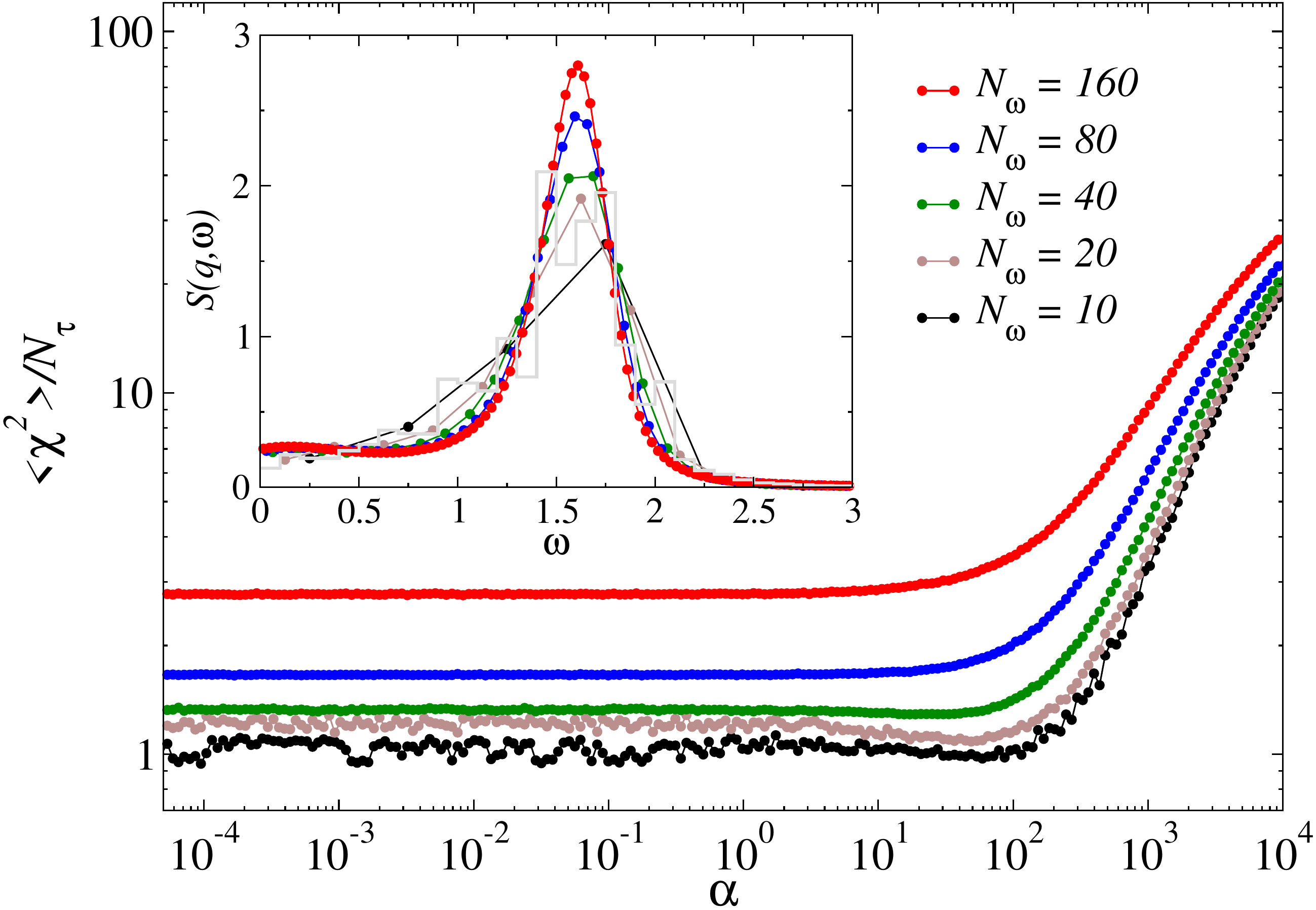}
\caption{Goodness of fit vs the entropy weighting parameter obtained by sampling the probability Eq.~(\ref{pmaxent2}) with spectra defined on grids
$\{ \omega_1,\ldots,\omega_{N_\omega}\}$ with $\omega_{N_\omega}=5$ and $N_\omega=10,20,40,80,160$. The system parameters are the same
as for the $L=16$ results in Fig.~\ref{x2_me}. The inset shows the averaged spectral functions at $\alpha \approx 50$. The dim gray histogram bins represent
the exact-diagonalization result as before (Fig.~\ref{sw1}). Note that the sampling with small $N_\omega$ is very slow (as demonstrated with
SAC in Fig.~\ref{nconv}), which causes the noisy behavior of $\langle \chi^2\rangle/N_\tau$ for $N_\omega=10$ and $20$.}
\label{x2_alpha}
\end{figure}

The main graph in Fig.~\ref{x2_alpha} shows $\langle \chi^2\rangle/N_\tau$ versus $\alpha$ and the inset shows the corresponding spectral functions together with
the histogram based on  exact diagonalization as, e.g., in Fig.~\ref{sw_me}(a). The behavior overall versus $N_\omega$ is similar to what we found previously with
$\Theta=1$ SAC in Fig.~\ref{ndep}; for given $\alpha$, $\langle \chi^2\rangle$ increases and eventually grows beyond the statistically acceptable bound. Interesting,
for the smallest $N_\omega$ cases (up to at least $N_\omega=40$) shallow minima can be seen between $\alpha=50$ and $\alpha=100$. While the histogram is very coarse
for $N_\omega=10$, those for $N_\omega=20$ and $40$ capture the overall shape quite well. For $N_\omega=160$ there is no $\chi^2$ minimum, and a small spurious
maximum has formed in the spectrum at low frequency, similar to the SAC results in Fig.~\ref{sw1} for the larger goodness-of-fit values. One could argue that
$\alpha$ should be set based on the $\chi^2$ minimum where possible, but it should be noted that $\chi^2$ is somewhat too high already with the small values of
$N_\omega$ for which a minimum exists. We have also not investigated this behavior in detail for other systems and spectral functions.

These results confirm our expectation that the probability distribution that forms the basis of the ME method, Eq.~(\ref{pme}), is flawed when a histogram
with a large number of bins is used. There is an implicit assumption, that has not been explicitly stated in this context as far as we are aware, that the
number of bins (or in general some number of parameters) must be rather small, so that the inherently extensive entropy of the configuration space does not
come significantly into play. In a recent application \cite{kora18}, where also $\alpha$ was sampled, indeed $\langle \chi^2\rangle$ is also marginally higher
than would be expected for good statistical fits, though the effects on the spectral function is likely only marginal and unlikely impacts any conclusions.
It should also be noted that a good default model reduces the detrimental effects of the extensive sampling entropy.

In principle, the entropy increase with $N_\omega$ can be counteracted by introducing a factor similar to $N_\omega/\theta$ in the SAC distribution in
Eq.~(\ref{psac}), but then both $\alpha$ and $\theta$ have to be optimized. Eventually for large $N_\omega$ the sampled spectrum would again become
identical to that maximizing the probability as discussed above in Sec.~\ref{sec:sacme}, in this case with the mixed entropy. Alternatively, $N_\omega$
should just not be too large, so that the entropic effect does not yet affect the result. In practice, it is easier to use SAC or the maximum-probability
version of the ME method, with the mixed entropy if so desired. Pathologies of the ME probability distribution are further discussed in Sec.~\ref{sec:pathology}.

\section{Conclusions and future prospects}
\label{sec:discussion}

An overarching theme of this work has been the critical role of the sampling space (parametrization) of the spectral function within the SAC approach to analytic
continuation. We have emphasized the fact that all parametrizations are associated with their different inherent entropic pressures under Monte Carlo sampling.
Thus, while good underlying QMC data will ensure that the main spectral features are well reproduced, details of the results will depend significantly on the
parametrization when the imaginary-time data are of statistical quality typical for QMC simulations. We have identified the main problems and advantages with
three different basic parametrizations, using $\delta$-functions on a grid or in the frequency continuum [Figs.~\ref{fig:spec}(a)--(c)].

Following studies of unrestricted sampling, we introduced a number of constraints intended to eliminate or reduce entropic effects that otherwise cause
smearing of sharp spectral features. We focused on the lower frequency bound, which physically reflects signatures of the effective conventional
quasi-particles or fractionalized excitations that are of primary interest in experimental and theoretical studies of quantum matter. Our results demonstrate
a remarkable fidelity of the methods to locate edges and determine their properties---even widths of narrow quasi-particle peaks and exponents governing
power-law singularities. The proper treatment of an edge also translates into significant fidelity enhancement at higher frequencies, since edge distortions
appearing in the absence of constraints (with all conventional analytic continuation methods) propagate and cause
distortions of the spectrum away from the edge. The new methods open opportunities to study quantitative aspects of spectral functions that
have been out of reach until now. To illustrate the power of these methods, we presented (Sec.~\ref{sec:ladders}) new results for the dynamic structure
factor of 2- and 3-leg $S=1/2$ Heisenberg ladders, uncovering previously unexplored features. 

While the main focus of our work has been on analytic continuation within the SAC framework, along the way we have also obtained new insights into the
ME method and its relationship to SAC. We have found that the Bayesian-based probability distribution underlying the ME method has a flaw in
that it describes divergent fluctuations in the continuum limit, though the most probable spectrum is still well defined. We have also revised the
formal relationship between the ME and SAC methods, finally achieving a rigorous mapping between the two. We demonstrated the relationship concretely by
comparing SAC calculations with different parametrizations with ME calculations with entropic priors corresponding to those parametrizations. These
insights and tests prove that the conventional Shannon information entropy used in the traditional ME method is not universal, and that generalized
ME methods with other entropic priors may produce better resolution.

Since the treatise is extensive, with progress reported on several interrelated fronts, we here concisely summarize and remark further on some of the key
developments and discoveries, not necessarily in the order presented previously but with references to relevant sections, equations, and figures.
In Sec.~\ref{sec:conca1} we focus on the developments of the SAC method itself, while in Sec.~\ref{sec:conca2} we discuss the more formal entropic aspects
of the sampling approach, our new insights into the ME method, and the relationships between the SAC and ME methods. These two sections do not merely
review material from the previous sections but include substantial commentary and additional conclusions from synthesizing different results.
In Sec.~\ref{sec:conc3} we discuss future prospects, including further developments of constraints (which can be regarded as generalized default models),
the possibility to use machine learning to augment SAC or ME calculations, as well as potential advantages of including a small fraction of negative
spectral weight in SAC sampling.

\subsection{Stochastic analytic continuation}

\subsubsection{Optimal $\Theta$ and $\Theta \to 0$ limit}
\label{sec:conca1}

An important aspect of SAC is how to determine the optimal sampling temperature $\Theta$. We have here expanded on the statistical motivations for
the previously suggested (in our work with collaborators in Refs.~\cite{qin17} and \cite{shao17}) simple criterion in Eq.~(\ref{eq:chi2}), which relates
$\langle \chi^2\rangle$ to the optimal $\Theta$. By an exact calculation of the configurational entropy in one of the parametrizations
(Sec.~\ref{sec:entropy1}), we demonstrated that $\Theta \sim 1/N_\omega$ is necessary, with $N_\omega$ being the number of $\delta$-functions in the
sampling space. This scaling behavior, which follows from the fact that the entropy is extensive in $N_\omega$ while $\chi^2$ acts as an unusual
intensive energy in a statistical-mechanics analogy (see \ref{app:statmech}), is realized automatically through our procedures.

We have investigated the $\Theta \to 0$ limit (in detail in \ref{app:lowtheta}), which defines the best goodness-of-fit $\chi^2_{\rm min}$ that also
enters in the $\Theta$-criterion. We explain why the spectrum in this limit consists of a small number $N_\delta$ of $\delta$-functions (about 2-6 for QMC data of
typical quality), and how these define an effective number of fitting parameters $N_{\rm para}=2N_\delta$ (from the position and amplitude of each
$\delta$-function) that can be collectively produced by a positive definite spectrum, given the noise in the imaginary-time data (which corresponds
to some violations of positive definiteness). The identification of the number $N_{\rm para}$, along with $\chi^2_{\rm min}$ serving as a proxy for
the number of effective degrees of freedom of the fit, closes the circle on the formal statistical applicability of the $\Theta$ criterion. These insights
also enable an a posteriori check of the statistical soundness of the covariance matrix for the QMC data set, in the form of the expected relationship
Eq.~(\ref{chi2expected}) between $\chi^2_{\rm min}$ and $N_{\rm para}$.

The designation of $N_{\rm para}$ based on the best-fit spectrum also provides a possible resolution to a mystery that most practitioners of numerical analytical
continuation have undoubtedly pondered: When improving the quality of the imaginary-time data, existing peaks and other features typically evolve somewhat (as
we have seen many examples of here, e.g., in Fig.~\ref{syntcomp}). However, it is commonly very difficult to achieve large qualitative improvements, i.e., to
observe the emergence of new previously unresolved spectral features (e.g., peaks or minima),
even when increasing the data quality substantially. This slowly evolving behavior, possibly
followed by ultimately large changes in the limit of very good data, is indeed what should be expected in light of the smallness of $N_{\rm para}=2N_\delta$.

If we consider the low-$\Theta$ behavior, a spectrum with a very small number $N_\delta$ of spikes, say $N_\delta=2$ or $N_\delta=3$, is typically very stable,
with $N_\delta$ itself as well as the locations and amplitudes of the spikes are not very sensitive to variations in $\bar G(\tau)$. For a given observed
$N_\delta$, it may take orders-of-magnitude improvements in the data quality to increase $N_\delta$ even by $1$, considering that the problem is exponentially
hard in the noise level. It is likely only when $N_\delta$ increases that additional significant spectral features also emerge in the sampled spectrum
with $\Theta$ in the optimal range. We have not yet systematically investigated this aspect of the analytic continuation problem but it would be very
interesting to pursue further. In this regard, we also point to the possibility that a very small fraction of negative weight in the spectrum may in some
cases lead to larger $N_\delta$ (\ref{app:low2}), which would suggest that sampling with negative weight can perhaps improve the resolution of SAC in some cases.

We have also demonstrated that a previously often used $\Theta$ criterion based an a maximum log derivative of $\chi^2(\Theta)$ in general results in a spectrum
with suboptimal fit to the data, while our method guarantees a good fit by construction. Moreover, our criterion is much easier to use in practice, because the
derivative is much noisier (i.e., requires longer sampling) than just $\langle \chi^2\rangle$ and $\chi^2_{\rm min}$. Our criterion is also much easier to
implement than the Baysian approach suggested in Ref.~\cite{fuchs10}, which we have not tested.

\subsubsection{Resolving sharp spectral edges}
\label{sec:conca2}
              
Constraining the sampling space in some way implies that certain additional pieces of information have been provided to the SAC process, in addition to the QMC data.
Thus, some specific aspect of the spectrum is built in, and this aspect should ideally correspond exactly to a known feature in the spectrum sought, e.g.,
the existence of a sharp quasi-particle peak or a power-law singularity at the lower edge of the spectrum. The method can also be regarded as a form of hypothesis
testing. In some cases, a treatment without constraint may already hints at a sharp peak or edge, and it is then worth trying constrained parametrizations
to further explore possible functional forms at the edge.

An imposed constraint often involves an optimized parameter that provides quantitative information not supplied
as input. Such information is frequently not just the optimized parameter itself, but additional quantitative features that can be reliable extracted from the
optimal spectrum. For instance, when optimizing the amplitude of a quasi-particle peak, also its location can be extracted at a level of
precision that would otherwise be impossible.

We stress that the amount of information provided when imposing a constraint seemingly is very limited, but the implications in terms of improved resolution and
extraction of quantitative not-supplied information can be dramatic. We also point out that it would be very difficult to construct generically useful fitting
functions with the degree of flexibility offered by our most generic constrained SAC parametrizations, e.g., a power-law edge followed by an arbitrary
continuum (Fig.~\ref{fig:mixed} and further generalizations). What our constrained  methods provide is a generic machinery for optimizing spectra of specific
types based on statistical criteria and minimal information provided.

Functions with close to the same degree of flexibility as SAC would still require a large number of parameters; say tens of parameters. Though this
would be much less than typically $N_\omega$ in SAC, the parameters would still be numerous enough to prohibit optimization in conventional ways, for similar
reasons as why $\chi^2$ minimization does not work with the parametrizations used here. The forms would have to be regularized in some way, e.g., by
sampling parameters, and then the task starts to resemble SAC with parametrizations beyond those considered here. Further complications may arise
from the likely complicated inherent entropic pressures of such forms, because of the way different parameters can affect the function very
differently in general. The parametrizations with large $N_\omega$ used here have the advantage of similarity to statistical mechanics, including
well-defined generalized thermodynamic limits.

Following our initial work on spectra with sharp quasi-particle peaks in Ref.~\cite{shao17}, we have clarified (Sec.~\ref{sec:delta2}) how the quasi-particle
weight converges toward its correct value with increasing data quality. We have also established
the statistical signatures of a quasi-particle peak of finite width when a $\delta$-function is incorrectly imposed in the parametrization, i.e., the method
can detect the inapplicability of the constraint (to an extent set by the data quality). As an application of this parametrization and optimization
to a problem with still open questions, we investigated the dynamic structure factor of the 2-leg Heisenberg spin-1/2 ladder (Sec.~\ref{sec:hladd2}), where,
at momenta close to $q=\pi$, there is an isolated $\delta$-peak at the gap $\Delta$, originating from the triplon quasi-particle, and a second gap
$3\Delta$ is also expected. We built in the second gap as a further constraint, which revealed new spectral features from three-triplon excitations.

We have further generalized the $\delta$-peak method to a quasi-particle peak of finite width, using a parametrization, Fig.~\ref{peakfig}(c), where the peak
weight is split up among $N_p$ $\delta$-functions, the number of which is much smaller than the total number $N_\omega$ (and their amplitudes conversely
are much larger). Both $N_p$ and the total peak weight $A_0$ are parameters that have to be optimized, for which we have proposed a reasonably efficient
scheme. We have presented promising test results for synthetic spectra with both narrow and broad peaks, showing that narrow peaks can be reproduced at a level
far beyond what can be achieved with conventional SAC or ME methods.

With an SAC parametrization specifically tailored to power-law singularities---the distance-monotonic $\delta$-functions depicted in Fig.~\ref{fig:spec}(e)---we
have demonstrated unprecedented fidelity in reproducing various synthetic spectral functions as well as the dynamic structure factor of the Heisenberg chain,
which hosts a spinon continuum with edge divergence. It should be noted that, as in the case of the quasi-particle peak discussed above, prior knowledge of
the location of the edge of the spectrum is not required, though if available such information can be used as input as well. Otherwise the sampling process
equilibrates to a stable edge very close to the correct location, typically to better than $1\%$ error in our tests (see, e.g., Fig.~\ref{fig:hbergsdens}).

With the basic distance-monotonic constraint, we can reproduce continuous spectra with strongly divergent edge followed by a monotonic decay. The divergence
can be quenched by introducing an optimized constraint on the minimum $\delta$-function separation. With a further generalization of the parametrization
involving varying amplitudes, the exponent governing a divergent or convergent power-law edge can also be extracted by optimizing a parameter (as in the
example in Fig.~\ref{p33scan}). In our most complete modeling of a spectrum with arbitrary continuum and power-law edge, we combine unrestricted SAC sampling
with the edge parametrization discussed above. Following promising tests with a synthetic spectrum (Fig.~\ref{fig:mixed} and other cases not presented here),
this parametrization was used to study the dynamic structure factor of the 3-leg Heisenberg ladder in Sec.~\ref{sec:hladd3}.

While there is still limited ability to resolve more than one or two peaks in the continuum, the constrained SAC methods offer unprecedented access to
spectral edges originating from quasi-particle fractionalization. We note that there is really no other way to reliably extract sharp edges. The most
competitive alternative method is time-dependent DMRG, which in some cases can deliver spectral functions with much more structure (more peaks) than what is
possible in practice with analytic continuation of QMC data. However, sharp edges pose a problem for DMRG calculations, and we have also pointed out apparent
difficulties with resolving the three-triplon contributions to the dynamic structure factor of the 2-leg Heisenberg ladder (which we did detect in
Sec.~\ref{sec:hladd2}). While spectral weight above
the isolated $\delta$-peak at the gap $\Delta$ has been detected in DMRG work \cite{schmidiger12}, this weight does not extend down to $\omega = 3\Delta$
but, for reasons that remain unresolved, exists only at higher energy. 

There are many other applications of time dependent DMRG and related methods directly targeting the frequency space. In relation to models discussed in this
work, the dynamic structure factor of the standard Heisenberg 1D chain at low temperatures has been studied at both high an low temperatures \cite{barthel09},
where in the latter case the sharp edge is still rounded by $T>0$ effects. There are also effects of the approximations made in the time evolution.

A method based on matrix product states and Chebyshev polynomials (in frequency space) at $T=0$ also leads to rounding of the edge \cite{xie18}, partially because
of the need to smoothen a rather small number of $\delta$-functions that appear in the exact (or well approximated) finite-size spectrum (see also
Ref.~\cite{wang19}). Even for the system size $L=500$ that we have considered for $T=0$ Heisenberg tests here, the ultimately discrete spectrum may
still come into play in some way. However, the sharp-edge SAC method also implicitly provides a natural smoothing mechanism for the continuum, while still
preserving the sharp edge, in a way that is fully appropriate for reaching the thermodynamic limit with increasing system size.

In the presence of trimerization, a more intricate spectral function forms in the Heisenberg chain as a consequence of coexistence of spinons with propagating
internal trimer excitations. Similar spectral features have been resolved using both QMC-SAC \cite{cheng22} and DMRG \cite{bera22} (and were also recently
observed experimentally \cite{bera22}), but the larger periodic systems accessible with QMC-SAC makes it possible to resolve the low-energy spinons,
in particular, to a much higher degree even with unrestricted sampling.

We also point to a DMRG study of a chain with long-range interactions \cite{yang21}, where the sharp (expected) magnon peak is likely broadened by the finite
time evolution, and the continuum exhibits (likely) artificial oscillations, at least partially because of true finite-size structure. It will be interesting
to apply the method discussed above for a quasi-particle of finite width to this case.

The DMRG and related matrix product methods \cite{aristov10,xie18,shu18}
are also largely limited to 1D systems (though some impressive 2D results have also been obtained recently \cite{verresen18,sherman22}),
while the methods discussed here are applicable to higher-dimensional systems as well (of course within the limitations set by the QMC sign problem). Here we
point to Ref.~\cite{ma18}, where unconstrained SAC was applied to a 2D quantum-critical system with deconfined spinon excitations. It would be interesting
to study the same model (and other models exhibiting spinon deconfinement) with the power-law edge parametrization of the spinon spectrum. 

\subsection{Maximum entropy methods}
\label{sec:conc2}

\subsubsection{Equivalence between sampling and maximum probability}

We have derived the exact entropy of spectra collected in histograms in SAC with equal-amplitude $\delta$-functions in the frequency continuum
(Sec.~\ref{sec:entropy}). The result, Eq.~(\ref{eea}), is the well known Shannon entropy, apart from the prefactor $N_\omega$ that we
discussed above in Sec.~\ref{sec:conca2}. With the temperature $\Theta$ scaled as $1/N_\omega$ in order to reach good $\chi^2$ values for the sampled spectra,
we define the intensive temperature $\theta=\Theta/N_\omega$. Then, if we identify $\theta=\alpha$, the probability distribution, Eq.~(\ref{psac}),
is the same as that in the ME method, Eq.~(\ref{pme}), apart from the overall factor $N_\omega/\theta$ in the exponential. This factor is of
critical importance, as it makes the ``free energy'' minimum increasingly deep as $N_\omega$ increases, thus effectively enforcing the minimization
of the functional $\chi^2(A)/2-\theta E(A)$ when $N_\omega \to \infty$. The average spectrum in SAC then becomes exactly the maximum-probability ME
spectrum with $\alpha=\theta$. In \ref{app:statmech} we further discuss the reason why the fluctuations are completely suppressed in this case,
unlike in conventional statistical mechanics.

An important consequence of the effective enforcement of maximum probability with increasing $N_\omega$ is that the sampling  fluctuations of the SAC
spectrum for finite $N_\omega$ cannot be used as reliable statistical errors on the average spectrum (as is some times implied \cite{beach04}). We will
discuss error calculations further below in Sec.~\ref{sec:errors}.

We have confirmed explicitly (Sec.~\ref{sec:sacme})
that SAC with different parametrizations correspond to the ME method with different entropies in the prior probability;
the Shannon entropy is generated only in the above discussed case of sampling the frequencies of equal-amplitude $\delta$-functions. With amplitudes
sampled on a grid, the entropy Eq.~(\ref{egk}) previously calculated by Ghanem and Koch \cite{ghanem20a} applies. When both the frequencies and the
amplitudes are sampled, we have conjectured, based on additivity of entropies, that a mixed entropy, Eq.~(\ref{emx}), applies to a good approximation.
This form can interpolate smoothly between the Shannon and GK entropies by tuning of the factor $\gamma$. To match the actual entropy generated by
sampled frequencies and amplitudes, we have proposed a simple heuristics method of setting $\gamma$ to the inverse width of the spectrum,
suitably defined (e.g., using frequency moments). Interestingly, the mixed entropy is also generated when the conventional ME probability
distribution is sampled stochastically (similar to SAC, but with prior-entropy weighting included), instead of solving for the most probable
spectrum (Sec.~\ref{sec:maxent2}).

Though entropy calculations in previous works had pointed to formal relationships between SAC and the ME method \cite{beach04,bergeron16,ghanem20a,ghanem20b},
they had never been pursued to the level of the full equivalence demonstrated here. Our results also differ in important respects from some of the previous
suggestions, as we discussed in Sec.~\ref{sec:maxent} and address further in Sec.~\ref{sec:pathology}.

The equivalence between SAC with different sampled parametrizations and the maximum-probability ME method with a corresponding entropy is fully
manifested when directly matching the goodness-of-fit of results obtained with the two methods, $\langle \chi^2_{\rm SAC}(\Theta)\rangle = \chi^2_{\rm ME}(\alpha)$,
and we have also developed similar methods to determine the optimal $\alpha$ or $\Theta$ value. Given this equivalence between the two methods, a relevant question
is which one is better in practice. While the ME spectrum probably can be optimized
faster in many cases, the average spectrum also can be sampled very efficiently with the continuous-frequency representations when $N_\omega$ is large
(despite the fact that the computational effort of the sampling algorithm in principle also scales as $N_\omega$). The high efficiency is in part
related to the fact that the variance of the fluctuations relative to the average spectrum decays as $1/N_\omega$. Based on our tests, with the
continuum representations it is also normally easier to extract the best goodness-of-fit, $\chi^2_{\rm min}$, which is needed to fix the value of
$\Theta$ or $\alpha$.

Another aspect to consider is that the entropy is not exactly known in the case of both amplitudes and frequencies sampled, and this, we have
argued, is the best parametrization with SAC. Still, the mixed entropy with the simple $\gamma$ fixing produces similar results---and in general
better results than the conventional Shannon entropy. Overall, it is essentially a matter of taste which method to use for unconstrained spectral
functions and flat default models. We have not investigated any other default models in our work reported here, and it would be interesting, in particular,
to further explore the mixed entropy, Eq.~(\ref{emxw}), with other default models.

We note here that regularizing functionals different from the Shannon entropy have also previously been discussed in the context of the generic
ME method \cite{bryan86} (though not in the QMC context as far as we are aware), including a form like $E_{\rm GK}$. However, those proposed alternative
forms were not derived from stochastic processes. Though $E_{\rm SH}$ can be considered superior for suppressing certain correlations,
it has also been recognized that it does not necessarily produce the best solutions \cite{bryan86}. Our finding that the mixed entropy may be the best
option therefore does not contradict any fundamental notion of the ME framework, though these methods have been completely dominated by the use
of $E_{\rm SH}$.

Constrained SAC can also in some cases be translated directly to the ME framework, e.g., the optimized plain lower frequency bound or
a $\delta$-function edge. Parameters would be optimized by minimizing the goodness of the fit at constant $\alpha$, in direct analogy with
fixed $\Theta$ in SAC. The multi-$\delta$ peak parametrizations in Fig.~\ref{peakfig} can perhaps be converted to the ME formalism by
considering two components of the spectrum with different relative spectral weight, parametrized by the (relative) peak weight $A_0$ (and
corresponding continuum weight $1-A_0$), with the continuum and peak parts mutually constrained in some way analogous to Fig.~\ref{peakfig}.
Instead of $N_p$ and $N_c$ in SAC,
the peak and continuum parts should be given different (optimized) entropy factors $\alpha_p$ and $\alpha_c$. After setting $\alpha_c$ in the
same way as $\Theta$ in SAC, the peak weight and the ratio $\alpha_p/\alpha_c$ would be the adjustable parameters in the quasi-particle
ME scheme.

The monotonic distance constraint used for power-law edges does not appear to have any obvious counterpart in the ME method,
since it effectively introduces a completely different entropy and is not directly tied to any conventional default model. In Sec.~\ref{sec:pdm},
we will discuss how to formulate this constraint, and generalizations of it, as a new type of default model. It is still not obvious how to translate
these concepts to the maximum-probability ME scheme, however.

\subsubsection{Pathological probability distribution}
\label{sec:pathology}

The ME method has been the standard analytical continuation tool for more than 30 years,
since it was adapted \cite{silver90,gubernatis91} from its more general use in statistics
\cite{gull84,bryan90} to extracting spectral functions from QMC data. However, it is apparent from the relationships that we have derived
between SAC and ME method in Sec.~\ref{sec:maxent} that the underlying probability distribution, Eq.~(\ref{pmaxent2}), has a fundamental flaw. While
the maximum-probability distribution, i.e., the spectrum minimizing the functional $\chi^2(S)/2-\alpha E(S)$, is well defined, the fluctuations about
this well-defined optimal spectrum are not. This problem arises because the mean spectrum in the continuum limit is defined as a functional integral over all possible
spectral functions, and when parametrizing the spectrum in some way (typically as a histogram), the entropy is extensive in the number of parameters
(histogram bins) $N_\omega$. Thus, if the spectrum is actually sampled, as it is in some implementations of the ME method \cite{boninsegni96,kora18},
it will approach the default model with increasing $N_\omega$, thus maximizing the entropy and leading to a typically very large $\langle \chi^2(S)\rangle$ value,
regardless of the choice of $\alpha$.

We observed this entropic catastrophe by sampling within the ME method in Sec.~\ref{sec:maxent2}. The same mechanism affects the SAC method as well, but
in that context it was realized early on \cite{sandvik98,beach04} that the sampling temperature $\Theta$ can be adjusted so that a statistically good fit is always
obtained. As discussed above, $\Theta$ must be chosen to achieve balance between the extensive entropy and the intensive goodness-of-fit $\chi^2$ (which
corresponds to the energy in statistical mechanics, even though its form is unusual; see further discussion in \ref{app:statmech}). In Sec.~\ref{sec:entropy2}
we demonstrated that our $\Theta$ criterion, Eq.~(\ref{eq:chi2}), indeed satisfies the expected scaling form $\Theta \sim 1/N_\omega$.

In the absence of an adjustable temperature, the maximum-probability ME spectrum is still well defined, because minimizing the functional only relies
on the explicitly imposed entropic prior and is not affected by the entropy inherent in the entire space of spectral functions. In a sense, the
probability-maximizing ME spectrum is a mean-field solution (as realized by Beach \cite{beach04}, though we partially disagree with some important details
of his conclusions) which is unstable once the fluctuations are taken into account. Hence, it is only the fluctuations about the most probable spectrum that
are problematic, which may not be that serious in practice (except for aspects discussed below in Sec.~\ref{sec:errors}) but is still of interest for
properly understanding the method and its relationship to SAC.

The original ME method \cite{gull84} has been extremely successful in many areas of statistics, and the pathological probability distribution
has not been noted before, as far as we are aware. The root cause of the problem in the context of spectral functions, i.e., the extensive entropy, most
likely does not arise in many other contexts where ME methods are applied. It arises here because of the large number of degrees of freedom $N_\omega$ of
the spectrum---infinite in the limit of continuous curves. In other applications of the ME method, the number of degrees of freedom of the solution would
typically be finite, and with no need to approach a continuum there would be no entropic catastrophe. Thus, there is nothing wrong with the ME method
per se, but the conditions under which it is applied to spectral functions lead to a pathological distribution.

As we have discussed (in detail in \ref{app:lowtheta}), in analytic continuation the problem is further exacerbated by the very small number $N_{\rm para}$
of effectively independent fitting parameters that a positive definite spectrum can realize, given the noisy and highly correlated QMC data points. The fact
that the number of effective parameters is very small has of course been recognized before, e.g., Jarrell and Gubernatis \cite{jarrell96} discuss a different
but in spirit similar concept of a number of ``good measurements'', which is determined using the eigenvalues of the covariance matrix and which we expect
to be close to $N_{\rm para}$.

The underdetermination of the spectrum is the well known fact that numerical analytic continuation is an ``ill posed'' problem. However, it should be noted that
the formally divergent fluctuations of the probability distribution Eq.~(\ref{pme}) in the limit of a continuous spectral function is a feature of the ME method,
not of the analytic continuation problem inherently (though the ill-posedness is of course inherent). It has been regarded as essentially a matter of taste whether
to use the maximum-probability spectrum or the average spectrum defined by the distribution \cite{jarrell96}. What we we have shown here is that the non-trivial
maximum-probability spectrum exists in the continuum limit, but the average spectrum flows toward the imposed defaul model.

Given that the maximum-probability ME solution is still well defined (and equivalent to the SAC average spectrum in the large-$N_\omega$ limit) even
in the continuum limit, one may ask why the pathologies of the probability distribution even matter. One can argue that the ME method with the standard
probability distribution Eq.~(\ref{pme}) is perfectly fine also for sampling (as in many cases it is \cite{kora18}), as long as the number of sampling
degrees of freedom is not taken too large, i.e., a histogram with, say, $N_\omega=100$ bins (and also a good default model will help in this regard).
Then the distribution can still be sampled, if so desired, with an acceptable $\langle \chi^2\rangle$ value, and with or without sampling, an error analysis
based on fluctuations (discussed further below in Sec.~\ref{sec:errors}) can be performed. However, the results will to some extent depend on $N_\omega$, and
it also seems fundamentally unsatisfying to work with a distribution that does not have the desired continuum limit. It is in this sense that we categorize
the ME probability distribution as pathological. 

\subsubsection{Fluctuations and statistical errors}
\label{sec:errors}

In practice, the main problem with the ME probability distribution is that it is not possible to use it for computing meaningful
statistical errors of the spectrum when the fluctuations are divergent.
Error analysis has been carried out \cite{jarrell96,linden96} by formally analyzing fluctuations about the maximum-probability spectrum. It was noted \cite{jarrell96},
that the correlation function $\langle \delta A_i \delta A_j\rangle$ of the deviations of the amplitudes (in a histogram) from their mean values is proportional
to elements of an inverse matrix $\Gamma^{-1}_{ij}$. The matrix $\Gamma$ (the definition of which we do not need here) is expected to
have some small eigenvalues, i.e., there are directions in amplitude space where the probability density is very flat, and it was pointed out that this could
lead to large fluctuations \cite{jarrell96}.

For a finite discretization, the fluctuations defined by $\Gamma$ are still well defined and can formally be used to compute error bars, as was done in examples
in Ref.~\cite{jarrell96}. It was also correctly pointed out that well-defined (finite) error bars can only be computed in this way for the spectral weight
integrated over some frequency window. What we have emphasized here (and which was also noted in Refs.~\cite{sandvik16,ghanem20a}) is that the
entropy of the spectrum increases with increasing density of grid points, to the extent that the mean spectrum flows toward the default model when
$N_\omega \to \infty$. A leading-order analysis of the fluctuations about the most probable spectrum in an essentially completely flat space then is
insufficient, and it is doubtful whether results for finite discretization have any well-defined statistical meaning, since the fluctuations about
the optimal spectrum also in finite frequency windows will grow with $N_\omega$.

In SAC, the situation is the opposite, in the sense that the factor $N_\omega$ in the exponent of the probability distribution Eq.~(\ref{psac}) causes diminishing
fluctuations about the average spectrum when $N_\omega \to \infty$ (and, as we summarized above, this average spectrum becomes the same as the most
probably ME spectrum). Thus, the method of directly using spectral weight fluctuations during the sampling process to estimate statistical errors
\cite{beach04} (as has also been done in the sampling version of the ME method \cite{kora18}) is also flawed, as those statistical errors depend on
$N_\omega$ and vanish when $N_\omega \to \infty$ (and in the ME case the sampled average also deteriorates).

In both SAC and ME calculations, the purely statistical errors propagated to the spectrum from the noise in $\bar G(\tau)$ can be calculated by bootstrapping.
The analytic-continuation procedures are then carried out also for some number (at least ten or more) of different bootstrap realizations of $\bar G(\tau)$,
to compute the statistical fluctuations about the spectrum obtained with the original full data set. However, in our experience, such statistical errors are
often much smaller than remaining errors originating from various sources of bias in the analytic continuation process. Thus, it remains difficult to estimate
all the errors in some absolute way, and it is important to consider the evolution of extracted quantities with the data quality. Benchmark studies, based on
exact model solutions or synthetic data are very useful, as we have seen in many examples in this work (e.g., the convergence tests illustrated in
Fig.~\ref{broadened-1}). Our main focus here has been on edge features, and we have seen that they are amenable to quantitative analysis. Statistical errors
and bias related to completely unresolvable (given typical QMC data quality) features above the edge, e.g., multiple sharp peaks, can likely never be estimated
properly in the absence of further information.

\subsection{Future prospects}
\label{sec:conc3}

We foresee a wealth of applications of the methods discussed here to quantum-many body systems accessible to QMC simulations, including  not only
spin Hamiltonians and other quantum lattice models, but also dynamical mean-field calculations \cite{georges96,song20}. Analytic continuation of
simulation data is also a key aspect of lattice field theory that is growing in importance \cite{ding18,aarts21,horak21}, and the improved SAC and ME techniques
may find applications there as well. The methods may also potentially  be useful for analytic continuation of imaginary-time DMRG data \cite{linden20}.
We do not discuss specific applications here, but outline possible future technical developments.

\subsubsection{Default peak structures and profile default models}
\label{sec:pdm}

\begin{figure}[t]
\centering
\includegraphics[width=75mm]{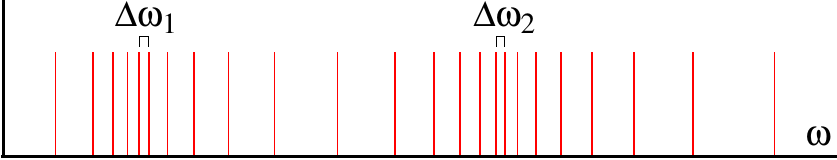}
\caption{Example of a DPS constrained to have two peaks (density maxima), with optional minimum distances controlled by
$\Delta\omega_1$ and $\Delta\omega_2$.}
\label{dps}
\end{figure}

The distance-monotonic parametrization, Fig.~\ref{fig:spec}(e), can be regarded as a special case of a more general {\it default peak structure}
(DPS), where density maxima are built in by requiring a certain number $n$ of local minima in the separation $d_i = \omega_{i+1}-\omega_i$ between the
$\delta$-functions. In the case in Fig.~\ref{fig:spec}(e), there is only one minimum, at the left edge of the sampled spectrum, but in general the minima
can be located anywhere. An example with $n=2$ is shown in Fig.~\ref{dps}. This type of DPS will clearly entropically favor $n$ sharp peaks when the
number of $\delta$-functions is large, and, by a generalization of the edge regulator we discussed in Sec.~\ref{sec:triangle}, the asymptotic inverse-square
divergences when $N_\omega \to \infty$ can be quenched by setting bounds $\Delta\omega_i$ on the minimum separations. These bounds can either be optimized,
which would be difficult and time consuming if
different values $\Delta\omega_i$ are considered for several peaks, or they can just be imposed without optimization to avoid excessively narrow peaks. The locations
of the minima should be allowed to migrate during the sampling process, keeping $n$ fixed (otherwise entropy would lead to large $n$---a proliferation of
peaks). Thus, one of the peaks may (if the imaginary-time data so dictate) migrate to the lower edge of the spectrum and form the kind of one-sided
singularity that we have investigated in detail.

We have already tried $n=2$ (Fig.~\ref{dps}) for a synthetic spectrum with two peaks. As in the case of the edge
singularity, the sampling adapts to the correct bounds of the spectrum and the peaks appear in the correct locations. We did not optimize $\Delta\omega_1$
and $\Delta\omega_2$, and the peaks were much sharper than the Gaussians used in our tests. Spectra with several sharp peaks are also
common and have been the subject of recent methods specifically adapted to resolving a number of such peaks \cite{linden19}. The DPS approach appears to be
competitive for such spectra. When the number of peaks is unknown, $n$ can be gradually increased until a satisfactory $\langle \chi^2\rangle$ value
can be obtained, thus signaling that the data at the present error level does not support additional peaks. Compared to the method with a sharp edge and
generic continuum, Sec.~\ref{sec:monomixed}, the DPS with $n>1$ should be better for reproducing multiple sharp peaks, while the generic continuum should be
better when the spectrum is rather featureless beyond the edge (or a single peak somewhere in the middle of the spectrum).

We envision a number of further developments of the DPS parametrization. For instance, it could be combined with amplitude sampling to better model
the behavior between the peaks---perhaps also to better morph the peaks into their correct shapes, though that may be more difficult because of the large
entropic pressures toward sharp peaks, unless the $\Delta\omega$ regulators are imposed as well. In this context, it may be useful to adopt the method of not
updating the amplitudes but only swapping them, $A_i,A_j \leftrightarrow A_j,A_i$, after initiating a set of varying amplitudes \cite{qin17}, e.g.,
$A_i \propto i$ or $A_i \propto {\rm constant} + i$. Then, to improve the fit in case the DPS by itself is sub-optimal, the sampling can distribute the available
amplitudes optimally among the frequencies. This method of including
some amplitude fluctuations, but not fully sampling the space of all possible amplitudes, may be preferable here because there are less entropic pressures
of the amplitudes themselves favoring sharper peaks (as in our tests, e.g., in Fig.~\ref{sw1}). Of course the DPS can also in principle be combined with
the amplitudes adapted for specific exponents of the singularities, as in Sec.~\ref{sec:monop}, though this approach may become too complicated if
there are several peaks to optimize.

We emphasize that there is an important difference between default models used in the ME method (and often also
within SAC \cite{beach04,fuchs10,ghanem20a,ghanem20b})
and constrained SAC parametrizations, specifically the DPS proposed here but also other cases that we have discussed (such as the $\delta$-peak).
A conventional default model
is locked to the frequency axis, i.e., it dictates where specific structures are favored.
In contrast, the DPS is self-adapting, in that it does not correspond to a well-defined
spectrum in the absence of data (in which case it completely spreads out over the entire frequency space) but finds the frequency bounds inherent in the imaginary-time
data once $\chi^2$ is used in the sampling. In the case of $n=1$ with the density maximum at the edge [Fig.~\ref{fig:spec}(e)], we found this self-adaptation
of the bounds to be remarkably precise, and we expect that to be the case also for $n>1$.

There are also other potential ways to achieve self-adaptation of the bounds with generalized default models.
A DPS can be regarded as an extreme case of hard-core interactions between the $\delta$-function ``particles'' (in this case a rather odd interaction that
enforces $n$ density maxima). Other, ``softer'' particle-particle potentials could be imposed instead---imagine the spectral amplitudes residing on a string
of particles connected by some kind of nonlinear springs with tunable properties. Such potentials would change the entropic pressures and lead to different
native spectral-weight distributions. The native shape can be regarded as a {\it profile default model} (PDM), and it should be noted that a PDM does not
define the location of the spectrum in frequency space. However, unlike the DPS, the shape can be defined also in the absence of data if the interactions
overall form a finite equilibrium shape, with bounds that can be regulated by a single optimized parameter (the overall mean distance between the particles).
Once $\chi^2$ is included in the sampling, the shape can be further modified from the native shape, similar to how the optimal spectrum in the ME method
is distorted relative to the default model. This overall approach of introducing potentials bears some resemblance to the way of combining $\chi^2$ fitting
with a gradient-squared term imposed on the spectrum \cite{white89}, but in continuous frequency space instead of the grid and with the possibility of more
general potentials between the particles, as well as the fact that the locations $\omega_i$ (and possible also the amplitudes $A_i$) are sampled,
not just statically optimized.

\subsubsection{Machine learning}

Machine learning (ML) methods are making inroads on many fronts in quantum many-body physics, including applications to the analytic continuation
problem \cite{arsenault17,fournier20,yoon18,song20,huang22,yao21,zhang22}. Without going into details, the attempts so far typically use a large data set of known
spectral functions $S(\omega)$ and their associated imaginary-time correlation function $\bar G(\tau)$. The neural network is trained to deliver $S(\omega)$ given
$\bar G(\tau)$ (with noise if appropriate for the intended application). These methods show some promise, e.g., it was claimed that they can resolve 
a Gaussian peak better than the ME method \cite{fournier20,zhang22}, and examples of spectra with sharp-edged features have also been presented \cite{yao21}.
However, mostly what has been produced so far can also be achieved with standard ME or SAC methods, and much better results can likely be produced with
the improved methods developed here. 

There are reasons to be skeptical about ML as a universal unbiased analytic continuation tool. While it is certainly possible to consider some
class of spectral functions, e.g., sums of small numbers of Gaussians \cite{zhang22}, more generally the possibilities are endless and it is
difficult to imagine constructing a complete training set. A specific problem with analytic continuation is that a given set of noisy imaginary-time
data would normally be consistent with a vast number of different spectral functions, as we have seen plentifully in the examples studied here.
While regularization can be built in at the stage of training (i.e., only including ``regular'' spectral functions), still many spectra regarded
as regular would be equally consistent with the data, and some additional criterion has to be applied to select the ``best'' spectrum. It is
not clear why this would be any better than using SAC or ME methods (or even function fitting with a relatively small number of parameters),
though of course one could decide to categorize different types of training spectra, e.g., with sharp edge features, and then select which set
to use for a given application, in analogy with selecting constraints for SAC.

We propose that ML could be used in a different way, in combination with SAC or ME results. The basic idea is that input to the neural network can
go beyond imaginary-time data $\bar G(\tau)$. As an example, looking at all our results in Fig.~\ref{sw1} (and for an even wider range of
$\langle \chi^2\rangle$ values), is it possible to determine which of the many spectral functions is the closest to the correct result? Thus,
we envision that the neural network is not trained just with a set of spectra $S(\omega)$ and corresponding $G(\tau)$ data, but that a whole series
of spectra, produced for a range of $\Theta$ ($\alpha$) values by SAC (ME) runs with different parametrizations (different entropy forms). Importantly,
results for $\langle \chi^2(\Theta)\rangle$ or $\chi^2(\alpha)$ could also be included in the training and decision making, as these functions may
contain information that could also improve the outcome, i.e., resulting in the best spectrum among those with acceptable $\langle \chi^2\rangle$ values.

The imaginary-time data set of course implicitly already contains all available information and, strictly speaking, a whole set of spectra produced under
different conditions does not contain any new information beyond $\bar G(\tau)$ (and its covariance matrix).
Coming in the form of a rather small set of floating-point numbers, the information on $S(\omega)$ is contained in $\bar G(\tau)$ in a rather subtle manner,
however, in the details of these numbers that are not ideal for ML to recognize. With pre-processing by SAC or ME procedures, information ``expanded'' from
the original $\bar G(\tau)$ data sets is produced. This new representation of the same information should be easier for ML
handle. The training would have to be done at different levels of noise added to the $G(\tau)$ data,
and later used at noise level corresponding to given QMC input data. Here a complication is the fact that the QMC data are correlated, but the majority
of these correlations can likely be modeled just by an autocorrelation time, as we do for synthetic data, Eq.~(\ref{corrnoise}).

Going beyond basic unrestricted parametrizations, a series of training spectra can further be produced with various constraints imposed, and, along
the same lines as above, the ML process would determine which constraint is the most appropriate. Characteristic ways in which output spectra
depend on the parameters in SAC would very likely be helpful for discriminating between various spectral features used
in constrained sampling. Thus, for a single spectral function to be included in a training set, SAC runs would be carried out for several
different parametrizations and values of constraining parameters. After training with spectra with different types of edge features (or other
prominent features), the ML process may be able to identify the proper constraint and the optimal value of any associated parameter $p$ based
on just just curves $\langle \chi^2(p)\rangle$ generated by SAC for different types of constraints. These ideas could be initially explored,
e.g., by testing whether ML could distinguish between a power-law edge singularity and a sharp quasi-particle peak. This problem is important
in the context of fractionalization of excitation, where cross-overs between these two types of spectral edges can be expected in realistic
situations \cite{shao17,zhou21}.

As an alternative to full-fledged ML approaches, simpler validation methods developed in statistical learning theoory \cite{mehta21} may be
useful for optimizing parameters and for quantifying the suitability of various parametrizations and constraints. Here the original data set
is divided into two groups, and in the present context the SAC would be run with one of these subsets and the other subset used to statistically
validate the results. A simple form of such a cross-validator has already been tested in the SAC context, with promising results \cite{efremkin21}
that motivate further explorations along these lines.

\subsubsection{Negative sampling weights}
\label{sec:prospnegative}

The imposed positive definiteness is responsible for the anomalies appearing in the spectrum when sampling at very low $\Theta$
(corresponding to $\chi^2$ minimization), i.e., a small number of spikes that would become true $\delta$-functions if the $\Theta \to 0$
limit could be realized in practice (\ref{app:lowtheta}). The entropy at $\Theta > 0$ (specifically for $\Theta$ in a range satisfying our
statistical criterion) is what removes these anomalies and leads to a dense spectrum. We have further demonstrated this point by relaxing the
constraint, introducing a small fraction of negative spectral weight in the sampling (\ref{app:low2}). Even a very small negative relative
weight (we tested $10^{-4}$ and $10^{-3}$) leads to a dramatic lowering (by orders of magnitude) of the $\Theta$ value at which the spike anomalies appear.
Thus, negative spectral weight is another mechanism that can supply entropy to the system and remove anomalies.

Our criterion for fixing $\Theta$, Eq.~(\ref{eq:chi2}), is intended to remove the anomalies by thermal fluctuations. In the presence of a non-thermal
smoothing mechanism---negative spectral weight---not only can $\Theta$ be lowered further but $\langle \chi^2\rangle$ remains low and can even be
below the lowest acceptable value attainable with only positive weight without causing overfitting (Fig.~\ref{fig:nega}). Thus, the negative weights
appear to be a mechanism for taking into account those noise features in $\bar G(\tau)$ that cause anomalies. The negative weight is of course itself
an unphysical feature, but at a small fraction its presence is inconsecuential and can be neglected. Though we have not yet constructed a suitable
criterion for fixing both the fraction of negative spectral weight and $\Theta$, it is possible that the resolution of the SAC method can be further
improved in this way.

In particular, the resolution may be improved at high physical temperatures $T$ (small $\beta=1/T$), where the imaginary-time correlations are available only
in a small range $\tau \in [0,\beta/2]$. The effective number of parameters is then typically very small, often $N_{\rm para}=4$, i.e., two spikes appear in the
$\Theta \to 0$ spectrum (as in the case of the $L=16$ Heisenberg chain at $\beta=2$, which we have used in many tests). The limited information is still
sufficient to resolve the rough distribution of spectral weight, but we have also seen (Fig.~\ref{sw1}) that the details of the profile depend
significantly on the parametrization used. Some of these details are very important, e.g., the $\omega \to 0$ limit dictates the spin-lattice relaxation
rate in NMR experiments (see Ref.~\cite{shu18} for an SAC application in this context). If the number of effective parameters can be increased by just
one step by sampling with some negative spectral weight, e.g., from $N_{\rm para}=4$ to $N_{\rm para}=6$, as is in fact the case in the test reported in
\ref{app:low2}, the reliability of the spectrum for small $\omega$ may be improved significantly. The small fraction of unphysical
negative weight typically appears at high frequencies (at least for gapless spectra) and would not directly be included in the final result.

For spectra where $N_{\rm para}$ is larger, the impact of negative weights is likely in general smaller. In addition to the tests reported in Fig.~\ref{fig:nega},
we have investigated the impact of negative weights with the synthetic spectrum in Fig.~\ref{syntcomp}. In this case the number of low-$\Theta$ spikes is five
with all-positive amplitudes and remains the same when a small fraction of negative spectral weight is included. We plan more extensive studies of SAC with
negative amplitudes.

\section*{Acknowledgments}
  We would like to thank Kevin Beach, Khaldoon Ghanem, and Erik Koch for their detailed comments on the manuscript.
  We also thank Adrian Feiguin for discussions about the time dependent DMRG method and Markus Holzmann for
  pointing out the cross-validation method. H.S. was supported by the
  National Natural Science Foundation of China under Grants No.~12122502, No.~11904024, and No.~11734002, and by National Key 
  Projects for Research and Development of China under Grant No.~2021YFA1400400.
  A.W.S. was supported by the the Simons Foundation under Simons Investigator Award No.~511064. Most of the computations
  were performed using the Shared Computing Cluster administered by Boston University’s Research Computing Services. 

\appendix

\section{Asymptotic best-fit spectrum}
\label{app:lowtheta}

Here we discuss the $\Theta \to 0$ limit of SAC in some more detail. This limit of course corresponds to pure $\chi^2$ fitting, and should,
in principle, be independent of how $\chi^2$ is minimized. We will demonstrate that the spectrum consisting of a few narrow peaks at low $\Theta$,
observed in Figs.~\ref{sw1} and \ref{sw2}, is directly related to the positive definiteness enforced in SAC. As an example, we here discuss
results for the $L=500$ Heisenberg chain at $\Theta$ much lower than in Fig.~\ref{sw2}, where the appearance of a small number of $\delta$-functions
becomes clear. 

\begin{figure}[t]
\centering
\includegraphics[width=75mm]{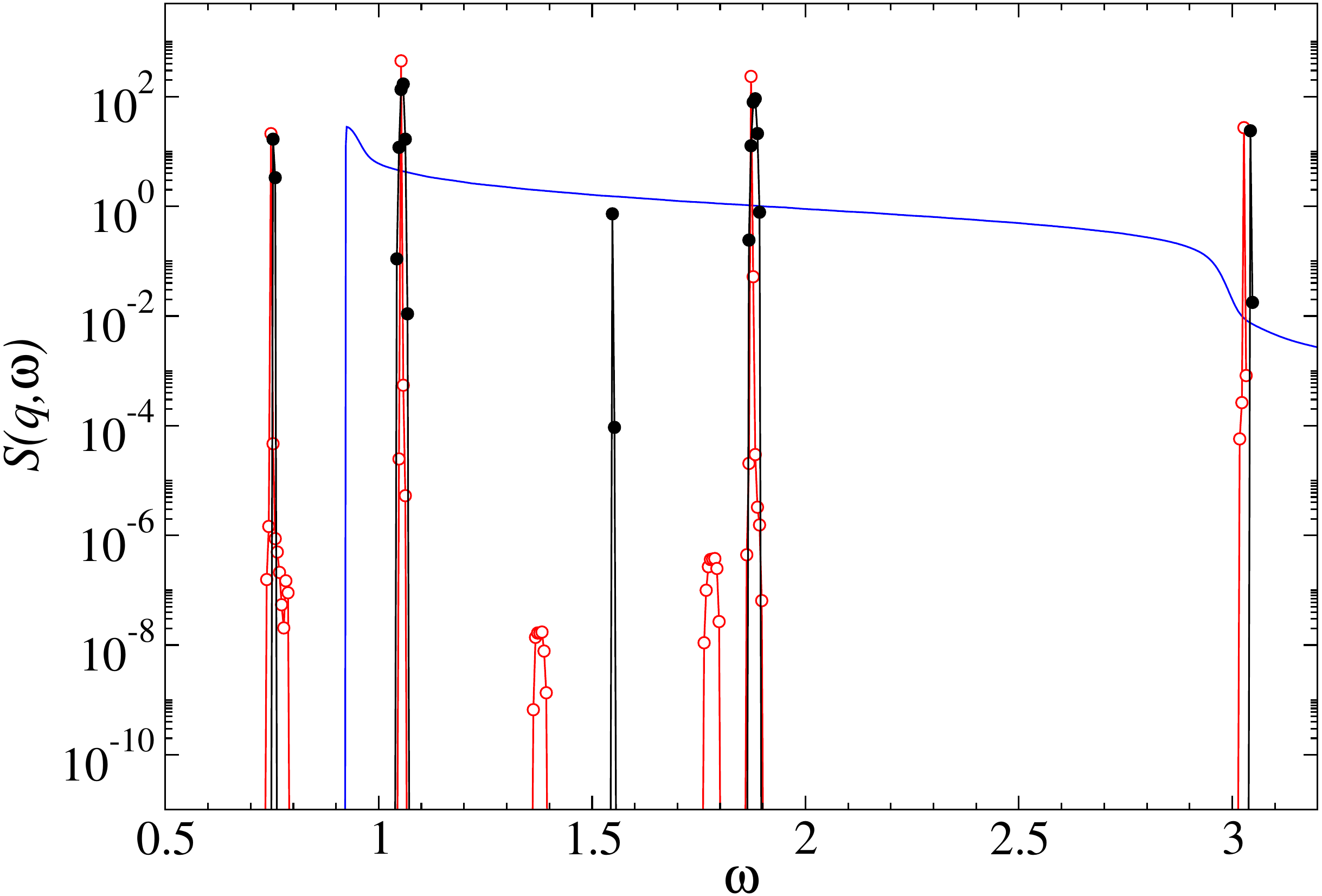}
\caption{Spectra with the lowest $\chi^2$ values obtained in the same SAC Heisenberg $S(q,\omega)$ runs as in Fig.~\ref{sw2}(b)
  and \ref{sw2}(c) at much lower $\Theta$, graphed on a logarithmic scale.
  The result of only frequency sampling (black solid circles) has $\chi^2/N_\tau=0.619093$, while $\chi^2/N_\tau=0.619035$
when also amplitude moves were employed (open red circles). The BA spectrum \cite{caux05a,cauxdata} is shown as the blue curve; note that this is a continuous
(broadened) representation, while the exact finite-size spectrum consists of a number  of discrete $\delta$-functions (likely around 100 based on
a rough extrapolation from results in Ref.~\cite{wang19}).}
\label{fig:final}
\end{figure}

Figure \ref{fig:final} shows the best converged results at the lowest $\Theta$ reached in the annealing runs with continuous frequency sampling
with the same data as in Fig.~\ref{sw2}(b) (fixed equal amplitudes) and \ref{sw2}(c) (including amplitude updates). The four dominant peaks contain
$99.9\%$ of the spectral weight in the case of the fixed amplitudes and very close to $100\%$ when the amplitudes are sampled as well. With only frequency
sampling, there is a small fifth peak at $\omega \approx 1.55$, which accounts for the remaining $0.1\%$ of the weight, while the amplitude sampling instead
produced two even smaller peaks at $\omega \approx 1.4$ and $\omega \approx 1.9$. Here it should be noted that the $0.1\%$ peak in the fixed-amplitude
representation corresponds exactly to one unit of spectral weight; a single $\delta$-function out of the 1000 sampled. If, ultimately, there indeed is
some smaller spectral weight at $\omega \approx 1.5$ it cannot be correctly reproduced within this fixed-amplitude parametrization
unless $N_\omega$ is increased further. With the amplitude updates
there is no such limitation, but given that the two peaks on either side of $\omega=1.5$ in this case are very small indeed, they may eventually also
shrink away in an even slower annealing process to lower $\Theta$. Most likely, the four dominant peaks would then also approach $\delta$-functions.

The low-$\Theta$ results in Fig.~\ref{fig:final} show little resemblance to the BA result, apart from the fact that the dominant peak is rather close to the
edge singularity---though it cannot even be used as a good lower spectral edge because there is also a significant peak much below the true edge. This effect of
overfitting leading to sharp peaks of little use was observed in early attempts of analytic continuation by $\chi^2$ minimization
\cite{schuttler85a,schuttler85b}, and the
results in Fig.~\ref{fig:final} solidify the claim \cite{sandvik98} that the limiting form is a small set of $\delta$-functions. Though the behavior may appear
bizarre, it can be understood as a consequence of the enforced positive definiteness of the spectrum.

From Eq.~(\ref{contrel1}) it is easy to see that a semi-positive definite spectrum (i.e., a sum or integral of $\delta$-functions with no negative coefficients)
leads to a monotonically decaying $G(\tau)$ whose $\tau$ derivatives are likewise all monotonically decaying. The statistically sampled $\bar G(\tau)$ will
to some extent violate these constraints, and for a perfect fit, i.e., $\chi^2=0$, some  negative spectral weight would have to appear. With negative weights
disallowed, the $\chi^2$ minimization does, in a sense, the ``next best thing'' and produces zeroes to the maximum extent possible.

A more quantitative way to understand a small number of $\delta$-functions is to imagine fitting $\bar G(\tau)$ at $T=0$ to a number $n$ of exponentials
$A_i {\rm e}^{-\omega_i\tau}$ (or hyperbolic cosines at $T>0$) with all positive coefficients, corresponding to a spectrum with the same number of $\delta$-functions
with positive definite amplitudes $A_i$. Increasing $n$ starting at $n=1$, initially each added $\delta$-function will improve the best-fit $\chi^2$ value. However,
after some $n=p$ (larger $p$ for better data quality), the optimal non-negative coefficient $A_{p+1}$ for the next $\delta$-function will be zero.
To improve the goodness of the fit $\chi^2(p+1)$ over $\chi^2(p)$, a negative coefficient $A_{p+1}$ will be required.

Once any amount of negative weight is required to further lower $\chi^2$, no amount of additional positive weight can be substituted for it
with $\chi^2=\chi^2(p+1)$ maintained [or even to reach a value below $\chi^2(p)$].
For instance, imagine that the first negative weight is not optimized but is added with
infinitesimal weight, $A_{p+1}=-\epsilon$ (and its location $\omega_{p+1}$ is optimized along with re-optimization of the locations $\omega_i$ and still
positive coefficients $A_i$ with $i \le n$). With the infinitesimal negative weight present, $\chi^2(p+1)$ can be infinitesimally below $\chi(p)$. The
negative weight contributes positively to the derivatives of $G(\tau)$, which cannot be accomplished by any added positive contributions to the spectrum,
and it is these effects of negative weight that enable a closer fit to the noisy data (as explicitly demonstrated in \ref{app:low2}).

The above reasoning strongly suggests that the optimal positive definite spectrum is a number $p$ (in practice a small number) of $\delta$-functions.
Going beyond the infinitesimal negative coefficient $A_{p+1}$, fully optimizing $A_{p+1}<0$ could in principle also cause some of the previously frequencies
$\omega_{\le i}$ and weights $A_{i\le p}$ to change completely, even with some of those amplitudes becoming negative as well. In other words, allowing a significant
fraction of negative spectral weight may completely change the best-fit spectrum. To eventually achieve $\chi^2=0$, a large number of both positive and
negative amplitudes should be expected. With improving QMC data quality, the number $p$ should increase, so that a continuous positive-definite spectrum
can apply when $G(\tau)$ is exact.

In practice, a number of Gaussian-like peaks appear at the end of a long but not fully converged SAC annealing procedure with enforced positive definiteness,
as in the rightmost panels of Figs.~\ref{sw1} and \ref{sw2}. The number of peaks with significant weight (noting that there may be some small scattered
peaks present unless the annealing is very slow) should be equal to the number $N_\delta$ (called $p$ above) of $\delta$-functions that form eventually
when $\Theta \to 0$, as discussed above.

We can take this reasoning further and propose that each significant peak at low $\Theta$ corresponds to two effective fitting parameters (frequency
and amplitude, while the peak widths ideally should vanish). We then expect $\chi^2_{\rm min} \approx N_{\rm dof} = N_\tau-2N_\delta$. This prediction implies
$\chi^2_{\rm min} \approx 8-2\times 2=4$ in Fig.~\ref{sw1} and $\chi^2_{\rm min} \approx 33-2\times 4=25$ in Fig.~\ref{sw2}, corresponding to $0.50$ and $0.76$,
respectively, after normalizing by $1/N_\tau$. Considering the variance $2N_{\rm dof}$ of the distribution, both the actual $\chi^2_{\rm min}/N_\tau$ values
reached, $0.48$ and $0.62$ for $L=16$ and $L=500$, respectively, are consistent with the statistical expectations.

These arguments can also serve as a criterion for statistical soundness of the SAC procedures: In our experience, excessively large $\chi^2_{\rm min}$ values,
according to the above criteria based on $N_\tau$ and $N_\delta$, can always be traced back to a poorly converged covariance matrix, which may be cured by
reducing the number of $\tau$ points or improving the data by performing additional QMC runs. 

We stress that detailed studies of slow annealing to very small $\Theta$ values would not normally be necessary for determining a sufficiently good
$\chi^2_{\rm min}$ value. The difference between the lowest $\langle \chi^2\rangle$ in Fig.~\ref{sw2} (or the lowest $\chi^2$ reached, which we
also keep track of) is only very marginally above the $\chi^2_{\rm min}$ values mentioned in the caption of Fig.~\ref{fig:final}. Given that the
sampled spectrum is not very sensitive to the exact value of $\Theta$, e.g., with $a=1$ or $a=0.5$ in the criterion Eq.~(\ref{eq:chi2}),
a slightly too high $\chi_{\rm min}$ estimate is also inconsequential. With typical $\bar G(\tau)$ noise levels, a good $\chi^2_{\rm min}$ can typically
be found by annealing for a few minutes.

Similar few-peak structures have been studied previously using non-negative least squares fitting \cite{Ghanemthesis}, where it was also proposed
that averaging these best-fit results over many instances of the QMC data with noise added might be a way to regularize the spectrum. Some results
were presented that indeed reproduced some aspects of synthetic spectra, though not as well as other methods. Noise may indeed lead to effects similar
to sampling at non-zero $\Theta$, but the method lacks the direct translation into statistical-mechanics language. Conventional Monte Carlo sampling
is also likely much more efficient.

\section{Sampling with negative spectral weight}
\label{app:low2}

As we argued above in \ref{app:lowtheta}, when $\chi^2$ has been minimized by a spectrum consisting of a number $N_\delta$ of $\delta$-functions,
a following $\delta$-function added with negative weight can still reduce $\chi^2$. Here we will relax the condition of positive definiteness slightly
and present SAC results showing dramatic effects of increased entropy when even a very small fraction of negative spectral weight is included
in the sampling process.

\begin{figure*}[t]
\centering
\includegraphics[width=120mm]{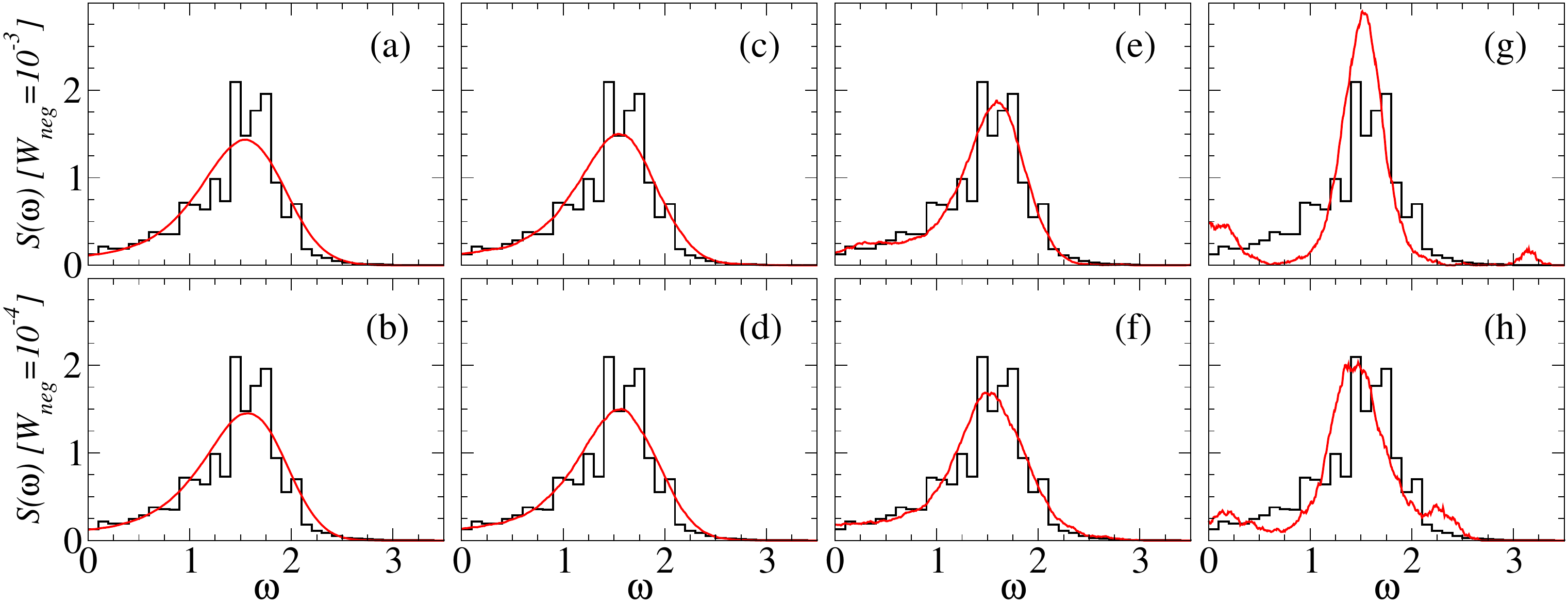}
\caption{SAC spectra obtained with $N_\omega=1000$, subdivided into 990 positive and 10 negative amplitudes (with equal amplitudes
within the group). The total negative contribution, defined in Eq.~(\ref{wnegdef}), is $10^{-3}$ for the panels in the upper row and $10^{-4}$
for those in the lower row. The sampling temperatures are $\Theta=10^{-3}$ in (a) and (b), $\Theta=10^{-4}$ in (c) and (d), $\Theta=10^{-5}$
in (e) and (f), and $\Theta=10^{-6}$ in (g) and (h). For comparison, the histogram from exact diagonalization is shown in black in all panels.}
\label{fig:nega}
\end{figure*}

\begin{figure*}
\centering
\includegraphics[width=100mm]{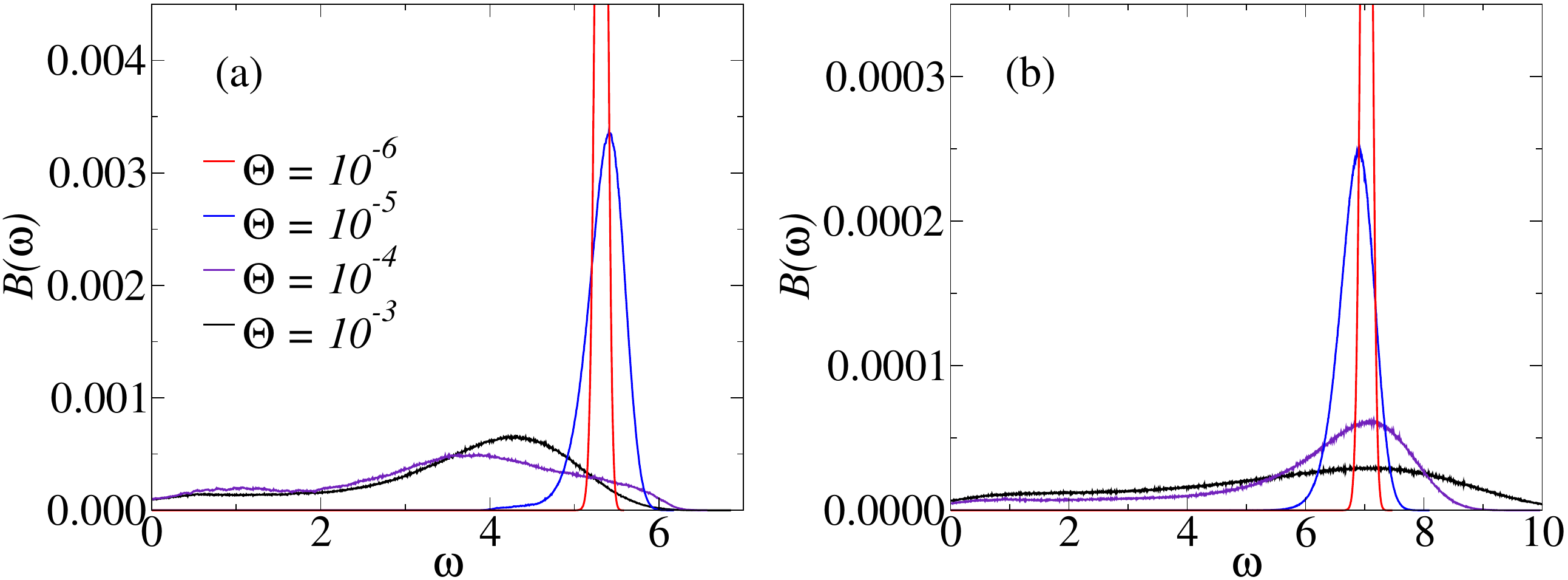}
\caption{The negative spectral weight distribution in the spectral functions of Fig.~\ref{fig:nega}; with negative fraction $W_{\rm neg}=10^{-3}$ in (a) and
$W_{\rm neg}=10^{-4}$ in (b). Here $B(\omega)$ is the mean spectral density obtained from the amplitudes of the $\delta$-functions $i=1$ through $i=M$, which are
added to the total spectrum with negative weight in Eq.~(\ref{awithnegative}).}
\label{fig:wnega}
\end{figure*}

To maintain normalization in the presence of negative spectral weight, the positive contributions of course have to increase. To achieve a perfect
fit, $\chi_{\rm min}=0$, a large number of $\delta$-functions with significant negative weight would have to be added. Even if we allow only a small amount
of negative spectral weight, so that $\chi^2_{\rm min}$ only changes marginally from its value for a positive definite spectrum, the entropy at low sampling
temperature $\Theta$ should increase substantially, because combinations of positive and negative contributions can be much more easily re-arranged with
little net effect on $G(\tau)$ than can purely positive contributions. The question addressed here is whether the additional entropy can counteract the tendency of a
spectrum to split into ``artificial'' narrow peaks, thereby perhaps allowing a better resolution of the shape of the spectrum by reducing $\Theta$ below the
point where otherwise precursor distortions start to appear.

We introduce mixed weights by defining the spectrum in continuous frequency as a sum of $N_\omega$ $\delta$-functions as in Fig.~\ref{fig:spec}(b)
or \ref{fig:spec}(c), with the first $M$ ones contributing with minus signs;
\begin{equation}
A(\omega) = - \sum_{i=1}^{M}A_i\delta(\omega-\omega_i) + \sum_{i=M+1}^{N_\omega}A_i\delta(\omega-\omega_i),
\label{awithnegative}
\end{equation}
where all $A_i\ge 0$ and this spectral function is normalized to unity as before. We define the ratio of negative to positive spectral weight as
\begin{equation}
W_{\rm neg} = \frac{\sum_{i=1}^{M}A_i}{\sum_{i=M+1}^{N_\omega}A_i} = \frac{\sum_{i=1}^{M}A_i}{1+\sum_{i=1}^MA_i}. 
\label{wnegdef}
\end{equation}
In order to achieve $\chi^2_{\rm min}=0$ (which is not our goal here), we expect that $W_{\rm neg}$ close to $1$ (i.e., almost equal positive and negative
weight) would be required, in which case sampling the spectrum with a good outcome would not be feasible. We will here only consider very small
ratios, $W_{\rm neg}=10^{-3}$ and $W_{\rm neg}=10^{-4}$, which, as we will see, still can have dramatic effects on the $\Theta \to 0$ behavior.

As an example, we consider the same $L=16$ Heisenberg chain as in Fig.~\ref{sw1}, using 1000 $\delta$-functions whereof 10 have negative
amplitudes. The positive and negative weights are divided equally over their respective groups.
The results do not change appreciably when 20 or 40 negative $\delta$-functions are used instead of 10. Fig.~\ref{fig:nega} shows average spectra
collected at four low sampling temperatures $\Theta=10^{-3}$, $10^{-4}$, $10^{-5}$, and $10^{-6}$, in each case for both $W_{\rm neg}=10^{-3}$ and
$W_{\rm neg}=10^{-4}$. For comparison, in the rightmost panel of Fig.~\ref{sw1}(b), the two-peak structure in the case of all positive amplitudes
has formed clearly when $\langle \chi^2\rangle/N_\tau = 0.48$, which corresponds to $\Theta \approx 2\times 10^{-5}$. The peak starts to split already at
$\Theta \approx 2\times 10^{-4}$. In contrast, in Fig.~\ref{fig:nega} there is still an intact single broad peak even at $\Theta= 10^{-5}$, and
for both $W_{\rm neg}=10^{-3}$ and $W_{\rm neg}=10^{-4}$ the goodness-of-fit at this $\Theta$ value is $\langle \chi^2\rangle/N_\tau \approx 0.477$,
which is marginally lower than the $\Theta \to 0$ limit witout negative weights. The shapes of the spectra at $\Theta= 10^{-5}$ and $\Theta= 10^{-6}$
are different from any shape realized versus $\Theta$ with all positive weight---interestingly, those at $\Theta=10^{-5}$ are closer to the spectrum at
$\langle \chi^2\rangle/N_\tau=0.6$ in Fig.~\ref{sw1}(c), obtained when also the amplitudes were sampled and which we argued is the best of the
three parametrizations.

Fig.~\ref{fig:wnega} shows the distribution of the negative amplitudes, defined as a spectrum $B(\omega)$ without the negative sign. At 
high $\Theta$, the negative weight covers the entire range $\omega \in [0,10]$ imposed for the sampling in this case. As $\Theta$ is lowered,
a sharp peak gradually forms, at a location that depends on the total negative weight $W_{\rm neg}$. It is clear that a $\delta$-function would
eventually form as $\Theta \to 0$. Given that the actual spectrum should be positive definite and that this negative $\delta$-function is
completely noise induced and has very small weight, it can be neglected in the output in practice.

The $\Theta= 10^{-6}$ results in Figs.~\ref{fig:final}(g) and \ref{fig:final}(h) are still far from the ultimate $\Theta \to 0$ limit, which is hard
to reach in practice. Results at $\Theta= 10^{-7}$ indicate that the large central peak and two smaller side peaks seen in Figs.~\ref{fig:nega}(g) and
\ref{fig:nega}(h) continue to narrow and likely develop into three distinct $\delta$-functions. Thus, even in the presence of these very small negative
weights, the best-fit spectrum changes from two positive $\delta$-functions to three substantial positive ones and a very small one carrying the full negative
spectral weight at much higher frequency. The number of effective parameters should be $N_{\rm para}=6$ in this case (counting only the positive contributions)
versus $N_{\rm para}=4$ before. This increase of $N_{\rm para}$ could possibly impact the potential of the SAC procedure to resolve more spectral details,
using lower $\Theta$ (and $\langle \chi^2\rangle$) than what is feasible with only positive weight. We plan to investigate this issue in more detail
in the future (as we discuss further
in Sec.~\ref{sec:prospnegative}).

\section{Fluctuations}
\label{app:fluct}

It is instructive to compare the native fluctuations of the three basic $\delta$-function parametrizations illustrated in
Figs.~\ref{fig:spec}(a)--(c). It is commonly believed that the fluctuations of spectral weight distribution can be used directly to compute statistical
errors \cite{beach04} (also when sampling within the ME approach \cite{kora18}, for which our findings also will apply), which is one reason for addressing
the fluctuations here. Insights gained here are also useful for other reasons.

It is easy to compute the fluctuations of spectral weight within windows of arbitrary size in the case of purely
entropic sampling. Here we consider a range of frequencies $\omega \in [0,1]$ and compute the fluctuations (with the total weight normalized to $1$)
within a window of size $\Delta$.

With $N$ equal-amplitude $\delta$-functions, the binomial distribution can be applied directly. In the limit of small $\Delta$, the
standard deviation of the weight $\sigma_\Delta$ in the window, normalized by the mean amplitude in the window (which here also equals $\Delta$), is
\begin{equation}
\frac{\sigma_\Delta}{\Delta} = \frac{c}{\sqrt{N\Delta}}, 
\label{fluctw}
\end{equation}  
with the coefficient $c=c_1 \equiv 1$.

\begin{figure}[t]
\centering
\includegraphics[width=75mm]{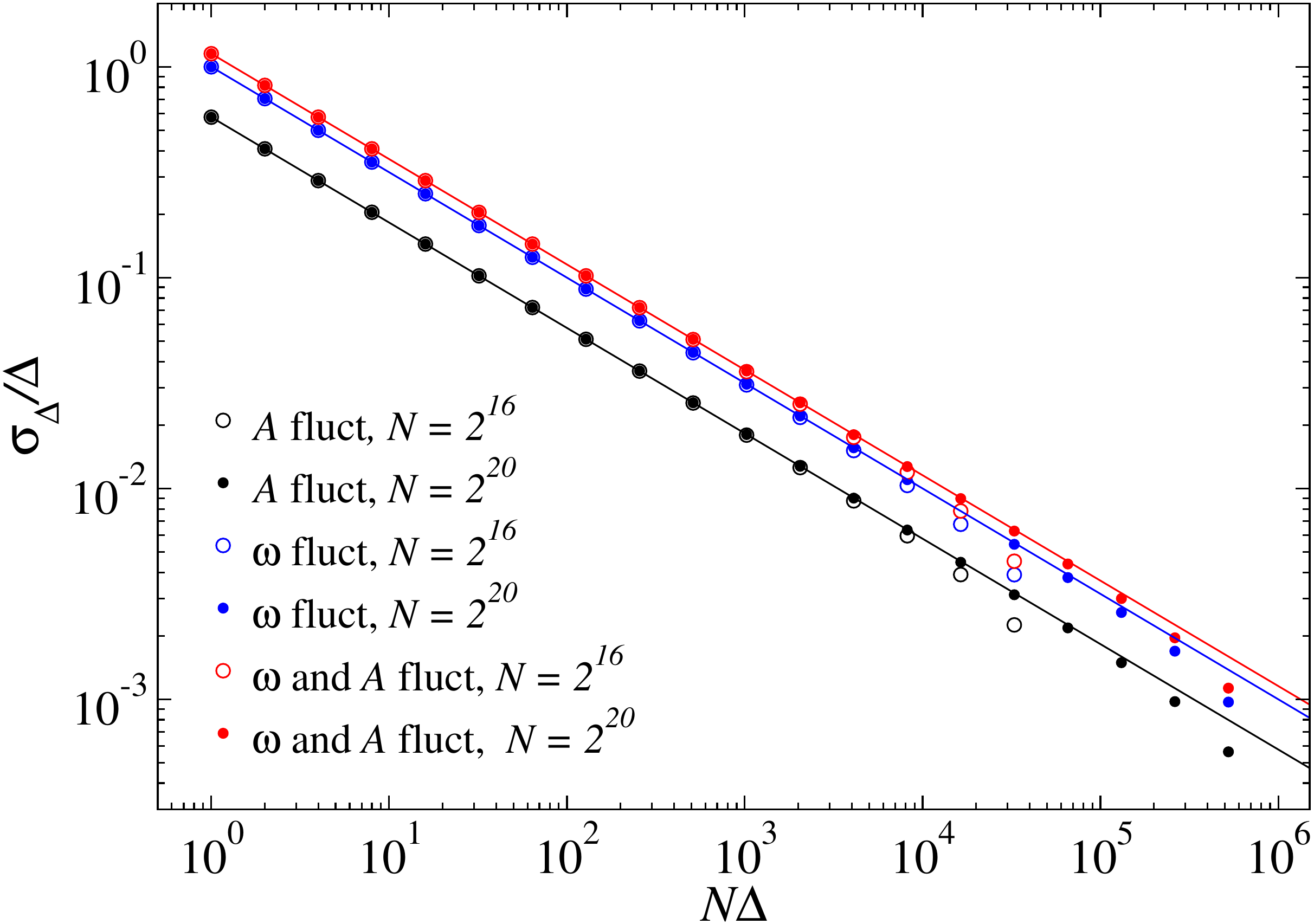}
\caption{Normalized standard deviation of the spectral weight vs  the size $\Delta$ of the frequency  window when sampling purely entropically with
  $N$ $\delta$-functions in the range $\omega \in [0,1]$ in three different ways; frequencies $\omega_i$ only (equal amplitudes), amplitudes $A_i$ only
  (on a uniform frequency grid), as well as both frequencies and amplitudes. Results are shown vs $N\Delta$ for $N=10^{16}$ (open symbols) and $N=10^{20}$
(solid symbols). The lines show Eq.~(\ref{fluctw}) with the predicted coefficients $c=(1/3)^{1/2}$, $1$, and $(4/3)^{1/2}$.}
\label{fluct}
\end{figure}

In the case of the uniform fixed grid, the smallest window is the grid spacing, $\Delta_1=1/N$, and it is also not difficult to compute the fluctuations
of this weight by dividing the total weight into windows separated by points $x_1,x_2\ldots,x_{N}$ corresponding to the partially summed weights;
$x_n=\sum_{i=1}^n A_i$ and $x_N=1$. Keeping $x_1$ fixed and integrating over the other ordered points gives the probability $P(A_1)=P(x_1)$, from which
the standard deviation $\sigma_{\Delta_1}/\Delta_1=(1/3)^{1/2}$ is obtained. For larger windows, with sizes $\Delta$ taken as multiples of $\Delta_1$, the
fluctuations can be regarded as independent and the variances from all the amplitudes in the window can be added, as long as $\Delta$ remains small.
Then the form is again Eq.~(\ref{fluctw}) but with $c=c_2 \equiv (1/3)^{1/2}$.

The case of sampling both frequencies and amplitudes at first sight appears more difficult. However, we can simply combine the above two results,
since we have two independent sources of fluctuations and their variances can be added. Thus, we predict the same form as Eq.~(\ref{fluctw}), but now
with $c=c_3 \equiv (c_1^2+c_2^2)^{1/2}=(4/3)^{1/2}$.

Figure \ref{fluct} shows the predicted forms of $\sigma_\Delta/\Delta$ together with simulation results for two large values of $N$. We observe perfect
agreement with the predictions when $\Delta \ll 1$ and expected deviations when $\Delta \to 1$. In particular, the additivity of frequency and amplitude
fluctuations is borne out by these results. We invoke a similar additivity when arguing for the mixed entropy form in Sec.~\ref{sec:entropy}.

While the overall $N^{-1/2}$ reduction of the fluctuations is of course completely expected from elementary statistics, this behavior has implicitly
been overlooked in some previous works on SAC, where it has been assumed that the fluctuations of the sampled spectra can be used directly to assign
error bars to the spectral weight accumulated in finite frequency windows. If the behaviors demonstrated here for purely entropic sampling carry over
when $\chi^2$ is included for weighting (which indeed is the case, as discussed further in \ref{app:statmech}),
the decay of fluctuations with $N$ implies that they cannot be used as
statistical uncertainties on the spectrum; at least not in the $N \to \infty$ limit, and then also the meaning of the fluctuations as $N$-dependent
statistical errors for finite $N$ becomes suspect. The vanishing fluctuations are also of relevance in the discussion of the relationships between
the SAC and ME methods, discussed in Sec.~\ref{sec:maxent}.

\section{Statistical mechanics}
\label{app:statmech}

Here we discuss the statistical mechanics aspects of SAC in some more detail, explaining why fluctuations of the spectrum in the SAC case
cannot drive the solution away from the maximum-probability spectrum when $N_\omega$ is taken to infinity. In contrast, in conventional statistical
mechanics fluctuations are of course very important and can drive systems away from their mean-field configurations, especially at phase transitions.

Consider a system like an Ising model, with $N_\sigma$ microscopic (lattice) degrees of freedom $\sigma({\bf r}) = \pm 1$ at lattice coordinates ${\bf r}$.
Its partition function at temperature $T$ is
\begin{equation}
Z = \sum_{\{\sigma\}} {\rm e}^{-H(\{\sigma\})/T},
\label{zsum}
\end{equation}
where $H$ is the Hamiltonian, which we do not need to specify here. We can perform a coarse graining of the microscopic degrees of freedom to a
real-valued magnetization density $m({\bf r}) \in [-1,1]$, as well as an energy density $e({\bf r})$.

A given coarse-grained
magnetization profile $m({\bf r})$ represents many possible microscopic spin configurations; thus there is an entropy $S(m)$ that depends on the
magnetization profile. We can also define a volume-normalized entropy density $s(m)=S(m)/V$ (where $V \propto N_\sigma$). With these definitions, the partition
function Eq.~(\ref{zsum}) can be converted into a functional integral over all magnetization profiles;
\begin{equation}
Z = \int_m dm W(m),
\label{zint}
\end{equation}
where the probability distribution $P(m)=W(m)/Z$ is given by
\begin{equation}
P(m) \propto {\rm e}^{-V[e(m) - Ts(m)]/T},
\label{pmag}
\end{equation}
where $e(m)$ is the total energy normalized by the volume, which represents an average over all microscopic spin configurations with magnetization
profile $m$.

Naively, when the volume $V$ is taken to infinity in Eq.~(\ref{pmag}), it would appear that we just have to minimize the free-energy (density) functional
\begin{equation}
f(m) = e(m) - Ts(m),
\end{equation}
and do not need to worry about fluctuations about the mean. We know that this is not correct in general, because of the high density of fluctuations
about the minimum that can overcome the volume factor $V$ in the exponential (\ref{zint}). In particular, at phase transitions non-trivial behaviors
are generated from contributions from magnetization profiles that are completely different from the flat ${\bf r}$ independent (zero or non-zero)
mean-field solution. The most striking of these behaviors is the form of the order parameter, e.g., the uniform value of $\langle m\rangle$ in a
ferromagnet, which acquires a power-law form with an exponent not explainable by mean-field theories.

We have claimed that fluctuations do not play the same crucial role in the case of SAC, where the probability distribution in Eq.~(\ref{psac})
looks very similar to Eq.~(\ref{zint}). For convenience, we here repeat Eq.~(\ref{psac}) with slightly different notation, using $s(A)$ for the
intensive SAC entropy instead of $E_{\rm X}(A)$;
\begin{equation}
P(A) \propto {\rm e}^{-N_\omega[\chi^2(A)/2 - \theta s(A)]/\theta}.
\label{paw}
\end{equation}
Recall that $\theta$ here should be regarded as $N_\omega$ independent, $\theta=\Theta N_\omega$ when $\Theta \propto N_\omega^{-1}$ is enforced,
and plays the same role as $T$ in Eq.~(\ref{pmag}). We further identify the magnetization profile $m({\bf r})$ with the spectral function $A(\omega)$,
the volume $V$ of the Ising model with the number $N_\omega$ of degrees of freedom ($\delta$-functions in our case) of the spectrum, and $\chi^2(A)/2$
with the energy density $e(m)$. We should also use appropriate forms for the intensive entropies, which we do not need to further specify here.

As in the Ising
model, it may seem like we also cannot neglect the fluctuations in the SAC case when $N_\omega \to \infty$. What is then the reason why, in fact, the
average spectrum becomes identical to the maximum-probability ME spectrum, as we have argued and shown explicitly with numerical comparisons in
Sec.~\ref{sec:maxent}?

To answer this question, we first consider the standard statistical mechanics setup (e.g., the Ising model). The system is defined on a lattice with
lattice constant set to $1$ (or any fixed value). Note that we are not allowed to imagine the system as squeezed into a finite box with lattice constant
vanishing ($V \to {\rm constant}$) as $N_\sigma \to \infty$ (which we mention for reasons that will become clear below), because we have assumed intensive
energy and entropy densities in Eq.~(\ref{pmag}), i.e., we must have $V \propto N_\sigma$.

When the volume increases, energetically inexpensive long-wave-length fluctuations can form in $m({\bf r})$, and these fluctuations are entropy favored
because of their vast number. Depending on the value of $T$, there can be a proliferation of non-uniform magnetization profiles with probability extremely close
to the one with formally maximum probability, and the volume factor in Eq.~(\ref{pmag}) cannot quench these fluctuations because the space of fluctuations (long
wave lengths) also increases. To really prove and quantify this behavior is of course a formidable task, eventually requiring the renormalization group, but
is a well established fact that we can invoke. In particular, we know that some systems undergo a phase transition between an ordered state with
$\langle |m| \rangle > 0$ (with low fluctuations) and one with $\langle m\rangle=0$ as $T$ increases and the entropy (fluctuations) takes over.

Now consider the situation in SAC. Here the first thing to note is that the spectrum is effectively confined between lower and upper frequency bounds,
which do not change much when $N_\omega$ is increased above some value. Thus, the ``volume'' $V_\omega$ of the spectrum is finite and does not scale in proportion
to $N_\omega$. This fact is critical, because the ``energy density'' $\chi^2$ when $\theta \sim 1/N_\omega$ nevertheless does not not grow in proportion to
$N_\omega$ (i.e., as the density of the degrees of freedom increases), because of its unusual form (if viewed from the standard statistical mechanics perspective).
It depends on the spectrum $A(\omega)$ through a correlation function $\bar G(\tau)$ available for a fixed number $N_\tau$ of points (i.e., independent of $N_\omega$
when $N_\omega$ is large). There can be no low-energy fluctuations corresponding to long wave-lengths in $\omega$-space, because spectral weight introduced
outside the de facto spectral bounds would be very expensive in terms of increasing $\chi^2$ (i.e., the volume is kept effectively finite by $\chi^2$). All
fluctuations are therefore confined within a finite volume $V_\omega$ in the infinite-density limit when $N_\omega \to \infty$ and should be regarded as
ultraviolet in the statistical mechanics language. Moreover, short-wave length fluctuations are associated with large ($\propto N_\omega$) reduction of
entropy (in contrast to long-wavelength fluctuations of the spin system) and are suppressed (in a way similar to the sampling without $\chi^2$ discussed
in \ref{app:fluct}). Thus, the SAC spectrum lacks exactly those ingredients that allow fluctuations to entropically proliferate (at transitions out of
ordered phases) in conventional statistical mechanics in the thermodynamic limit of large volume and constant density.

An analogy with phase transitions has been used \cite{beach04,bergeron16} to describe the way the SAC spectrum changes from a smooth data-fitting one
to a noise dominated ``glassy'' one. However, there is no true phase transitions with singularities in the ``free energy'' functional, for the reasons
outlined above. In statistical mechanics, the fluctuations also affect expectation values away from a phase transition, if the correlation length exceeds
the lattice spacing. In the SAC case, when $N_\omega \to \infty$ any fluctuation in a histogram of essentially fixed width will be too expensive entropically,
and fluctuations bring spectral weight outside the $\chi^2$ favored bounds will be energetically expensive. In the absense of fluctuations, there can also
be no correlations in frequency space and no deviations from the maximum-probability solution, which indeed also can be regarded as a mean-field
solution \cite{beach04}.

While the above arguments are not completely rigorous mathematically, they are fully borne out by the numerical demonstrations in Sec.~\ref{sec:sacme}.

\end{document}